\documentclass[12pt]{article}
\usepackage[top=2.5cm,bottom=2.5cm,left=2.55cm,right=2.55cm]{geometry}

\usepackage{enumitem}

\usepackage{bbm}
\usepackage{overpic}
\usepackage{caption}
\usepackage{mathtools}
\usepackage{latexsym} 
\usepackage{verbatim}  
\usepackage{tikz}
\usetikzlibrary{matrix}
\usepackage{color}
\usepackage{graphicx,amssymb,amsfonts,amsmath,amssymb,amscd,amstext, mathrsfs}
\usepackage{graphicx}
\usepackage{dsfont}
\usepackage{xcolor}
\usepackage{amsmath}
\usepackage{empheq}
\usepackage{cite}
\usepackage{tikz}
\usetikzlibrary{shapes.misc, positioning,decorations.pathreplacing,angles,quotes}
\usetikzlibrary{decorations.pathmorphing}

\numberwithin{equation}{section}

\newlength\dlf

\newcommand{\bw}{\begin{widetext}}
\newcommand{\ew}{\end{widetext}}
\newcommand{\bea}{\begin{eqnarray}}
\newcommand{\eea}{\end{eqnarray}}

\renewcommand{\bar}[1]{\overline{#1}}
\renewcommand{\tilde}[1]{\widetilde{#1}}

\newcommand{\<}{\langle}
\renewcommand{\>}{\rangle}

\renewcommand{\cal}{\mathcal}

 \definecolor{themecolor}{RGB}{0, 64, 168}
\definecolor{myRed}{RGB}{150,22,22}
\definecolor{myRedL}{RGB}{255,239,237}
\definecolor{myGrayL}{RGB}{220,220,220}
\definecolor{blul}{RGB}{51, 153,255}
\definecolor{greenl}{RGB}{76,153,0}

\DeclareFontShape{OT1}{cmr}{mx}{n}{<->cmr10}{}
\newcommand{\titlefont}{\fontseries{mx}\selectfont}

\def\frac#1#2{{#1\over #2}}

\usepackage{subcaption}
\usepackage{graphicx}
\usepackage{mymacroMOD}
\usepackage{jheppub}

\usepackage{braket}
\usetikzlibrary{decorations.pathmorphing}
\tikzset{snake it/.style={decorate, decoration=snake}}
\usetikzlibrary{arrows.meta}
\makeatletter
\newlength\lrvec@height
\newlength\lrvec@width
\newif\iflrvec@same@height
\def\lrvec{\@ifstar\slrvec@\lrvec@}
\newcommand{\slrvec@}[2][.4ex]{
  \lrvec@same@heighttrue
  \mathpalette\lrvec@@{{#1}{#2}}
}
\newcommand{\lrvec@}[2][.4ex]{
  \lrvec@same@heightfalse
  \mathpalette\lrvec@@{{#1}{#2}}
}
\def\lrvec@@#1#2{\lrvec@@@#1#2}
\def\lrvec@@@#1#2#3{%
  \iflrvec@same@height
    \settoheight{\lrvec@height}{$\m@th#1 \mathbf{T}#3$}
  \else
    \settoheight{\lrvec@height}{$\m@th#1#3$}
  \fi
  \settowidth{\lrvec@width}{$\m@th#1#3$}
  \kern.08em
  \raisebox{#2}{\raisebox{\lrvec@height}{\rlap{%
    \kern-.05em
    \begin{tikzpicture}[<-> /.tip={To[width=.4em, length=.2em]}]
      \draw [<->] (-.05em,0)--(\lrvec@width+.05em,0);
    \end{tikzpicture}%
  }}}%
  #3
  \kern.08em
}
\makeatother

\usepackage{simpler-wick}
\graphicspath{ {./figure/} }

\newcommand\blfootnote[1]{%
  \begingroup
  \renewcommand\thefootnote{}\footnote{#1}%
  \addtocounter{footnote}{-1}%
  \endgroup
}

\begin{document}
\pagestyle{myplain}
\begin{titlepage}

\begin{flushright} 
\end{flushright}

\begin{center} 

\vspace{0.35cm}

{\fontsize{20.5pt}{25pt}
{\titlefont  
Holography and Regge Phases with U(1) Charge
}}

\vspace{1.6cm}  

{{Giulia Fardelli$^\dagger$\blfootnote{${}^\dagger$\href{mailto:fardelli@bu.edu}{\tt fardelli@bu.edu}}, A. Liam Fitzpatrick$^\ddagger$\blfootnote{${}^\ddagger$\href{mailto:fitzpatr@bu.edu}{\tt fitzpatr@bu.edu}},  Wei Li$^{\text{\S}}$\blfootnote{${}^{\text{\S}}$\href{mailto:weili17@bu.edu}{\tt weili17@bu.edu}}
}}

\vspace{1cm} 

{{\it
Department of Physics, Boston University, 
Boston, MA  02215, USA
}}\\
\end{center}
\vspace{1.5cm}

{\noindent 
We use holography to study the large spin $J$  limit of the spectrum of low energy states with charge $Q$  under a U(1) conserved current in CFTs in $d>2$ dimensions, with a focus on $d=3$ and $d=4$.  For $Q=2$, the spectrum of such states is known to be universal and properly captured by the long-distance limit of holographic theories, regardless of whether the CFT itself is holographic.  We study in detail the holographic description of such states at $Q>2$, by considering the contribution to the energies of $Q$ scalar particles  coming from single photon and graviton exchange in the bulk of AdS; in some cases,  scalar exchange and bulk contact terms are also included.  For a range of finite values of $Q$ and $J$, we numerically diagonalize the Hamiltonian for such states and examine the resulting spectrum and wavefunctions as a function of the dimension $\Delta$ of the charge-one operator and the central charges $c_\CT, c_\CJ$ of the stress tensor and U(1) current, finding multiple regions in parameter space with qualitatively different behavior.  We discuss the extension of these results to the regime of parametrically large charge $Q$, as well as to what extent such results are expected to hold universally, beyond the limit of holographic CFTs.  We compare our holographic computations to results from the conformal bootstrap for the $3d$ O(2) model at $Q = 3$ and $Q = 4$ and find excellent agreement.
}

\end{titlepage}

\tableofcontents

\newpage

\newpage

\section{Introduction and Summary} 

A single Conformal Field Theory (CFT), despite the absence of an intrinsic length scale, can  have a rich phase diagram with qualitatively different emergent behavior in a range of different regimes parameterized by the conserved charges of the theory. At a minimum, these conserved charges include angular momentum $J$ and total energy $E$ measured relative to the volume of space, but often include the charges of internal symmetries as well.  Mapping out the full phase diagram of a CFT as a function of such charges paints a beautiful picture of how physics of a single model changes qualitatively when pushed to different extremes, and sheds light on collective phenomena in Quantum Field Theory more generally.

One can ask specifically about CFTs with a global U(1) current, and consider the phase diagram as a function of global U(1) charge $Q$ and total angular momentum $J$, at large $Q$ and $J$.  Already in this scenario there are several different phases that can arise depending on the relative size of the charges $Q$ and $J$, and in the specific case of the $3d$ O(2) model a nearly complete classification of the different possible phases has been proposed \cite{Hellerman:2015nra,Cuomo:2017vzg,Cuomo:2022kio,Su:2022ysc,Cuomo:2023vvd,Banerjee:2017fcx,Cuomo:2023mxg}.  In this paper we will specifically be interested in the Regge limit, where the large $J$ limit is taken first, in a precise sense that we will define below.  This limit has a useful dual holographic interpretation, where in AdS$_{d+1}$ the large angular momentum $J$ causes the state to rotate rapidly around the center of AdS so that its centrifugal force pushes it outwards toward the boundary of AdS.   This limit has been studied in detail in the case  $Q=2$, where the lightest primary state is a `double-twist' state with spin $J$ that consists of two partons (or `blobs') in AdS separated by a large distance.  Moreover, this holographic picture is very general, and applies to all unitary CFTs in $d\ge 3$, regardless of whether or not the CFT has a sparse holographic AdS dual \cite{Komargodski:2012ek,Fitzpatrick:2012yx,Fitzpatrick:2014vua,Alday:2007mf}.  The basic reason for this generality is that, even if interactions in AdS are not intrinsically small, their strength nevertheless decays and becomes small at large distances, and the leading interactions are due to long-distance exchange of the lowest-twist neutral states, which can be systematically included at large spin \cite{Alday:2015eya,Alday:2015ewa,Alday:2016njk,Simmons-Duffin:2016wlq,Fitzpatrick:2015qma,Li:2015rfa,Li:2017lmh}.   When $Q$ is taken to be larger, however, the picture becomes more complicated, and it is not completely obvious if a holographic description can still be used in general.  A simple dynamical reason that a weakly coupled bulk description might still be valid at $Q>2$ is if at large  spin $J$ the lowest-energy states tend to fall apart into $Q$ partons each with spin approximately $J/Q \gg 1$, all far apart from each other in AdS.  More generally, if some partons are close together and others are far apart, then one can at least model the long-range interactions using perturbation theory in AdS.

With this picture in mind, we will begin our analysis with a thorough treatment of the scenario of an AdS model with a bulk scalar weakly coupled to gravity and electromagnetism.  The key technical assumption of this treatment is that we only need to include tree-level exchange of bulk photons and bulk gravitons, and the analysis of the Regge limit will be a purely perturbative bulk computation.  Nothing about these results will require taking $Q$ large, and in fact we will spend considerable time studying various small values of $Q$.  Because photon exchange is repulsive whereas graviton exchange is attractive in the limit of long distances, the qualitative behavior of the Regge limit will depend on a competition between their effects, and there are two distinct phases that we will call the `attractive' phase and the `repulsive' phase, depending on whether gravity or gauge interactions, respectively, are stronger at large distances.  Equivalently, if $\Phi$ is the lightest charged state, which we assume has $Q=1$, then the attractive vs repulsive phases are defined by the sign of the anomalous dimension of the double-twist operator $[\Phi, \Phi]_J$ at large $J$:
\begin{equation}
\lim_{J \rightarrow \infty} J^{d-2} \gamma_{[\Phi, \Phi]_J} \textrm{ is } \quad \begin{array}{cc} \textrm{positive} \\  \textrm{negative} \end{array}  \quad \Leftrightarrow \quad  \begin{array}{cc} \textrm{repulsive} \\ \textrm{attractive} \end{array}  .
\end{equation}

In the repulsive phase, the dominant photon interactions keep the individual $Q$ partons as far away from each other as possible, so that the lightest state at large $J$ is $Q$ partons spinning around the center of AdS, separated by angular intervals of $2\pi/Q$.  In fact, because the distance between such partons is large, we expect the large spin results of the simple holographic model hold more generally than the holographic setup used to derive it, and apply to any unitary $d\ge 3$ CFT where lowest-twist exchange is dominantly  a U(1) current.

By contrast, in the attractive phase, the dominant graviton interactions tend to pull the individual partons towards each other, so that they group into  `blobs', where the $Q$ total partons all fall into one or another blob.  Then, there is a nontrivial question as to the description of these individual blobs such that the energy of the total state is minimized.  In the holographic limit it is possible to answer this question with our perturbative analysis.  In our numeric analysis, we will consider $Q=2,3,4,5$, and we will see that the lowest-energy state consists of two blobs rotating around each other in AdS.  The detailed description of the blobs depends on the parameters of the bulk model.  For instance, when both gauge and gravity interactions are turned on, the details of the blobs  depend not only on which of the two interactions is larger at large $J$ but also on their quantitative relative strength at finite values of $J$.  We will see that in some sense, the attractive phase contains multiple distinct smaller phases, distinguished by the internal spins of the  blobs.  
 
Finally, we will consider what happens in the attractive phase away from the holographic limit, specifically in the O(2) model. Because the description of the blobs is a strongly coupled problem, we will have to inject some nonperturbative information into our analysis.  At small values of $Q$, specifically at $Q=3$ and $Q=4$, the blobs are made up of a small number of partons and so there are precise nonperturbative results on the energies of all possible blob states.  For instance, at $Q=3$ a blob can contain at most $Q=2$ partons, and the spectrum of the lowest-twist $Q=2$ states is known fairly well.  This $Q=2$ data can be reproduced well by our bulk computations with the simple addition of an extra bulk scalar dual to the $\Delta \approx 1.511$ `$\epsilon$' operator, together with a bulk $\phi^4$ contact term; one reason that this description works well is that the Lorentzian inversion formula \cite{Caron-Huot:2017vep} converges down to $J=2$, so one gets a fairly good prediction by including the exchange of the low-twist operators $\CT_{\mu\nu}, \CJ_\mu$ and $\epsilon$, together with a contact term to account for the $J=0$ value.  We compare this simple bulk model to  $Q=3$ data from the conformal bootstrap and find excellent agreement.  We also find reasonable agreement in a similar comparison at $Q=4$.  Moreover, this agreement is an improvement over previous analyses using the inversion formula, because our bulk computation lets us efficiently include mixing between a very large number of primary operators that arise as $Q$ and $J$ increase;  in principle, we expect the inversion formula could reproduce these results, but would require considering larger systems of mixed correlators than has been done in the past.  
 
 At very large values of $Q$, it is no longer practical to inject nonperturbative data into our bulk model using the method just described.  Instead, in this regime we will use a combination of  intuition from holography together with the conjectured large $Q$ phase structure of the theory when $J \ll Q^2$ and the respective large charge EFTs for these phases.  In particular,  we will argue that at $J \gg Q^2$ the lowest-energy state can still be described holographically as a multi-blob state, where now it is energetically favorable for each of the individual blobs to have $q$ partons inside them for some value of $q$ determined dynamically.  The relevant dynamics is to minimize the twist-to-charge ratio of these blobs among all possible values of their charge $q$ and internal spin $\ell$.  Existing numeric results for the case $q=2$ put an upper bound on the minimum possible twist-to-charge ratio, which is just small enough that the gravitational attraction between blobs is now {\it weaker} than the repulsive force of gauge interactions, so that the blobs repel each other at large distance.

The paper is organized as follows.  In section  \ref{sec:DT}, we review the bulk setup for the case $Q=2$ as a warm-up before going to higher $Q$. In section \ref{sec:largeJBulk}, we discuss some analytic results for the holographic description at large $J$ and show how to extend these to general values of $Q$, culminating in an analytic expression for the lowest-energy state at large $J$ for any $Q$ in the repulsive phase. In section \ref{sec:FiniteQJ}, we turn to our numeric bulk analysis at finite value of $Q$ and $J$, and study in detail the resulting eigenvalues and wavefunctions.  In section \ref{sec:LargeQ}, we discuss taking the limit of large $Q$ in various phases of the theory.  We also include in this section a detailed comparison with the conformal bootstrap results  in the O(2) model at $Q=3, 4$ (where one already starts to see some qualitative similarity to the `large $Q$' limit). Finally in section \ref{sec:Future}, we conclude with a discussion of possible future directions.

\section{Double-Twist Warm-up $(Q=2)$} 
\label{sec:DT}

We begin with a review of the case of double-twist operators, \textit{i.e.} $Q=2$, at large spin $J$.  This case has been studied extensively,\footnote{For a partial list of recent works, see \cite{Kaviraj:2015cxa,Alday:2016njk,Dey:2016zbg,Pal:2022vqc,Bertucci:2022ptt,Henriksson:2022rnm,Albayrak:2019gnz}.} and starting with the four-point function of $Q=1$ operators $\Phi$, one can prove for any unitary CFT in $d\ge 3$ that at large $J$ the lightest $Q=2$ state is a `double-twist' state $[\Phi, \Phi]_J$ that is described, qualitatively and quantitatively, by two $\phi$ particles far away from each other, spinning around the center of AdS \cite{Fitzpatrick:2012yx,Komargodski:2012ek,Fitzpatrick:2014vua}, and that the energy of this state is given by long-distance exchange of the lowest-twist neutral states.  More precisely, a $\phi$ `particle' in AdS is the state dual to the CFT operator $\Phi$, and is not necessarily a weakly coupled point-like particle; we will refer to this state in AdS as a $Q=1$ `parton'.  Assuming there are no neutral scalars with dimension less than $d-2$, the stress tensor and current saturate the unitarity lower bound $\tau \ge d-2$ on twist, and so are the lowest-twist neutral states.  Because the stress tensor and  current are dual to the AdS graviton and photon, respectively, this means the leading AdS interaction is just long-distance gauge and gravity exchange.  Moreover, the relevant couplings with $\phi$ are fixed by the Ward identity (equivalently, by diffeomorphism and gauge invariance in AdS), and so at large $J$ one can fully determine the leading large $J$ correction to the dimension of $[\Phi, \Phi]_J$:
\begin{equation}
\begin{aligned}
\Delta_{[\Phi, \Phi]_J} &= 2 \Delta + J + \frac{ \Gamma(d) \Gamma^2(\Delta)}{\Gamma^2(\frac{d}{2}) \Gamma^2(\Delta-\frac{d-2}{2})} \left( \frac{1}{c_\CJ} - \frac{2d(d+1)\Delta^2}{(d-1)^2 c_\CT} \right) \frac{1}{J^{d-2}} + \dots.  
\label{eq:Q2AnomDim}
\end{aligned}
\end{equation}
In this equation, $\Delta$ is the dimension of $\Phi$, and $c_\CJ, c_\CT$ are the central charges of the current $\CJ$ and stress tensor $\CT$, respectively.  Because the two $\phi$ partons have equal charge, the gauge force between the partons is repulsive while that of the gravitation force is attractive, so that  their energy contributions are positive and negative, respectively.  This relative sign will play an important role in our later analysis at $Q>2$.  At $Q=2$, however, the two partons in AdS are held apart from each other by their total angular momentum, regardless of the sign of the force between them.

In the following, we review how to derive the formula (\ref{eq:Q2AnomDim}) from the AdS description, which will allow us to cover some necessary techniques and  intuition, as well as to establish our conventions.

\subsection{Bulk Setup}

The only degrees of freedom that play a role on the AdS side are the gauge and gravitation fields, as well as the complex scalar $\phi$ which sources them.  The action for these fields in AdS is
\begin{equation}
S = \frac{1}{\kappa^2} \int d^{d+1}x \sqrt{-g} \left( \frac{R-\Lambda}{2} - \frac{1}{4} F_{\mu\nu} F^{\mu\nu} - (D^\mu \phi)^\dagger (D_\mu \phi) - m^2 \phi^\dagger \phi \right), \label{action}
\end{equation}
where we have defined the covariant derivative
\begin{equation}
D_\mu = \nabla_\mu - i g_{\rm U(1)} A_\mu.
\end{equation}
For now, we do not include any additional contact terms (\textit{e.g.} $\phi^4$) in the action.  We will discuss the effect of adding such terms later, but for now the main point is that their effect is negligible at long distances.
The state-operator correspondence, where the boundary CFT metric is that of $\mathbb{R} \times S^{d-1}$, is most transparent on the AdS side in global coordinates, 
\begin{equation}
ds^2 = \frac{1}{\cos^2 \rho} (- dt^2 + d \rho^2 + \sin^2 \rho d \Omega^2) = - \cosh^2 \chi dt^2 + d \chi^2 + \sinh^2 \chi d \Omega^2.
\end{equation}
The mass $m$ of the complex scalar in AdS is related to the dimension $\Delta$ of the dual CFT operator by the usual relation:
\begin{equation}
m^2 = \Delta(\Delta-d).
\end{equation}
At long distances, the interactions between bulk $\phi$ partons become weak and we can perturbatively evaluate the energy corrections due to photon and graviton exchange.  At leading order, where we neglect such interactions, a large spin double-twist state in the CFT is just a two-particle state in AdS, which we can fix uniquely by the conditions of being primary and minimal twist.  The full set of single-particle  modes for $\phi$ are determined by solving its free bulk equations of motion, with the following well-known result~\cite{Fitzpatrick:2010zm}:
\begin{equation}
\phi_{n \ell L} = \frac{1}{N_{\Delta,n,\ell}}  e^{i E_{n,\ell} t} Y_{\ell, L}(\Omega) \sin^\ell \rho \cos^\Delta \rho \  {}_2F_1(-n,\Delta+\ell+n, \ell+\frac{d}{2}, \sin^2 \rho),
\label{eq:BulkDescendantWvFcn}
\end{equation}
where $N_{\Delta, n, \ell}$ is a normalization factor, $Y_{\ell,L}$ are $S^{d-1}$-spherical harmonics with eigenvalue $-\ell(\ell+d-2)$ and $L$ is a collective index for the quantum numbers of the rotation group, and the energy of the mode is
\begin{equation}
E_{n,\ell} = \Delta+2n+\ell.
\end{equation}
The free bulk field can then be expanded in terms of these wavefunctions times creation and annihilation operators:
\begin{equation}
\phi(X) = \sum_{n \ell L} \phi_{n \ell L}(X) a_{n \ell L} + \phi^*_{n \ell L}(X) b^\dagger_{n \ell L}.
\label{eq:BulkFieldDecomp}
\end{equation}
Next, we rescale the fields as 
\eqna{
(\phi, A_{\mu}, h_{\mu \nu})=\kappa (\phi, A_{\mu}, h_{\mu \nu})\, ,
}[]
such that the action takes the form
\eqna{
S&\sim \int d^{d+1} x \sqrt{-\tilde{g}} \Big( -\nabla^\mu \phi^\dagger \nabla_\mu \phi -m^2\phi^\dagger \phi -\frac{1}{2} F_{\mu \nu} \nabla^{\mu} A^\nu +\kappa A^\mu \CJ_{\mu}\\
&\quad \, -\frac{1}{4} h_{\mu \nu} \Delta_{T}^{\mu\nu, \rho \sigma} h_{\rho \sigma }+\frac{1}{2}\kappa \lsp  h^{\mu\nu} \CT_{\mu \nu}\Big)+\CO(\kappa^2)\, ,
}[]
where we have expanded to order $\kappa^2$ and $\tilde{g}$ is the pure AdS background metric ($g_{\mu\nu}=\tilde{g}_{\mu\nu}+h_{\mu\nu}$). We have defined  the current and the stress-tensor to be
\twoseqn{
\CJ_{\mu}&=i g_{\mathrm{U(1)}} \left( \phi \nabla_\mu \phi^\dagger -\phi^\dagger \nabla_{\mu}\phi\right)\,,
}[]
{\CT_{\mu \nu}&=\nabla_{\mu}\phi^\dagger \nabla_{\nu}\phi+\nabla_{\nu}\phi^\dagger \nabla_{\mu}\phi-\tilde{g}_{\mu\nu} \left(\nabla^{\sigma} \phi^\dagger  \nabla_{\sigma} \phi+m^2\phi^\dagger \phi\right)\, .
}[][JTfromPhi]
Now, following \cite{Fitzpatrick:2011hh}, the idea is to integrate out the photon and the graviton classically, leaving an effective potential for the scalar, using the linearized equation of motions\footnote{We define $\Delta_T^{\mu\nu, \rho\sigma} h_{\rho \sigma}$ as the linearized perturbation of $R_{\mu \nu}-\frac{1}{2}g_{\mu\nu} (R-\Lambda)$.}
\twoseqn{
\nabla_{\mu}F^{\mu\nu}[\phi, \phi^\dagger]&=-\kappa \lsp \CJ^{\nu}[\phi, \phi^\dagger]\, ,
}[]
{
\Delta_T^{\mu\nu, \rho\sigma}h_{\rho \sigma}[\phi, \phi^\dagger]&=\kappa \lsp  \CT^{\mu\nu}[\phi, \phi^\dagger]\, ,
}[][EoM]
such that
\eqna{
S_{\text{eff}}&=\int d^{d+1} x \sqrt{-\tilde{g}} \Big( -\nabla^\mu \phi^\dagger \nabla_\mu \phi -m^2\phi^\dagger \phi -V_{\mathrm{eff}}[\phi, \phi^\dagger] \Big)\, ,\\
V_{\mathrm{eff}}[\phi, \phi^\dagger]&=-\frac{1}{2}\kappa\, A^\mu[\phi, \phi^\dagger] \CJ_{\mu}[\phi, \phi^\dagger]-\frac{1}{4}\kappa \lsp h^{\mu \nu}[\phi, \phi^\dagger]\CT_{\mu\nu}[\phi, \phi^\dagger]\, ,
}[VeffBulkAction]
where $A^\mu[\phi, \phi^\dagger]$ and $h^{\mu\nu}[\phi, \phi^\dagger]$ are the solutions to~\eqref{EoM}.  Performing canonical quantization, one can show that  the leading order interaction Hamiltonian density is  just given by $V_{\text{eff}}$~\cite{Fitzpatrick:2011hh}.  Therefore, the leading order correction to the binding energies of a generic state $\ket{\Psi}_J $ with spin $J$ and charge $Q$ is given by\footnote{Notice that the integration is only in space and not in time. }
\eqna{
\gamma^{(d,Q)}({J})=\int d^{d}x \sqrt{-\tilde{g}} \, {}_J\!\bra{\Psi } V_{\mathrm{eff}} [\phi, \phi^\dagger] \ket{\Psi}_J\equiv {}_J\!\bra{\Psi } \CV_{\mathrm{eff}} [\phi, \phi^\dagger] \ket{\Psi}_J\, .
}[anomInt]
To make contact with the CFT result, we  need to relate the AdS coupling to the CFT three-point function. For a general bulk interaction of the form 
\eqna{
g_{\phi^\dagger\phi V} \int d^{d+1}x\sqrt{-g} \left(\phi^\dagger \nabla_{\mu_1}\cdots \nabla_{\mu_s}\phi  \right)V^{\mu_1 \cdots \mu_s}\, ,
}[]
where the bulk coupling is related to the  CFT OPE coefficient $C_{\phi^\dagger \phi V}$ as~\cite{Costa:2014kfa} 
\eqna{
C_{\phi^\dagger \phi V}&=\frac{f(\Delta_\phi, \Delta, s)}{\sqrt{\CC_{\Delta_\phi,0}^2 \CC_{\Delta, s}}} g_{\phi^\dagger\phi V}\, ,\qquad \qquad \CC_{\Delta, s}=\frac{\pi ^{-\mathit{d}/2} \Gamma (\Delta ) (\Delta +s-1)}{2 (\Delta -1) \Gamma
   \left(-\frac{\mathit{d}}{2}+\Delta +1\right)}\\
f(\Delta_\phi, \Delta, s)&=\CC_{\Delta_\phi,0}^2 \CC_{\Delta, s}\frac{\pi ^{\mathit{d}/2}  \Gamma \left(\frac{s+\Delta }{2}\right)^2 \Gamma \left(\frac{s-\Delta
   }{2}+\Delta_{\phi }\right) \Gamma \left(\frac{1}{2} (s-\mathit{d}+\Delta )+\Delta_{\phi }\right)}{2^{1-s}\Gamma (\Delta_{\phi })^2 \Gamma (s+\Delta )}\,,
}[]
where in this expression it is assumed that all two-point functions are unit normalized.  Because we are considering  conserved currents, the OPE coefficient are related by Ward Identities to current and stress-tensor two-point function normalization 
\eqna{
C_{\phi^\dagger \phi J}&=\frac{1}{\sqrt{c_\CJ}}\, , \qquad  \qquad && g_{\phi^\dagger \phi J}=g_{\mathrm{U(1)}}\kappa\, ,\\
C_{\phi^\dagger \phi T}&=- \frac{\Delta \lsp d}{(d-1)\sqrt{c_\CT}}\, , \qquad  \qquad &&g_{\phi^\dagger \phi T}=-\kappa\, .
}[]
Putting everything together one obtains the following relation between bulk and boundary parameters:
\twoseqn{
(g_{\mathrm{U(1)}}\kappa)^2&=\frac{2 \pi ^{\mathit{d}/2} (\mathit{d}-2) \Gamma (\mathit{d})}{\Gamma
   \left(\frac{\mathit{d}}{2}\right)^3} \frac{1}{c_\CJ}\, ,
}[]
{
\kappa^2&= \frac{4 \pi ^{\mathit{d}/2} \Gamma (\mathit{d}+2)}{(\mathit{d}-1) \Gamma
   \left(\frac{\mathit{d}}{2}\right)^3}\frac{1}{c_\CT}\, .
}[][]

\subsection{Construction of Primaries} \label{subsec:construct primaries}

We will generally only be interested in the lowest-energy states at fixed twist $E-\ell$ and charge $Q$, so in all cases we will only be considering modes with $n=0$ and we will restrict to the symmetric traceless representations of SO($d$). We can also focus on the highest weight component of each spin representation, which fixes $L$.\footnote{In terms of CFT operators,  these highest-weight components must be constructed from  null derivatives $\partial_-$ and  $Q$ factors of $\Phi$.  Under these assumptions, in $d=3$ we are constructing all lowest-twist operators, since all SO$(3)$ representations are symmetric traceless.  In $d=4$, instead, more general representations are possible for $Q\ge 3$.
These more general operators were analyzed, for $Q=3$, in~\cite{Buric:2021kgy,Harris:2024nmr}, studying the lightcone limit of  higher-point conformal blocks.  In their notation, one classifies  SO(4) representations  with two labels $(\ell, \kappa)$, and  the symmetric traceless representations correspond to $\kappa=0$.} So, from now on, we will drop the $n$ and $L$ quantum labels, unless we want to explicitly emphasize this choice.  In this case, the wavefunctions simplify (since ${}_2F_1(0, \dots)=1$) and the normalization factor is 
\begin{equation}
N_{\Delta,\ell} \equiv N_{\Delta,0,\ell} = \sqrt{ \frac{\Gamma(\ell+ \frac{d}{2}) \Gamma(\Delta- \frac{d-2}{2})}{\Gamma(\Delta+\ell)}}.
\label{norm}
\end{equation}
A general spin $J$ (minimal twist, highest weight) state made from two $\phi$s is a linear combination of all states of the form $b^\dagger_\ell b^\dagger_{J-\ell} | {\rm vac}\>, \ell = 0, \dots, J$.  In general, we can define the `monomial' state
\begin{equation}
|\ell_1, \ell_2, \dots, \ell_n \> \equiv b_{\ell_1}^\dagger  b_{\ell_2}^\dagger \dots b_{\ell_n}^\dagger | {\rm vac} \>.
\label{eq:monomialstate}
\end{equation}
We can also represent the creation operators in terms of the boundary CFT operator $\Phi$, and vice versa.  Taking the limit of (\ref{eq:BulkFieldDecomp}) as $X$ approaches the boundary of AdS, we obtain the mode decomposition of $\Phi$ in radial quantization:
\begin{equation}
(\Phi)_{\mathbb{R} \times S^{d-1}} = \lim_{\rho \rightarrow \frac{\pi}{2}} \frac{1}{\cos^\Delta \rho}  \phi(X) = \sum_{n, \ell, L} \frac{1}{N_{\Delta, n ,\ell}} Y_{\ell, L}(\Omega)(e^{i E_{n, \ell} t}  a_{n \ell L} + e^{-i E_{n, \ell} t} b_{n \ell L}^\dagger).
\end{equation}
Alternatively, we can obtain the modes $b_{n \ell L}^\dagger$ by taking derivatives of $\Phi$ at the origin in Cartesian coordinates.  The state $b_{0,0 0}^\dagger | {\rm vac}\>$ is the primary state $\Phi(0) | {\rm vac}\>$, and all other modes can be obtained by acting multiple times with the translation operator $P_\mu$ on $\Phi(0)$.  We are mainly interested in the lowest-twist descendants, and the index-free formulation of \cite{Costa:2011mg} is a particularly convenient  device for keeping track of the corresponding descendants operators.  The main ingredient is to introduce an auxiliary null vector $z^\mu$ such that $z \cdot z =0$, so that contracting $z^\mu$ with all the free indices of a tensor $T^{\mu_1 \dots \mu_n}$ projects out all traces.  The lowest-twist descendants of $\Phi$ can then be written simply as the following operators:
\begin{equation}
(z \cdot P)^\ell \Phi = z^{\mu_1} \dots z^{\mu_\ell} (P_{\mu_1} \dots P_{\mu_\ell}) \Phi.
\end{equation}
With a little algebra, one can calculate the two-point function of this operator to determine the normalization of the corresponding state.  Since the states $b^\dagger_\ell  | {\rm vac}\>$ have unit norm by definition, this fixes the following relative coefficient:
\begin{equation}
(z \cdot P)^\ell \Phi(0) | {\rm vac} \> = \sqrt{2^\ell \ell! (\Delta)_\ell } b_\ell^\dagger | {\rm vac} \>,
\label{eq:KetFromZP}
\end{equation}
where $(a)_n$ denotes the Pochhammer symbol.

A general multi-particle lowest-twist state can be written as a linear combination of products of $b_\ell^\dagger$ acting on the vacuum:
\begin{equation} \label{eq:multibasis}
\begin{aligned}
|\Psi\>_{J;Q} &=   \sum_{\ell_1 + \dots + \ell_Q = J} C_{\ell_1 \dots \ell_Q} : (z \cdot P)^{\ell_1} \Phi(0) \dots (z \cdot P)^{\ell_Q} \Phi(0)  : | {\rm vac} \>  \\
 &=\sum_{\ell_1 + \dots + \ell_Q = J} C_{\ell_1 \dots \ell_Q}\sqrt{2^{\ell_1} \ell_1! (\Delta)_{\ell_1} } \dots \sqrt{2^{\ell_Q} \ell_Q! (\Delta)_{\ell_Q} }| \ell_1, \dots , \ell_Q \>,
\end{aligned}
\end{equation}
where $::$ indicates normal-ordering.
We can require the states to be primary and to be unit normalized,
\begin{equation}
(z' \cdot K) | \Psi\>_{J;Q} = 0, \qquad {}_{J;Q} \< \Psi | \Psi\>_{J;Q},
\end{equation}
with the following conventions for the conformal algebra:
\begin{equation}
\begin{aligned}
[M_{\mu\nu}, P_\rho] &= i (\eta_{\mu\rho} P_\nu - \eta_{\nu\rho} P_\mu), \qquad &&[ P_\mu, K_\nu] = -2(\eta_{\mu\nu} D + i M_{\mu\nu}), \\
[M_{\mu\nu}, K_\rho] &= i (\eta_{\mu \rho} K_\nu - \eta_{\nu \rho}K_\mu), \qquad &&\,\,  \lsp [D, P_\mu]  = P_\mu, \\
[M_{\mu\nu}, D] &=0, \qquad && \,\lsp [D, K_\mu]  = - K_\mu.
\end{aligned}
\end{equation}
Imposing the primary state condition is straightforward, and facilitated by the following commutation relation:
\begin{equation}
[z' \cdot K, (z \cdot P)^\ell ] \Phi = 2 \ell (\Delta+\ell-1)( z \cdot z') (z \cdot P)^{\ell-1} \Phi.
\end{equation}
Then, the condition $(z' \cdot K) | \Psi\>_{J;Q}=0$ translates into relations that must be satisfied by the coefficients $C_{i_1, \dots, i_Q}$.  For the special case of $Q=2$, these relations take the form of recursion relations that fix $C_{\ell, J-\ell}$ in terms of $C_{\ell+1, J-\ell-1}$, and so for any $J$ there is a unique solution, corresponding to the fact that for $Q=2$ at any $J$ there is a unique lowest-twist primary operator.  This recursion relation was solved in \cite{Penedones:2010ue}, and the solution is
\begin{equation}
\begin{aligned}
|\Psi\>_{2,J} &=  \frac{1}{\sqrt{{\cal N}_{J;2}}} \sum_{\ell=0}^J \frac{(-1)^\ell }{\sqrt{(J-\ell)! \ell! \Gamma(\Delta+J -\ell)\Gamma(\Delta+\ell)}} | \ell, J-\ell\>, \\
{\cal N}_{J;2} & = (1+(-1)^J) \frac{ \Gamma(2\Delta+2J-1)}{ \Gamma(J+1) \Gamma^2(\Delta+J) \Gamma(2\Delta+J-1)}.
\end{aligned}
\label{eq:DoubleTwistPrim}
\end{equation}
The sum vanishes for odd $J$, due to Bose symmetry of $\Phi$.

\subsection{Energy Correction from Graviton and Photon Exchange}\label{sec:energycorrection}
Since we will restrict to two-particle interactions, no matter the number of particles we will consider, the relevant integral we need to compute is the integral of the matrix element 
\eqna{
 \bra{\ell_2\ell_4} V_{\mathrm{eff}}[\phi, \phi^\dagger]\ket{\ell_1 \ell_3}= -2 \Bigg( &\frac{(\kappa g_{\mathrm{U(1)}})^2}{2} \wick{ \langle \c1 \ell_2 \c2 \ell_4 \vert  A[\c1 \phi, \c1 \phi^\dagger ] \cdot  \CJ[\c2 \phi, \c2 \phi^\dagger ] \vert \c1 \ell_1 \c2 \ell_3}\rangle\\
  &\, +\frac{\kappa^2}{4} \wick{ \langle \c1 \ell_2 \c2 \ell_4 \vert  h[\c1 \phi, \c1 \phi^\dagger ] \cdot  \CT[\c2 \phi, \c2 \phi^\dagger ] \vert \c1 \ell_1 \c2 \ell_3}\rangle + \ell_2 \leftrightarrow \ell_4  \Bigg)\, ,
}[]
where we have pulled out the explicit coupling dependence. More explicitly 
\twoseqn{
\CI_\CJ(\ell_1, \ell_3, \ell_2, \ell_4)&\equiv \int d^d x \sqrt{-\tilde{g}} \wick{ \langle \c1 \ell_2 \c2 \ell_4 \vert  A[\c1 \phi, \c1 \phi^\dagger ] \cdot  \CJ[\c2 \phi, \c2 \phi^\dagger ] \vert \c1 \ell_1 \c2 \ell_3}\rangle\\ \nonumber
&=  \int d^d x \sqrt{-\tilde{g}} \,  A[\phi^*_{0\ell_2 \ell_2}, \phi_{0\ell_1\ell_1}]\cdot \CJ[\phi^*_{0\ell_4 \ell_4}, \phi_{0\ell_3 \ell_3}]\, ,
}[]
{
\CI_\CT(\ell_1, \ell_3, \ell_2, \ell_4)&\equiv \int d^d x \sqrt{-\tilde{g}} \wick{ \langle \c1 \ell_2 \c2 \ell_4 \vert  h[\c1 \phi, \c1 \phi^\dagger ] \cdot  \CT[\c2 \phi, \c2 \phi^\dagger ] \vert \c1 \ell_1 \c2 \ell_3}\rangle\\ \nonumber 
&=  \int d^d x \sqrt{-\tilde{g}}\,   h[\phi^*_{0\ell_2 \ell_2}, \phi_{0\ell_1\ell_1}]\cdot \CT[\phi^*_{0\ell_4 \ell_4}, \phi_{0\ell_3 \ell_3}]\, ,
}[][IJIT]
with $\phi_{n\ell L}$ and $\phi^*_{n \ell L}$ as in~\eqref{eq:BulkFieldDecomp}.

In order to evaluate the matrix elements, we first need to solve the classical equation of motion defining the expressions for $A_\mu$ and $h_{\mu\nu}$. For a general bra state $\< \ell | $ and ket state $|\ell'\>$, the source terms $\CJ[\phi, \phi^\dagger]$ and $\CT[\phi, \phi^\dagger]$ are not invariant under the AdS isometries and a brute force approach involves solving a complicated set of coupled partial differential equations.  However, for any fixed bra and ket states, we can always perform a conformal transformation that maps them to the single-particle primary state, in which case the source terms are manifestly more symmetric and the resulting background fields $A_\mu$ and $h_{\mu\nu}$ can be determined in simple closed form \cite{Fitzpatrick:2011hh}. Therefore,  the background fields for any particular external states should be obtainable from the appropriate conformal transformation on this simple solution.  In practice, we can formalize this procedure by first writing the gauge and gravity backgrounds as a function of the boundary CFT operators in position space, and then pick out specific external states by using~\eqref{eq:KetFromZP}).  

More precisely, we can think of the background gauge field due to the external states $\langle \ell |$ and $| \ell' \rangle$ as
\begin{equation}
 \wick{ \langle \c1 \ell  \vert  A^\mu(X)[\c1 \phi, \c1 \phi^\dagger ]  \vert \c1 \ell'}\rangle  \propto (z \cdot P_1)^\ell (z' \cdot P_2)^{\ell'} \< \Phi(x_1) A^\mu(X) \Phi^*(x_2)\> ,
\end{equation}
and since $\< \Phi(x_1) A^\mu(X) \Phi^*(x_2)\>$ is a three-point function (with two points at the boundary and one point in the bulk of AdS) it must be covariant under conformal transformations.  The constraints of conformal invariance on such correlators are most transparent in embedding space, and we will take $P_1$ and $P_2$ to be the embedding space lifts of the boundary points $x_1$ and $x_2$, while $A^M(X)$ will be the embedding space lift of the bulk gauge field.  In fact, up to gauge-dependent pieces that can be discarded, we can read off the position-dependence for 
\begin{align} \begin{aligned}
A^M(X;P_1, P_2) &= \< \Phi(P_1) A^M(X) \Phi^*(P_2)\>\, , \\
  h^{MN}(X;P_1, P_2) &= \< \Phi(P_1) h^{MN}(X) \Phi^*(P_2)\>
  \end{aligned}
\end{align}
from the results in~\cite{DHoker:1999mqo}; the uplifting of their results from real space to embedding space is straightforward.  
\subsubsection{Photon matrix element} \label{subsec:PhotonMatrixEl}
We begin by translating the results in~\cite{DHoker:1999mqo} for the background gauge field to embedding space: 
\twoseqn{
A^M(X;P_1,P_2) &\propto \frac{1}{(-2 P_1\cdot P_2)^\Delta} \left(\frac{  P_2^M }{X \cdot P_2}-\frac{P_1 ^M}{X \cdot P_1} \right) f\left(\frac{- P_1\cdot P_2}{2 X\cdot P_1 \lsp  X\cdot P_2}\right)\, ,}[]
{f(t)&=\begin{cases}
-\frac{t^{\Delta }-t}{2 (\Delta -1) (t-1)} & \quad d=4\, , \\
\frac{t^{\Delta } \, _2F_1\left(1,\Delta ;\Delta +\frac{1}{2};t\right)}{2 \Delta -1}-\frac{\sqrt{\pi }
   \sqrt{t} \Gamma \left(\Delta -\frac{1}{2}\right)}{2 \sqrt{1-t} \Gamma (\Delta )} &\quad d=3\, .
\end{cases}}[DhokerSolnGaugef]
[DhokerSolnGauge]
Moreover for $\Delta$ integer (half-integer) in $d=4$ ($d=3$), $f(t)$  can be written as a finite sum:
\eqna{
f(t)&=-\sum_{k=\frac{d-2}{2}}^{\Delta-1}{\frac{\Gamma (k) \Gamma \mleft(\Delta-\frac{\mathit{d}}{2} +1\mright)}{2 \Gamma (\Delta ) \Gamma
   \mleft(k-\frac{\mathit{d}}{2}+2\mright)}}t^k\, .
}[]
More generally, when the sum does not truncate, it can still be written as the difference of two infinite sums, $f(t) = \sum_{k=\Delta}^\infty a_k t^k -\sum_{k=\frac{d-2}{2}}^\infty a_k t^k$ with $a_k = \frac{\Gamma (k) \Gamma \mleft(\Delta-\frac{\mathit{d}}{2} +1\mright)}{2 \Gamma (\Delta ) \Gamma
   \mleft(k-\frac{\mathit{d}}{2}+2\mright)}$.

As discussed above, to extract the spin $\ell_1$, $\ell_2$ component we can take 
\eqna{
(z_1 \cdot \nabla_1 )^{\ell_1}(z_2 \cdot \nabla_2 )^{\ell_2} A_{\mu}(X;P_1,P_2)\, , 
\label{eq:Aderivs}
}[]
with $\nabla_M=\frac{\partial}{\partial P^M}$ and then expand around the points  $x_1=0$ and $x_2=\infty$, which in embedding space corresponds to $P_1=(1,0,\vec{0})$ and $P_2=(0,1,\vec{0})$.  An efficient way to take the derivatives $(z_1 \cdot \nabla_1)^\ell$ and $(z_2 \cdot \nabla_2)^\ell$  is to restrict to the following coordinates:
\eqna{
P_1^{M}=(1,0, \vec{0}_{d-3}, \frac{z_1+\zb_1}{2},\frac{z_1-\zb_1}{2i},0)\, , \quad P_2^{M}=(0,1, \vec{0}_{d-3}, \frac{z_2+\zb_2}{2},\frac{z_2-\zb_2}{2i},0) .
}[PPar]
With this choice of coordinates for $P_1, P_2$, the derivatives in (\ref{eq:Aderivs}) at $x_1=0, x_2 = \infty$ reduce to
\eqna{
 A_\mu [\phi^*_{0 \ell_2 \ell_2}, \phi_{0 \ell_1  \ell_1}]\propto\left[ (\partial_{\zb_1})^{\ell_1}(\partial_{z_2})^{\ell_2} A_\mu (X;P_1, P_2) \right]_{\zb_1=0, z_2=0}\, ,
}[]
so we can actually  set directly  $z_1=0$ and $\zb_2=0$ in~\eqref{PPar}. We can fix the normalization factor by requiring that $A_\mu$ satisfies the differential equation \eqref{EoM}.

Finally, to obtain the anomalous dimensions, we substitute these solutions for the background field back into the integrals \eqref{IJIT}. For generic values of $\Delta$, we evaluate the anomalous dimensions by Taylor series expanding $f(t)$ from (\ref{DhokerSolnGauge}) in powers of $t$, and evaluating the integrals term by term; for any finite value of $J$, only a finite number of terms in the Taylor series contribute, due to (\ref{eq:Aderivs}).\footnote{For any finite value of $J$, the sum over terms in this series converges.  However, there is a subtlety that if one tries to expand around infinite $J$ in powers of $1/J$, this limit does not commute with the sum in the series.  We discuss this in more detail in Appendix~\ref{app:SubtleScalar}.} 

For integer (half-integer) values of $\Delta$ in $4d$ ($3d$), because the series expansion of $f(t)$ is finite, we can evaluate the integrals in closed form for arbitrary $J$. In $4d$, this sum takes the following explicit but somewhat unwieldy form for integer $\Delta$:
\eqna{
&\CI_{\CJ, \Delta \in \mathbb{Z}}^{(4d)}=-\frac{\pi ^2  \Gamma (\Delta )^2 \tilde{N}_{\Delta, \ell_1, \ell_2}^{(4d)}\tilde{N}_{\Delta, \ell_3, \ell_4}^{(4d)}}{\Gamma (\Delta +\ell_1) \Gamma (\Delta +\ell_2)}\! \sum_{j=j_0}^{\ell_2}\!\frac{ \ell_1! \ell_2! \Gamma (j+\ell_4) \Gamma
   (-j+\Delta +\ell_2-1)}{\Gamma (-j+\ell_2+1) 
   \Gamma (j+\Delta +\ell_4+2)} \times\\
  &\Bigg(\! j(j+\ell_{12}){}_4F_3 \left( \!\!\begin{array}{c}
  j+1, \quad  j+\ell_{12}+1, \quad 2-\Delta \quad\Delta \\
  2, \quad j-\ell_2-\Delta+2,\quad j+\ell_4+\Delta+2
\end{array}  ; 1 \right)
  \\ 
&+  (j\Delta-\ell_4(\ell_{12}-\ell_4-\Delta)){}_4F_3 \left(\!\! \begin{array}{c}
  j+1, \quad j+\ell_{12}+1, \quad 2-\Delta \quad\Delta \\
 1, \quad j-\ell_2-\Delta+2,\quad j+\ell_4+\Delta+2
\end{array}  ; 1 \right)\!\!\Bigg) 
}[IJ4d]
   where $\ell_{12}\equiv \ell_1-\ell_2$,  $j_{0}=\mathrm{max}(0, \ell_2-\ell_1)$ and momentum conservation fixes $\ell_3=\ell_2+\ell_4-\ell_1$.  We have defined
   \eqna{
   \tilde{N}_{\Delta, \ell_1, \ell_2}^{(4d)}=\frac{\sqrt{(\ell_1+1)( \ell_2+1)}}{2\pi^2 N_{\Delta, \ell_1}N_{\Delta, \ell_2}}\, ,
   }[Ntilde4d]
   with $N_{\Delta, \ell}$ in~\eqref{norm}.  Although \eqref{IJ4d} is  complicated, for some specific $\Delta$ it  greatly simplifies;  for example, for $\Delta=2$ one has
   \eqna{
   \CI_{\CJ, \Delta=2}^{(4d)}=-\pi^2 \tilde{N}_{2, \ell_1, \ell_2}^{(4d)}\tilde{N}_{2, \ell_3, \ell_4}^{(4d)} \begin{cases}
  \frac{ (\ell_2+2) (\ell_2+\ell_4+1)+(\ell_4+1) (-\ell_1+\ell_2+\ell_4)}{(\ell_1+1) (\ell_4+1) (\ell_2+\ell_4+1) (\ell_2+\ell_4+2) (\ell_2+\ell_4+3)} & \ell_1\geq \ell_2\, , \\
  \text{same with } \ell_1 \leftrightarrow \ell_3,\, \ell_2 \leftrightarrow \ell_4 & \ell_1 < \ell_2\,.
   \end{cases}
   }[]
Similarly in three dimensions for half-integer $\Delta$ the explicit form of the contribution to the anomalous dimension from any specific external state is
   \eqna{
  & \CI_{\CJ,\Delta \in \mathbb{Z}-\frac{1}{2}}^{(3d)}=-\frac{\pi  \Gamma(\Delta)^2 \tilde{N}_{\Delta, \ell_1, \ell_2}^{(3d)}   \tilde{N}_{\Delta, \ell_3, \ell_4}^{(3d)}}{\Gamma(\Delta+\ell_1)\Gamma(\Delta+\ell_2)}\sum_{j=j_0}^{\ell_2} \frac{\ell_1! \ell_2!\Gamma \left(j+\frac{1}{2}\right) \Gamma \mleft(j+\ell_{12}+\frac{1}{2}\mright)\Gamma (j+\ell_4)}{j! (j+\ell_{12})!  (-j+\ell_2)! } \times  \\
   & \frac{  \Gamma \mleft(-j+\Delta +\ell_2-\frac{1}{2}\mright)}{\Gamma
   \mleft(j+\Delta +\ell_4+\frac{3}{2}\mright)}\Bigg ( 2\Delta j(j+\ell_{12}){}_4F_3 \left(\!\! \begin{array}{c}
  j+\frac{1}{2}, \quad j+\ell_{12}+\frac{1}{2}, \quad \frac{3}{2}-\Delta \quad\Delta \\
 \frac{3}{2}, \quad j-\ell_2-\Delta+\frac{3}{2},\quad j+\ell_4+\Delta+\frac{3}{2}
\end{array}  ; 1 \right)\\
&+(j\Delta-\ell_4(\ell_{12}-\ell_4-\Delta)){}_4F_3 \left(\!\! \begin{array}{c}
  j+\frac{1}{2}, \quad j+\ell_{12}+\frac{1}{2}, \quad \frac{3}{2}-\Delta \quad\Delta \\
 \frac{1}{2}, \quad j-\ell_2-\Delta+\frac{3}{2},\quad j+\ell_4+\Delta+\frac{3}{2}
\end{array}  ; 1 \right) \Bigg)
   }[]
   \eqna{
    \tilde{N}_{\Delta, \ell_1, \ell_2}^{(3d)}=\frac{2^{-(2+\ell_1+\ell_2)}\sqrt{(2\ell_1+1)!( 2\ell_2+1)!}}{\pi \lsp \ell_1!\ell_2! N_{\Delta, \ell_1}N_{\Delta, \ell_2}} 
   }[Ntilde3d]
   \subsubsection{Graviton matrix element}
   The graviton matrix element is computed analogously to the previous results. First we write the three-point function in embedding space
   \eqna{
h^{MN}(X, P_1, P_2)\propto \frac{\Delta}{(-2 P_1 \cdot P_2)^\Delta}   \left( \frac{P_1^M P_1^N}{(X\cdot P_1)^2}-\frac{\eta^{M N}}{d-1} \right)f\left(\frac{-P_1\cdot P_2}{2X\cdot P_1\, X \cdot P_2} \right)\, ,
   }[hMNsoln]
   where we have discarded gauge-dependent terms and the function $f(t)$ is exactly the one defined  in~\eqref{DhokerSolnGaugef}.  The full expressions for the graviton matrix element are quite cumbersome, so we relegate all the details to Appendix~\ref{Appendix:gravitonMatrix}. Here we just report a few examples in four and three dimensions for some specific external dimensions and spins.  In four dimensions, for $\ell_i=0$ we find
   \eqna{
   \CI_\CT^{(4d)}(0,0,0)=\frac{4 \pi ^2 (\Delta -2) \Delta ^2}{3 (\Delta -1) (2 \Delta -1)} (\tilde{N}_{\Delta, 0,0}^{(4d)})^2\, ,
   }[]  
which is consistent with the results in~\cite{Fitzpatrick:2011hh,Andriolo:2022hax}. For $\Delta=2$ and generic spins
\eqna{
\CI_{\CT,\Delta=2}^{(4d)}&=\frac{16\pi^2}{3} \tilde{N}^{(4d)}_{2,\ell_1,\ell_2}\tilde{N}^{(4d)}_{2,\ell_3,\ell_3} \begin{cases}
\frac{\ell_2^2+2 (\ell_4+2) \ell_2+\ell_4^2+\ell_4-3 \ell_1 (\ell_4+1)}{(\ell_1+1) (\ell_4+1) (\ell_2+\ell_4+1) (\ell_2+\ell_4+2) (\ell_2+\ell_4+3)}& \ell_1\geq \ell_2\, , \\
 \text{same with } \ell_1 \leftrightarrow \ell_3,\, \ell_2 \leftrightarrow \ell_4 & \ell_1 < \ell_2\,.
\end{cases}
}[]
Passing to three dimensions, we have for all vanishing spins
\eqna{
\CI_\CT^{(3d)}(0,0,0)=\frac{2 \pi ^{3/2} \Delta ^2 (2 \Delta -3) \Gamma \mleft(2 \Delta
   -\frac{3}{2}\mright)}{\Gamma (2 \Delta )} (\tilde{N}_{\Delta,0,0}^{(3d)})^2\,.
}[]
\subsubsection{Final Result}
So far, we have computed the contribution to the anomalous dimensions when the bra state $\< \ell_2, \ell_4|$ and ket state $|\ell_1, \ell_3\>$ in \eqref{IJIT} have arbitrary spins.  The final step is to take specific linear combinations of these states to form primary wavefunctions with definite total spin $J$.  At $Q=2$, there is only one such state, $[\Phi, \Phi]_J$, at each $J$.  For any $\Delta$ and $J$ it is possible to explicitly perform all the necessary sums with the following result for $J>0$:
\twoseqn{ 
&\begin{aligned} \bra{[\Phi, \Phi]_J} \CV_{\mathrm{eff}} \ket{[\Phi, \Phi]_J}&= \frac{\Gamma \left(\frac{\mathit{d}}{2}\right) \Gamma (\Delta )^2}{2\pi ^{\frac{\mathit{d}}{2}}  \Gamma
   \mleft(\frac{2-\mathit{d}}{2}+\Delta \mright)^2} \left( \frac{(g_{\mathrm{U(1)}}\kappa )^2}{d-2}-\frac{\Delta^2 \kappa^2}{d-1}\right) \gamma_{[\Phi, \Phi]_J}^{(d)} \\
   &=\frac{ \Gamma (\mathit{d}) \Gamma (\Delta )^2}{\Gamma \mleft(\frac{\mathit{d}}{2}\mright)^2 \Gamma
   \mleft(\Delta -\frac{\mathit{d}}{2}+1\mright)^2} \left( \frac{1}{c_\CJ}-\frac{2d(d+1)\Delta^2}{(d-1)^2 c_\CT}\right) \gamma_{[\Phi, \Phi]_J}^{(d)}\,,
   \end{aligned}
}[]
{ &\gamma_{[\Phi, \Phi]_J}^{(4d)} =
   \frac{1}{(J +1) (2 \Delta +J -2)} \, , \qquad  \quad \quad \, \,
  \gamma_{[\Phi, \Phi]_J}^{(3d)}  = \frac{\Gamma (J +1) \Gamma \mleft(J +2 \Delta -\frac{3}{2}\mright)}{\Gamma \mleft(J
   +\frac{3}{2}\mright) \Gamma (J +2 \Delta -1)} \, . 
}[][DoubleTwistLargeJ]
The results for $J=0$ differs from the ones above, signaling a non-analyticity,
\eqna{
\bra{[\Phi, \Phi]_0} \CV_{\mathrm{eff}} \ket{[\Phi, \Phi]_0}&=\frac{ \Gamma (\mathit{d}) \Gamma (\Delta )^2 \Gamma \mleft(2 \Delta  +1-\frac{\mathit{d}}{2}\mright)}{\Gamma \mleft(\frac{\mathit{d}}{2}\mright)^3 \Gamma (2 \Delta ) \Gamma
   \mleft(\Delta +1-\frac{\mathit{d}}{2}\mright)^2}\left( \frac{1}{c_\CJ}-\frac{4  (\mathit{d}^2+d) \Delta ^2 (\mathit{d}-2 \Delta )}{(\mathit{d}-1)^2 (\mathit{d}-4
   \Delta )\lsp c_\CT} \right)\, ,
}[] 
where notice the pole at $\Delta=\frac{d}{4}$.  This will be relevant in the discussion in Sec.~\ref{Subsec: gravity}.\\
In Appendix \ref{app:FourPointAnomDim}, we review the standard computation of these anomalous dimensions from a conformal block expansion of the four-point function $\langle \Phi(x_1) \Phi(x_2) \Phi^\dagger(x_3) \Phi^\dagger(x_4)\>$, which gives the same result.

Notice that, as expected,  at large $J$ the anomalous dimensions decay with $J$ as a power law~\cite{Fitzpatrick:2012yx,Komargodski:2012ek}.   The power is determined by the lowest-twist exchanged operator, which in this case is $d-2$,  
\eqna{
\gamma_{[\Phi, \Phi]_J}^{(d)} \to \frac{1}{J^{d-2}}\, .
}[]

A particularly important qualitative feature is that the gauge boson  and graviton contributions have different signs, since gauge interactions between equally charged particles are repulsive whereas gravity is attractive at long distances. Therefore, there is an interesting qualitative transition in behavior depending on which of the two terms in parentheses is larger:
\begin{equation}
\frac{1}{c_\CJ} \lessgtr \frac{2d(d+1) \Delta^2}{(d-1)^2 c_\CT} ,
\label{eq:RepulsiveVsAttractive}
\end{equation}
where if the LHS is larger then there is a tendency for particles in AdS to move apart from each other whereas if the RHS is larger then by contrast they tend to collapse together.  
For $Q=2$, at each $J$ there is only a single primary state and so even if gravity is the strongest of the two interactions, the particles are held apart from each other by their angular momentum.
When we turn our attention to higher particle number states, however, there are many primary states at each $J$ (for large enough $J$) and it is a dynamical questions which of them has the smallest total energy.  Therefore, the physics of the Regge limit is naturally divided into two qualitatively different regimes, which we will call the `attractive' phase and the `repulsive' phase, depending on the relative size of the two terms in (\ref{eq:RepulsiveVsAttractive}).

From the perspective of the AdS computation, the repulsive phase is under much better perturbative control.  This is because the dynamical tendency of particles to repel each other typically will cause the lowest-energy states at fixed $Q$ and large enough $J$ to be described by widely-separated particles rotating around the center of AdS, and this large separation leads to weak interactions that can be treated perturbatively even if the couplings themselves are not small and the bulk spectrum is not sparse.  Therefore, we expect the results of the AdS computation in the repulsive Regge phase to hold beyond the holographic setting in which we perform the computation.  

By contrast, the computations we will perform for the lowest-energy states in the Regge limit of the attractive phase are not obviously reliable outside of holographic settings.  Nevertheless, in Sec.~\ref{sec:LargeQ} we will argue that with a few reasonable dynamical assumptions and a small amount of nonperturbative data as inputs, there is again a reliable holographic computation that can be applied in the limit of large spin.

\subsection{Contact Terms}\label{Subsec:contactTerms}

In addition to particle exchanges, the bulk description may contain local contact term interactions like $\phi^4$.  In real space, such terms are $\delta$ functions (or derivatives of $\delta$ functions), and so their contributions vanish in the limit of large spin/large separation. There is an interesting question of how many contact terms one should include, since from an EFT perspective there are an infinite number of them.  On the other hand, for the case $Q=2$, the inversion formula indicates that double-twist anomalous dimensions should be analytic down to $J=2$, so if one really had access to the full tower of bulk primary operators being exchanged (in other words, knowledge of the full set of conformal blocks exchanged in the cross-channel of the four-point function), then among the quartic-interactions only the coefficient of $\phi^4$ should be a free parameter.  
Such an application of the inversion formula would be extracting the double-twist anomalous dimensions from the exact four-point function and thus working with the full UV theory, which has healthier high-energy behavior in the bulk of AdS than an EFT~\cite{Caron-Huot:2017vep,Alday:2017gde,Maldacena:2015waa}.  A closer analogy is the use of dispersion relations, and the fact that they have to be supplemented by only a finite number of subtraction terms.  Of course, we rarely have the exact correlator, but rapid OPE convergence means that typically using a few conformal blocks is a good approximation.  Indeed, in later sections, we will turn to the $3d$ O(2) model as an application, and here one can see explicitly that just from exchange of low-twist fields in AdS, one can obtain a good match to the data for $Q=2$ double-twist dimensions all the way down to $J=2$, but not for $J=0$. We interpret this to mean that, even if it is the case that higher-derivative contact terms should be included in principle, in practice their coefficients are very small and can be neglected, though we think that making this statement sharper is an important open question.    These considerations motivate us to include at the quartic level only those local interactions that do not affect spins greater than or equal to 2:
\eqna{
V_{\mathrm{eff}}\to V_{\mathrm{eff}}+V_{\mathrm{contact}}\, , \quad  V_{\mathrm{contact}}=\frac{\kappa^2\lambda_0}{R_{\mathrm{AdS}}^{4-d}} \lsp (\phi \phi^\dagger)^2+\frac{\kappa^2\lambda_1}{R_{\mathrm{AdS}}^{2-d}} \lsp \phi \phi^\dagger \partial_\mu \phi \partial^\mu \phi^\dagger\, .
}[Vcontact]
The corresponding matrix elements read
\twoseqn{
&\begin{aligned}
\CI_{\lambda_0}(\ell_1, \ell_3, \ell_2, \ell_4)&=\int d^d x \sqrt{-\tilde{g}} \wick{ \langle \c1 \ell_2 \c2 \ell_4 \vert  \c1 \phi \c1 \phi^\dagger   \c2 \phi \c2 \phi^\dagger  \vert \c1 \ell_1 \c2 \ell_3}\rangle\\
&=\tilde{N}^{(d)}_{\Delta,\ell_1, \ell_2}\tilde{N}^{(d)}_{\Delta,\ell_3, \ell_4}\frac{\pi ^{\mathit{d}/2} \Gamma \mleft(2 \Delta -\frac{\mathit{d}}{2}\mright) \Gamma (\ell_2+\ell_4+1)}{\Gamma (\ell_2+\ell_4+2 \Delta )}
\end{aligned}
}[lambda0El]
{&\begin{aligned}
\CI_{\lambda_1}(\ell_1, \ell_3, \ell_2, \ell_4)&=\int d^d x \sqrt{-\tilde{g}} \wick{ \langle \c1 \ell_2 \c2 \ell_4 \vert  \partial_\mu \c1    \phi \lsp  \partial^\mu \c1  \phi^\dagger   \c2 \phi, \c2 \phi^\dagger  \vert \c1 \ell_1 \c2 \ell_3}\rangle\\
&=\tilde{N}^{(d)}_{\Delta,\ell_1, \ell_2}\tilde{N}^{(d)}_{\Delta,\ell_3, \ell_4}\frac{\Delta \Gamma \mleft(2 \Delta -\frac{\mathit{d}}{2}\mright) \Gamma (J)}{\pi ^{-\mathit{d}/2}\Gamma (J+2 \Delta+1 )}(\ell_1(\ell_2-\ell_4)\left(4\Delta-d\right)\\
&\quad\,-J(\ell_2 (3 \Delta -\mathit{d})-\Delta  (\mathit{d}-2 \Delta +\ell_4)) )\, ,
\end{aligned}
}[][]
where we have defined $J=\ell_2+\ell_4=\ell_1+\ell_3$.
For two-particle primaries, the contribution to the anomalous dimension simplifies to
\eqna{
\bra{[\phi, \phi]_J} \CV_{\mathrm{cont}} \ket{[\phi, \phi]_J}&=( \tilde{N}^{d}_{\Delta,0,0})^2 \frac{2\Gamma \left(2 \Delta -\frac{\mathit{d}}{2}\right)}{ \pi ^{-\mathit{d}/2} \Gamma (2 \Delta )} \lsp \delta_{J,0} \kappa^2\!\left(\! \lambda_0-\lambda_1 \Delta \left( \Delta-\frac{d}{2}\right)\! \right)\, .
}[]
Because we  consider states made from products of $\phi$, but not $\phi^\dagger$, contributions from both these operators are proportional to each other  and we lose nothing by setting $\lambda_1=0$.  

When we pass to $Q= 3$, there are two different kinds of effects to consider.  The first is that there will be sextet contact terms $\sim \phi^6$ that directly contribute to the low spin triple-twist anomalous dimensions.  The second is that the $\phi^4$ interaction will affect $Q=3$ states with any finite value of spin, because $Q=3$ states contain components of the form $[\Phi, [\Phi, \Phi]_{\ell=0}]_J$  where an internal two-particle blob has spin-0 and is therefore affected by $\phi^4$ --- see Appendix~\ref{App:contact} for a dedicated analysis at large spin. Similar effects are present at all higher values of $Q$ as well. Thus, even if we are only interested in large $J$, we are forced to consider the effect of contact terms.  

\section{Large $\ell$ Bulk Approximation} \label{sec:largeJBulk}

In this section we will consider various analytic approximations for the wavefunctions at large $\ell$, in order to determine the behavior of the anomalous dimensions for $Q\ge 2$ states in this limit.  
Recall that the wavefunction of a lowest-twist large spin descendant state, from (\ref{eq:BulkDescendantWvFcn}), at the equator is simply 
\begin{equation}
\phi_{\ell} \propto e^{i \Delta t+ i \ell (t+\varphi)} \sin^\ell \rho \cos^\Delta \rho = e^{i \Delta t + i \ell (t+\varphi)} \frac{\tanh^\ell \chi }{\cosh^\Delta \chi},
\end{equation}
where we have restricted to highest-weight spin components $L=\ell$, and the proportionality factor above depends on the other angles on the sphere besides $\varphi$.  The magnitude of the wavefunction is maximized at $\chi=\chi_*$ where
\begin{equation}
\begin{aligned}
\sinh \chi_* &= \sqrt{\frac{\ell}{\Delta}} \quad \Rightarrow \quad \chi_* = \frac{1}{2} \log \frac{4 \ell}{\Delta} + O(\ell^{-1}), 
\label{eq:ChiVsLPart}
\end{aligned}
\end{equation}
and away from this maximum the wavefunction is
\begin{equation}
|\phi_{\ell}|  = |\phi_\ell(\chi_*)| e^{  -\Delta f(\chi-\chi_*)+O(1/\ell)}, \qquad f(\delta) = \frac{1}{2} (e^{-2\delta}-1) + \delta .
\end{equation}
Therefore at large spin $\ell$ the state gets pushed out toward the boundary of AdS, with width roughly $\sim \sqrt{\Delta}$ in $\chi$, though as one can see from this expression the wavefunction is not exactly Gaussian for finite $\Delta$ and there is a power-law tail at large $\chi$. 

The descendant states $|\ell\>$ are by definition eigenfunctions of rotation generators and so are spread out over the sphere. However, the linear combinations of them that arise in primary multiparticle states generally form wavepackets that have some width $\sim 1/\ell$ in the angular directions.  We can already see this explicitly in the double-twist primary states, where the specific linear combinations in (\ref{eq:DoubleTwistPrim}) at large $J$ are approximately Gaussians of the form
\begin{equation}
|\Psi\>_{2,J} \sim  \sum_{\ell=0}^J (-1)^\ell e^{-\frac{2}{J} (\ell-\frac{J}{2})^2 } | \ell, J-\ell\>,
\end{equation}
that is, they are peaked at $\ell = \frac{J}{2}$ with width $\sim \sqrt{J}$. Consequently, in real space they are Gaussians in the angle $\varphi$ with width $\sim J^{-1/2}$. 

These localized wavefunctions for individual particles in AdS can also be understood by taking the primary state wavefunction $\phi_0$, which describes a particle at rest in the center of AdS, and performing an AdS isometry to map it to a particle rotating around the center of AdS at fixed radial distance.  This transformation is most easily done using embedding space coordinates:
\begin{equation}
X_0 = \cos t \cosh \chi, \quad X_{d+1} = \sin t \cosh \chi, X_i = \sinh \chi ~\Omega_i.
\end{equation}
The trajectory for a particle at rest at $\chi=0$ is manifestly 
\begin{equation}
(\chi=0): \quad X_0 = \cos t, \quad X_{d+1}=\sin t, \quad X_i =0.
\label{eq:XAtRest}
\end{equation}
By contrast, a particle at fixed $\bar{\chi}$, rotating in $\varphi$ around the equator at the speed of light is
\begin{equation}
\begin{aligned}
(\chi=\bar{\chi}, \varphi= \varphi_0 + t): \quad & X_0 = \cosh \bar{\chi} \cos t , \quad && X_{d+1} =  \cosh \bar{\chi} \sin t, \\
 & X_1 = \sinh \bar{\chi} \cos (\varphi_0+t), \quad && X_2 = \sinh \bar{\chi} \sin(\varphi_0+t) , \\
 & X_3 = \dots = X_d =0.
 \label{eq:XRotating}
\end{aligned}
\end{equation}
Without loss of generality, we can always rotate $X_1$ and $X_2$ into each other using the generator $J_{12} = M_{12}$, to shift $\varphi_0=0$ for the trajectory of a single particle. Then, to map the $\chi=\bar{\chi}$ trajectory to the $\chi=0$ trajectory, we can perform the following SO$(d-1,2)$ transformation: 
\begin{equation}
\begin{aligned}
 &\left( \begin{array}{c} X_0 \\ X_{1} \end{array} \right) \rightarrow R(\bar{\chi}) \left( \begin{array}{c} X_0 \\ X_{1} \end{array} \right)\, , \qquad \qquad
 \left( \begin{array}{c} X_{d+1} \\ X_{2} \end{array} \right) \rightarrow R(\bar{\chi}) \left( \begin{array}{c} X_{d+1} \\ X_{2} \end{array} \right)\, , \,  
\end{aligned}
\label{eq:XFrameChange}
\end{equation}
with
\eqna{
R(\beta)  &= \left( \begin{array}{cc}  \cosh \beta & - \sinh \beta \\ - \sinh\beta & \cosh \beta \end{array} \right)
}[]
It is easy to verify that this transformation maps (\ref{eq:XAtRest}) to (\ref{eq:XRotating}) with $\varphi_0=0$.  

Now, consider the  transformation $R(\bar{\chi})$ acting on the primary state wavefunction $\phi_0$:
\begin{equation}
\phi_0 \propto e^{i \Delta t} \cos^\Delta \rho = (X_0 - i X_{d+1})^{-\Delta}.
\end{equation}
Acting with $R(\bar{\chi})$, we obtain the wavefunction for the spinning particle:
\begin{equation}
\begin{aligned}
\phi_R &\propto  (X_0' - i X'_{d+1})^{-\Delta} =  \left( X_0 \cosh \bar{\chi} - X_1 \sinh \bar{\chi}- i X_{d+1} \cosh \bar{\chi} + i X_2 \sinh \bar{\chi}\right)^{-\Delta} \\
 & = \left( e^{-i t} \cosh \bar{\chi} \cosh \chi - e^{- i \varphi} \Omega_3 \sinh \bar{\chi} \sinh \chi \right)^{-\Delta},
\end{aligned}
\end{equation}
where $\Omega_3 = \sin \theta_1 \sin \theta_2 \dots \sin \theta_{d-2}$ carries the dependence on the other angles on the sphere and is equal to 1 at the equator $\theta_i=\frac{\pi}{2}$.
At $t=0$, this wavefunction is now peaked at $\chi=\bar{\chi}$ and $\varphi=0$: 
\begin{equation}
\phi_R|_{t=0} \approx\! \exp\! \left(\!\! - \frac{\Delta}{2} \Big( (\chi-\bar{\chi})^2 + \sinh^2(\bar{\chi})\!\!\sum_{i=1}^{d-2} (\theta_i - \frac{\pi}{2})^2 \Big)\! - \frac{\Delta \sinh^2(2\bar{\chi})}{8} \varphi^2  + i \Delta \varphi  \sinh^2 \bar{\chi} \! \right),
\label{eq:WvFcnGaussianApprox}
\end{equation}
where we have shown terms up to quadratic order in $\varphi, (\chi-\bar{\chi})$ for the real part and up to linear order for the imaginary part of the exponent.  Note that we can again see $\ell = \Delta \sinh^2 \bar{\chi}$ by acting with the rotation generator $M_{12} = -i \partial_\varphi$.  

\begin{figure}[t!]
\begin{center}
\includegraphics[width=0.4\textwidth]{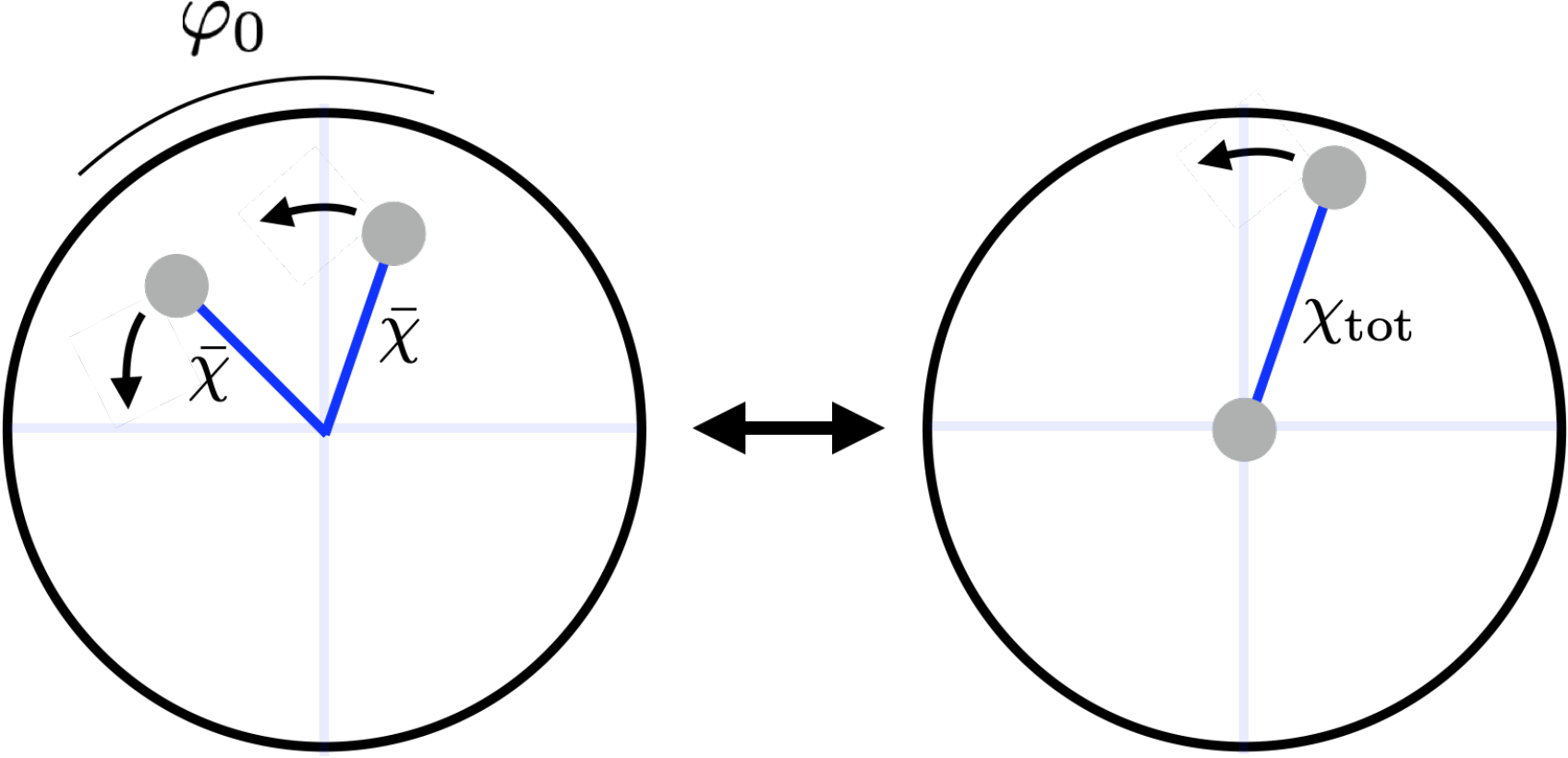}
\caption{Depiction of the trajectories of two particles in the frame where ({\it left}) they orbit AdS at the same distance $\bar{\chi}$ from the center of AdS, separated by an angle $\varphi_0$, and in the frame where ({\it right}) one of them is at the center of AdS and the other circles it at distance $\chi_{\rm tot}$.}
\label{fig:AdSFrames}
\end{center}
\end{figure}

The main scenario in which we will apply these formulas is to the case where we have multiple partons rotating around the center of AdS at some fixed radial distance $\chi$, separated from each other in the angular direction by some angle $\varphi_0$.  We can simplify the analysis of this configuration by performing the transformation that maps one of the partons to $\chi=0$, and then the other parton will be mapped to a new trajectory circling the center of AdS but at a new radial distance (see Fig.~\ref{fig:AdSFrames}). Because the wavefunctions are highly peaked, we can focus on the classical trajectory of this peak in the following discussion, but the AdS isometry $R(\bar{\chi})$ maps the  wavefunction of the partons to the new configuration and therefore relates the leading-order binding energy between the two partons to the energy of the quantum two-body problem;  this is also a convenient way of making manifest the symmetries of the problem of calculating the background gauge and gravity field sourced by an individual parton, as we cover in appendix \ref{app:singleparton}.  

 To see what happens to the particle originally rotating with offset $\varphi_0$, we take the trajectory in (\ref{eq:XRotating}) and act with the transformation (\ref{eq:XFrameChange}).  The new trajectory of the second parton has
\begin{equation}
\begin{aligned}
X_0 &\rightarrow  \cosh^2 \bar{\kappa} \cos t- \sinh^2 \bar{\kappa}  \cos (\varphi_0 +t), \\
X_{d+1} &\rightarrow \cosh^2\bar{\kappa}  \sin t - \sinh^2\bar{\kappa} \sin (\varphi_0+t).
\end{aligned}
\end{equation}
Let $\chi_{\rm tot}$ be the distance from the center of AdS of the second parton in this new frame.  We can read off the value of $\chi_{\rm tot}$ from the fact that, by inspection of (\ref{eq:XRotating}) for a generic rotating trajectory, $\cosh^2 \chi_{\rm tot} = X_0^2 + X_{d+1}^2$.  Therefore, the new trajectory of the second parton is at distance $\chi_{\rm tot}$ given by
\begin{equation}
\cosh^2 \chi_{\rm tot} = X_0^2 + X_{d+1}^2 = \frac{1}{4} \left( \cosh(4 \bar{\chi}) (1- \cos \varphi_0) + \cos \varphi_0 +3\right).
\end{equation}
In the limit of large $\bar{\chi}$ and $\chi_{\rm tot}$, this simplifies to the approximate relation
\begin{equation}
e^{ \chi_{\rm tot}} \approx e^{2 \bar{\chi}} \left|\sin(\varphi_0/2)\right|.
\label{eq:ChiTotVsChiVarPhi}
\end{equation}
For the special case where $\varphi_0=\pi$, the two particles in the original frame orbit the center of AdS at antipodal points, and the total geodesic distance between them is $\chi_{\rm tot} = 2 \bar{\chi}$, as one would expect.\footnote{For generic angles $\varphi_0$, however, the geodesic distance between the particles on a fixed time slice $t=0$ is not the same in the two frames, because the change in frame distorts the constant time slices.}  

Because we will mainly be considering photon and graviton exchange between two particles, and the anomalous dimension for a $Q=2$ primary is already known from the work we have reviewed in previous sections, it will often be useful to take the trajectories of two partons and determine the total spin $J_{\rm prim}$ of the corresponding state in the center-of-mass frame for the two-parton system, which is the frame in which they form a two-particle primary state.  In their center-of-mass frame, they are separated by an angle $\varphi_0=\pi$, so $\chi_{\rm tot}=2 \bar{\chi}$ as above.  Moreover, the total spin $J_{\rm prim}$ in the center-of-mass frame is just the sum $\ell+\ell = 2 \Delta \sinh^2 \bar{\chi} \approx \frac{\Delta}{2} e^{2 \bar{\chi}}$ of the spins  of the two partons.  Therefore, we obtain the following relation between $\chi_{\rm tot}$ and $J_{\rm prim}$ \cite{Fitzpatrick:2014vua}:
\begin{equation}
\frac{2J_{\rm prim} }{\Delta}
 \approx  e^{\chi_{\rm tot}}.
 \label{eq:JprimVsChiTot}
\end{equation}
Now we can use this formula to infer the contribution to the anomalous dimension from photon and graviton exchange between the more general configuration of two particles separated by a generic angle $\varphi_0$.  The idea is to map the trajectories at equal $\chi$ and fixed angular separation to the trajectory with one particle at the origin and another rotating at radial distance $\chi_{\rm tot}$, and then use (\ref{eq:JprimVsChiTot}) to find the anomalous dimension of the corresponding two-particle primary with spin $J_{\rm prim}$.  Using our result (\ref{eq:ChiTotVsChiVarPhi}) above, we obtain
\begin{equation}
J_{\rm prim} \approx \frac{ \Delta}{2} e^{\chi_{\rm tot}}  \approx \frac{ \Delta}{2} e^{ 2 \bar{\chi}} \left| \sin(\varphi_0/2)\right| \approx 2\ell \left| \sin(\varphi_0/2)\right|,
\label{eq:JprimVsVarphi}
\end{equation}
where we have used (\ref{eq:ChiVsLPart}) to relate $\bar{\chi}$ to the angular momentum $\ell$ of the individual partons in the original frame (left plot in Fig.~\ref{fig:AdSFrames}).

Consider the full anomalous dimension of a $Q$-parton state when the interactions are weak and to leading order we can sum over all pairwise exchanges. We assume the particles are evenly spread out in angles, so $\varphi_0$ takes the values  $\frac{2\pi}{Q} n$  with $n=0, 1, \dots, Q-1$, and the angular momentum $\ell$ of the individual partons is $J/Q$.  The anomalous dimension $\gamma^{(d,Q)}(J)$ of the $Q$-parton state at spin $J$ is related to the anomalous dimension $\gamma^{(d,Q=2)}(J)$ of 2-parton state  by
\begin{equation}
{\gamma^{(d, Q)}( J) \approx \frac{Q}{2} \sum_{n=1}^{Q-1} \gamma^{(d, Q=2)}(2\frac{J}{Q} \left| \sin ( \pi n/Q )\right|)\, ,}
\label{eq:MultQAnomDim}
\end{equation}
which is the main result of this section. Substituting $\gamma^{(d,Q=2)}(J)$ from (\ref{DoubleTwistLargeJ}), one obtains an analytic expression for the lowest-dimension state at large $J$ for any $Q$ in the repulsive phase.  We will use this in later sections to compare to numeric results at finite $Q$ and to explore the large $Q$ limit.

\section{Finite ${Q}$ and $J$ Exact Analysis}
\label{sec:FiniteQJ}
In this section we will turn to a numerical analysis of the eigenstates of our system at finite values of $Q$ and $J$, where we will use the perturbative bulk description that arises from tree-level gauge and graviton exchange, as well as tree-level contributions from contact terms, where relevant. 

Alternative approaches to extract large spin data of multi-trace operators can be found in~\cite{Antunes:2021kmm,Kaviraj:2022wbw, Harris:2024nmr}. These works take a purely CFT perspective and extract the large spin information by studying the lightcone limit of multi-point correlators as a direct generalization of the $Q=2$ results in~\cite{Komargodski:2012ek,Fitzpatrick:2012yx}.
\subsection{Primary states}
We have already discussed the general structure of lowest-twist primary states with charge $Q$, in Sec.~\ref{subsec:construct primaries}.  
Recall that we obtain the primary state basis as the null space of the operator $(z' \cdot K)$ acting on the space of `monomials' (\ref{eq:monomialstate}), where it takes the form of a finite-dimensional matrix. 
In the case  $J\leq3$, we have exactly one primary state at each level. Explicitly,
\eqna{
\ket{\Psi}_{0;Q}&=\frac{1}{\sqrt{Q!}} \ket{\vec{\boldsymbol{0}}_Q}\, ,\\
\ket{\Psi}_{1;Q}&=0\, , \\
\ket{\Psi}_{2;Q}&=\frac{1}{\sqrt{\Gamma(Q-1)(Q \Delta+1)}}\left( \sqrt{\Delta}\ket{2, \vec{\boldsymbol{0}}_{Q-1}}-\sqrt{\frac{\Delta+1}{2}}\ket{1,1, \vec{\boldsymbol{0}}_{Q-2}} \right)\, ,\\
\ket{\Psi}_{3;Q}& =
\frac{  \Delta \ket{3, \vec{\boldsymbol{0}}_{Q-1}}-\sqrt{3\Delta  (\Delta +2)}\ket{2,1, \vec{\boldsymbol{0}}_{Q-2}}+\sqrt{\frac{2(\Delta +1) (\Delta +2)}{3}}\ket{1,1, 1,\vec{\boldsymbol{0}}_{Q-3}}}{\sqrt{ \Gamma (Q-2)(\Delta  Q+2) (\Delta  Q+4)}} \, ,
}[smallSpinsPrimaries]
with $\vec{\boldsymbol{0}}_{Q}$ a vector with $Q$ zero entries.
In general for $Q>3,J>3$, we will have multiple primary states  for fixed $J$ and $Q$.  As an example,  suppose we would like to construct the primary state basis at  $Q=4,J=4$ with external dimensions $\Delta=3$.   Recall that $(z' \cdot K)$ decreases the total spin by one, so  to construct the matrix of our example, we need to know the `monomial' basis for the spins  $J=4$ and $J=3$ 
\eqna{
&\{|4,0,0,0\rangle,~|3,1,0,0\rangle,~|2,2,0,0\rangle,~|2,1,1,0\rangle,~|1,1,1,1\rangle\} && \qquad \text{for }J=4\, ,\\
&\{ |3,0,0,0\rangle,~|2,1,0,0\rangle~,|1,1,1,0\rangle\} && \qquad \text{for }J=3\, .
}[]
In this basis,  the matrix $(z'\cdot K)$ reads,
\begin{equation}
(z'\cdot K)=(z\cdot z') 
\begin{pmatrix}
4\sqrt{3} & \sqrt{6} & 0 &0 &0\\
0&\sqrt{30} & 8& 2\sqrt{6} &0\\
0&0&0&4&4\sqrt{6} 
\end{pmatrix} .
\end{equation}
The null space is made of two orthogonal vectors, which we identify with our primary states,
\eqna{
|v_1\rangle&=-\frac{1}{\sqrt{82}}|4,0,0,0\rangle+\frac{2}{\sqrt{41}}|3,1,0,0\rangle-\sqrt{\frac{5}{41}}|2,1,1,0\rangle+\sqrt{\frac{5}{246}}|1,1,1,1\rangle\, , \\
|v_2\rangle&=2\sqrt{\frac{3}{697}}|4,0,0,0\rangle-4\sqrt{\frac{6}{697}}|3,1,0,0\rangle+\frac{1}{2}\sqrt{\frac{41}{85}}|2,2,0,0\rangle\\
&\quad \, -11\sqrt{\frac{2}{10455}}|2,1,1,0\rangle +\frac{11}{3\sqrt{3485}}|1,1,1,1\rangle\, .
}[J4Prims]
The method we have described so far is very straightforward, but, unfortunately, becomes quite cumbersome for larger values of $J$ and $Q$, given the increasing size of the matrices we have to construct.  Luckily, in these cases, we can employ a more efficient way to construct the primaries,  inspired by the recursive method used in~\cite{Anand:2020gnn} which built on results from \cite{Mikhailov:2002bp,Penedones:2010ue,Fitzpatrick:2011dm}.\footnote{For even earlier recursive constructions of multi-twist operators  in  $\epsilon$ and large $N$ expansions, and applications to the large spin limit, see~\cite{Derkachov:2010zza} and reference therein.} The idea is that, at each $J$ and $Q$, we can construct a new primary as a double-twist state made of a primary, with spin $\ell$ and constructed at order $(Q-1)$, and $\Phi$ itself,  schematically
\eqna{
\CO_{Q, J}\sim \left[ \CO^{\prime}_{Q-1, \ell}, \Phi\right]\, .
}[]
More details can be found in Appendix~\ref{app:buildprimary}, where we also derive the number of degenerate primaries at fixed $Q$ and $J$, which we report here for simplicity:
\begin{equation}
\mathcal{N}(Q,J)=\sum_{k=0}^{J/Q}\mathcal{N}(Q-1,J-kQ)\, .
\end{equation}

\begin{figure}[t!]
\centering
\begin{subfigure}{0.48\textwidth}
    \includegraphics[width=\textwidth]{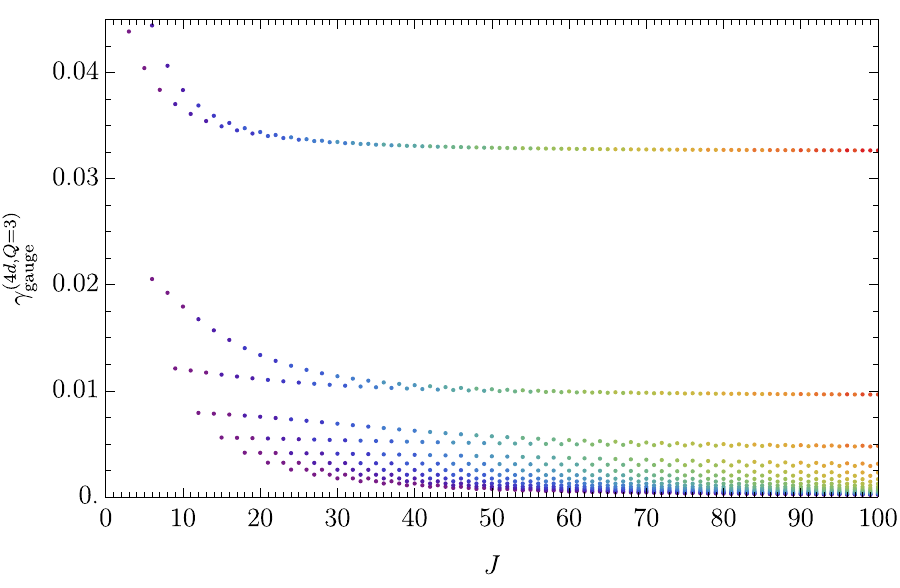}
    \caption{${}^{\quad \quad \! r}_{J;Q=3}\!\bra{\Psi}  \CV_{\CJ}\ket{\Psi}_{J;Q=3}^{s}$}\label{fig:q3AllEigGauge}
\end{subfigure}
\hfill
\begin{subfigure}{0.48\textwidth}
    \includegraphics[width=\textwidth]{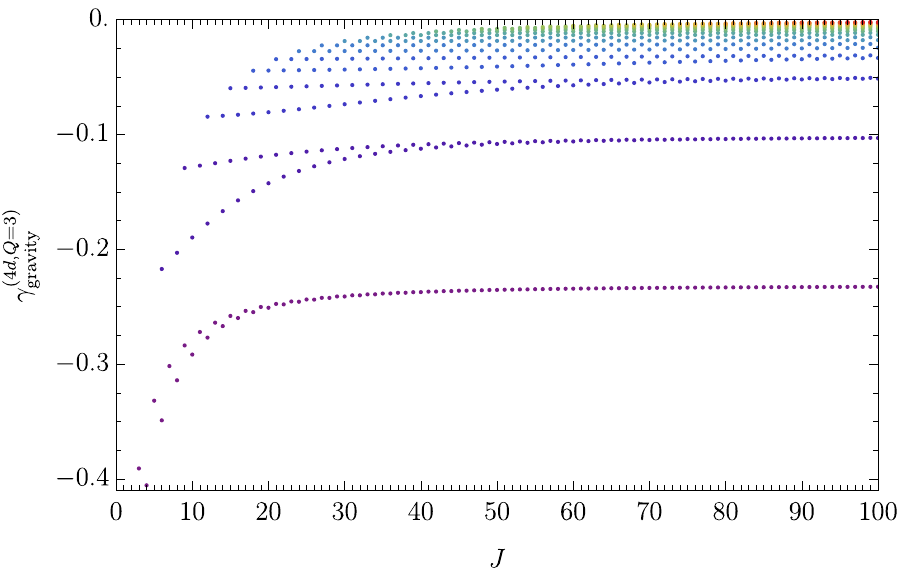}
    \caption{${}^{\quad \quad \! r}_{J;Q=3}\!\bra{\Psi}  \CV_{\CT}\ket{\Psi}_{J;Q=3}^{s}$}\label{fig:q3AllEigGravity}
\end{subfigure}
\caption{All eigenvalues of ${}^{\quad \quad \! r}_{J;Q=3}\!\bra{\Psi}  \CV\ket{\Psi}_{J;Q=3}^{s}$,  $\CV=\CV_\CJ$ (\textit{left}) and $\CV=\CV_\CT$ (\textit{right}), in $4d$ for $\Delta=4$ for different values of the spin. The gradient of color represents increasing energy eigenvalues at fixed $J$. }
\label{fig:q3D4AllEig}
\end{figure}

Given a  basis of primaries,  similarly to the two-particle analysis,  we can now compute the correction to their dimensions due to  photon and graviton tree-level exchange. The anomalous dimension  of the lightest state is given by the minimum of the eigenvalues of the matrix $V_{\mathrm{eff}}$, defined in~\eqref{anomInt}, evaluated in the primary state basis:
\eqna{
\gamma^{(d,Q)}&=\min  {}^{\quad r}_{J;Q}\!\bra{\Psi} \CV_{\mathrm{eff}} \ket{\Psi}_{J;Q}^{ s}\, , \qquad \quad r, s=1, \cdots,\,  \CN(Q, J)\\
&=\min  {}^{\quad r}_{J;Q}\!\bra{\Psi} (g_{\mathrm{U}(1)}\kappa)^2 \CV_{\CJ}+\kappa^2 \CV_\CT \ket{\Psi}_{J;Q}^{s}\, .
}[]
The restriction to two-particle interactions simplifies a lot the computation of this matrix. As an example,  consider $Q=3$
\eqna{
 {}^{\quad \quad \! r}_{J;Q=3}\!\bra{\Psi}  \CV_{\mathrm{eff}}\ket{\Psi}_{J;Q=3}^{s}&=\sum_{\substack{\ell_1+\ell_2+\ell_3=J\\ \ell_4+\ell_5+\ell_6=J}} \tilde{C}^r_{\ell_1\ell_2\ell_3}\tilde{C}^s_{ \ell_4\ell_5\ell_6} \bra{\ell_1\ell_2 \ell_3} \CV_{\mathrm{eff}} \ket{\ell_4 \ell_5\ell_6}\\&=\sum \tilde{C}^r_{\ell_1\ell_2\ell_3}\tilde{C}^s_{ \ell_4\ell_5\ell_6}\!\left( \delta_{\ell_1 \ell_4}\bra{\ell_2 \ell_3} \CV_{\mathrm{eff}} \ket{\ell_5 \ell_6} +\text{permutations}\right)\,,
}[]
where $\tilde{C}_{\ell_1 \cdots  \ell_Q}=C_{\ell_1 \dots \ell_Q}\sqrt{2^{\ell_1} \ell_1! (\Delta)_{\ell_1} } \dots \sqrt{2^{\ell_Q} \ell_Q! (\Delta)_{\ell_Q} }$ in~\eqref{eq:multibasis} and the matrix elements can be read off from Sec.~\ref{sec:energycorrection}.  Given the absence of  analytic expressions for the primary basis,  to extract the large spin behavior of the anomalous dimensions, we will compute this sum for various value of $J$,  then diagonalize the matrix and finally take the smallest eigenvalue.  To give some intuition of the degeneracy of the states and how the primary state eigenvalues are distributed, in  Fig.~\ref{fig:q3D4AllEig} we show the eigenvalues of ${}^{\quad \quad \! r}_{J;Q=3}\!\bra{\Psi}  \CV_{\CJ}\ket{\Psi}_{J;Q=3}^{s}$  and ${}^{\quad \quad \! r}_{J;Q=3}\!\bra{\Psi}  \CV_{\CT}\ket{\Psi}_{J;Q=3}^{s}$ separately at $Q=3$ and external dimension $\Delta=4$.\footnote{Certain large spin trajectories in $Q=3$ with interactions arising from bulk exchange can also be obtained exactly \cite{PetrInProgress}.}   It is evident that, in both cases, the lowest-energy eigenstate, which we are interested in,  has a qualitatively different behavior compared to the highest one. Moreover,  because the sign of the long-distance interaction for gravity is reversed compared to the gauge interactions, the highest-energy eigenvalues due to  a photon exchange (Fig.~\ref{fig:q3AllEigGauge}) are qualitatively similar to the graviton lowest-energy eigenvalues (Fig.~\ref{fig:q3AllEigGravity}), and vice versa.  In the next subsections, we will study in detail the large spin behavior of the lowest eigenvalues   together with the form of the corresponding eigenvectors. At  first we will  consider the contributions from gauge and gravity separately, and only at the very end will we sum them up and show the full dependence of the anomalous dimension on  $c_\CJ/c_\CT$.
\subsection{Gauge} \label{Sec:repulsivephase}
\begin{figure}[t]
\centering
\begin{subfigure}{0.48\textwidth}
    \includegraphics[width=\textwidth]{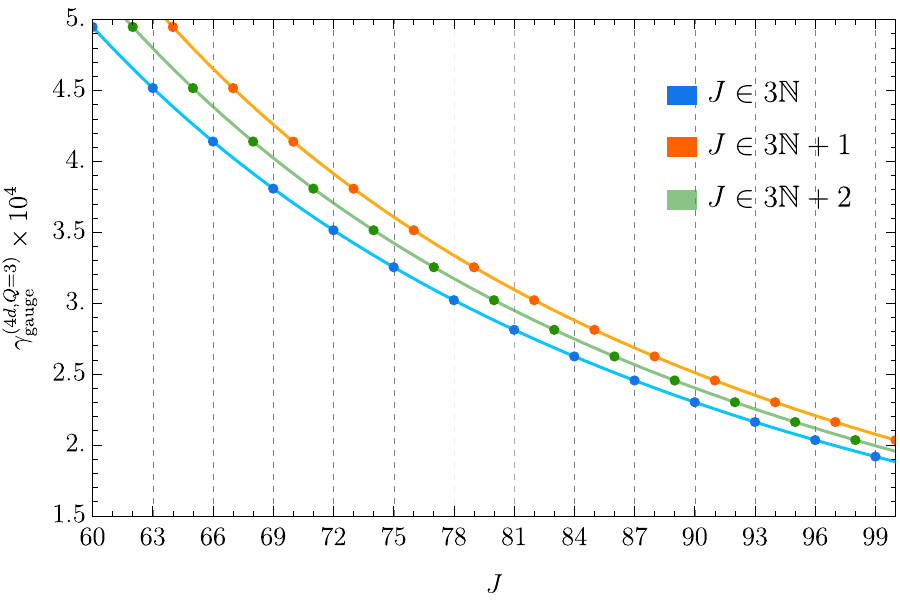}
    \caption{$Q=3$ $\Delta=4$ in $4d$}
    \label{fig:Q3D4}
\end{subfigure}
\hfill
\begin{subfigure}{0.48\textwidth}
    \includegraphics[width=\textwidth]{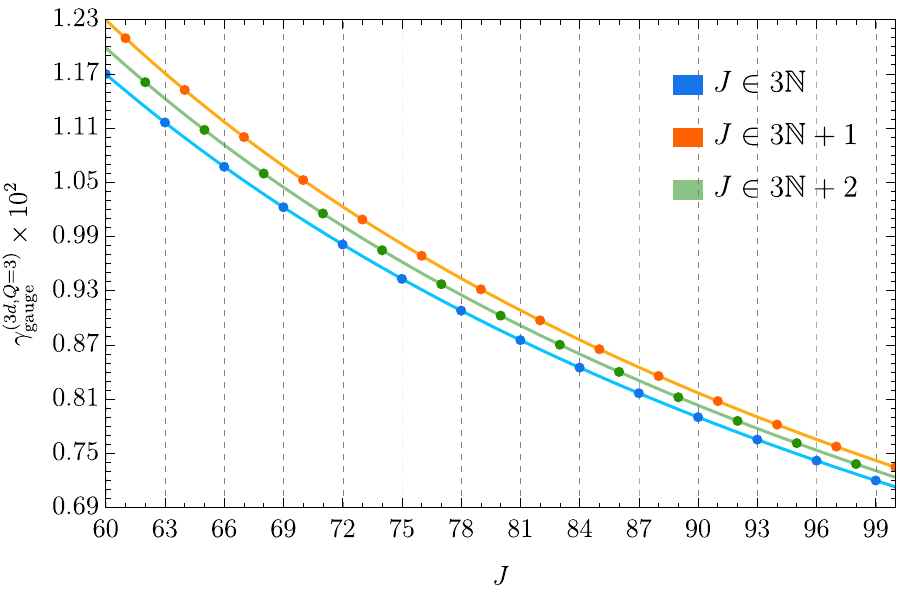}
    \caption{$Q=3$ $\Delta=\frac{5}{2}$ in $3d$}
    \label{fig:Q3D52}
\end{subfigure}
\caption{Anomalous dimensions corresponding to the lowest eigenvalue at each spin for 3-particle states due to U(1) gauge boson exchange; note that they fall into three separate trajectories.  The points correspond to the computed values while the solid lines are obtained by fitting the computed values to the function $\gamma=\frac{a_0}{J}+\frac{a_1}{J^2}+\frac{a_2}{J^3}$ in $3d$ and  $\gamma=\frac{b_0}{J^2+b_1J+b_2}$ in $4d$. The leading order coefficients $a_0,b_0$ for the three different trajectories are $a_0\approx (0.7307,0.7307,0.7307)$ and $b_0\approx (2.0516,2.0517,2.0518)$. The expected value from \eqref{gammagauge} are $a_0^{\text{exact}}=\frac{27\sqrt{3}}{64}\approx0.7307,~b_0^{\text{exact}}=\frac{81}{4\pi^2}\approx2.05175$.}
\label{fig:Q3}
\end{figure}
\begin{figure}[t!]
\centering
\begin{subfigure}{0.48\textwidth}
    \includegraphics[width=\textwidth]{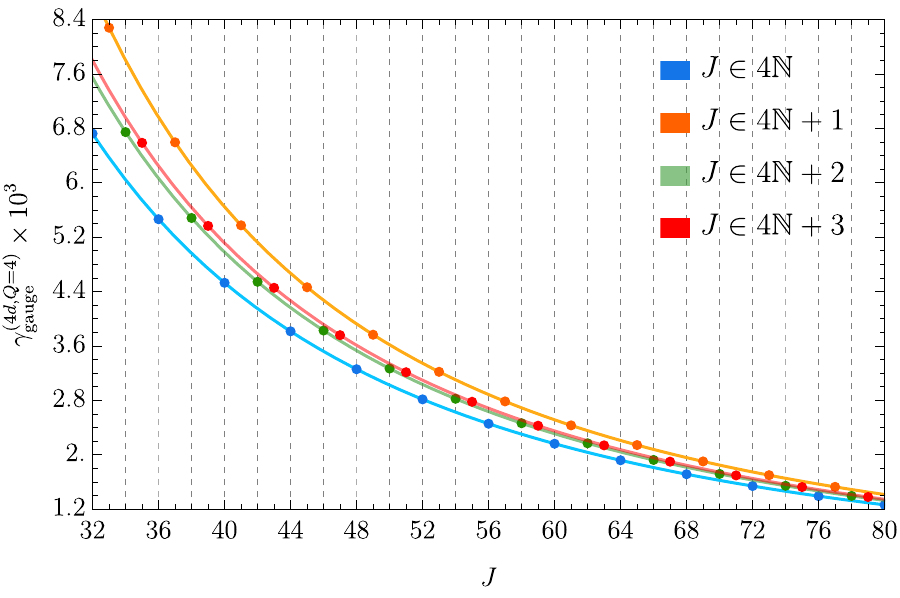}
    \caption{$Q=4$ $\Delta=4$ in $4d$}
    \label{fig:Q4D4}
\end{subfigure}
\hfill
\begin{subfigure}{0.48\textwidth}
    \includegraphics[width=\textwidth]{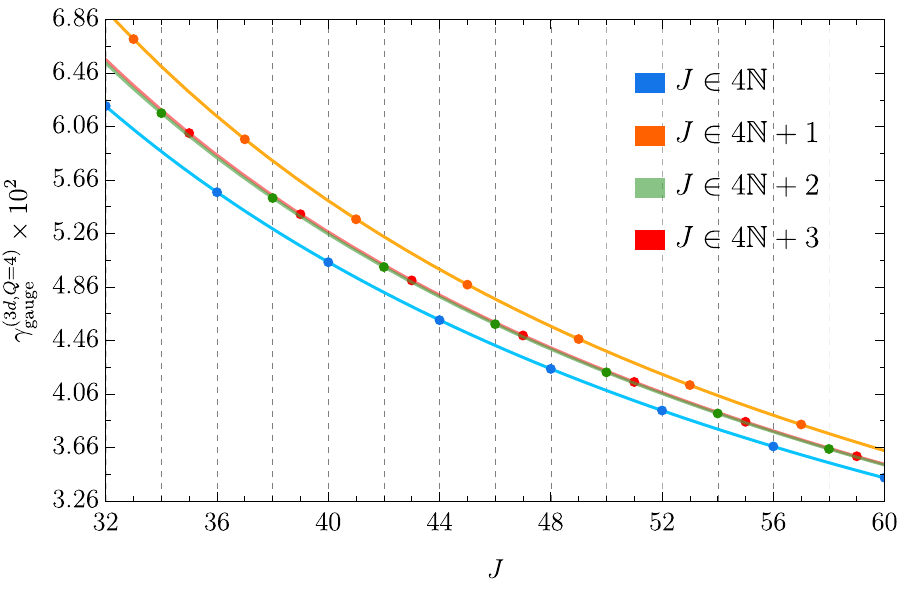}
    \caption{$Q=4$ $\Delta=\frac{5}{2}$ in $3d$}
    \label{fig:Q4D52}
\end{subfigure}
\caption{Anomalous dimensions corresponding to the lowest eigenvalue at each spin for 4-particle states due to U(1) gauge boson exchange. The points correspond to the computed values while the solid lines are obtained by fitting as described in  the caption of Fig.~\ref{fig:Q3}.  
The best-fit results for the parameters $a_0,b_0$ for the four different trajectories are $a_0\approx (2.152,2.152,2.152,2.152)$ and $b_0\approx(9.11,9.12,9.11,9.12)$, in agreement with the expected values from \eqref{gammagauge},   $a_0^{\text{exact}}=\frac{9(1+2\sqrt{2})}{16}\approx2.153,~b_0^{\text{exact}}=\frac{90}{\pi^2}\approx9.12$.}
\label{fig:Q4}
\end{figure}
\begin{figure}[t!]
\centering
\begin{subfigure}{0.48\textwidth}
    \includegraphics[width=\textwidth]{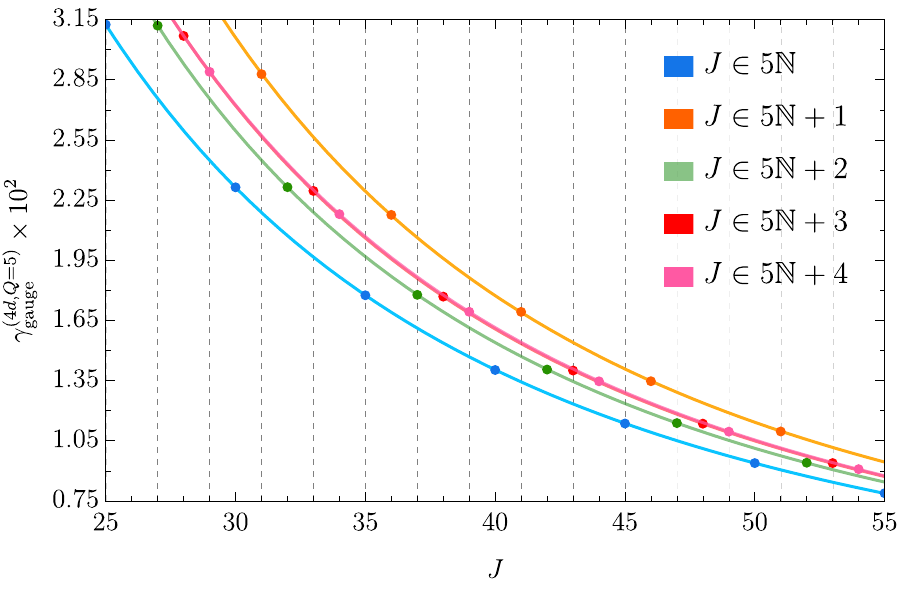}
    \caption{$Q=5$ $\Delta=4$ in $4d$}
    \label{fig:Q5D4}
\end{subfigure}
\hfill
\begin{subfigure}{0.48\textwidth}
    \includegraphics[width=\textwidth]{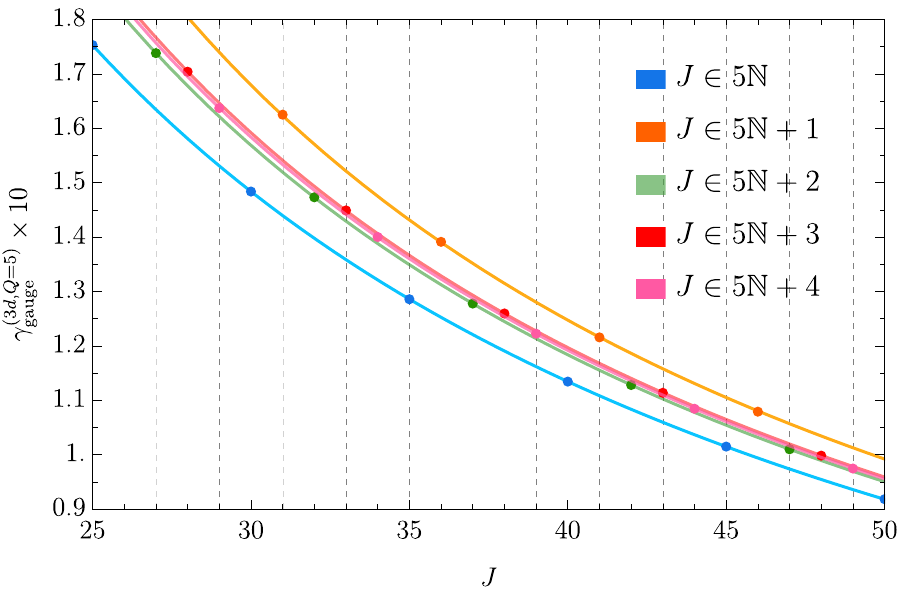}
    \caption{$Q=5$ $\Delta=\frac{5}{2}$ in $3d$}
    \label{fig:Q5D52}
\end{subfigure}
\caption{Anomalous dimensions corresponding to the lowest eigenvalue at each spin for 5-particle states due to U(1) gauge boson exchange. As in Fig.~\ref{fig:Q3} and \ref{fig:Q4}, the points correspond to the computed values while the solid lines are obtained by fitting. In this case the best-fit values for the five trajectories are
$a_0\approx (4.71,4.78,4.82,4.80,4.78)$ and $b_0\approx (27.7,28.1,28.1,27.8,28.3)$, in agreement with the expected values from \eqref{gammagauge}, $a_0^{\text{exact}}=\frac{45\sqrt{5(5+2\sqrt{5})}}{64}\approx4.84,~b_0^{\text{exact}}=\frac{1125}{4\pi^2}\approx28.5$.}
\label{fig:Q5}
\end{figure}
If we imagine turning off  gravity and focusing on gauge interactions only,  at large spin,  the $Q$ charged partons will evenly distribute around the center of AdS because of their mutual repulsion. This is exactly the configuration described in Sec.~\ref{sec:largeJBulk}, so we expect that the lowest-energy eigenvalue, defined as
\eqna{
\gamma^{(d, Q)}_{\mathrm{gauge}}\equiv \min  {}^{\quad r}_{J;Q}\!\bra{\Psi}  \CV_{\CJ}\ket{\Psi}_{J;Q}^{s}\, ,
}[gammaGaugeMatrix]
will behave at large spin as follows
\eqna{
J^{d-2}\gamma_{\mathrm{gauge}}^{(d, Q)} &\xrightarrow[\,J\to \infty\,]{}\frac{Q}{2} \sum_{n=1}^{Q-1}\left( \frac{Q}{2 \sin\! \frac{\pi n}{Q}}\right)^{d-2}\tilde{\gamma}_{\mathrm{gauge}}^{(d, Q=2)} \, ,\\ 
J^{d-2}\gamma_{\mathrm{gauge}}^{(d, Q=2)} &\xrightarrow[\,J\to \infty\,]{}\tilde{\gamma}_{\mathrm{gauge}}^{(d, Q=2)}=\frac{\pi ^{-\mathit{d}/2} \Gamma \mleft(\frac{\mathit{d}}{2}\mright) \Gamma (\Delta )^2}{2 (\mathit{d}-2)
   \Gamma \mleft(\Delta-\frac{\mathit{d}}{2} +1\mright)^2}\,.
}[gammagauge]
We can verify these expectations by numerically diagonalizing the matrix in~\eqref{gammaGaugeMatrix} for various values of the spin and by extracting the lowest eigenvalue at each $J$.  In Fig.~\ref{fig:Q3}, we show the result for $\gamma^{(d,Q=3)}_{\rm gauge}$, the anomalous dimension of the lightest three-particle primaries, in four and three dimensions for external dimensions respectively $\Delta=4$ and $\Delta=\frac{5}{2}$.\footnote{We checked that the same behavior persists for other values of $\Delta$.} Interestingly, the lowest-energy eigenvalue, as a function of $J$, falls on three separate trajectories depending on whether $J\in 3 \mathbb{N},  3\mathbb{N}+1,3 \mathbb{N}+2$.  A similar analysis can be done  for $Q=4$ (Fig.~\ref{fig:Q4}) and $Q=5$ (Fig.~\ref{fig:Q5}). The results are analogous for any $Q$: $\gamma_{\mathrm{gauge}}^{(d,Q)}$ is positive,  as expected for a repulsive interaction, and  its values for different $J$ are organized into $Q$ different trajectories and decay at large spin as prescribed by~\eqref{gammagauge}.

\begin{figure}[t!]
\centering
\begin{subfigure}{0.32\textwidth}
    \includegraphics[width=\textwidth]{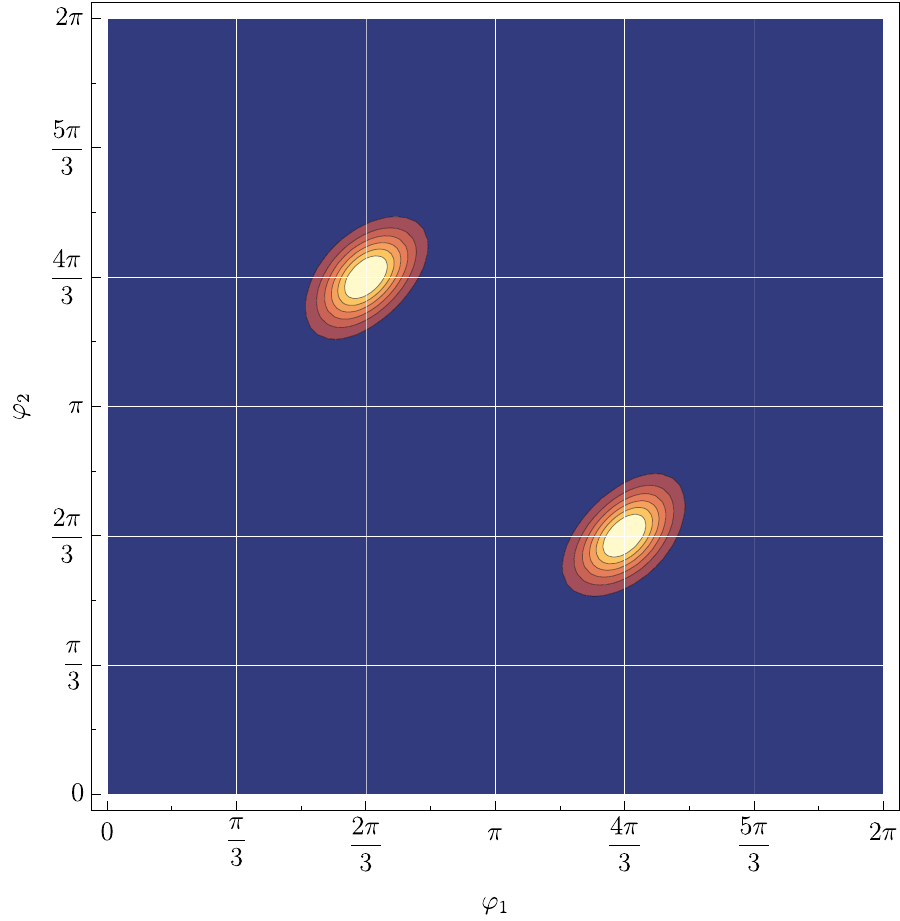}
    \caption{$J=99$}
\end{subfigure}
\hfill
\begin{subfigure}{0.32\textwidth}
    \includegraphics[width=\textwidth]{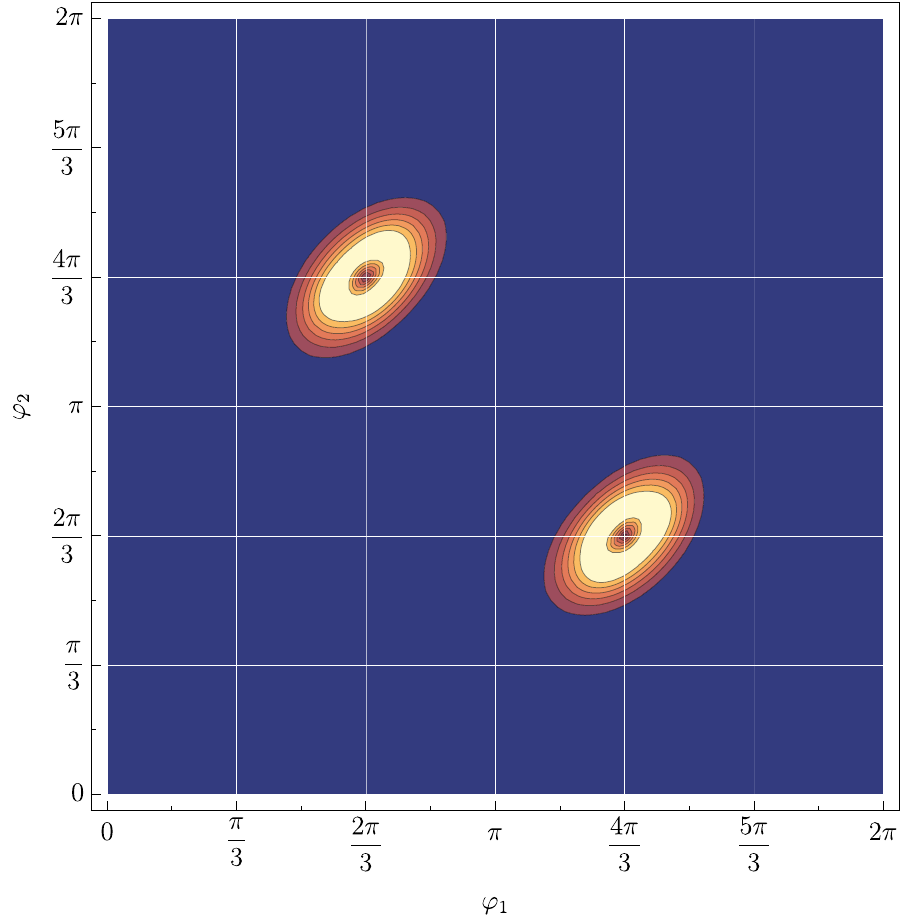}
    \caption{$J=98$}
\end{subfigure}
\hfill
\begin{subfigure}{0.32\textwidth}
    \includegraphics[width=\textwidth]{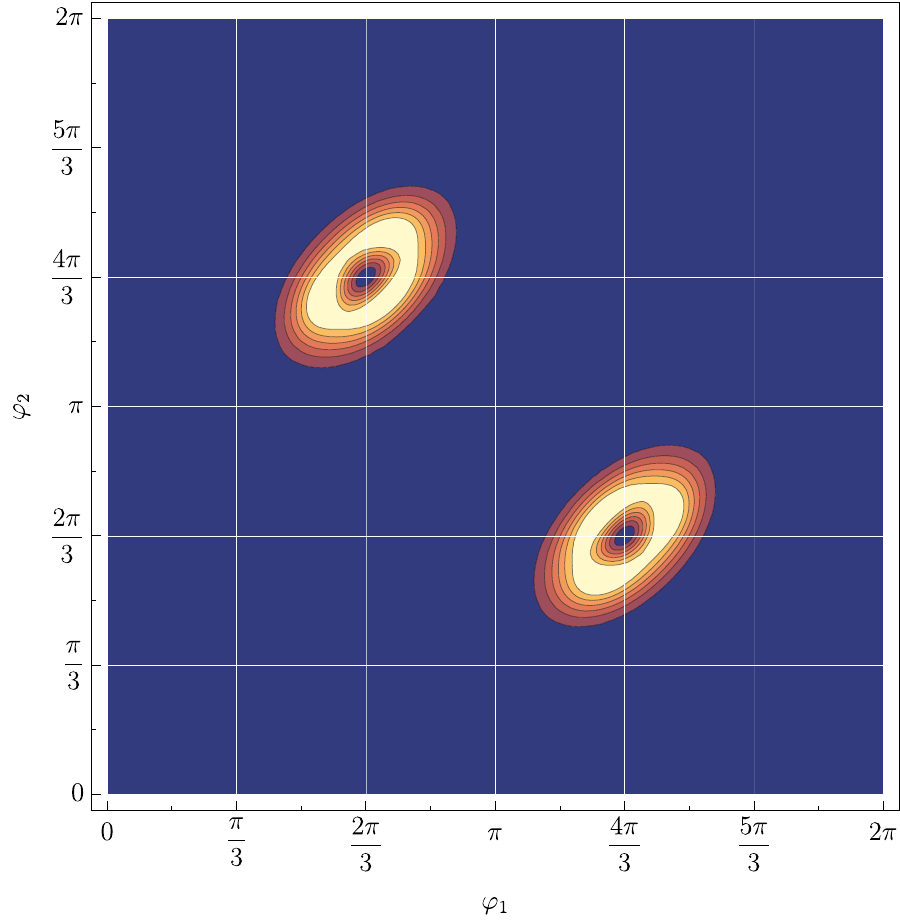}
    \caption{ $J=97$}
\end{subfigure}
\hfill
\begin{subfigure}{0.32\textwidth}
    \includegraphics[width=\textwidth]{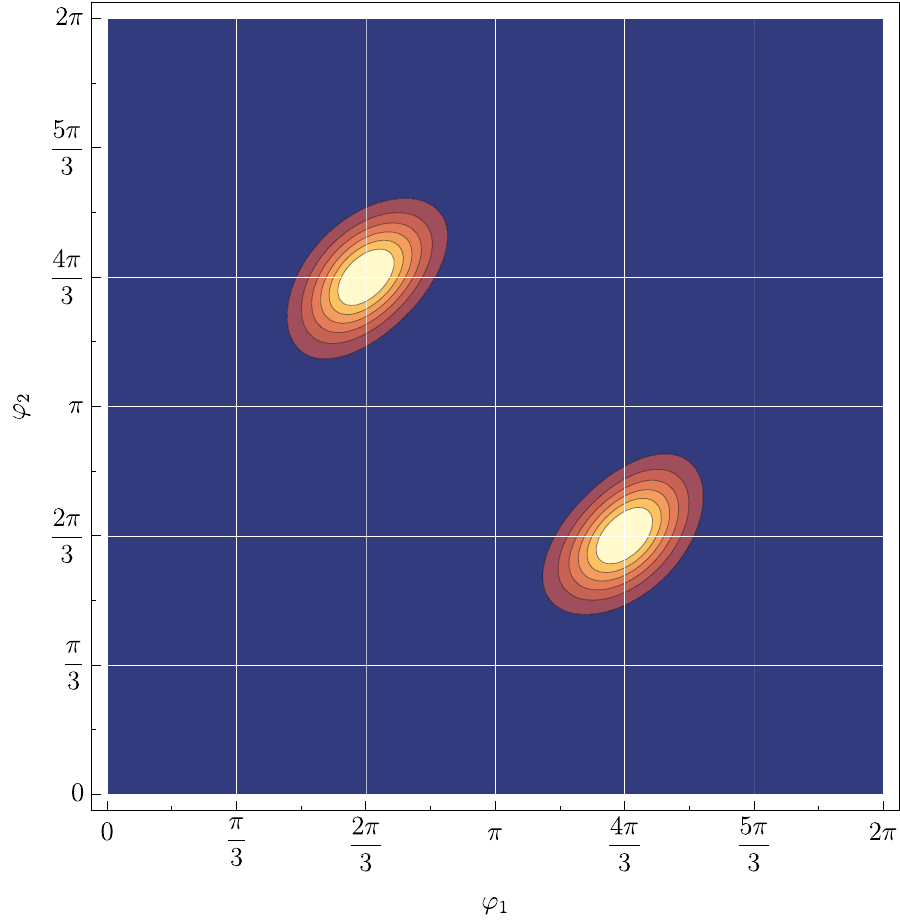}
    \caption{ $J=57$}
\end{subfigure}
\begin{subfigure}{0.32\textwidth}
    \includegraphics[width=\textwidth]{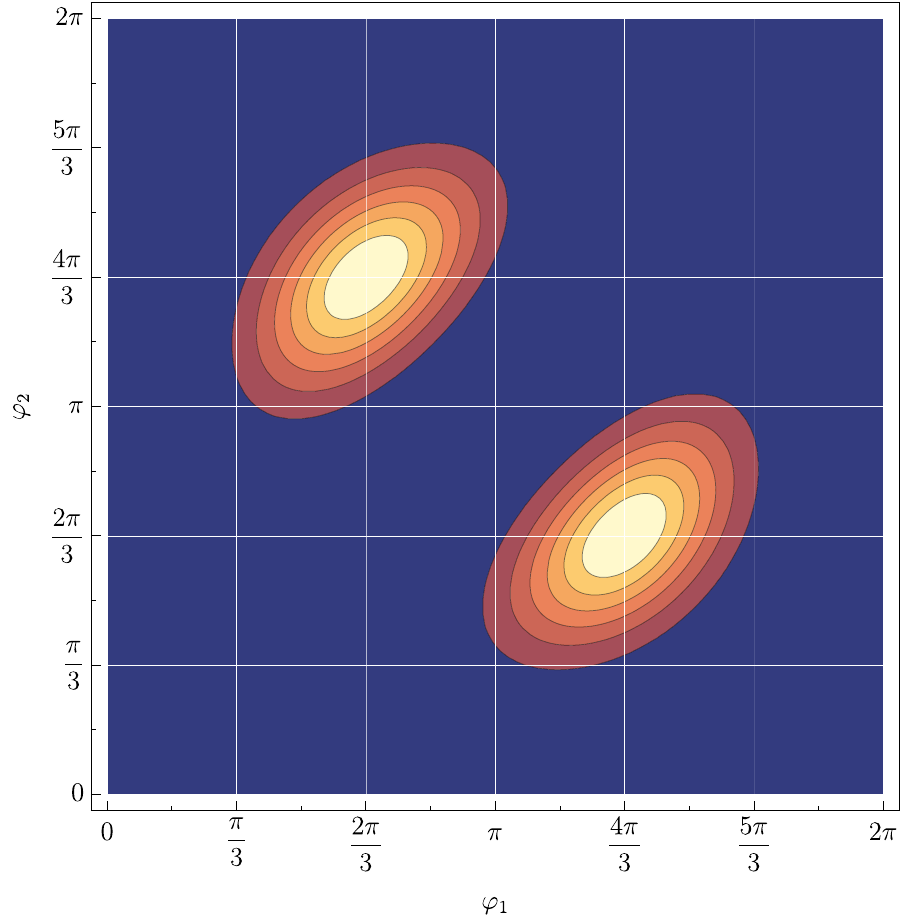}
    \caption{ $J=18$}
\end{subfigure} 
\begin{subfigure}{0.32\textwidth} \centering
\includegraphics[scale=1]{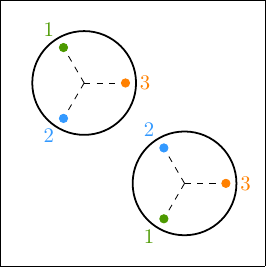}\hspace{-0.8cm} \vspace{0.6cm} 
  \caption{ Cartoon version} \label{fig:gaugeCartoon}
\end{subfigure}
\caption{$|\Psi_{\mathrm{min}}(\varphi_1, \varphi_2)|$ for the 3-particle state with $\Delta=\frac{5}{2}$ in $3d$ for different values of the spin $J$. We have chosen the first three values of $J=99,98,97$ to have one from each of our three trajectories, and have shown some smaller values of $J$ as well from the first trajectory to exhibit the $J$ dependence along a trajectory. In the last figure we have drawn a cartoon version of how the partons are distributed in AdS, each system corresponds to one of the two possible configurations.} 
\label{fig:Q3D52wavephi}
\end{figure}
\begin{figure}
\centering
\begin{subfigure}{0.4\textwidth}
    \includegraphics[width=\textwidth]{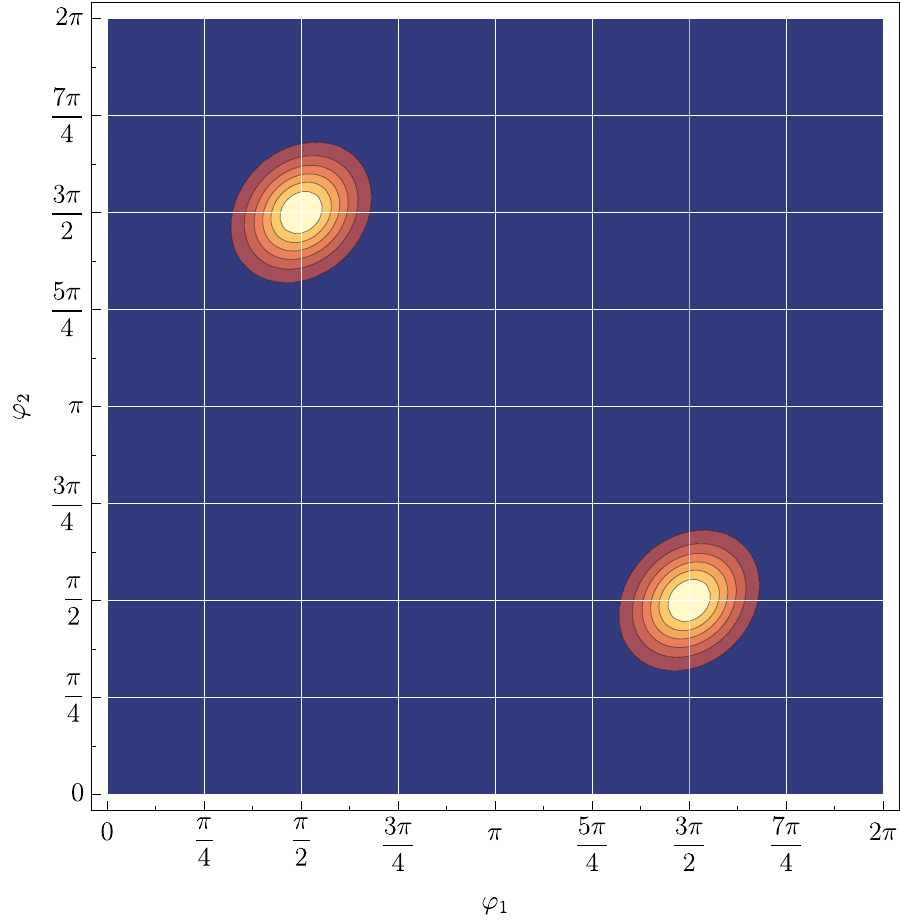}
    \caption{$\varphi_3=\pi$}
\end{subfigure}
\hfill
\begin{subfigure}{0.4\textwidth}
    \includegraphics[width=\textwidth]{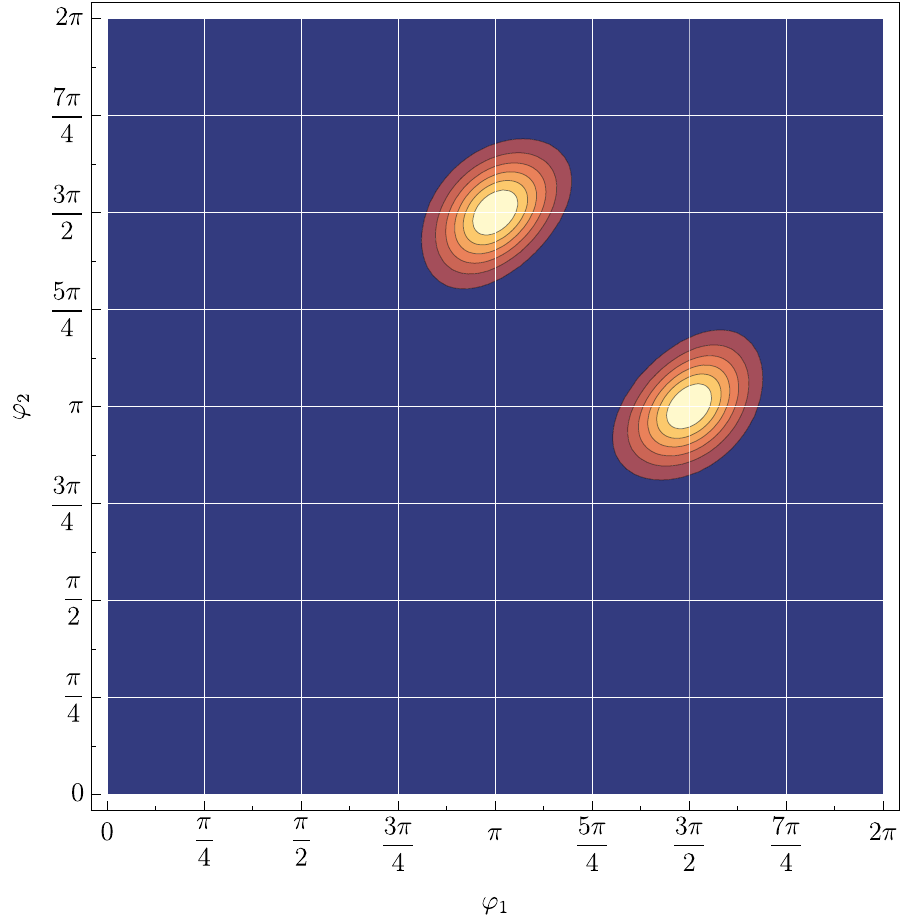}
    \caption{ $\varphi_3=\frac{\pi}{2}$}
\end{subfigure}
\caption{$Q=4$ $|\Psi_{\mathrm{min}}(\varphi_1,\varphi_2, \varphi_3,0)|$ for $\Delta=\frac{5}{2}$ in $3d$ and $J=60$.}
\label{fig:Q4D52phi}
\end{figure}
\begin{figure}
\centering
\begin{subfigure}{0.4\textwidth}
    \includegraphics[width=\textwidth]{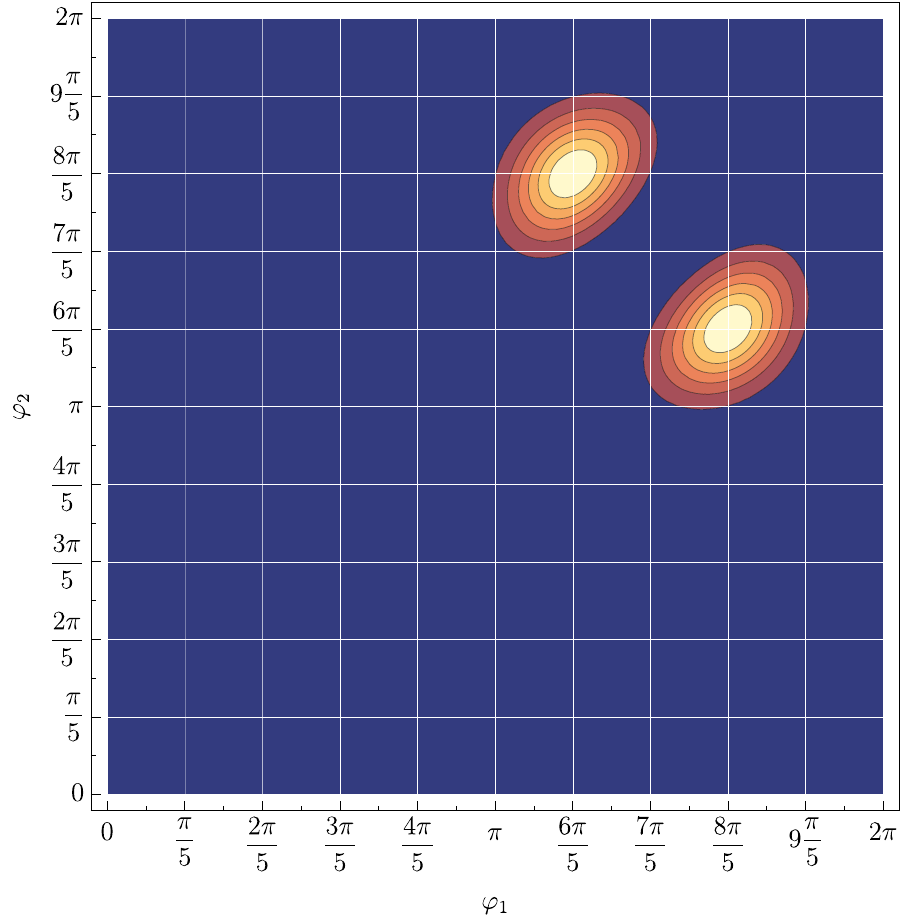}
    \caption{$(\varphi_3, \varphi_4)=\left(\frac{2\pi}{5},\frac{4\pi}{5}\right)$}
\end{subfigure}
\hfill
\begin{subfigure}{0.4\textwidth}
    \includegraphics[width=\textwidth]{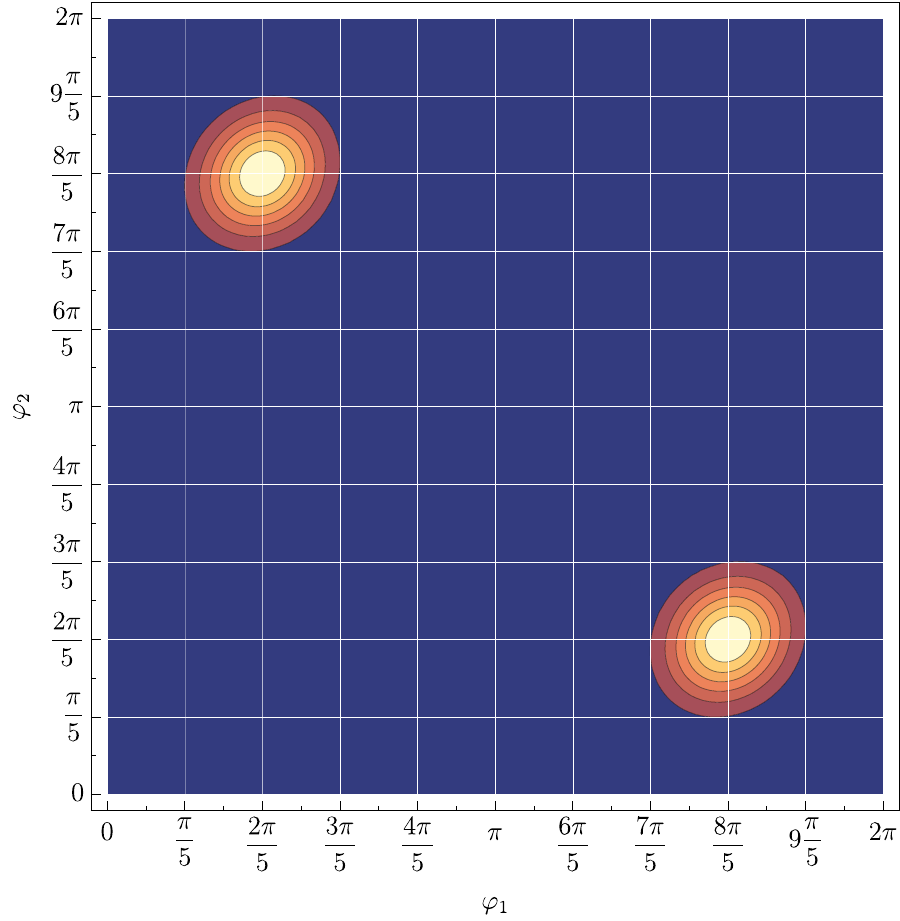}
    \caption{$(\varphi_3, \varphi_4)=\left(\frac{6\pi}{5},\frac{4\pi}{5}\right)$}
\end{subfigure}
\caption{$Q=5$ $|\Psi_{\mathrm{min}}(\varphi_1,\varphi_2, \varphi_3,\varphi_4,0)|$ for $\Delta=\frac{5}{2}$ in $3d$ and  $J=50$.}
\label{fig:Q5D52phi}
\end{figure}
\begin{figure}
 \includegraphics[width=\textwidth]{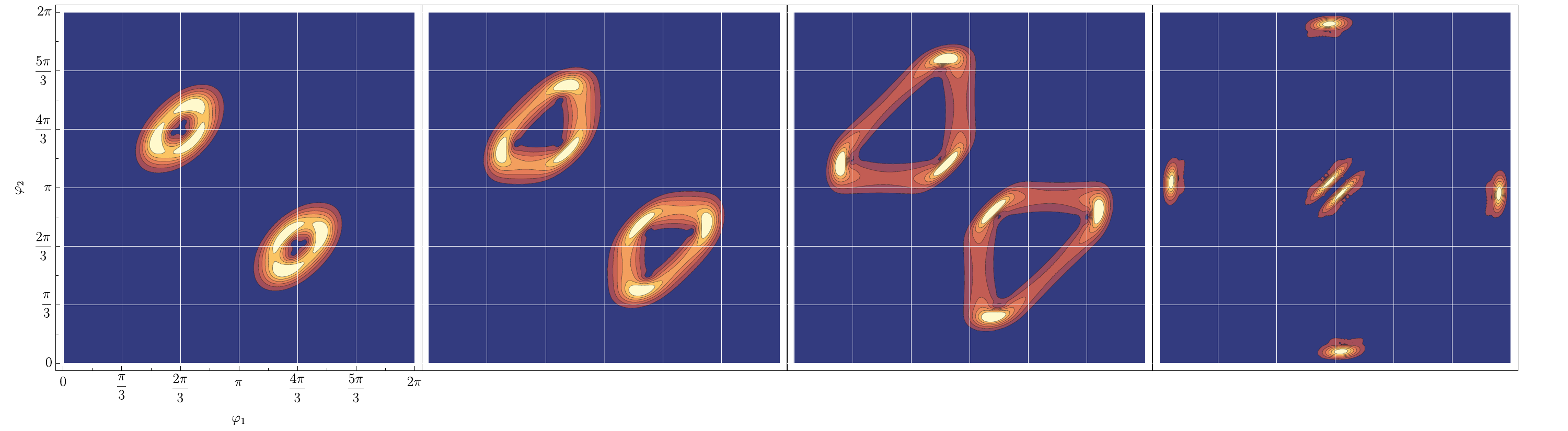}
 \caption{$|\Psi_r(\varphi_1, \varphi_2)|$ for the 3-particle state with $\Delta=\frac{5}{2}$ in $3d$ at $J=99$ for various eigenvalues. Recall that $r$ labels different primaries at a fixed spin , $r=1, \cdots, \CN(3,99)=17$. We plot, from left to right, $r=2,4,8,16$.}  \label{fig:q3D52spectra}
\end{figure}

The fact that our numeric eigenvalues reproduce the expression derived in Sec.~\ref{sec:largeJBulk} (see the captions of Fig.~\ref{fig:Q3}, \ref{fig:Q4}, \ref{fig:Q5} for the detailed numeric comparison)  assuming that the $Q$ partons were equally distributed around the equator  is a check of this assumption. To substantiate it even further and try to visualize this configuration, we can plot the eigenvectors in real space.  Let us define the diagonal matrix
\eqna{
D_{qp}=U^{\dag}_{qr} {}^{\quad r}_{J;Q}\!\bra{\Psi}  \CV_{\CJ}\ket{\Psi}^{s}_{J;Q} U_{sp}\, ,
}[]
where $U$ is the unitary matrix that diagonalizes $\CV_\CJ$ and $r, \, s$ label the $\CN(Q, J)$ degenerate primaries. 
Then the eigenvectors can be written
\eqna{
\Psi_r&=U^{\dag}_{rs}\ket{\Psi}_{J;Q}^{s} \, .
}[Psir]
Suppose we order them in ascending order based on the value of the corresponding eigenvalues (with the lowest one being $\gamma_{\mathrm{gauge}}^{(d,Q)}$). To project to real space, we evaluate the first-quantization wavefunctions $\< \phi(X_1) \dots \phi(X_Q) |\Psi\>_{J;Q}$.  In practice, this amounts to  trading each $\ket{\ell_1 \cdots \ell_Q}$ in $\ket{\Psi}_{J;Q}$ for the product of the corresponding wavefunctions $\phi_{0 \ell_i\ell_i}(X)$, as  in~\eqref{eq:BulkFieldDecomp}, evaluated at $t=0$. Moreover, since the particles are dominantly near the equator, we will only evaluate the wavefunctions at  $\theta=0$ ($\theta_1=\theta_2=0$) in $3d$ ($4d$).  Then,
\eqna{
\Psi_r (\varphi_i, \chi_i)=U^{\dag}_{rs}\sum_{\ell_1+\cdots+\ell_Q=J} \tilde{C}^s_{\ell_1 \cdots \ell_Q} \prod_{i=1}^Q \frac{1}{N_{\Delta,0, \ell_i}} e^{i \ell_i \phi_i} \frac{\tanh^{\ell_1} \chi_i}{\cosh^{\Delta} \chi_i}\, , 
}[PsiRealSpace]
with $\Psi_{\mathrm{min}}(\varphi_i, \chi_i)$ defined as the eigenvector corresponding to $\gamma_{\rm{gauge}}^{(d,Q)}$.  We can use the rotational symmetry to set $\varphi_Q=0$ and, suppressing for a second the $\chi$ dependence, we can study $|\Psi_{\rm min} (\varphi_1, \cdots, \varphi_{Q-1})|$.  For example, at $Q=3$ --- see Fig.~\ref{fig:Q3D52wavephi} for $\Delta=\frac{5}{2}$ in $3d$\footnote{For other $\Delta$'s and in $4d$ the plots of $|\Psi_{\rm min}(\varphi_1, \varphi_2)|$ are similar so we have not reported them.} --- $|\Psi_{\mathrm{min}}(\varphi_1, \varphi_2)|$ has a Gaussian behavior: it is more localized (yellow region in the plot) around two symmetric points $(\varphi_1, \varphi_2)=\left(\frac{2\pi}{3},\frac{4\pi}{3}\right)$ and $(\varphi_1, \varphi_2)=\left(\frac{4\pi}{3},\frac{2\pi}{3}\right)$ and it decays away from these values.  Since $\varphi_3=0$, these are exactly the two possible configurations we expect in AdS,  with the partons localized at $2\pi n/Q$ angles, as represented in the cartoon in   Fig.~\ref{fig:gaugeCartoon}. 
The same behavior persists at higher $Q$, as one can verify by looking at   Fig.~\ref{fig:Q4D52phi} (for $Q=4$) and Fig.~\ref{fig:Q5D52phi} (for $Q=5$), where for illustrative purposes we have restricted to sections with  fixed values of  $\varphi_3, \cdots \varphi_{Q-1}$.  
For different values of the spin the positions of the peaks of $|\Psi_{\rm min}|$ stay the same, but the shape itself changes. In Fig.~\ref{fig:Q3D52wavephi}, as an example, we show the absolute value of the wavefunction at $Q=3$ for various $J$. Notice in particular  how it varies when $J$ takes values in the three different trajectories ($J=99,98,97$ in the figure). This is likely  related to the fact that when $J$ is not a multiple of 3, the spin can not be exactly evenly distributed among the three different partons, so one will get eigenvectors that are linear combinations of different ways of distributing the spin.  Finally, we would like to remark that, as shown in  Fig.~\ref{fig:q3D52spectra}, for larger eigenvalues the distribution of the partons changes significantly.

\begin{figure}
\centering
\begin{subfigure}{0.38\textwidth}
    \includegraphics[width=\textwidth]{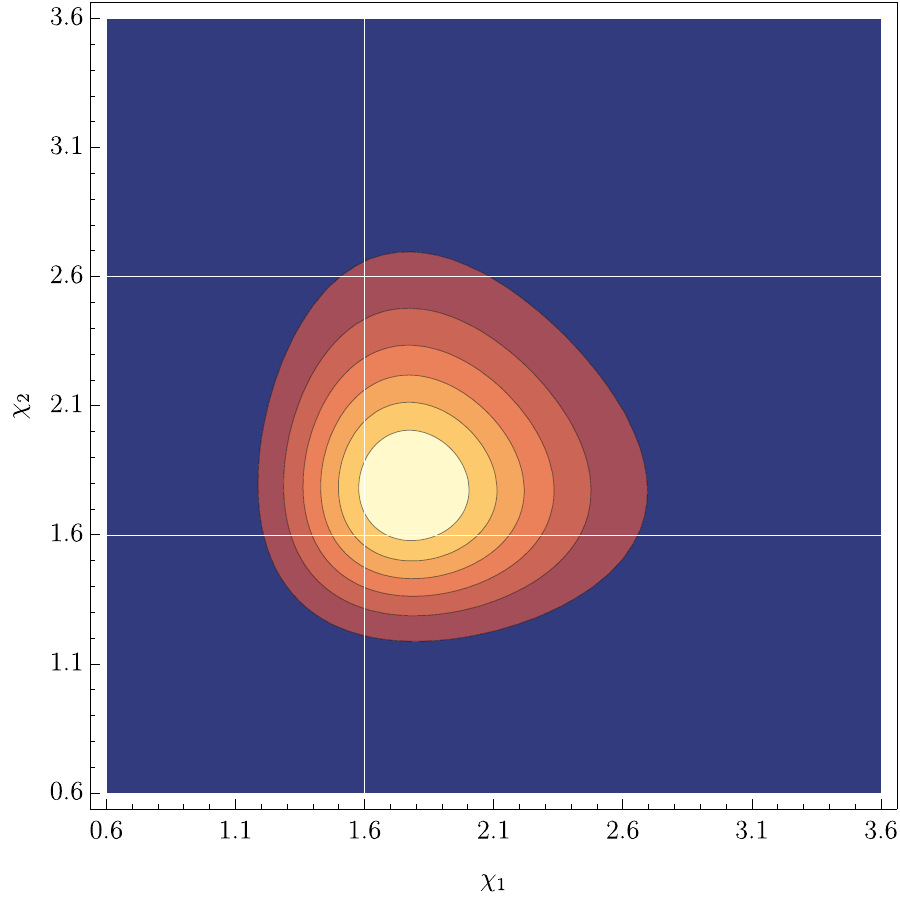}
    \caption{$\Delta=4$ and $\chi_3\sim 1.78$}
\end{subfigure}
\hfill
\begin{subfigure}{0.38\textwidth}
    \includegraphics[width=\textwidth]{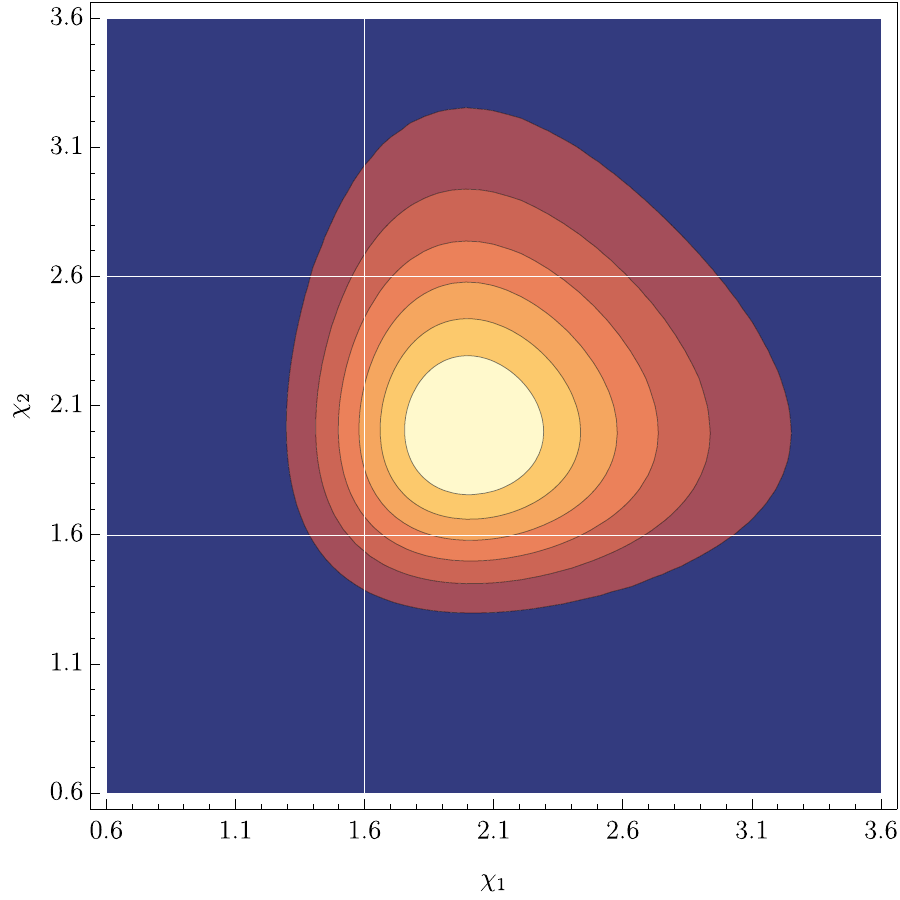}
    \caption{$\Delta=\frac{5}{2}$ and $\chi_3\sim 2$}
\end{subfigure}
\caption{$|\Psi_{\mathrm{min}}(\chi_1, \chi_2, \chi_3)|$ at $\varphi_1=\frac{2\pi}{3}$ and $\varphi_2=\frac{4\pi}{3}$ for the 3-particle state with $\Delta=4$ in $4d$ and $\Delta=\frac{5}{2}$ in $3d$ for $J=99$. For plotting purposes we have set $\chi_3$ to its maximum.   } 
\label{fig:Q3wavechi}
\end{figure}

To conclude,  let us discuss the $\chi$- dependence of $\Psi_{\mathrm{min}}$, namely how close or far away the partons are with the respect to the center of AdS ($\chi=0$) in the radial direction.  For concreteness, let us focus on $Q=3$ and set $\varphi_1$ and $\varphi_2$ to one of the two maxima discussed above.  In Fig.~\ref{fig:Q3wavechi}, we present these results for fixed total spin $J=99$ at $\Delta=4$ and $\Delta=\frac{5}{2}$.  Once again, the shape of $|\Psi_{\mathrm{min}}(\chi_1, \chi_2, \chi_3)|$ is approximately Gaussian: it is centered around $\chi^*=\mathrm{arcsinh} \left( \sqrt{\frac{J}{3\Delta}}\right)$ and it has standard deviation $\sigma \sim \frac{1}{2\Delta}$, as expected based on equations (\ref{eq:ChiVsLPart}) and  (\ref{eq:WvFcnGaussianApprox}).  

\subsection{Gravity} \label{Subsec: gravity}
\begin{figure}
\centering
\begin{subfigure}{0.32\textwidth}
    \includegraphics[width=\textwidth]{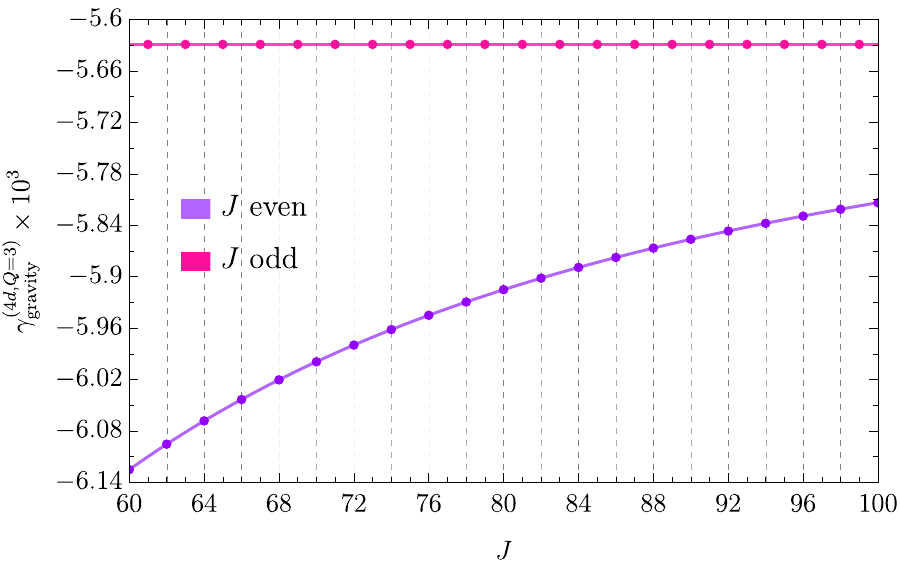}
    \caption{$Q=3$}
\end{subfigure}
\hfill
\begin{subfigure}{0.32\textwidth}
    \includegraphics[width=\textwidth]{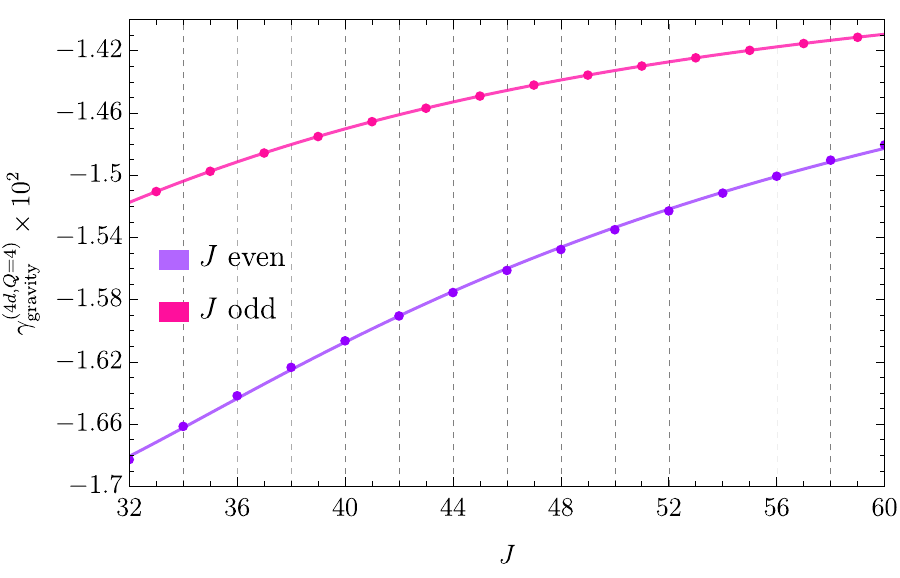}
    \caption{$Q=4$}
\end{subfigure}
\hfill
\begin{subfigure}{0.32\textwidth}
    \includegraphics[width=\textwidth]{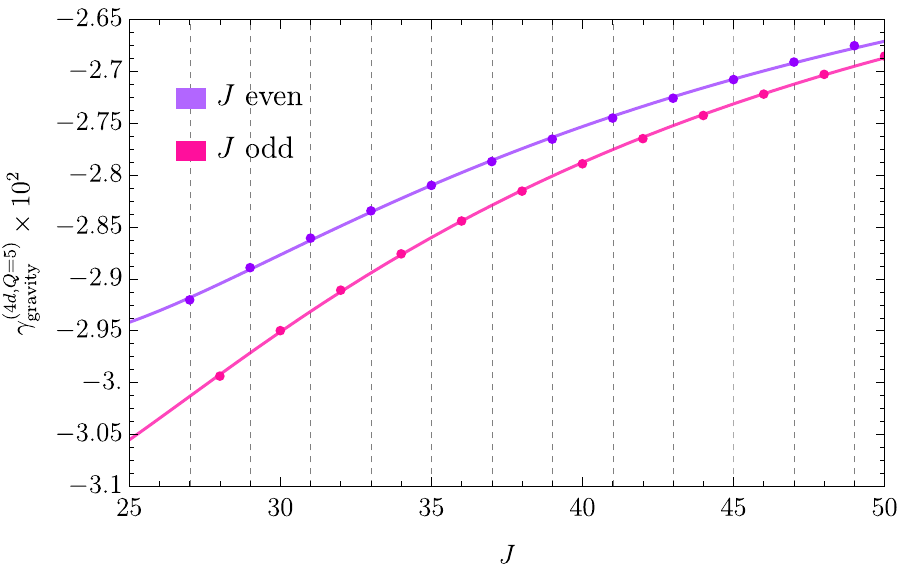}
    \caption{$Q=5$}
\end{subfigure}
\caption{Anomalous dimensions corresponding to the lowest eigenvalue at each spin for 3,4,5-particle states (left to right) due to graviton exchange for $\Delta=2$ in $4d$.  The points correspond to the computed values while the solid lines are obtained by fitting to the curve $a_0+\frac{a_1}{J^{d-2}}+\frac{a_2}{J^{d-1}}$.} 
\label{fig:QD2gravity}
\end{figure}

\begin{figure}
\centering
\begin{subfigure}{0.32\textwidth}
    \includegraphics[width=\textwidth]{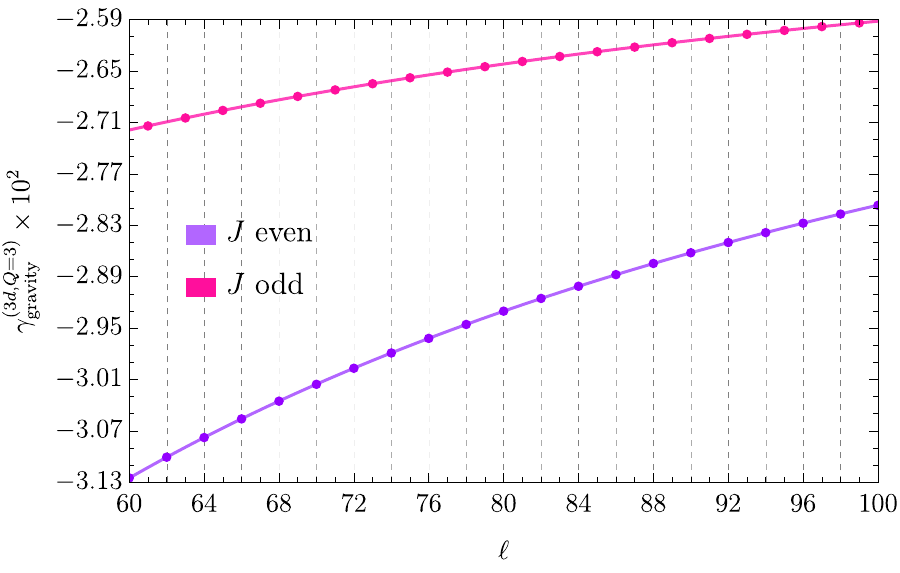}
    \caption{$Q=3$}
\end{subfigure}
\hfill
\begin{subfigure}{0.32\textwidth}
    \includegraphics[width=\textwidth]{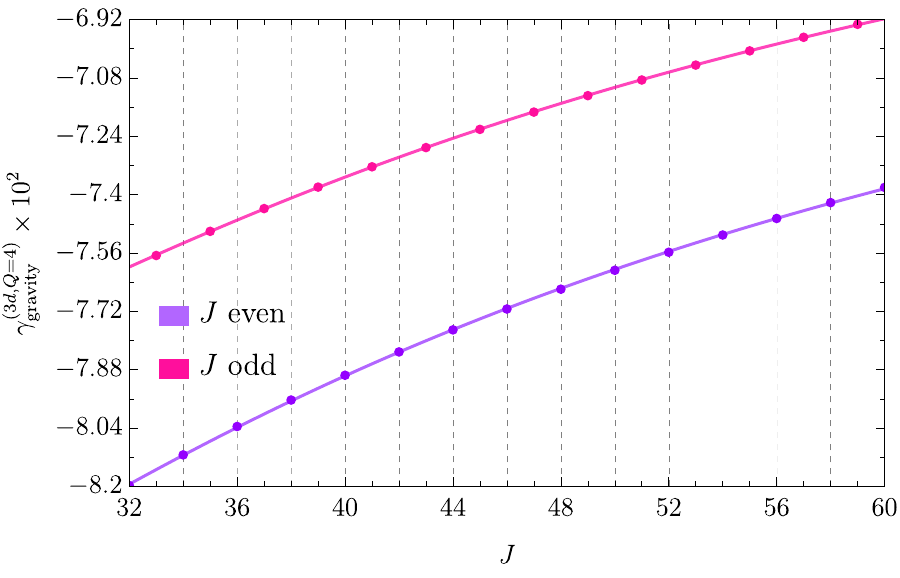}
    \caption{$Q=4$}
\end{subfigure}
\hfill
\begin{subfigure}{0.32\textwidth}
   \includegraphics[width=\textwidth]{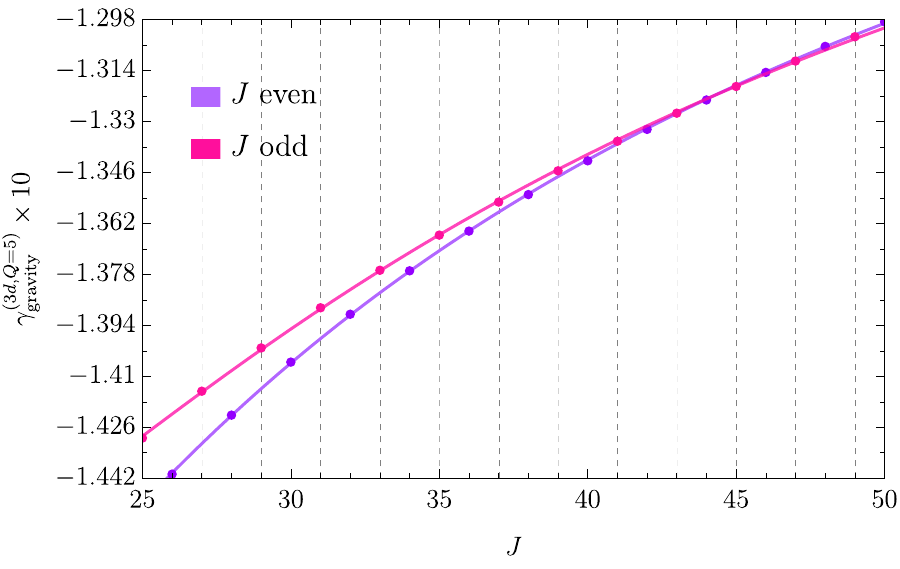}
    \caption{$Q=5$}
\end{subfigure}
\caption{Anomalous dimensions corresponding to the lowest eigenvalue at each spin for 3,4,5-particle states (left to right) due to graviton exchange for $\Delta=\frac{3}{2}$ in $3d$.  The points correspond to the computed values while the solid lines are obtained by fitting to the curve $a_0+\frac{a_1}{J^{d-2}}+\frac{a_2}{J^{d-1}}$.} 
\label{fig:QD32gravity}
\end{figure}

The analysis of the gravity contribution proceeds similarly to the gauge one, but the picture in AdS changes substantially.  Since gravity is attractive,  it is energetically favorable for the $Q$ partons to cluster together.  At large spin, they  reach a stable configuration in which we have $(Q-n)$ partons forming a `blob' on one side of AdS and another $n$-parton `blob' on the opposite side,  with these two `blobs' rotating around the center.\footnote{At sufficiently large $Q$, we expect it could be possible to form stable configurations with more than two blobs, as we discuss in Sec.~\ref{sec:LargeQ}.}  At large spin, we expect  the smallest anomalous dimension, defined as 
\eqna{
\gamma_{\mathrm{gravity}}^{(d, Q)} \equiv  \min  {}^{\quad r}_{J;Q}\!\bra{\Psi}  \CV_{\CT}\ket{\Psi}_{J;Q}^{s}\, ,
}[gammaMatrixGrav] 
to be
\eqna{
\gamma_{\mathrm{gravity}}^{(d, Q)} &\xrightarrow[\,J\to \infty\,]{} a_0^{(\rm{gravity})}(d, Q)+\frac{a_1^{(\rm{gravity})}(d, Q)}{J^{d-2}}+ \cdots\, .
}[gammaGravGen]
The leading constant term corresponds to the energy of  the configuration of $(Q-n)$ partons with the minimum twist
\eqna{
a_0^{\rm{gravity}}=\min_{n, \ell} \gamma^{(d, Q-n)}_{\mathrm{gravity}}(\ell)\, ,
}[]
where the twist is a function of the internal spin $\ell$ of the $(Q-n)$-`blob'.  To determine the optimal values of $n$ and $\ell$, we can numerically diagonalize~\eqref{gammaMatrixGrav} and extract the lowest eigenvalue.  
\begin{figure}
\centering
\begin{subfigure}{0.42\textwidth}
    \includegraphics[width=\textwidth]{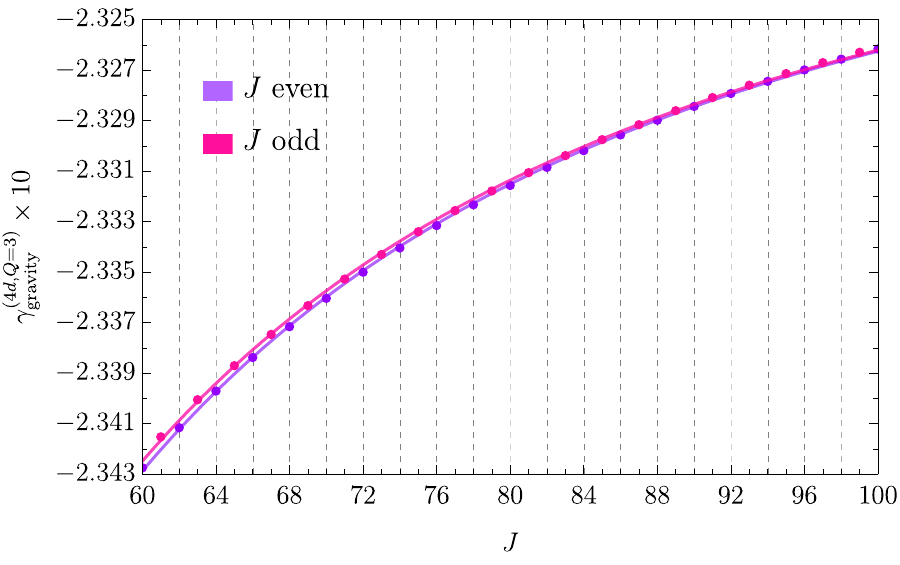}
    \caption{$Q=3$ $\Delta=4$ in $4d$}
\end{subfigure}
\hfill
\begin{subfigure}{0.42\textwidth}
    \includegraphics[width=\textwidth]{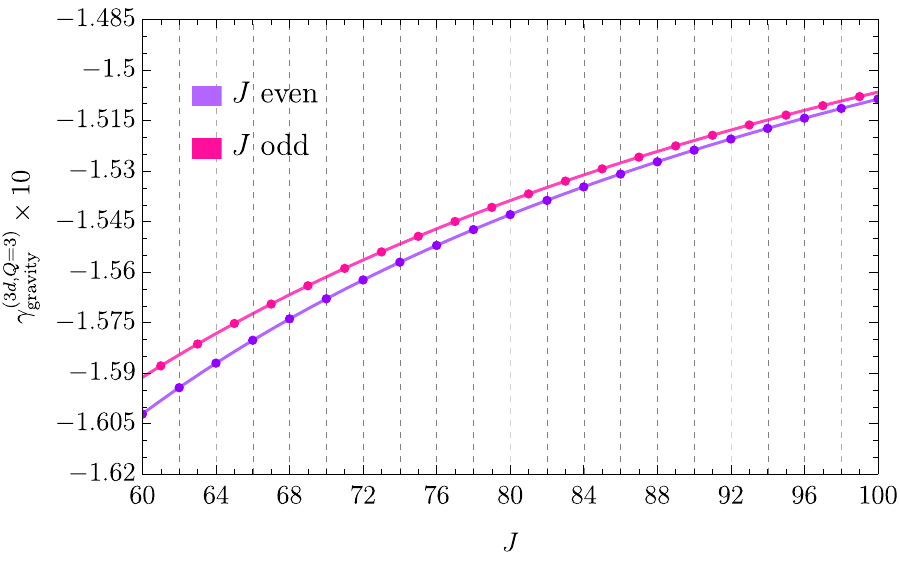}
    \caption{$Q=3$ $\Delta=\frac{5}{2}$ in $3d$}
\end{subfigure}
\caption{Anomalous dimensions corresponding to the lowest eigenvalue at each spin for 3-particle states due to graviton exchange. The points correspond to the computed values while the solid lines are obtained by fitting to the curve  $a_0+\frac{a_1}{J^{d-2}}+\frac{a_2}{J^{d-1}}$. The best-fit values for the leading order constant coefficients $a_0^{3d},a_0^{4d}$ for the two different trajectories are $a_0^{3d}\approx(-0.138,-0.137)$ and $a_0^{4d}\approx(-0.2316,-0.2316)$, in agreement with  the expected value from \eqref{gammagra}  $a_0^{3d,\text{exact}}=-\frac{1125}{8192}\approx-0.1373$ and $a_0^{4d,\text{exact}}=-\frac{16}{7\pi^2}\approx-0.2316$.}
\label{fig:Q3gravity}
\end{figure}
\begin{figure}
\centering
\begin{subfigure}{0.42\textwidth}
    \includegraphics[width=\textwidth]{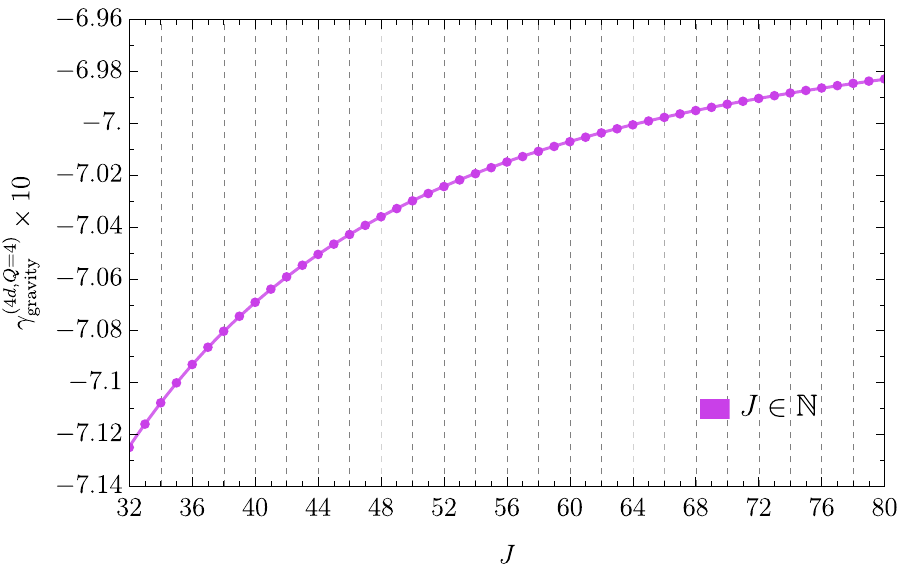}
    \caption{$Q=4$ $\Delta=4$ in $4d$}\label{fig:q4gravityA}
\end{subfigure}
\hfill
\begin{subfigure}{0.42\textwidth}
    \includegraphics[width=\textwidth]{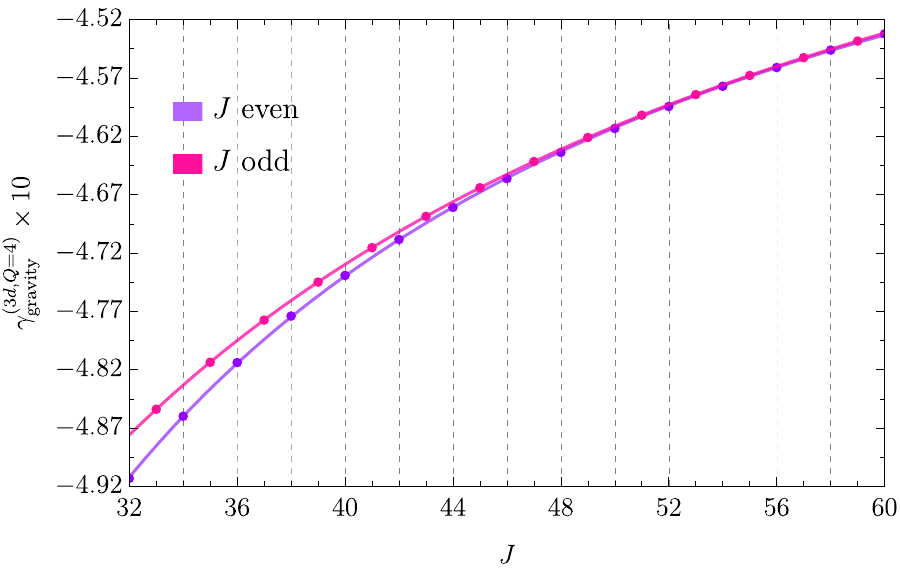}
    \caption{$Q=4$ $\Delta=\frac{5}{2}$ in $3d$}
\end{subfigure}
\caption{Anomalous dimensions corresponding to the lowest eigenvalue at each spin for 4-particle states due to graviton exchange. The points correspond to the computed values while the solid lines are obtained by fitting as described in the caption for Fig.~\ref{fig:Q3gravity}.   
The best-fit values for  $a_0^{3d},a_0^{4d}$ for the two different trajectories are $a_0^{3d}\approx (-0.4138,-0.4126)$ and $a_0^{4d}\approx(-0.6948,-0.6948)$, in agreement with the expected value from \eqref{gammagra}, $a_0^{3d,\text{exact}}=-\frac{3375}{8192}\approx-0.412,~a_0^{4d,\text{exact}}=-\frac{48}{7\pi^2}\approx-0.6947$.}

\label{fig:Q4gravity}
\end{figure}
\begin{figure}
\centering
\begin{subfigure}{0.42\textwidth}
    \includegraphics[width=\textwidth]{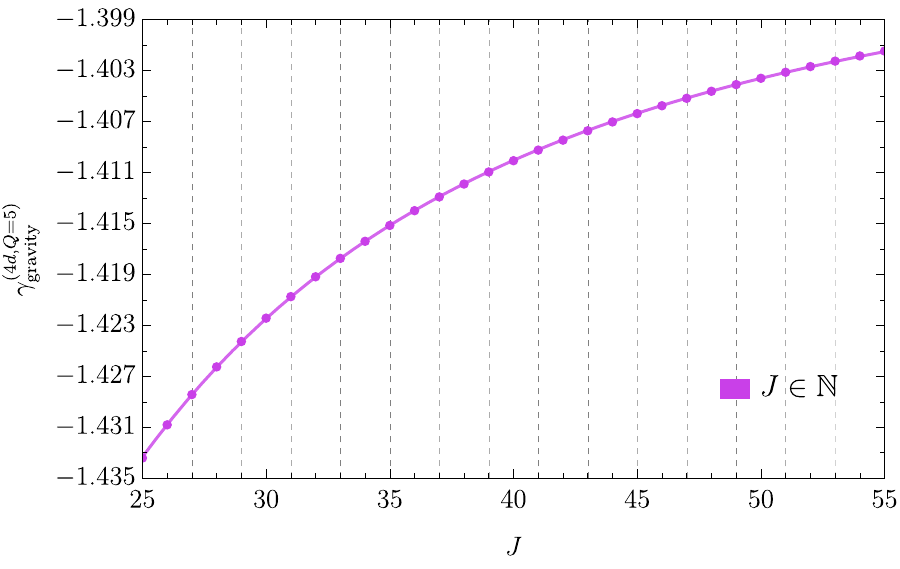}
    \caption{$Q=5$ $\Delta=4$ in $4d$}\label{fig:q5gravityA}
\end{subfigure}
\hfill
\begin{subfigure}{0.42\textwidth}
    \includegraphics[width=\textwidth]{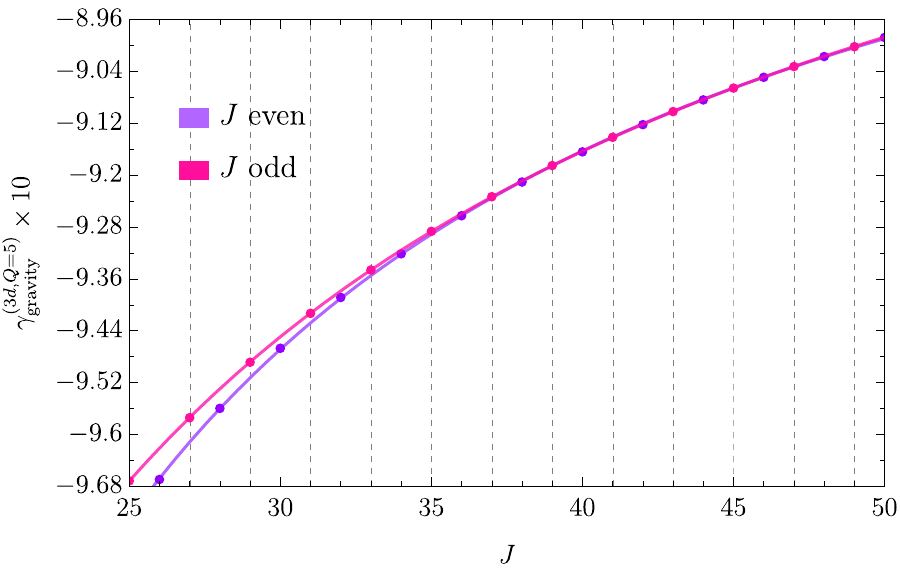}
    \caption{$Q=5$ $\Delta=\frac{5}{2}$ in $3d$}
\end{subfigure}
\caption{Anomalous dimensions corresponding to the lowest eigenvalue at each spin for 5-particle states due to graviton exchange. The points correspond to the computed values while the fits are performed as in Fig.~\ref{fig:Q3gravity} and \ref{fig:Q4gravity}. 
The best-fit values  
for $a_0^{3d},a_0^{4d}$ 
for the two different trajectories are $a_0^{3d}\approx (-0.826,-0.826)$ and $a_0^{4d}\approx(-1.3899,-1.3899)$, in agreement with the expected value from \eqref{gammagra}, $a_0^{3d,\text{exact}}=-\frac{3375}{4096}\approx-0.824,~a_0^{4d,\text{exact}}=-\frac{96}{7\pi^2}\approx-1.3896$.}
\label{fig:Q5gravity}
\end{figure}
In Fig.~\ref{fig:Q3gravity}-\ref{fig:Q5gravity}  we report these results for various external dimensions $\Delta$ in $3d$ and $4d$ at fixed $Q=3,4,5$.  By plotting the lowest-energy eigenvalues, first of all, we notice that  they form two trajectories, one for even spins and one for odd spins. And these two trajectories get closer and closer as $\Delta$ increases,  until they become practically indistinguishable --- see for instance Fig.~\ref{fig:q4gravityA} and~\ref{fig:q5gravityA} at $\Delta=4$. More importantly, we learn that the optimal configuration is the one with a $(Q-1)$-parton blob and an isolated parton ($Q=1$), therefore
\eqna{
a_0^{\rm{gravity}}=\min_{\ell} \gamma^{(d, Q-1)}_{\mathrm{gravity}}(\ell)\, .
}[]
The value of $\ell$ minimizing the twist depends on the dimension $\Delta$ of the external operator.  A more complete analysis reveals that it actually depends on whether $\Delta\lessgtr \frac{d}{2}$, where $\Delta=\frac{d}{2}$ is the point where the bulk scalar mass-squared saturates the Breitenlohner-Freedman (BF) bound~\cite{Breitenlohner:1982bm}.  
For the examples in Fig.~\ref{fig:Q3gravity}-\ref{fig:Q5gravity} and Fig.~\ref{fig:QD2gravity}-\ref{fig:QD32gravity} respectively, we find 
\twoseqn{
a_0^{\rm{gravity}}=\gamma^{(d, Q-1)}_{\mathrm{gravity}}(\ell=0)\, , \phantom{-1}\qquad \text{ if }\, \Delta>\frac{d}{2}\, ,
}[a0dGr]
{
a_0^{\rm{gravity}}=\gamma^{(d, Q-1)}_{\mathrm{gravity}}(\ell=Q-1)\, , \qquad \text{if }\, \Delta=\frac{d}{2}\, .
}[a0d2][gammagra]
We defer a detailed analysis of why these are the lowest-twist configurations, and how the $\Delta$-dependence emerges, to Sec.~\ref{Sec:smallSpinContact}, where we derive explicit expressions for  $\gamma^{(d, Q)}(\ell)$ at low values of $\ell$.

Going back to our original expression in~\eqref{gammaGravGen} for the large spin behavior of $\gamma^{(d, Q)}_{\mathrm{gravity}}$, we still have to comment on the origin of the sub-leading corrections in $1/J$.  We believe that these are due to a  combination of different factors. The most obvious contributions are corrections coming from tree-level graviton exchange between the $(Q-1)$ partons and the isolated parton at the opposite side of AdS. However, there is also a larger $1/J$ correction coming from the fact that the blob only settles down to a specific $Q-1$ primary state asymptotically at large $J$. 
This effect will hopefully be clearer when we study the combined effect of gauge and graviton exchanges in Sec.~\ref{sec:PhaseDiagram}.

\begin{figure}
\centering
\begin{subfigure}{0.32\textwidth}
    \includegraphics[width=\textwidth]{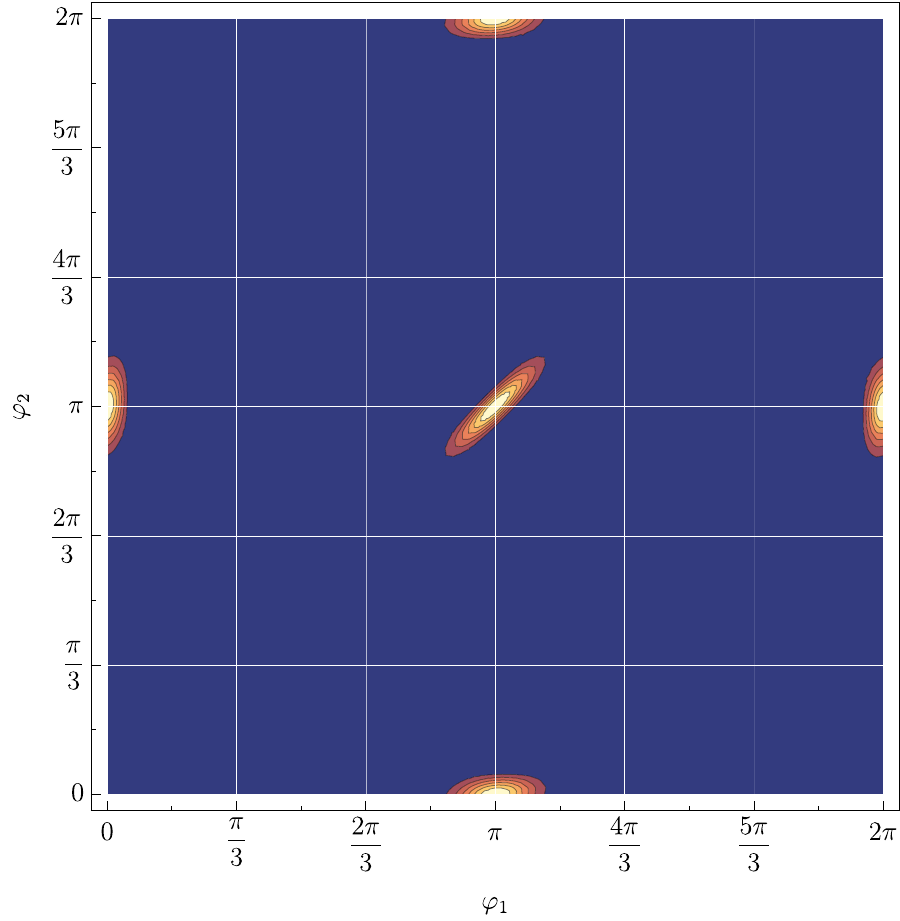}
    \caption{$J=99$}
\end{subfigure}
\hfill
\begin{subfigure}{0.32\textwidth}
    \includegraphics[width=\textwidth]{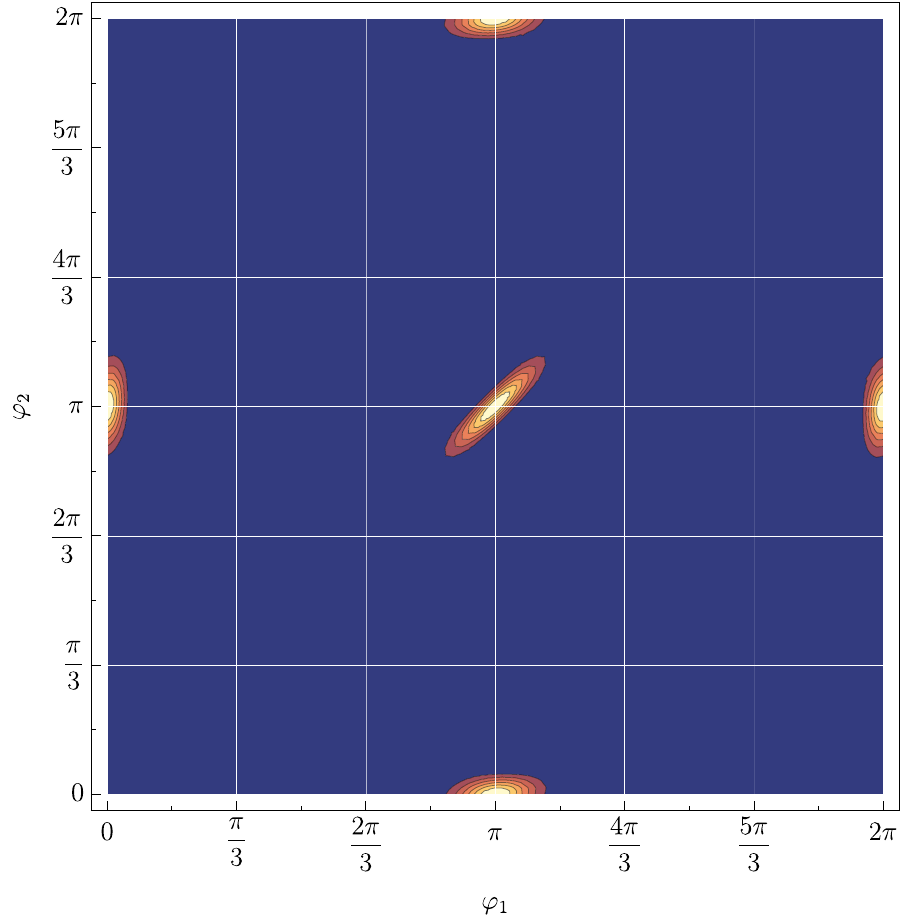}
    \caption{$J=98$}
\end{subfigure}
\hfill
\begin{subfigure}{0.32\textwidth}
    \includegraphics[width=\textwidth]{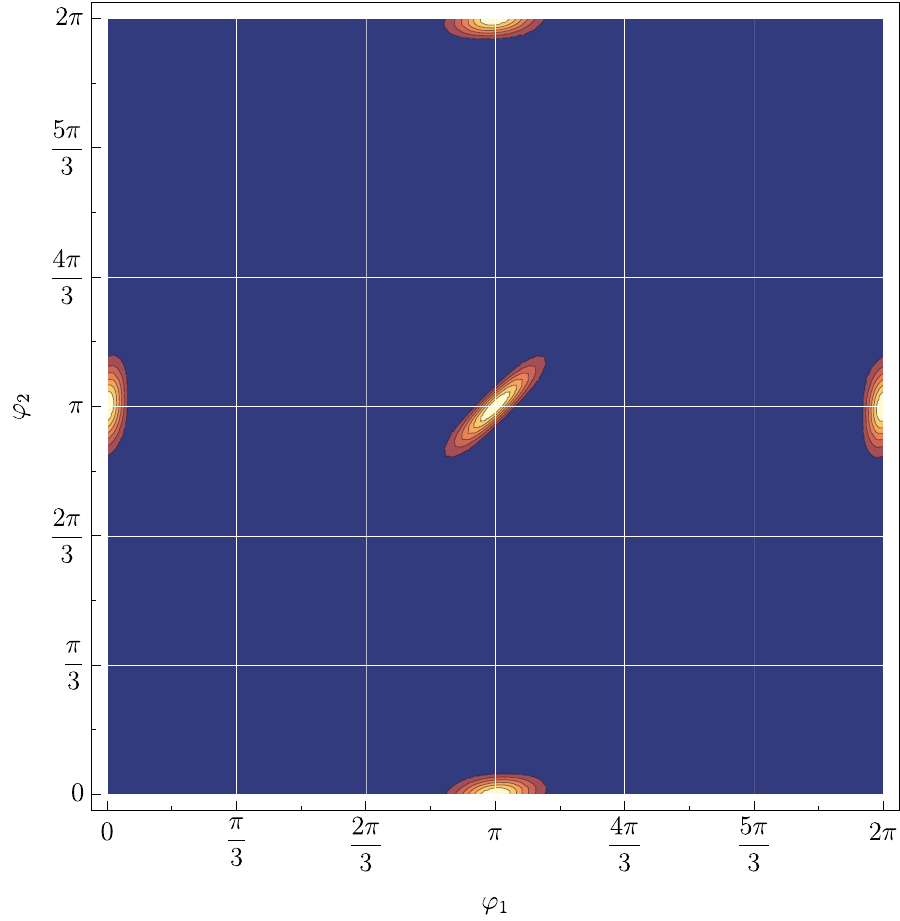}
    \caption{ $J=97$}
\end{subfigure}
\hfill
\begin{subfigure}{0.32\textwidth}
    \includegraphics[width=\textwidth]{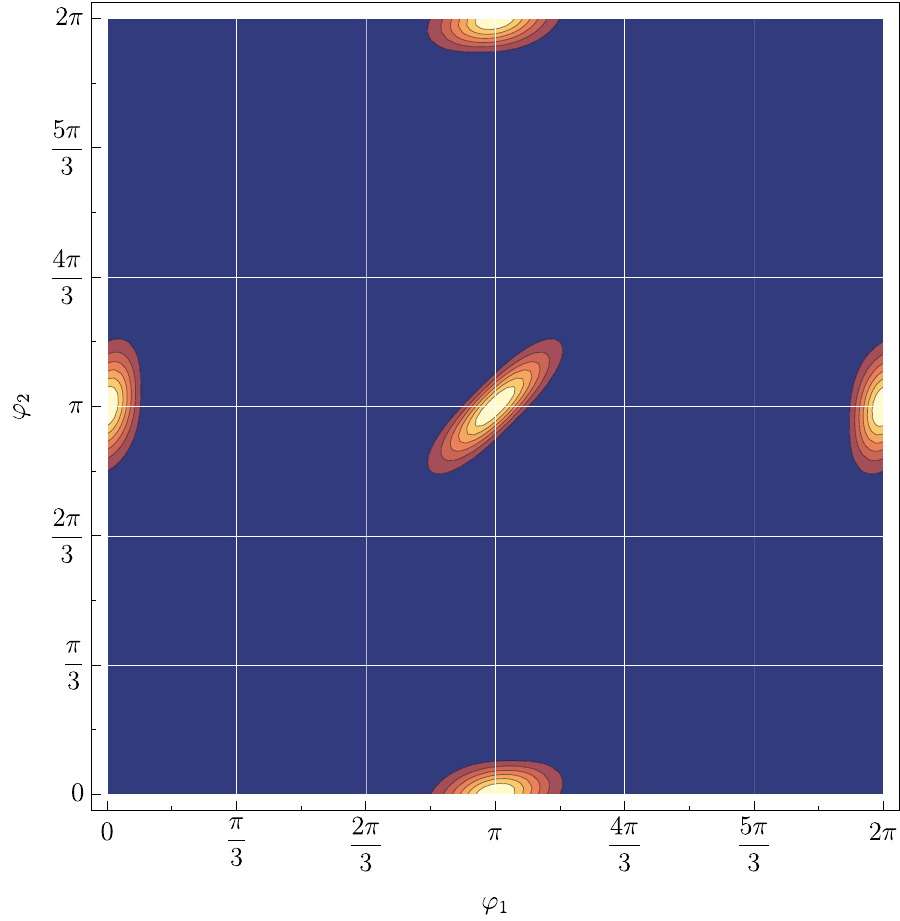}
    \caption{ $J=57$}
\end{subfigure}
\begin{subfigure}{0.32\textwidth}
    \includegraphics[width=\textwidth]{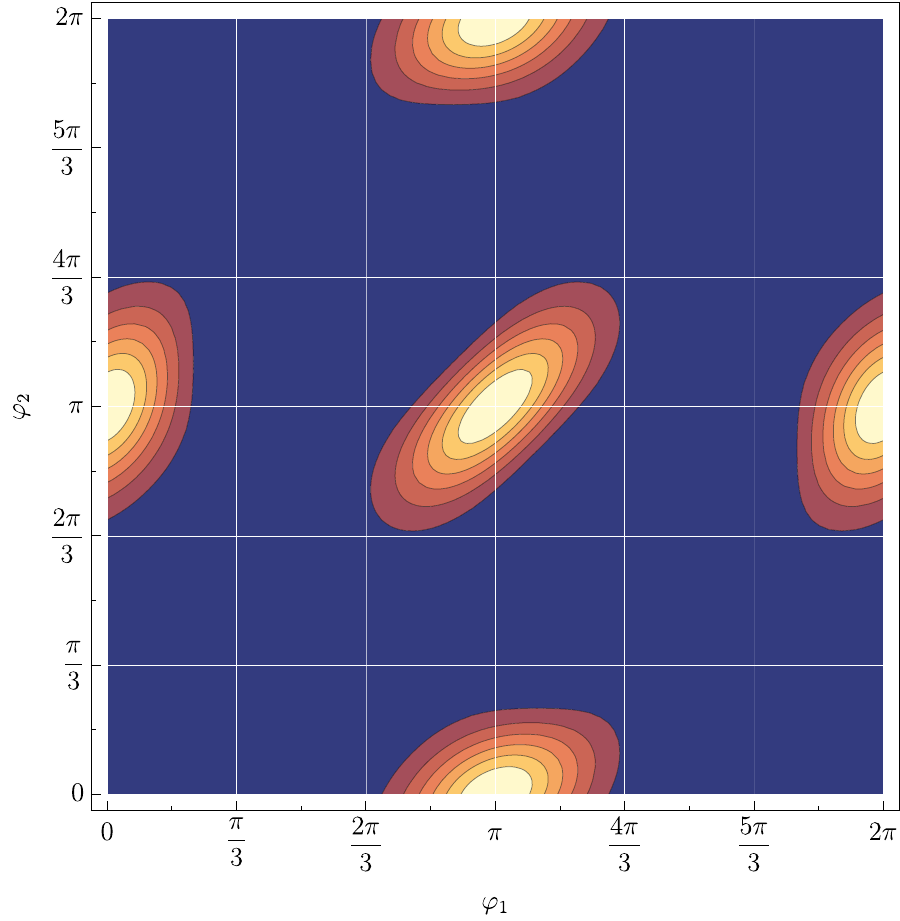}
    \caption{ $J=18$}
\end{subfigure}
\begin{subfigure}{0.32\textwidth}
\centering
\includegraphics[scale=1]{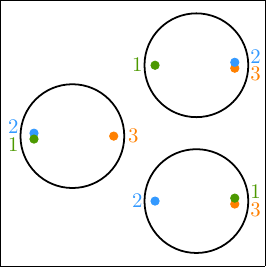}\hspace{-0.8cm} \vspace{0.6cm}  
  \caption{ Cartoon version} \label{fig:gravityCartoon}
\end{subfigure}
\caption{$|\Psi_{\mathrm{min}}(\varphi_1, \varphi_2)|$ for the 3-particle state with $\Delta=\frac{5}{2}$ in $3d$ for different values of the spin $J$.  For comparison, we have chosen the same values of $J$ as for the repulsive case discussed in Sec.~\ref{Sec:repulsivephase}. In the last figure we have drawn a cartoon version of how the partons are distributed in AdS.  The central configuration corresponds to $(\varphi_1,\varphi_2)=(\pi, \pi)$, which, in the $|\Psi_{\mathrm{min}}|$ plots, corresponds to the central yellow region.  The right top and bottom corner configurations correspond respectively to $(\pi, 0)$ and $(0, \pi)$. In  $|\Psi_{\mathrm{min}}|$ they represent the extreme vertical and horizontal regions. } 
\label{fig:Q3D52wavephigravity}
\end{figure}

\begin{figure}
\centering
\begin{subfigure}{0.4\textwidth}
    \includegraphics[width=\textwidth]{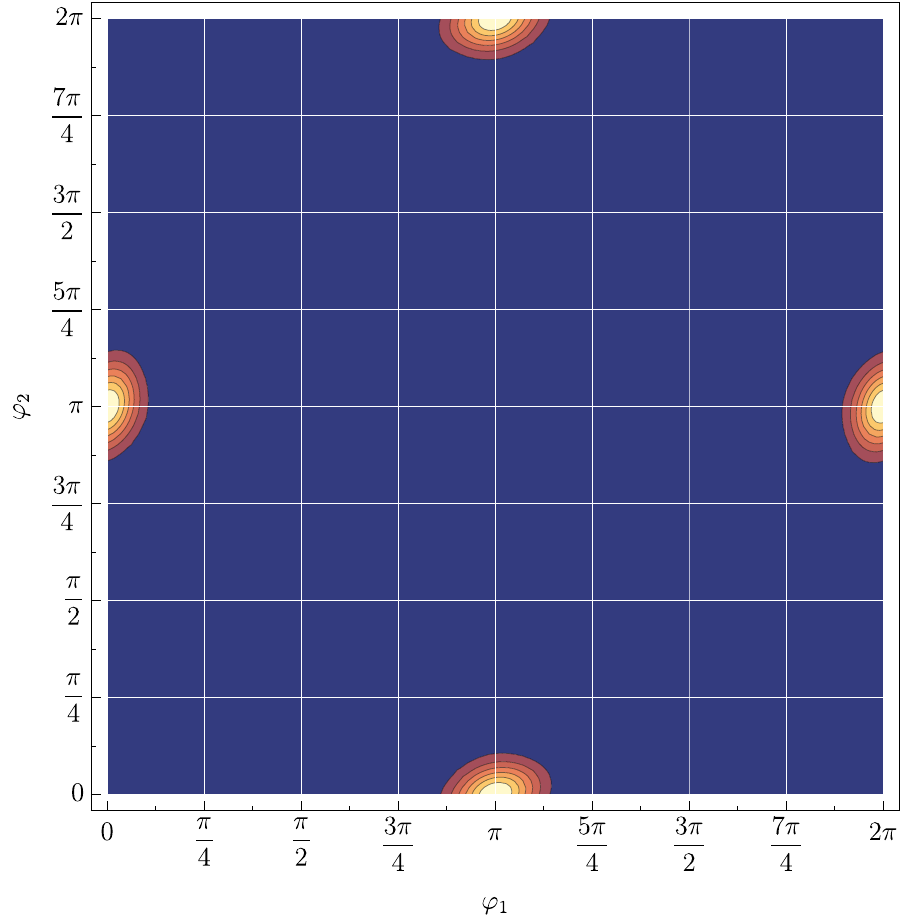}
    \caption{$Q=4$ $\Delta=4$ in $4d$ and $\varphi_3=0$}\label{fig:Q4D4phigravityA}
\end{subfigure}
\hfill
\begin{subfigure}{0.4\textwidth}
    \includegraphics[width=\textwidth]{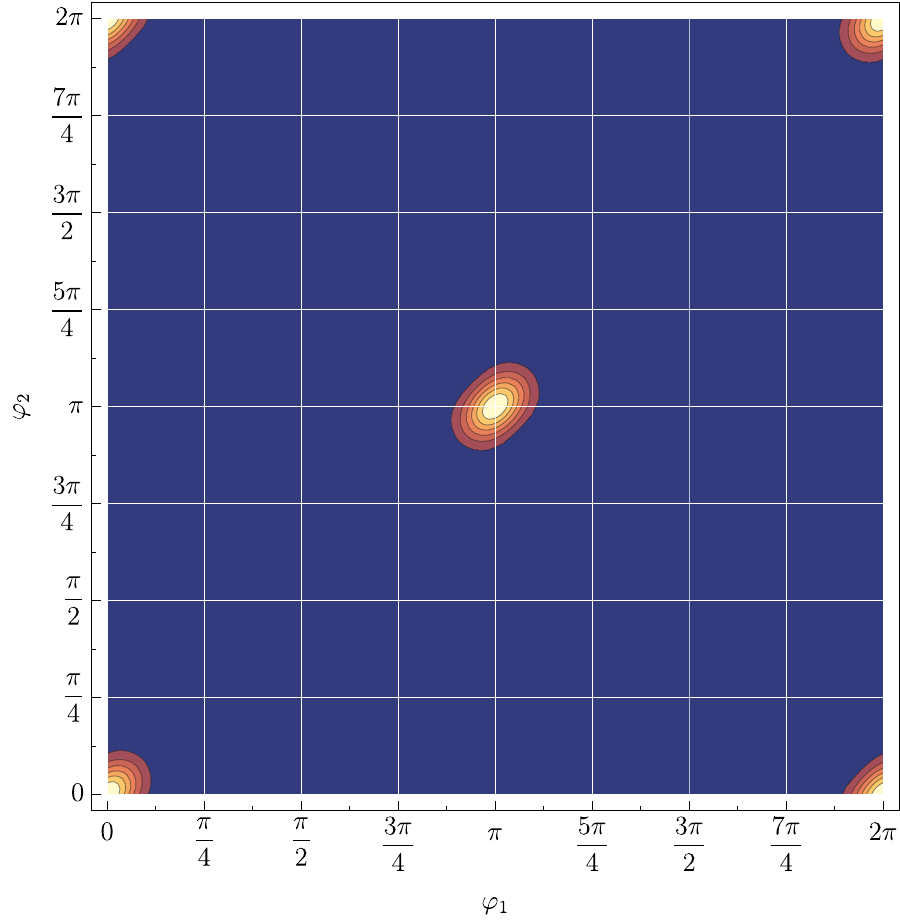}
    \caption{$Q=4$ $\Delta=4$ in $4d$  and $\varphi_3=\pi$}\label{fig:Q4D4phigravityB}
\end{subfigure}
\caption{$|\Psi_{\mathrm{min}}(\varphi_1,\varphi_2, \varphi_3,0)|$ for $\Delta=4$ in $4d$, $Q=4$ and $J=80$. As expected,  we find $3$ partons are together at either 0 or $\pi$ and one is isolated at the opposite angle.  \textit{Left}: 1,3,4 at 0 and parton 2 at $\pi$ (left and right yellow regions) or  $1\leftrightarrow 2$ (top and bottom yellow regions). \textit{Right}: parton 1,2,3 together at $\varphi_i=\pi$ and the fourth one at zero (central region),  or parton 1,2,4 at zero and parton 3 at $\pi$ on the diagonals. 
}
\label{fig:Q4D4phigravity}
\end{figure}

\begin{figure}
\centering
\begin{subfigure}{0.4\textwidth}
    \includegraphics[width=\textwidth]{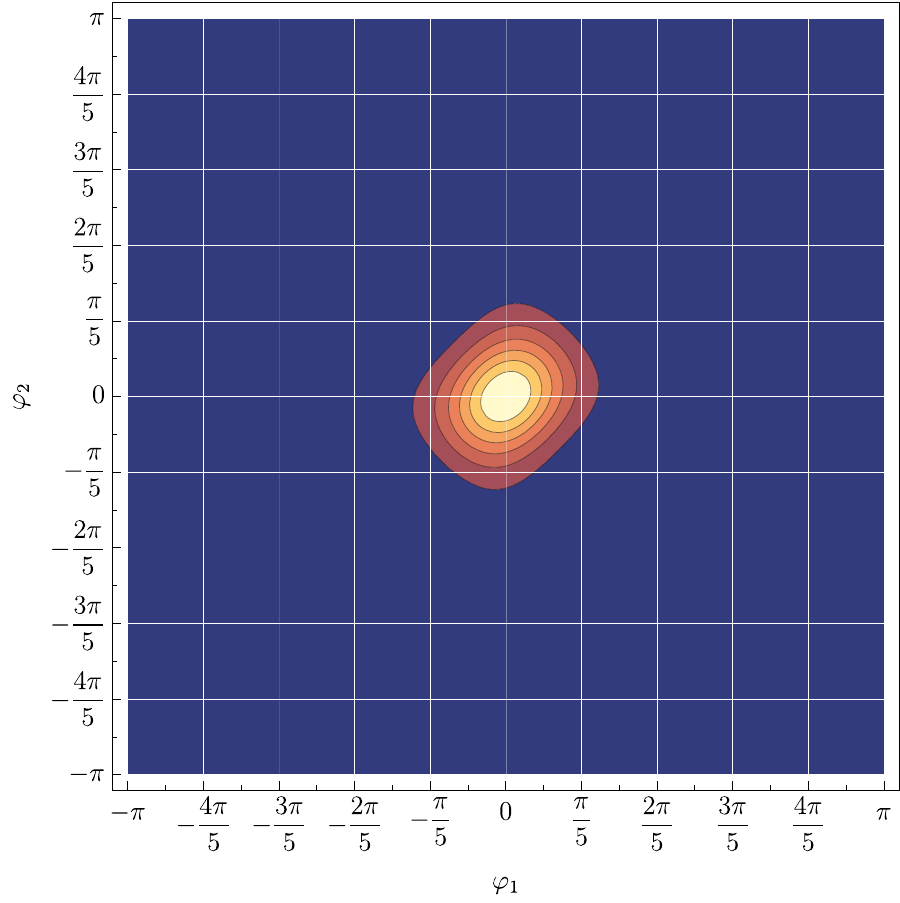}
    \caption{$Q=5$ $\Delta=4$ in $4d$ and $J=55$}
\end{subfigure}
\hfill
\begin{subfigure}{0.4\textwidth}
    \includegraphics[width=\textwidth]{q5D4gaugeMaxWaveL40phi30phi4Pi.pdf}
    \caption{$Q=5$ $\Delta=\frac{5}{2}$ in $3d$ and $J=50$}
\end{subfigure}
\caption{$|\Psi_{\mathrm{min}}(\varphi_1,\varphi_2, \varphi_3,\varphi_4, 0)|$ in $4d$ and $3d$ with $\varphi_3=0$ and $\varphi_4=\pi$. As expected,  we find $4$ partons (1,2,3,5) are together at zero, while parton 4 is at $\pi$.}
\label{fig:Q5phigravity}
\end{figure}

To further justify our intuitive picture of the  lowest-energy eigenstate, 
we can again look at the eigenstate in real space. 
Let us first focus on $Q=3$.  In Fig.~\ref{fig:Q3D52wavephigravity}, we plot $|\Psi_{\mathrm{min}}(\varphi_1, \varphi_2)|$ for $\Delta=\frac{5}{2}$ in $3d$ and fix $\varphi_3=0$ --- other values with $\Delta>\frac{d}{2}$ give similar results in $3d$ as well as in $4d$.  This absolute value shows a Gaussian behavior and it appears to be more localized (yellow region in the plot) around three possible configurations $(\varphi_1, \varphi_2)=(\pi, \pi), \, (0, \pi)$ and $(\pi, 0)$. This is consistent with our expectations, \textit{i.e.} two particles on top of each other placed at a $\pi$-angle with the respect to the third one. In Fig.~\ref{fig:gravityCartoon} we  sketch these three possible configurations in a cartoon version.  
Similar results are obtained at $Q=4$ (Fig. ~\ref{fig:Q4D4phigravity}) and at $Q=5$ (Fig.~\ref{fig:Q5phigravity}).

\begin{figure}
\centering
\begin{subfigure}{0.38\textwidth}
    \includegraphics[width=\textwidth]{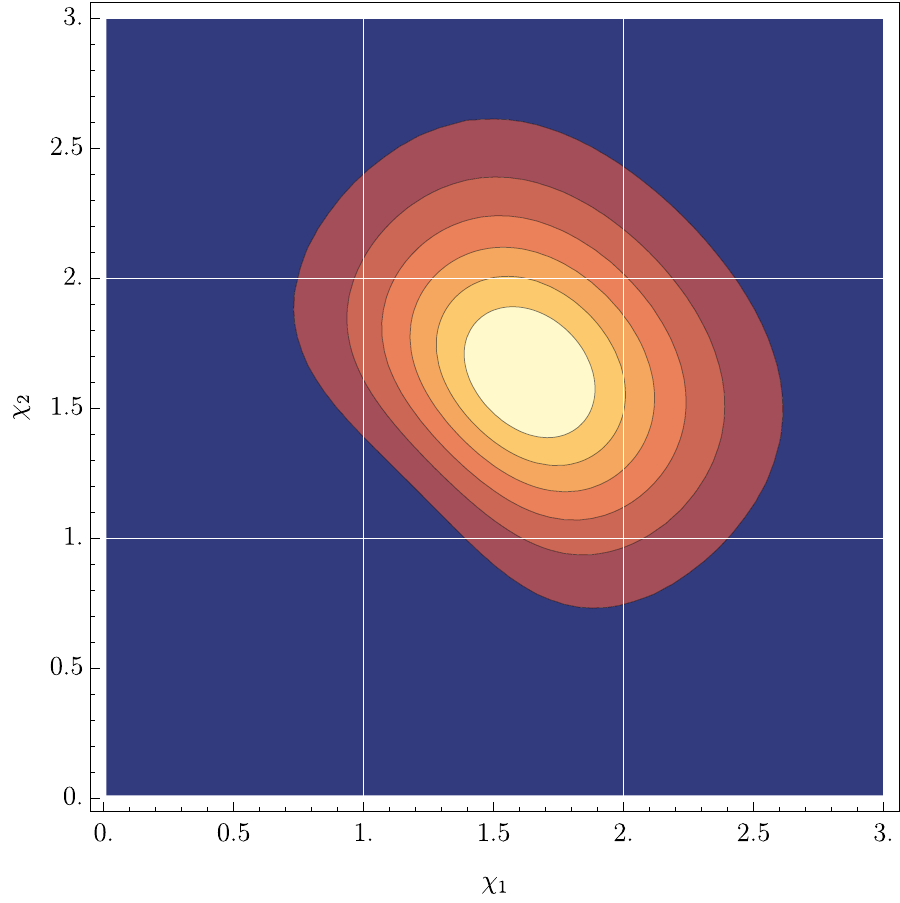}
    \caption{$\chi_3=1.98$, $(\varphi_1, \varphi_2)=(\pi, \pi)$} \label{fig:Q3D4wavechigravityA}
\end{subfigure}
\hfill
\begin{subfigure}{0.38\textwidth}
    \includegraphics[width=\textwidth]{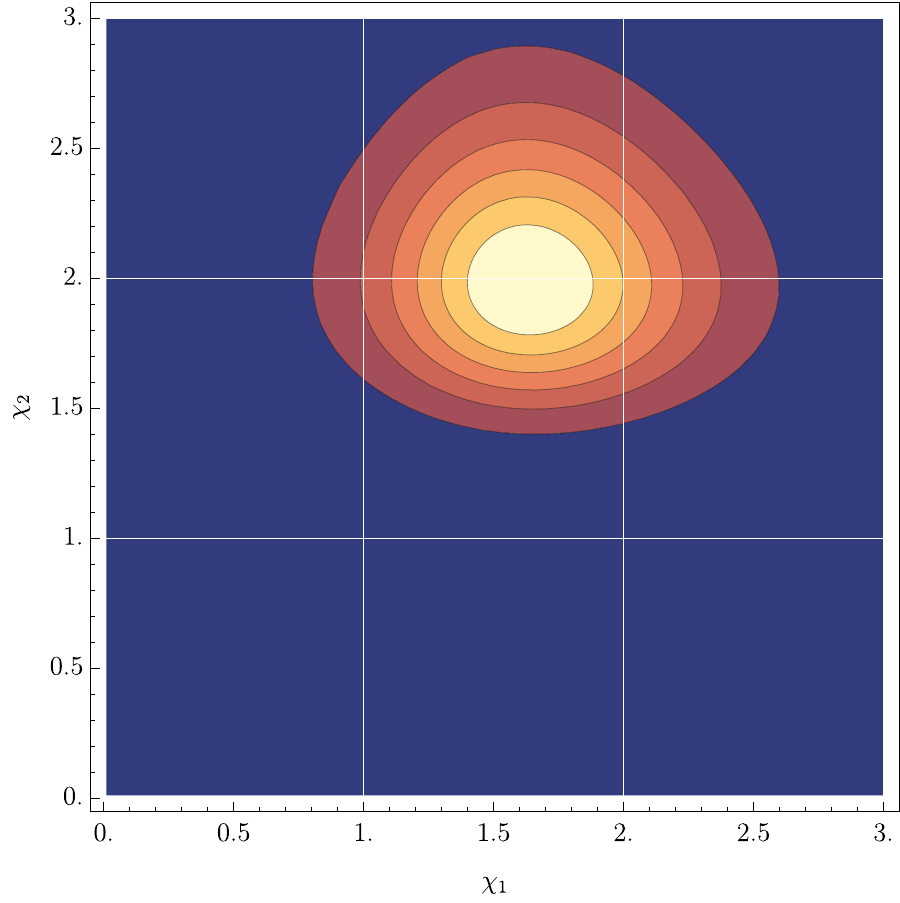}
    \caption{$\chi_3=1.63$, $(\varphi_1, \varphi_2)=(0, \pi)$}\label{fig:Q3D4wavechigravityB}
\end{subfigure}
\caption{$|\Psi_{\mathrm{min}}(\chi_1, \chi_2, \chi_3)|$ for the 3-particle state with $\Delta=4$ in $4d$ for $J=99$ at different values of $(\varphi_1, \varphi_2)$. $\chi_3$ it is fixed to its maximum.} 
\label{fig:Q3D4wavechigravity}
\end{figure}

Finally for fixed values of $\varphi_i$, we can study the $\chi$-dependence of the absolute value of the wave-function.  In Fig.~\ref{fig:Q3D4wavechigravity} we report $|\Psi_{\mathrm{min}}(\chi_1, \chi_2, \chi_3)|$ for $\Delta=4$ in $4d$.  Notice that the behavior is Gaussian and the maximum is located at $\chi_i^*=\mathrm{arcsinh}\sqrt{\frac{\ell_i^*}{\Delta}}$. The value of $\ell_i^*$ depends on the specific configurations: if parton 1 and 2 are on top of each other (Fig.~\ref{fig:Q3D4wavechigravityA}) then $\ell_1^*=\ell_2^*=\frac{J}{4}$ and $\ell_3^*=\frac{J}{2}$; if they lie on opposite sides (Fig.~\ref{fig:Q3D4wavechigravityB}), instead,  $\ell_2^* \leftrightarrow \ell_3^*$. 

\subsection{Small spin  and contact terms}\label{Sec:smallSpinContact}
So far, we have seen in these examples that a $Q$-parton state in the attractive phase typically contains  $q$-parton `blobs' with $q<Q$.  
These $q$-parton blobs  have some internal spin which is distinct from the spin of the full $Q$-parton state.
So to understand the behavior of the blobs, we have to also understand the behavior of the anomalous dimensions at small values of the spin.

At sufficiently small values of $J$, the free theory in AdS contains only a single primary state and in these cases it is straightforward to determine the analytic expressions for the anomalous dimensions.  In~\eqref{smallSpinsPrimaries}, we have listed the form of the primaries for spin $0,2,3$ for generic $Q$.  We consider each of these cases in turn.

\subsubsection*{$\boldsymbol{J=0}$}
\begin{figure}[t!]
\centering
\begin{subfigure}{0.32\textwidth}
    \includegraphics[width=\textwidth]{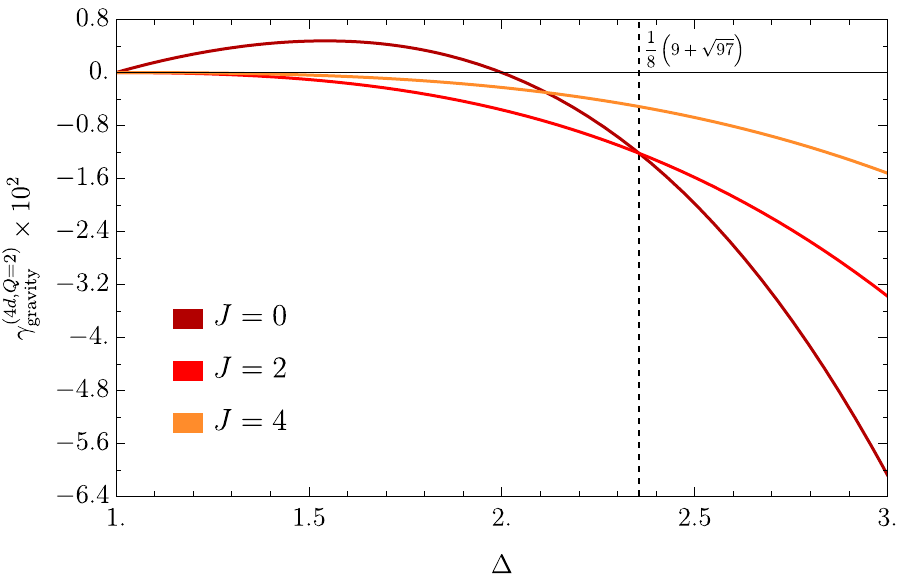}
    \caption{$Q=2$}
\end{subfigure}
\hfill
\begin{subfigure}{0.32\textwidth}
    \includegraphics[width=\textwidth]{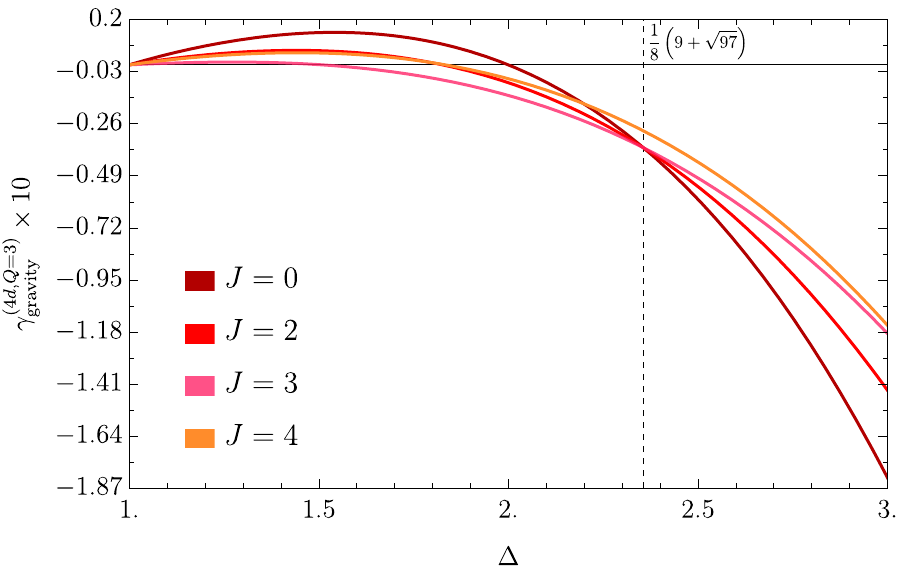}
    \caption{$Q=3$}
\end{subfigure}
\hfill
\begin{subfigure}{0.32\textwidth}
    \includegraphics[width=\textwidth]{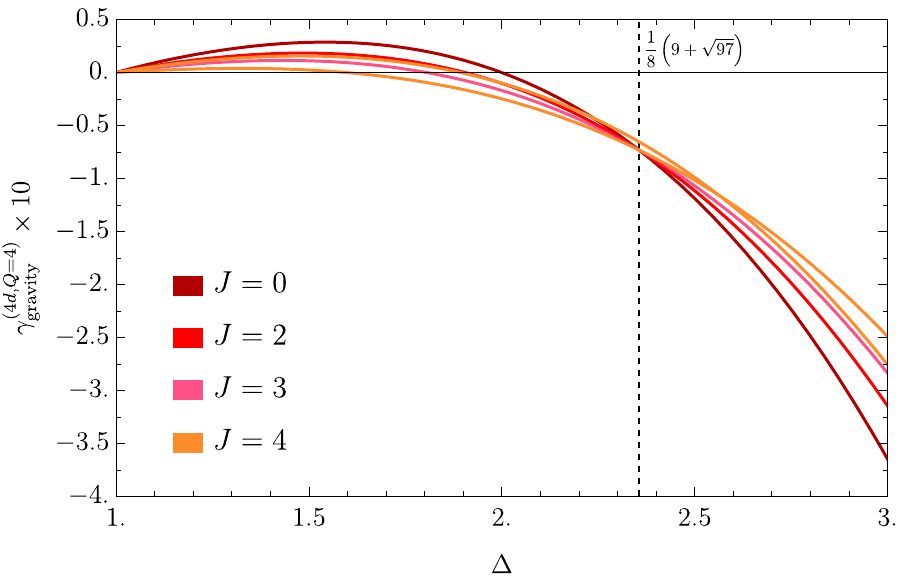}
    \caption{$Q=4$}
\end{subfigure}
\caption{$4d$ anomalous dimensions for $Q$-particle states due the graviton exchange as a function of the external dimensions $\Delta$ for $J=0,2,3,4$.  Notice that there are two lines at $Q=4$ due to the degeneracy of the primaries.} 
\label{fig:4dlowSpinDelta}
\end{figure}

\begin{figure}[t!]
\centering
\begin{subfigure}{0.32\textwidth}
    \includegraphics[width=\textwidth]{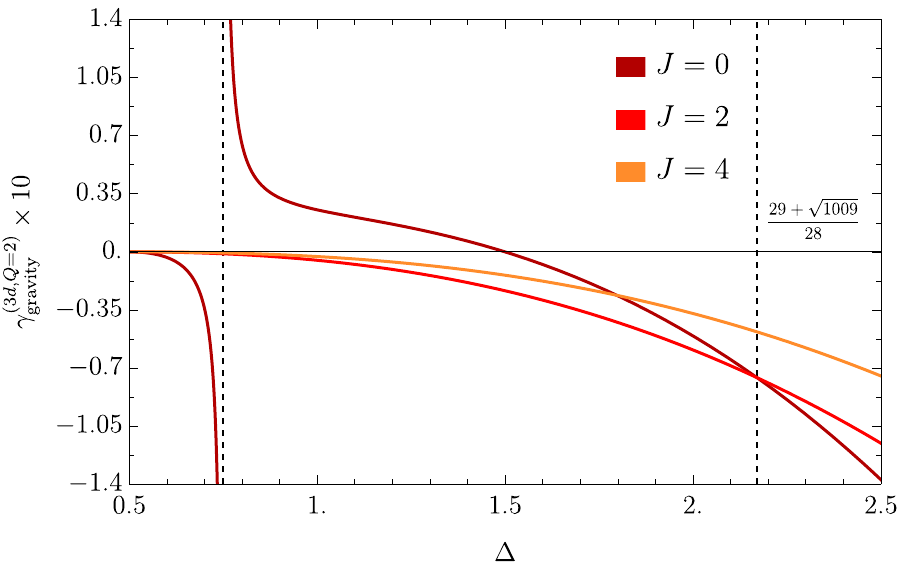}
    \caption{$Q=2$}\label{3dlowSpinDeltaQ2}
\end{subfigure}
\hfill
\begin{subfigure}{0.32\textwidth}
    \includegraphics[width=\textwidth]{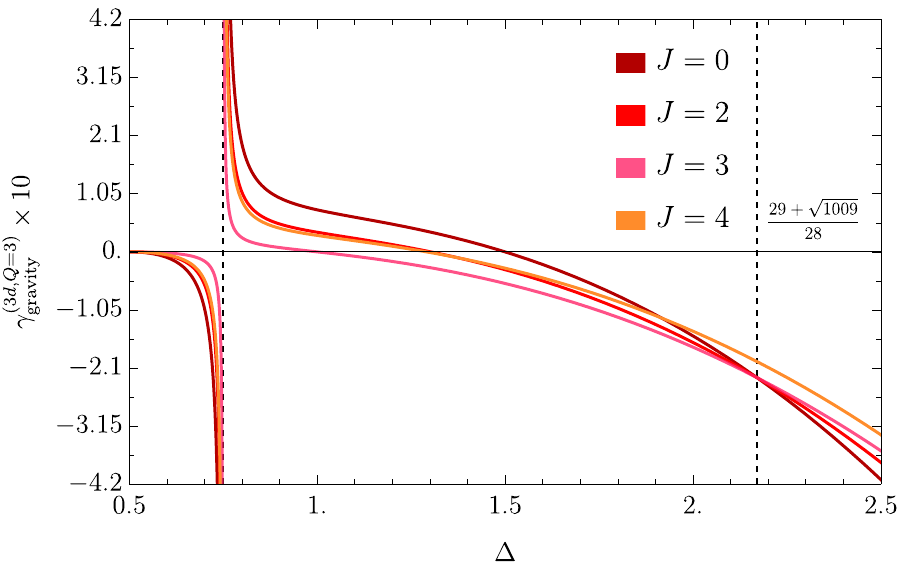}
    \caption{$Q=3$}
\end{subfigure}
\hfill
\begin{subfigure}{0.32\textwidth}
    \includegraphics[width=\textwidth]{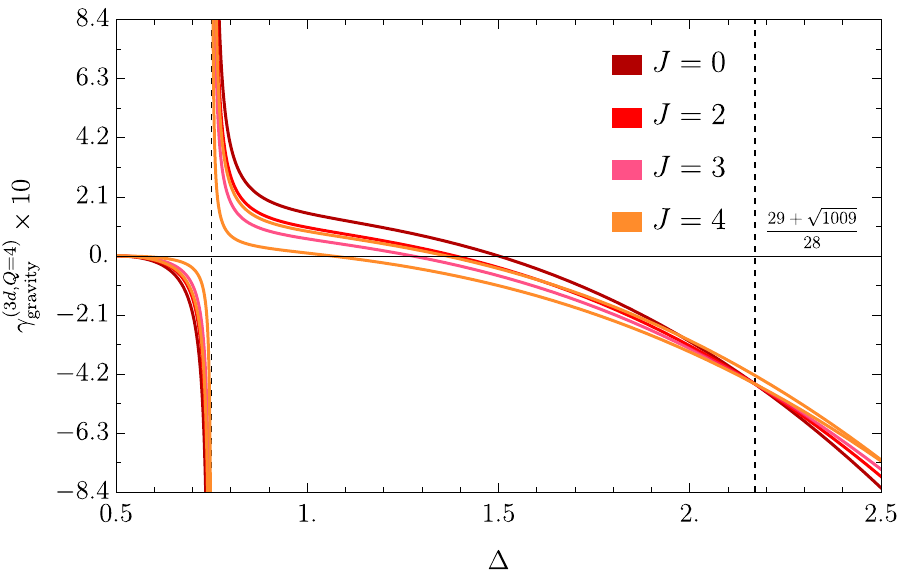}
    \caption{$Q=4$}
\end{subfigure}
\caption{$3d$ anomalous dimensions for $Q$-particle states due the graviton exchange as a function of the external dimensions $\Delta$ for $J=0,2,3,4$.  Notice that there are  two lines at $Q=4$ due to the degeneracy of the primaries. } 
\label{fig:3dlowSpinDelta}
\end{figure}

For $J=0$, the anomalous dimension at any value of $Q$ is
\eqna{
\gamma^{(d,Q)}|_{J=0}&= \frac{Q(Q-1)}{2}\gamma^{(d,Q=2)}|_{J=0}\, ,
}[]
with 
\eqna{
\gamma^{(d,Q=2)}_{\mathrm{gauge}}|_{J=0}&=\frac{\pi ^{-\mathit{d}/2} \Gamma (\Delta )^2 \Gamma \mleft(2 \Delta-\frac{d}{2}+1 \mright)}{2
   (\mathit{d}-2) \Gamma (2 \Delta ) \Gamma \mleft(\Delta-\frac{d}{2}+1 \mright)^2}\, , \\
 \gamma^{(d,Q=2)}_{\mathrm{gravity}}|_{J=0}&=-\frac{\pi ^{-\mathit{d}/2} \mleft(\Delta -\frac{\mathit{d}}{2}\mright) \Gamma (\Delta +1)^2 \Gamma \mleft(2
   \Delta -\frac{\mathit{d}}{2}\mright)}{ (\mathit{d}-1) \Gamma (2 \Delta ) \Gamma
   \mleft(\Delta-\frac{\mathit{d}}{2}+1 \mright)^2}\, .
}[gammaJzero]
 Notice that the sign of the gravity contributions depends on $\Delta\lessgtr \frac{d}{2}$.  Moreover, the gravity contribution in $4d$  has a zero at $\Delta=2$ and $\Delta=1$ (the unitarity bound), while in $3d$ it has a zero at $\Delta=\frac{3}{2}$ and $\Delta=\frac{1}{2}$ and a pole at $\Delta=\frac{3}{4}$ (see Figs.~\ref{fig:4dlowSpinDelta} and \ref{fig:3dlowSpinDelta}).  This pole may seem puzzling, but it can be removed by adding a contact term as we will discuss in more detail below.  Since the anomalous dimension is not analytic in $J$ at $J=0$, our attitude is that the contact terms should be chosen, in the language of EFT, by a `matching calculation'  where we take the physically measured anomalous dimension as input in order to set the coefficient of $\phi^4$.

\subsubsection*{$\boldsymbol{J=2,3}$}
For the spin-2 anomalous dimension we find
\twoseqn{
\gamma_{\mathrm{gauge}}^{(d,Q)}|_{J=2}&=\gamma_{\mathrm{gauge}}^{(d,Q=2)}|_{J=0}\begin{cases}
\frac{Q(Q-1)}{2}-\frac{(\Delta  \lsp Q+1)}{3}\frac{(4 \Delta +1) }{2 \Delta  (2 \Delta +1)} & \quad \qquad \, \,  \, 4d\, ,\\
\frac{Q(Q-1)}{2} -\frac{(\Delta  \lsp Q+1)}{3}\frac{(7 \Delta +2)}{5 \Delta  (2 \Delta +1)} &\quad \qquad \, \, \, 3d\, ,
\end{cases}
}[]
{
\gamma_{\mathrm{gravity}}^{(d,Q)}|_{J=2}&=\gamma_{\mathrm{gravity}}^{(d,Q=2)}|_{J=0}\begin{cases}
\frac{Q(Q-1)}{2}-\frac{(\Delta  \lsp Q+1)}{3}\frac{4 \Delta ^2-9 \Delta -1}{2 (\Delta -2) \Delta  (2 \Delta +1)}& \quad 4d\, ,\\
\frac{Q(Q-1)}{2} -\frac{(\Delta  \lsp Q+1)}{3}\frac{14 \Delta ^2-29 \Delta -3}{5 \Delta  (2 \Delta -3) (2 \Delta +1)}&\quad 3d\, ,
\end{cases}
}[][spin2Q]
and at $J=3$
\twoseqn{
\gamma_{\mathrm{gauge}}^{(d,Q)}|_{J=3}&=\gamma_{\mathrm{gauge}}^{(d,Q=2)}|_{J=0}\begin{cases}
\frac{Q(Q-1)}{2}-\frac{(\Delta  \lsp Q+2)}{2}\frac{(4 \Delta +1) }{2 \Delta  (2 \Delta +1)} &  \quad \qquad \, \,\, 4d\, ,\\
\frac{Q(Q-1)}{2} -\frac{(\Delta  \lsp Q+2)}{2}\frac{(7 \Delta +2)}{5 \Delta  (2 \Delta +1)} & \quad \qquad \, \,  \,  3d\, ,
\end{cases}
}[]
{
\gamma_{\mathrm{gravity}}^{(d,Q)}|_{J=3}&=\gamma_{\mathrm{gravity}}^{(d,Q=2)}|_{J=0}\begin{cases}
\frac{Q(Q-1)}{2}-\frac{(\Delta  \lsp Q+2)}{2}\frac{4 \Delta ^2-9 \Delta -1}{2 (\Delta -2) \Delta  (2 \Delta +1)}& \quad 4d\, ,\\
\frac{Q(Q-1)}{2} -\frac{(\Delta  \lsp Q+2)}{2}\frac{14 \Delta ^2-29 \Delta -3}{5 \Delta  (2 \Delta -3) (2 \Delta +1)}&\quad 3d\, .
\end{cases}
}[][spin3Q]
The gauge part of the anomalous dimension is always positive for $\Delta$ above the unitarity bound.  

The situation is rather different for the gravity contribution, where as we can see from the results at $J=0,2,3$ -- see also Fig.~\ref{fig:4dlowSpinDelta} and~\ref{fig:3dlowSpinDelta} --  the sign and the shape of $\gamma_{\mathrm{graviton}}^{(d, Q)}(J)$ is strongly affected by the value of $Q$ and $\Delta$.

\subsubsection*{$\boldsymbol{J \ge 4}$}

Starting at $J=4$, for generic values of $Q$ there are multiple degenerate primary states in the free theory.  For $Q=2,3$, however, there is still only a single $J=4$ state.  For the case $Q=4$, there are two degenerate primary states in the free theory, which were  given explicitly in the monomial basis in (\ref{J4Prims}).  In Figs.~\ref{fig:3dlowSpinDelta} and \ref{fig:4dlowSpinDelta}, both of the corresponding $Q=4, J=4$ eigenvalues of the gravity exchange contribution are presented.  For larger values of $J$ when the degeneracy becomes too large to find the eigenstates analytically, we diagonalize the anomalous dimension matrix numerically.

\subsubsection*{Minimal Twist at Small $\boldsymbol{Q}$ and small $\boldsymbol{J}$}
To understand the dynamically preferred configurations of the blobs that form inside states at large $J$, we need to determine for each $Q$ what is the value of $J$ that minimizes the twist.  In our perturbative computation, all the minimal-twist states we are considering have twist $Q \Delta + \gamma$ where $\gamma$ is the perturbative bulk anomalous dimension.

\begin{figure}
\centering
\begin{subfigure}{0.32\textwidth}
    \includegraphics[width=\textwidth]{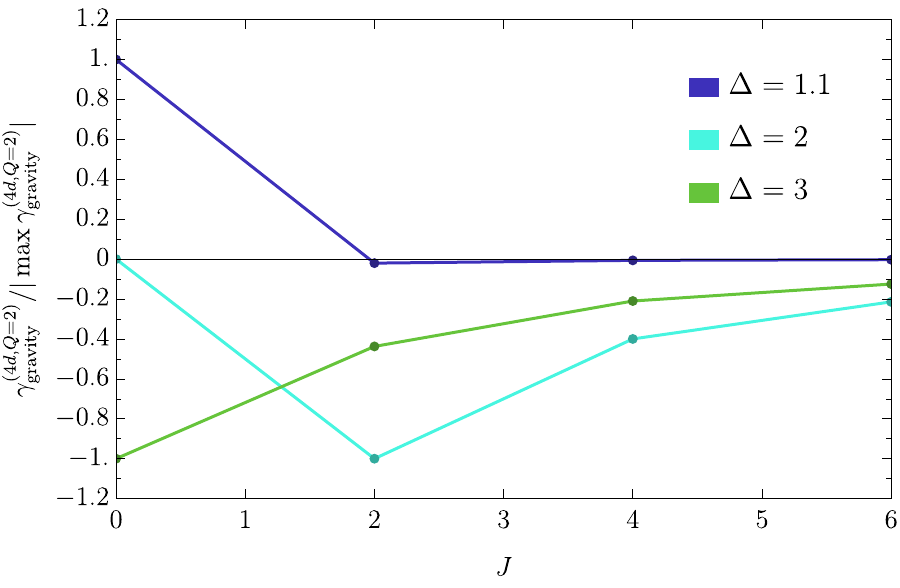}
    \caption{$Q=2$}
\end{subfigure}
\hfill
\begin{subfigure}{0.32\textwidth}
    \includegraphics[width=\textwidth]{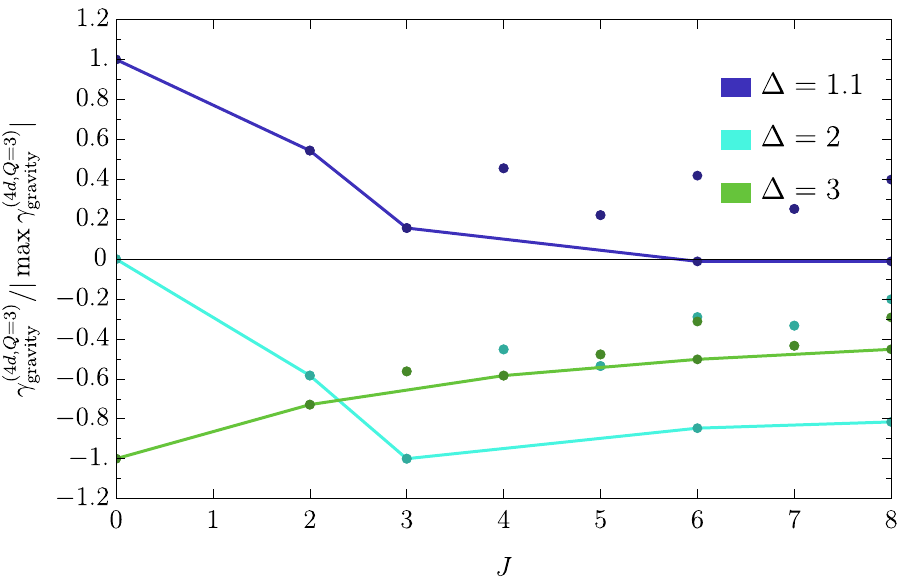}
    \caption{$Q=3$}
\end{subfigure}
\hfill
\begin{subfigure}{0.32\textwidth}
    \includegraphics[width=\textwidth]{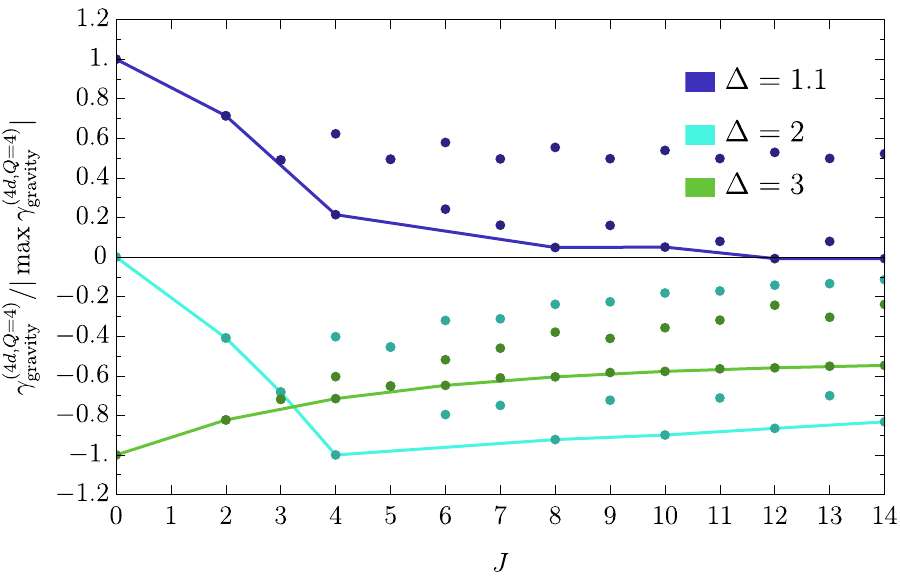}
    \caption{$Q=4$}
\end{subfigure}
\caption{$4d$ anomalous dimensions for $Q$-particle states due the graviton exchange as a function of $J$ for different external dimensions.   $\gamma^{(4d,Q)}_{\mathrm{gravity}}$ is normalized with the respect to its maximum value.} 
\label{fig:4ddeltaSpins}
\end{figure}

\begin{figure}
\centering
\begin{subfigure}{0.32\textwidth}
    \includegraphics[width=\textwidth]{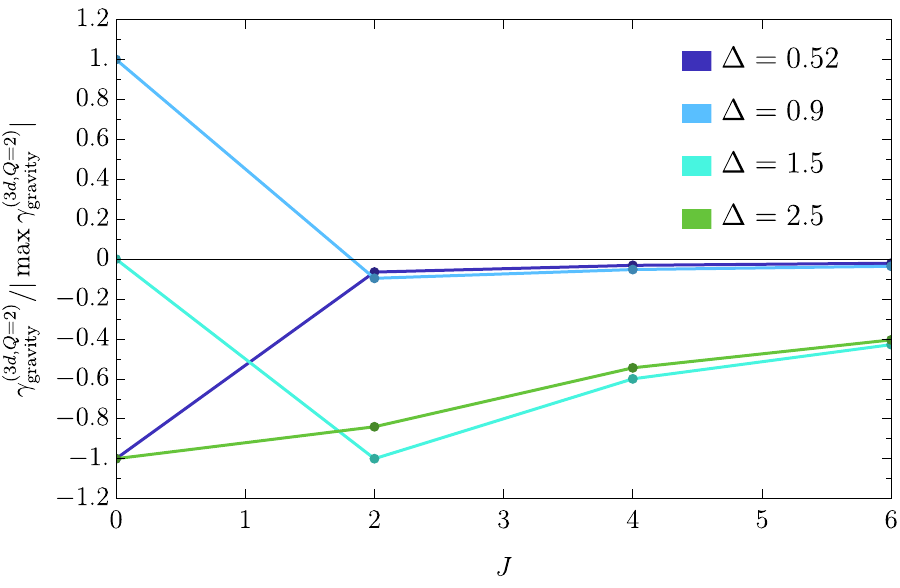}
    \caption{$Q=2$} 
\end{subfigure}
\hfill
\begin{subfigure}{0.32\textwidth}
    \includegraphics[width=\textwidth]{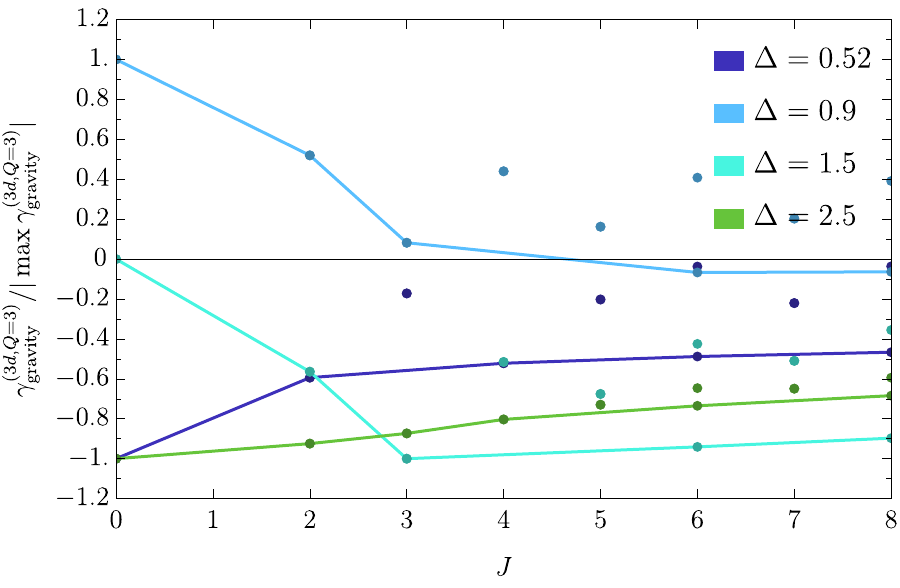}
    \caption{$Q=3$}
\end{subfigure}
\hfill
\begin{subfigure}{0.32\textwidth}
    \includegraphics[width=\textwidth]{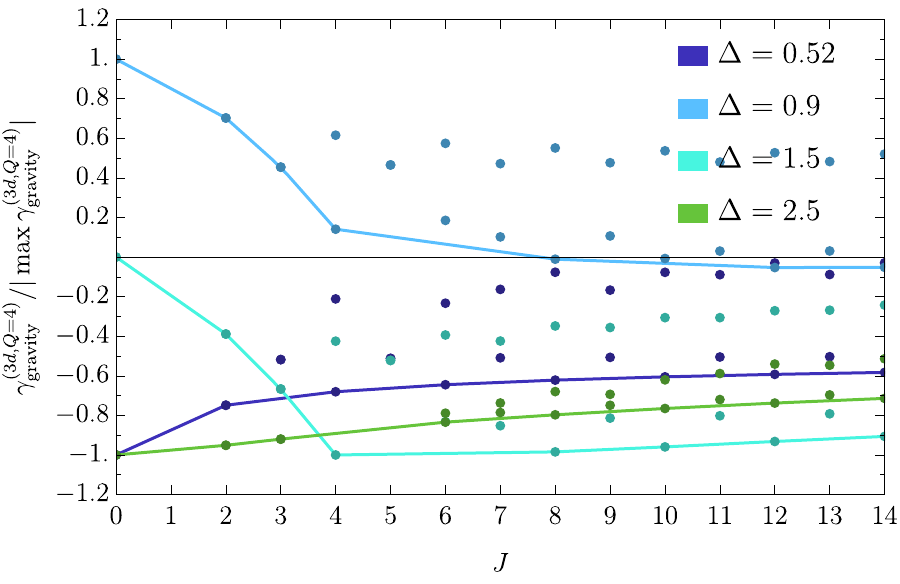}
    \caption{$Q=4$}
\end{subfigure}
\caption{$3d$ anomalous dimensions for $Q$-particle states due the graviton exchange as a function of $J$ for different external dimensions.  $\gamma^{(4d,Q)}_{\mathrm{gravity}}$ is normalized with the respect to its maximum value.}
\label{fig:3ddeltaSpins}
\end{figure}

In Fig.~\ref{fig:4ddeltaSpins} and~\ref{fig:3ddeltaSpins} we show how $\gamma_{\mathrm{graviton}}^{(d, Q)}(J)$ is a convex function whose minimum varies as we tune the value of $\Delta$.  Notice that for any $Q$ for $\Delta>\frac{d}{2}$, the minimum of $\gamma_{\mathrm{graviton}}^{(d, Q)}(J)$ is always at $J=0$, thus supporting our result in~\eqref{a0dGr}, while for $\Delta=\frac{d}{2}$ the minimum is at $J=Q$, in accordance with~\eqref{a0d2}.
We can exactly quantify for which value of $\Delta$ the behavior of $\gamma^{(d,Q)}_{\mathrm{gravity}}$ changes and its minimum changes as a function of $Q$.  Focusing in four dimensions first,   we can identify three different ranges of $\Delta$: starting from the unitarity bound ($\Delta=\frac{d-2}{2}$)
\begin{align}
\begin{aligned}
1\leq \Delta < 2&: \text{minimum at }J=Q(Q-1)\, , \\
2\leq \Delta < \frac{9+\sqrt{97}}{8}&: \text{minimum at }J=Q\, , \\
 \Delta  > \frac{9+\sqrt{97}}{8}&: \text{minimum at }J=0\, .
 \end{aligned}
\end{align} 
In three dimensions the situation is more complicated because of the presence of a pole at $\Delta=\frac{3}{4}$ 
\begin{align}
\begin{aligned}
\frac{1}{2} \leq \Delta <\frac{3}{4}&: \text{minimum at }J=0\, ,\\
\frac{3}{4}<\Delta < \frac{3}{2}&: \text{minimum at }J=Q(Q-1)\, , \\
\frac{3}{2}\leq \Delta < \frac{29+\sqrt{1009}}{28}&: \text{minimum at }J=Q\, , \\
 \Delta  > \frac{29+\sqrt{1009}}{28}&: \text{minimum at }J=0\, .
 \end{aligned}
\end{align}

As we have mentioned, the anomalous dimension can have nonanalytic contributions in $J$ at $J=0$, and so the cases where we see a minimum at $J=0$ must be reevaluated in light of such contributions.   

\subsubsection*{Including Contact Terms}

In the last part of this section, we will discuss the effect of adding contact terms to the anomalous dimensions.  In Sec.~\ref{Subsec:contactTerms}  we have discussed the contribution to the 2-particle anomalous dimensions due to the addition of the  interactions $(\phi \phi^\dagger)^2$ and $ \phi \phi^\dagger \partial_\mu \phi \partial^\mu \phi^\dagger$. Differently from the $Q=2$ case, for $Q\geq 3$, these interactions have an effect on the anomalous dimensions for any spin. For spin $J=0,2,3$ it is easy to find an explicit formula for them
\eqna{
\gamma^{(d,Q)}_{\mathrm{contact}}|_{J=0,2,3}&=\kappa^2\frac{\pi ^{-\mathit{d}/2} \Gamma (\Delta )^2 \Gamma \mleft(2 \Delta -\frac{\mathit{d}}{2}\mright)}{2
   \Gamma (2 \Delta ) \Gamma \mleft(\Delta-\frac{\mathit{d}}{2} +1\mright)^2}\Big \lbrace \frac{Q(Q-1)}{2} \left( \lambda_0 -{\lambda_1} \Delta  \left( \Delta-\frac{d}{2}\right)\right)\\
   &\quad\, -\frac{J   (\Delta\lsp Q+J -1)}{2 (2 \Delta +1)} \left(\lambda_0- \lambda_1 \Delta  \left(\Delta -\frac{\mathit{d}}{2}\right)\right)
 \Big\rbrace\, .
}[contactQ]
Recall that in three dimensions, the graviton anomalous dimension in~\eqref{gammaJzero} exhibits a pole for $\Delta=\frac{3}{4}$. 
We would like to remove this pole and we will show how to do so by  appropriately tuning the $\lambda_i$'s above.  
 To be as generic as possible, we will choose the contact term to satisfy the following condition for some value of $\alpha$:
\eqna{
\gamma^{(3d,Q)}_{\mathrm{gravity}}+\gamma^{(3d,Q)}_{\mathrm{contact}}|_{J=0} &=\alpha \kappa^2 \frac{Q(Q-1)}{2} \left(2\Delta-\frac{3}{2} \right)\frac{\Gamma (\Delta )^2 \Gamma \mleft(2 \Delta -\frac{3}{2}\mright)}{2 \pi ^{3/2} \Gamma \mleft(\Delta
   -\frac{1}{2}\mright)^2 \Gamma (2 \Delta )}\\
   &=\alpha\kappa^2 \frac{Q(Q-1)}{2}\frac{\Gamma (\Delta )^2 \Gamma \mleft(2 \Delta -\frac{1}{2}\mright)}{2 \pi ^{3/2} \Gamma \mleft(\Delta
   -\frac{1}{2}\mright)^2 \Gamma (2 \Delta )}\, ,
}[contactJ0]
The point is that in the first line, we impose that the sum of the two contributions produces an extra factor of $(2\Delta-3/2)$ so that the pole is cancelled for any finite value of $\alpha$.  This parameterization is convenient since the combination of the contact term and gravity contributions removes the pole simply by shifting  the argument of the $\Gamma$ function.  At the level of the couplings, this requirement translates to
\eqna{
\lambda_0=\alpha \left( 2\Delta-\frac{3}{2}\right)+\Delta \left( \Delta-\frac{3}{2}\right)\left(\Delta+\lambda_1 \right)\, .
}[]
Quite remarkably, this choice  produces the same factor $\left(2\Delta-\frac{3}{2} \right)$  also for $J \neq 0$, thus effectively removing the pole from the anomalous dimension in all spins,  as one can check using ~\eqref{contactQ} and~\eqref{spin2Q}, \eqref{spin3Q}.\footnote{We explicitly check up to $J=5$. }
Moreover notice that, by changing the value of $\alpha$ we can change the value of the spin-zero state with the respect to the other lowest spin states, thus modifying the behavior in Fig.~\ref{fig:3ddeltaSpins}. In particular, there is always a choice of $\alpha$ such that the spin-0 is the lowest-energy state
\eqna{
\alpha&>-\frac{1}{15} \Delta  (4 \Delta -1)\, ,  \qquad \text{for }\frac{1}{2}<\Delta<\frac{3}{4}\, , \\
\alpha&<-\frac{1}{15} \Delta  (4 \Delta -1)\, ,  \qquad \text{for }\Delta>\frac{3}{4}\,.
}[]

\begin{figure}
\centering
\begin{subfigure}{0.32\textwidth}
    \includegraphics[width=\textwidth]{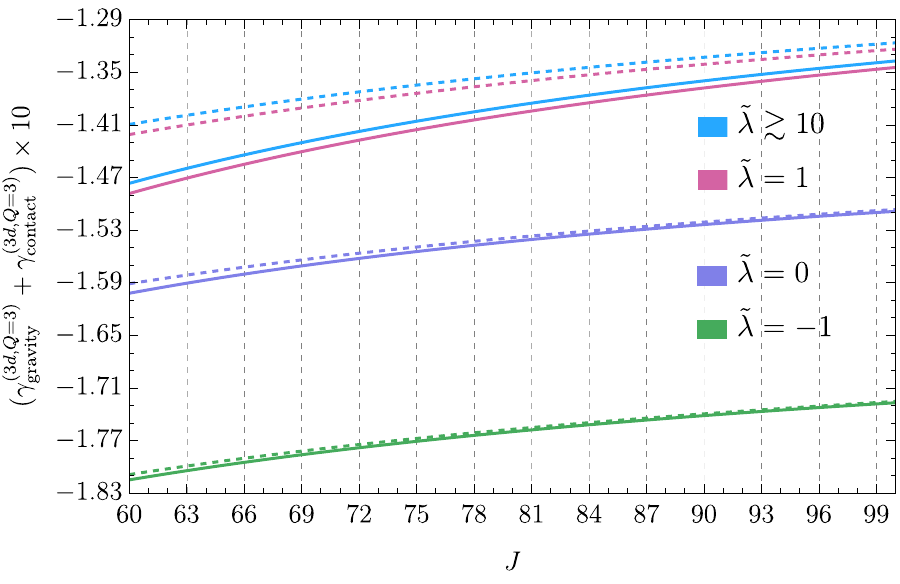}
    \caption{$Q=3$} 
\end{subfigure}
\hfill
\begin{subfigure}{0.32\textwidth}
    \includegraphics[width=\textwidth]{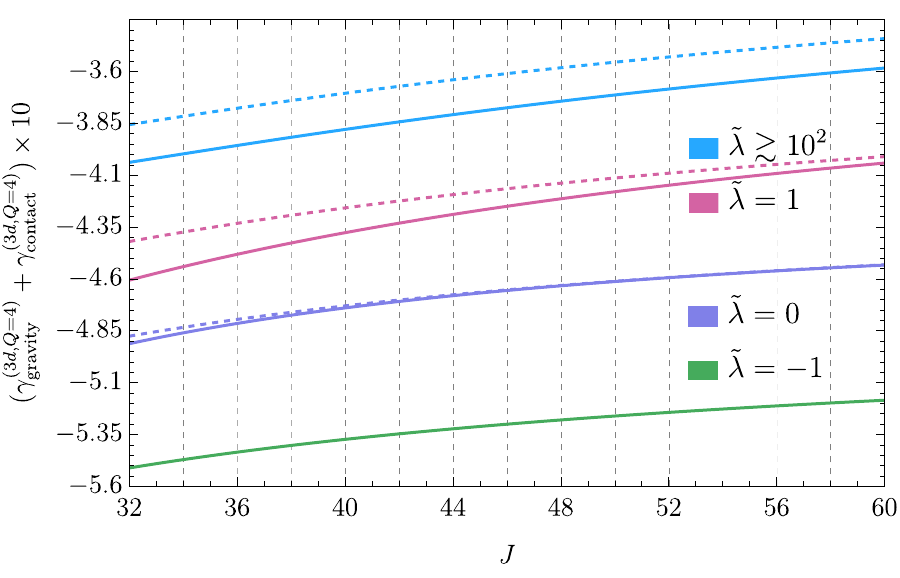}
    \caption{$Q=4$}
\end{subfigure}
\hfill
\begin{subfigure}{0.32\textwidth}
    \includegraphics[width=\textwidth]{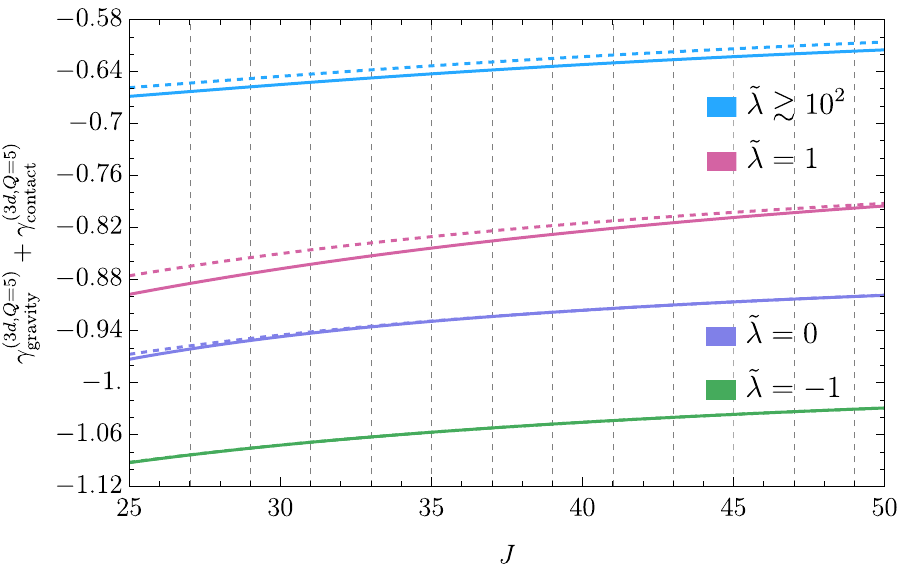}
    \caption{$Q=5$}
\end{subfigure}
\caption{$3d$ anomalous dimensions for $Q$-particle states due the graviton exchange and contact terms at $\Delta=\frac{5}{2}$ for different values of $\tilde{\lambda}$ ($\tilde{\lambda}=0$ coincides with $\gamma_{\mathrm{gravity}}^{(3d,Q)}$). Solid lines represent even spin, while dashed ones odd spins.} 
\label{fig:gravityContactd52}
\end{figure}

\begin{figure}
\centering
\begin{subfigure}{0.32\textwidth}
    \includegraphics[width=\textwidth]{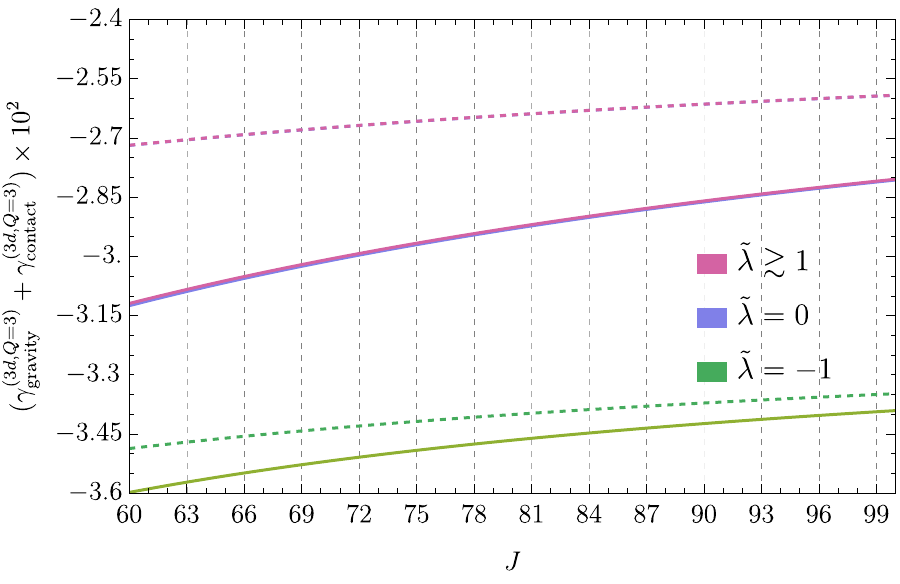}
    \caption{$Q=3$} 
\end{subfigure}
\hfill
\begin{subfigure}{0.32\textwidth}
    \includegraphics[width=\textwidth]{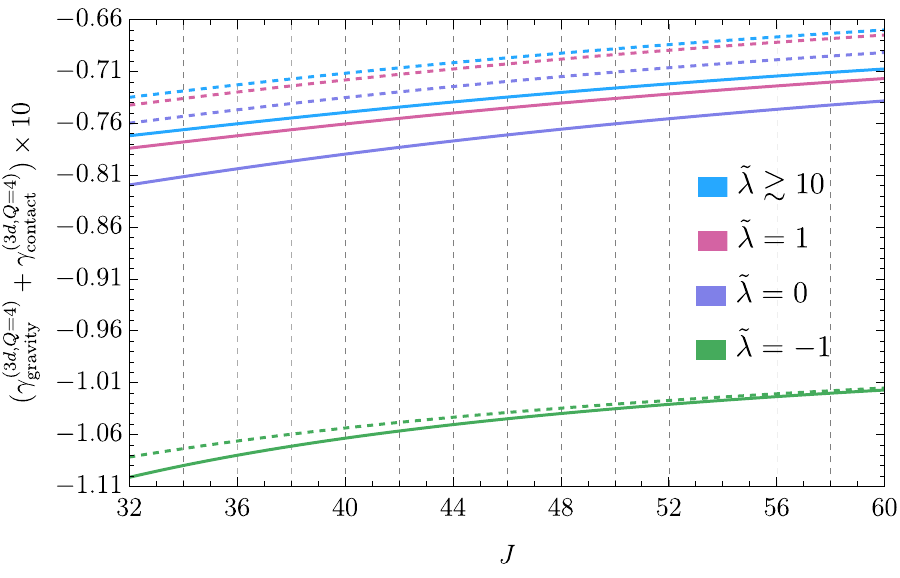}
    \caption{$Q=4$}
\end{subfigure}
\hfill
\begin{subfigure}{0.32\textwidth}
    \includegraphics[width=\textwidth]{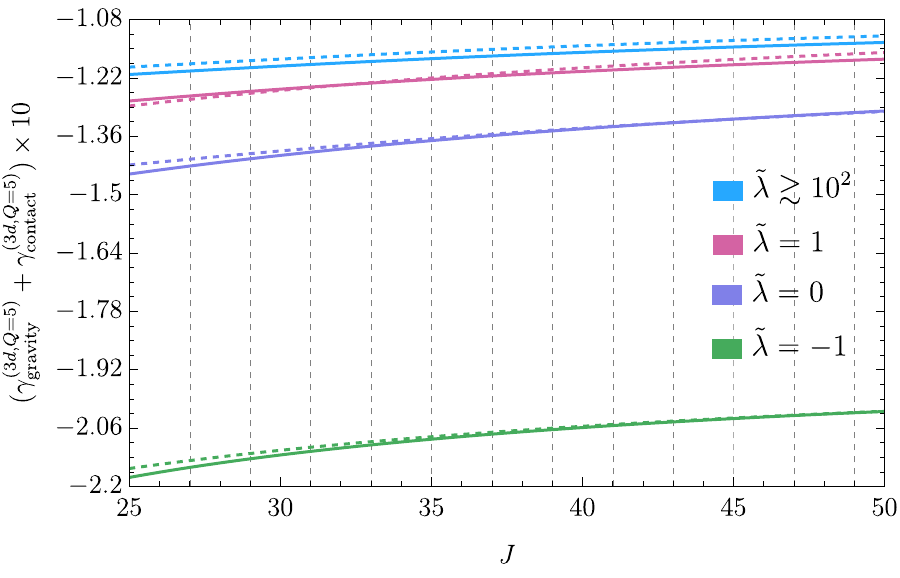}
    \caption{$Q=5$}
\end{subfigure}
\caption{$3d$ anomalous dimensions for $Q$-particle states due the graviton exchange and contact terms at $\Delta=\frac{3}{2}$ for different values of $\tilde{\lambda}$ ($\tilde{\lambda}=0$ coincides with $\gamma_{\mathrm{gravity}}^{(3d,Q)}$). Solid lines represent even spin, while dashed ones odd spins.} 
\label{fig:gravityContactd32}
\end{figure}

So far we have seen the effect of adding a contact term to the lower spin states, but what happens at large $J$? Since,  regardless of the value of $J$ and $Q$, the contributions from $(\phi \phi^\dagger)^2$ and $\phi\phi^\dagger \partial_\mu \phi \partial^\mu \phi^\dagger$ always appear in the same linear combination seen in~\eqref{contactQ},  we define
\eqna{
\tilde{\lambda} \equiv \lambda_0- \lambda_1 \Delta  \left(\Delta -\frac{\mathit{d}}{2}\right)\, .
}[]
As one can see in Fig.~\ref{fig:gravityContactd52} and~\ref{fig:gravityContactd32} in $3d$,\footnote{The results are analogous in $4d$, so we are not explicitly including them.}  adding some contact terms shifts the curves but  does not change the overall behavior as a function of $J$ as we have seen in~\eqref{gammaGravGen}. 
More precisely, as  $\tilde{\lambda}$ gets more and more negative, the $J=\infty$ asymptotic value of $\gamma^{(d,Q)}$ is pushed towards more and more negative values.  In the opposite direction, \textit{i.e.}  $\tilde{\lambda}>0$, instead,  the asymptotic value of $\gamma^{(d,Q)}$ increases as we dial $\tilde{\lambda}$ until it saturates a certain point,  where it does not change any more regardless the size of $\tilde{\lambda}$. Finally, notice how the effect of turning on a positive contact term has a less pronounced effect for $\Delta=\frac{d}{2}$.  This should be related to the fact that the $(Q-1)$ wavefunctions are not perfectly overlapping since the $(Q-1)$-`blob' has some internal spin, thus mitigating the size of contact term interactions, which, as the name suggest,  gets more relevant the closer the partons are to each others. 
\subsection{Phase diagrams}\label{sec:PhaseDiagram}
\begin{figure}
\centering
\begin{subfigure}{0.48\textwidth}
    \includegraphics[scale=0.4]{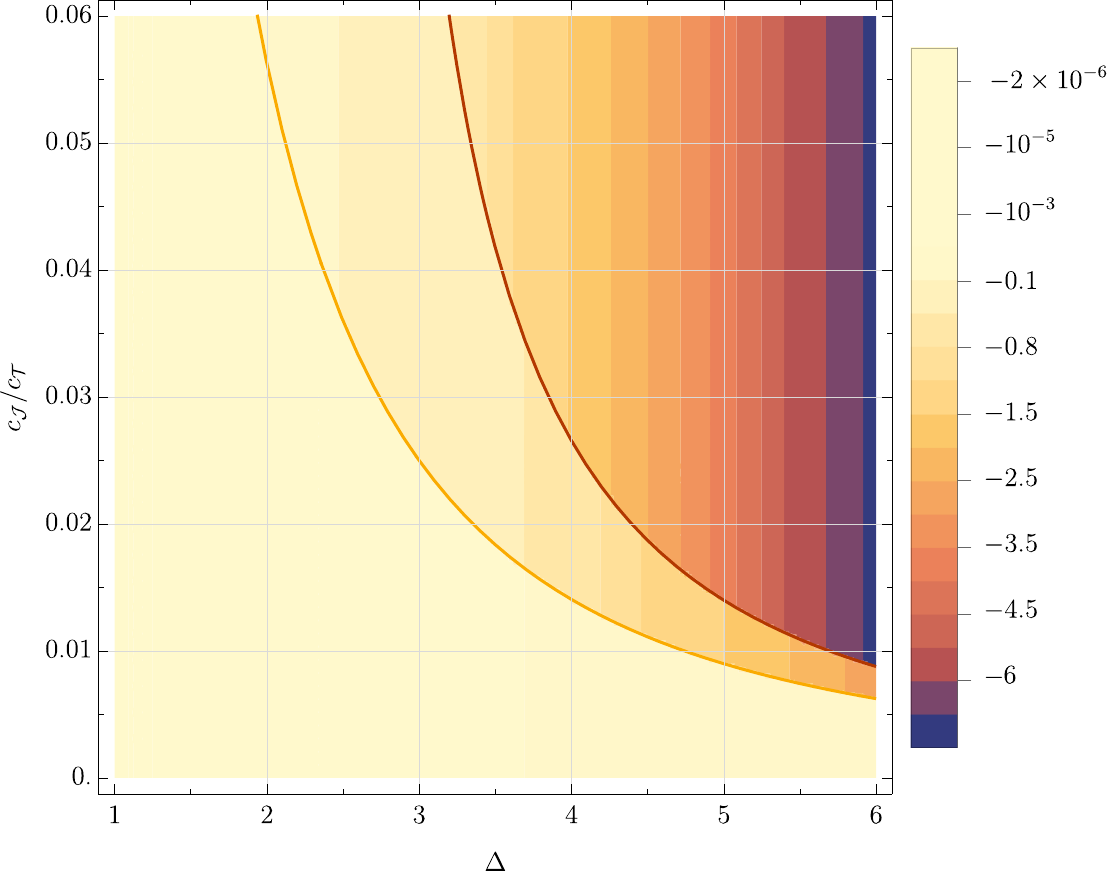}
    \caption{$4d$} 
\end{subfigure}
\hfill
\begin{subfigure}{0.48\textwidth}
   \includegraphics[scale=0.4]{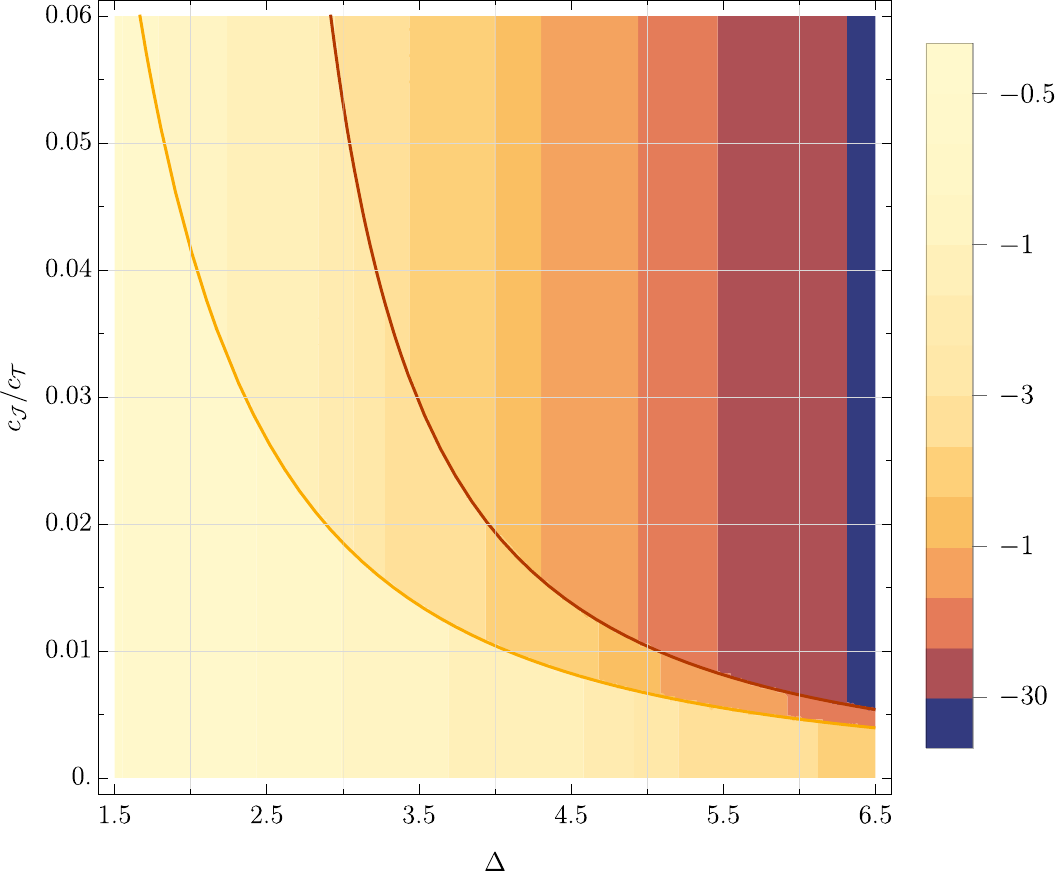}
    \caption{$3d$}
\end{subfigure}
\caption{$\frac{\partial \gamma^{(d, Q)}}{\partial (c_\CJ / c_\CT)}$ as a function of $\Delta$ and $ (c_\CJ / c_\CT)$ at fixed $J=100$ and $Q=3$ in four and three dimensions. The lines correspond to a function $ (c_\CJ / c_\CT)=\beta(\Delta)$ such that the system passes from the repulsive $\to$ attractive phase (orange line) and $a_0=\gamma^{(d, Q=2)}(c_\CJ / c_\CT)|_{\ell=2}$ $\to$  $a_0=\gamma^{(d, Q=2)}(c_\CJ / c_\CT)|_{\ell=0}$ (red line). } \label{fig:Q3phases}
\end{figure}
\begin{figure}
\centering
\begin{subfigure}{0.32\textwidth}
    \includegraphics[width=\textwidth]{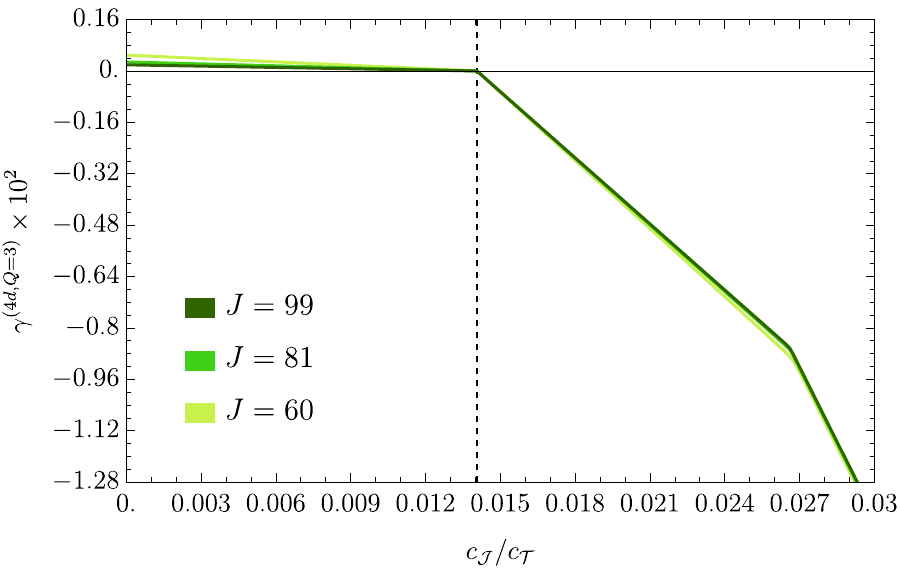}
    \caption{$Q=3$}  \label{fig:Q3phaseD4}
\end{subfigure}
\hfill
\begin{subfigure}{0.32\textwidth}
    \includegraphics[width=\textwidth]{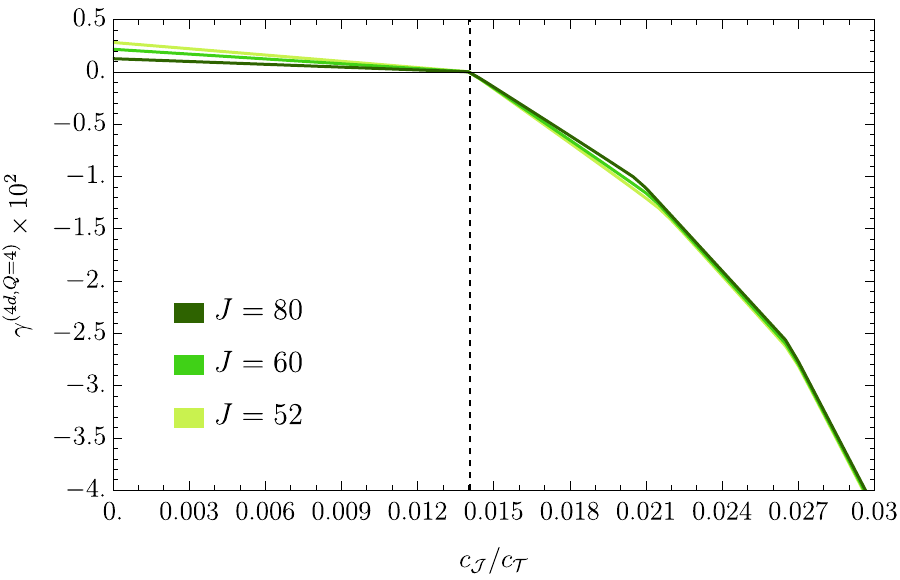}
    \caption{$Q=4$}\label{fig:Q4phasesD4}
\end{subfigure}
\hfill
\begin{subfigure}{0.32\textwidth}
    \includegraphics[width=\textwidth]{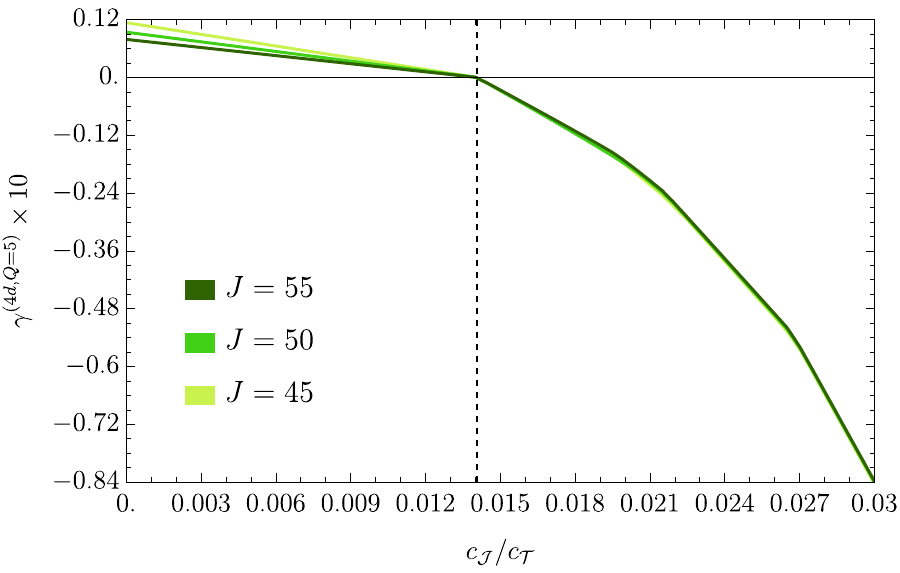}
    \caption{$Q=5$}\label{fig:Q5phasesD4}
\end{subfigure}
\caption{$4d$ anomalous dimensions for $Q$-particle states due the photon and  graviton exchange combined at $\Delta=4$ as a function of $c_\CJ/c_\CT$. At fixed $Q$, every line corresponds to a different value of the spin.  There are $Q$ different slopes corresponding to different configurations of the partons. } \label{fig:D4phase}
\end{figure}
\begin{figure}\centering
\includegraphics[scale=0.45]{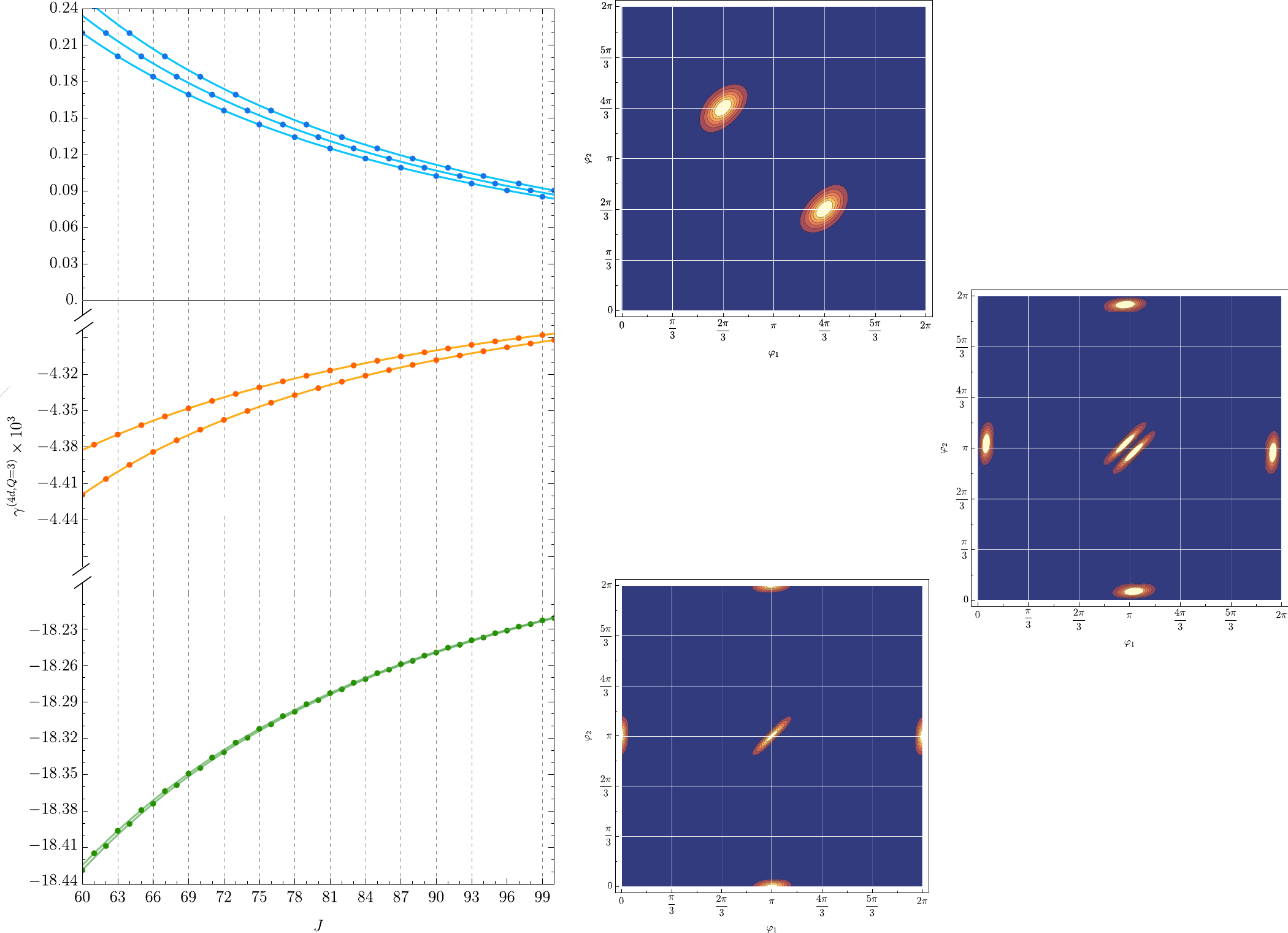}
\caption{Anomalous dimension and corresponding ground state angular wave-function (setting one particle at $\varphi_3=0$) $|\Psi_{\mathrm{min}}(\varphi_1, \varphi_2)|$ for $Q=3,\Delta=4$ for different values of $\frac{c_\CJ}{c_\CT}$.  \textbf{Top} (blue, $\frac{c_\CJ}{c_\CT}=\frac{1}{128}$) is in the repulsive phase and the three partons are evenly distributed around the equator with angular separation $2\pi/3$. \textbf{Middle} (orange, $\frac{c_\CJ}{c_\CT}=\frac{13}{640}$) has entered the attractive phase, with one parton on one side of AdS and  two nearby partons on the other side. The effect of the repulsive gauge interaction still visibly keeps the two nearby partons separated, with internal spin 2.  \textbf{Bottom} (green, $\frac{c_\CJ}{c_\CT}=\frac{21}{640}$) is deeper into the attractive phase, so that now the  internal spin of the two-particle blob is zero.}\label{fig:q3D4ctcJGW}
\end{figure}

In the previous sections we have discussed  the contribution to the anomalous dimensions of  the lowest-twist and lowest-energy $Q$-particle state from a single photon or graviton exchange separately.  Combining them together, we obtain the full anomalous dimension as a function of the central charges:
\eqna{
\gamma^{(d, Q)}_{\mathrm{tot}}=\frac{2 \pi ^{\mathit{d}/2} (\mathit{d}-2) \Gamma (\mathit{d})}{\Gamma
   \mleft(\frac{\mathit{d}}{2}\mright)^3} \frac{1}{c_{\CJ}} \gamma^{(d, Q)}_{\mathrm{gauge}}+\frac{4 \pi ^{\mathit{d}/2} \Gamma (\mathit{d}+2)}{(\mathit{d}-1) \Gamma
   \mleft(\frac{\mathit{d}}{2}\mright)^3} \frac{1}{c_{\CT}} \gamma^{(d, Q)}_{\mathrm{gravity}}\, . 
}[]
For ease of notation, we can multiply this expression by a  prefactor 
\eqna{
\gamma^{(d, Q)} (c_\CJ / c_\CT)\equiv \frac{  c_\CJ \Gamma
   \mleft(\frac{\mathit{d}}{2}\mright)^3}{2 \pi ^{\mathit{d}/2} (\mathit{d}-2) \Gamma (\mathit{d})} \gamma^{(d, Q)}_{\mathrm{tot}}=\gamma^{(d, Q)}_{\mathrm{gauge}}+\frac{2d(d+1)}{(d-2)(d-1)} \frac{c_\CJ}{c_\CT}\gamma^{(d, Q)}_{\mathrm{gravity}}\, ,
}[]
and study the behavior of this function as we vary the ratio $\beta\equiv c_\CJ/c_\CT$ and the external dimension $\Delta$.  In Fig.~\ref{fig:Q3phases},  we show the behavior of the slope $\frac{\partial \gamma^{(d, Q=3)}}{\partial \beta}$ for $Q=3$ in four and three dimensions.\footnote{To avoid dealing with the problems caused by the pole at $\frac{3}{4}$ discussed in Sec.~\ref{Sec:smallSpinContact}, in $3d$ we will restrict to $\Delta\geq \frac{3}{2}$.} To read this phase diagram, suppose to fix $\Delta$, for example $\Delta=4$, and move along the $y$-axis, increasing the value of $c_\CJ/c_\CT$.  To help the reader, in Fig.~\ref{fig:Q3phaseD4}, we  show the behavior of $\gamma(\beta)$ for this exact example.  The system starts in the repulsive phase, for $\beta=0$
\eqna{
\gamma^{(d, Q=3)}(0)=\gamma^{(d, Q=3)}_{\rm gauge} >0\, ,
}[]
and stays in this phase until we reach a critical value (orange line in Fig.~\ref{fig:Q3phases})
\eqna{
\gamma^{(d, Q)} >0 \quad \to \quad \gamma^{(d, Q)} <0 \qquad \text{for} \qquad \beta^*_{\mathrm{rep}\to \mathrm{att}}=\frac{(\mathit{d}-2) (\mathit{d}-1)}{\mathit{d} (\mathit{d}+1) \Delta ^2}\,. 
}[repToAttr]
This is  exactly the value of $\beta$  we see from \eqref{eq:RepulsiveVsAttractive},  describing the transition from the repulsive to the attractive phase for the double-twist anomalous dimensions.  It turns out that, since we are restricting to tree-level exchanges, this behavior is universal and  the transition occurs at $\beta^*_{\mathrm{rep}\to \mathrm{att}}$ for any value of $Q$.\footnote{To be precise, for $Q\geq 3$, the value of $\beta^*_{\mathrm{rep}\to \mathrm{att}}$ can deviate from this value and it can be moved towards zero if we add a $\lambda (\phi \phi^\dagger)^2$-contact term with a large enough $\lambda<0$. } Once we have entered the attractive phase, we know from our previous discussion that 
\eqna{
\gamma^{(d, Q)} (\beta)\sim a_0 (\beta)+O(1/J)\, , \qquad \forall Q\, .
}[]
As we can sense from the phase diagram, at $Q=3$,  $a_0(\beta)$ can change depending on the value of $\Delta$  and the system can undergo an additional phase transition (red line in Fig.~\ref{fig:Q3phases})
\eqna{
a_0(\beta)=\begin{cases}
\gamma^{(d, Q=2)}|_{\ell=2} \xrightarrow[]{\tilde{\beta}^*}\gamma^{(d, Q=2)}|_{\ell=0}\, ,  & \quad \text{ for }\, \Delta>\Delta^* \, ,\\
\gamma^{(d, Q=2)}|_{\ell=2} \, ,  & \quad \text{ for }\, \Delta<\Delta^* \, ,
\end{cases}
}[a0Q3phase] 
where the exact value of $\Delta^*$ can be read off from the expressions in Sec.~\ref{Sec:smallSpinContact}. The notation `$\gamma_1 \stackrel{\tilde{\beta}^*}{\rightarrow} \gamma_2$' here  indicates that as the value of $\beta$ crosses $\tilde{\beta}^*$, the constant term $a_0(\beta)$ in (\ref{a0Q3phase}) transitions from $\gamma_1$ to $\gamma_2$. 
The example $\Delta=4$ in $4d$ falls into the first case.  In the attractive phase, the gravitational interaction between the partons, tends to pull them together and form two blobs. $a_0$ corresponds exactly to the energy of the blob with the lowest twist.  Close to the transition between attractive and repulsive phases, the residual gauge repulsion causes the 2-parton blob with $\ell=2$ to be the energetically most favorable  configuration.  Eventually, at sufficiently large values of $\beta$,   $a_0$ settles to the value we have studied in Sec.~\ref{Subsec: gravity}.  As a visual summary, in Fig.~\ref{fig:q3D4ctcJGW}, we show the  anomalous dimensions of the lightest state and corresponding wave-functions for the example $\Delta=4$ for the different phases we have described: in order $\beta<\beta^*_{\mathrm{rep}\to\mathrm{att}}$, $\beta^*_{\mathrm{rep}\to\mathrm{att}}<\beta<\tilde{\beta}^*$ and $\beta>\tilde{\beta}^*$.  Notice in particular that in the intermediate phase, since the internal blob has some internal spin, the two partons are not exactly at $\varphi_{i}=0$ or $\pi$ but there is some separation.

\begin{figure}
\centering
\begin{subfigure}{0.32\textwidth}
    \includegraphics[width=\textwidth]{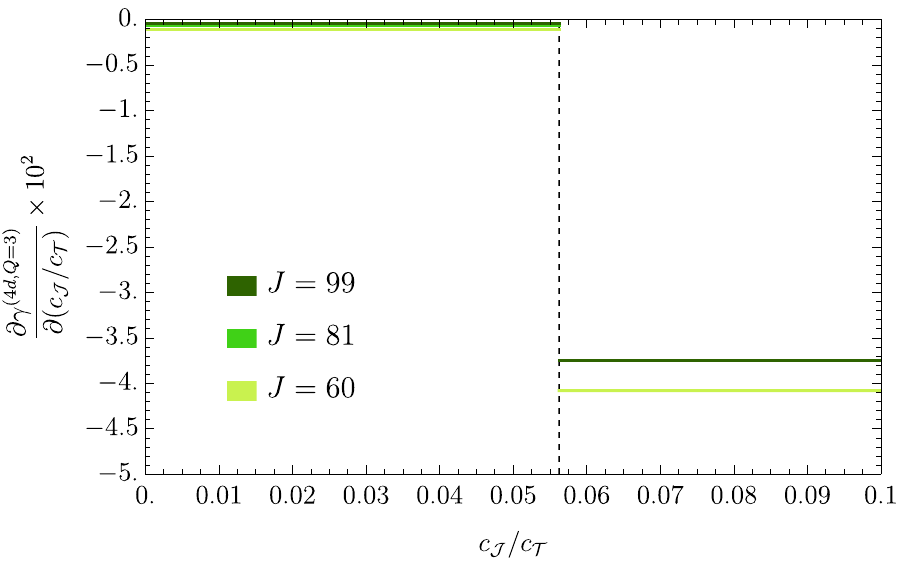}
    \caption{$Q=3$} \label{fig:Q3phasesD2}
\end{subfigure}
\hfill
\begin{subfigure}{0.32\textwidth}
   \includegraphics[width=\textwidth]{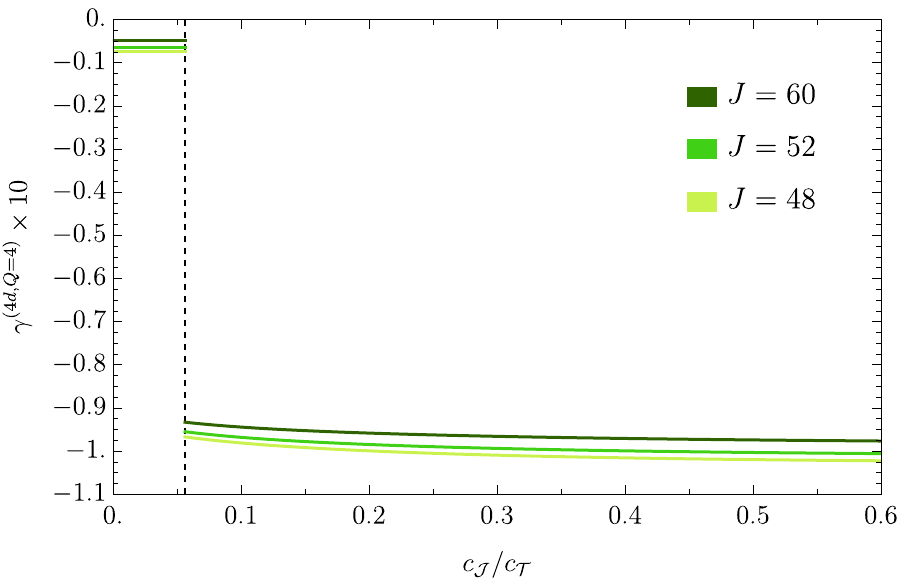}
    \caption{$Q=4$}\label{fig:Q4phasesD2}
\end{subfigure}
\hfill
\begin{subfigure}{0.32\textwidth}
   \includegraphics[width=\textwidth]{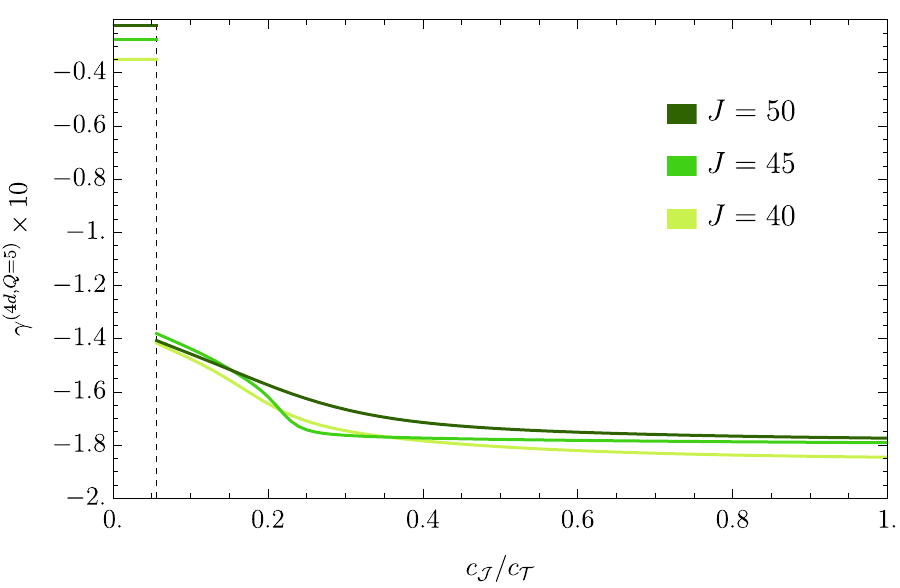}
    \caption{$Q=5$}\label{fig:Q5phasesD2}
\end{subfigure}
\caption{Derivative of the $4d$ anomalous dimensions for $Q$-particle states due the photon and  graviton exchange combined at $\Delta=2$ as a function of $c_\CJ/c_\CT$. At fixed $Q$, every line corresponds to a different value of the spin.  } \label{fig:D2phase}
\end{figure}

When $\Delta=\frac{d}{2}$, see for example $\Delta=2$ in $4d$ in Fig.~\ref{fig:Q3phasesD2},  the system falls in the second category listed in~\eqref{a0Q3phase}. 
In this case,  the system undergoes only the transition repulsive $\to$ attractive. After that, the anomalous dimension is given by~\eqref{a0Q3phase} with $a_0=\gamma^{(d, Q=2)}(c_\CJ / c_\CT)|_{\ell=2}$.  This is consistent with our previous discussion: for $\Delta=\frac{d}{2}$ at $Q=2$ the spin two is always the lowest twist configuration. 

\begin{figure}
\centering
\begin{subfigure}{0.48\textwidth}
    \includegraphics[scale=0.4]{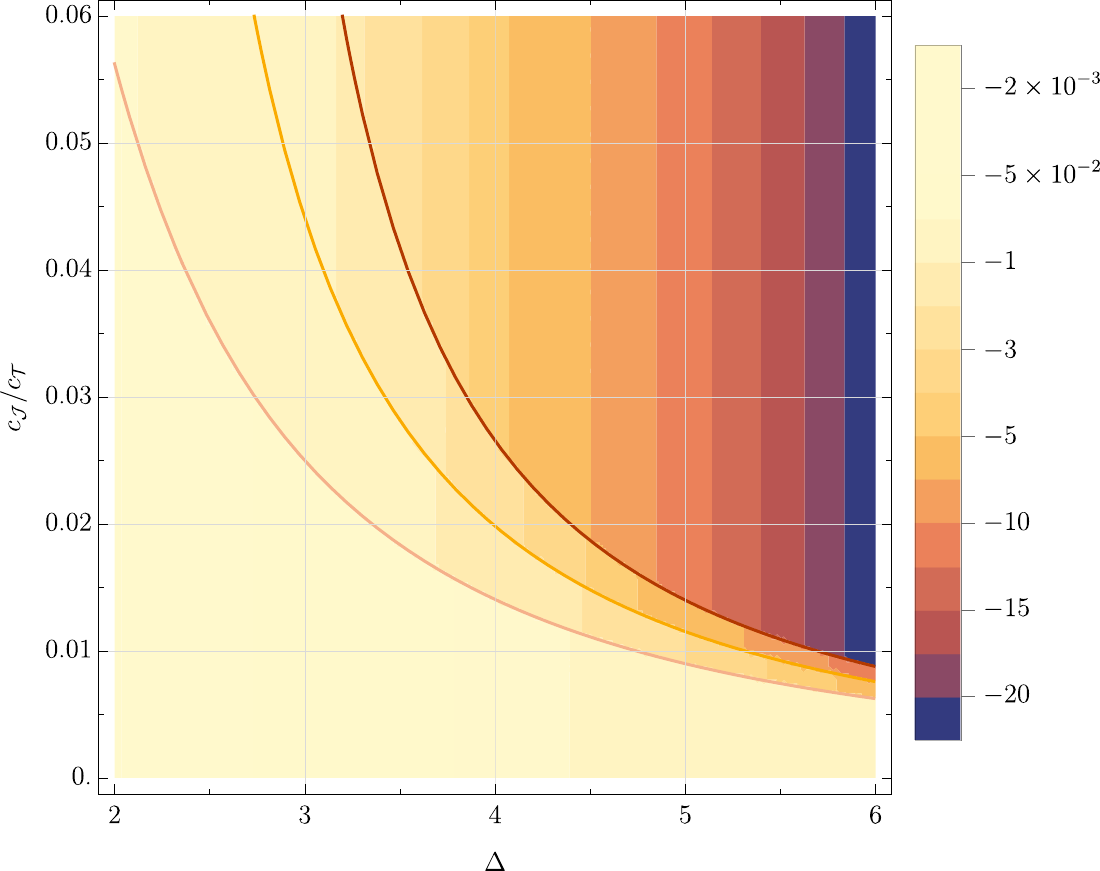}
    \caption{$4d$} 
\end{subfigure}
\hfill
\begin{subfigure}{0.48\textwidth}
   \includegraphics[scale=0.4]{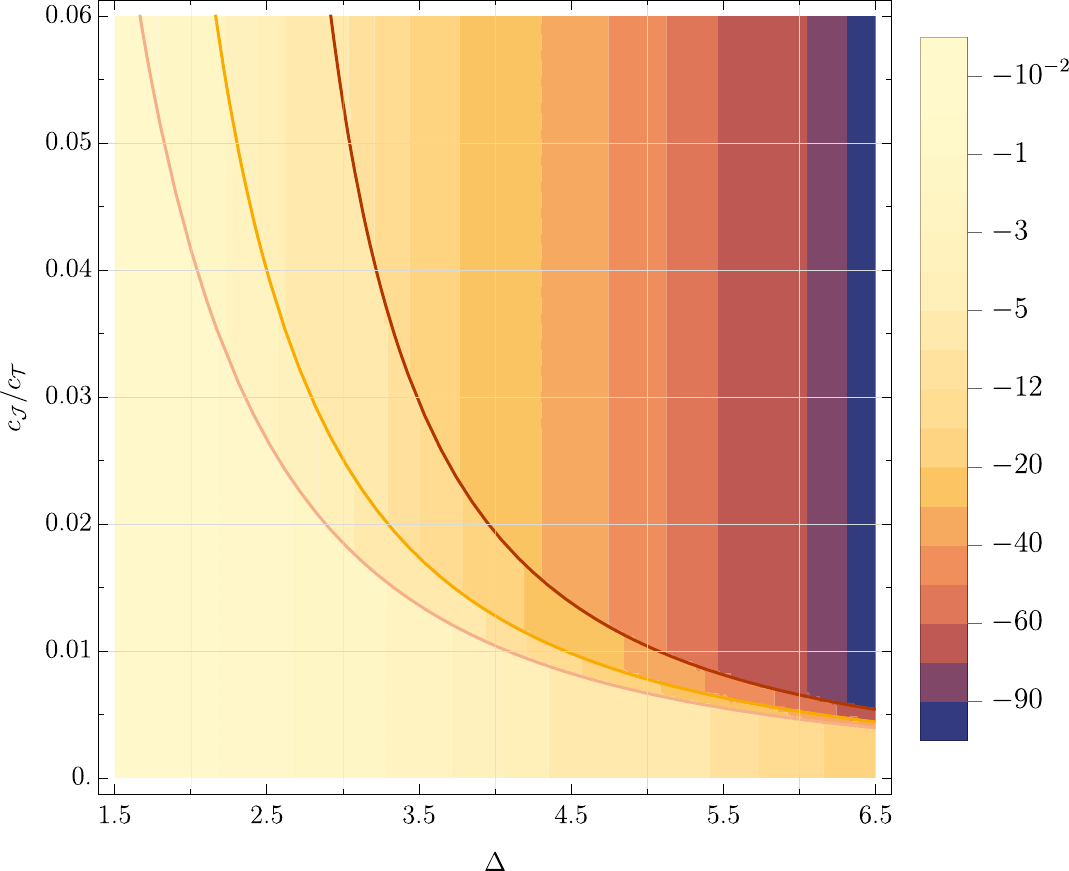}
    \caption{$3d$}
\end{subfigure}
\caption{$\frac{\partial \gamma^{(d, Q)}}{\partial (c_\CJ / c_\CT)}$ as a function of $\Delta$ and $ (c_\CJ / c_\CT)$ at fixed $J=100$ and $Q=4$ in four and three dimensions. The lines correspond to a function $ (c_\CJ / c_\CT)=f(\Delta)$ such that the system undergoes different transitions.  Referring to the notation in the main text, the pink line corresponds to $\beta^*_{\mathrm{rep}\to\mathrm{att}}(\Delta)$, the orange line corresponds to $\beta^*_1(\Delta)$ and the red line corresponds to $\beta^*_2(\Delta)$.} \label{fig:Q4phases}
\end{figure}

For larger values of $Q>3$, the description of the different phases becomes more complicated.  In Fig.~\ref{fig:Q4phases}, we show the behavior of the slope $\frac{\partial \gamma^{(d, Q=4)}}{\partial \beta}$ in $4d$ and $3d$ as a function of $\beta=c_\CJ/c_\CT$ and $\Delta\geq \frac{d}{2}$ at $Q=4$ --- for more concrete examples at fixed $\Delta$ and $Q=4,5$, see Fig.~\ref{fig:Q4phasesD4}, \ref{fig:Q5phasesD4} and Fig.~\ref{fig:Q4phasesD2}, \ref{fig:Q5phasesD2}.
As before, looking at the general phase diagram for four-particle primaries,  we can think of fixing $\Delta$  and moving on a vertical line.  Dialing from smaller to larger $\beta$, the system passes through the following phases:
\begin{align}
\beta &<\beta^*_{\mathrm{rep}\to\mathrm{att}}(\Delta)=\frac{(\mathit{d}-2) (\mathit{d}-1)}{\mathit{d} (\mathit{d}+1) \Delta ^2} \, \quad \text{repulsive}\\ \nonumber 
\beta &> \beta^*_{\mathrm{rep}\to\mathrm{att}}(\Delta) \, \quad \text{attractive:}\,  a_0=\begin{cases} 2\gamma^{(Q=2)}|_{\ell=2} \xrightarrow[]{\beta^*_1} \gamma^{(Q=3)}|_{\ell=3} \xrightarrow[]{\beta^*_2} \gamma^{(Q=3)}|_{\ell=0}  & \Delta>\Delta^*\, , \\
2\gamma^{(Q=2)}|_{\ell=2} \xrightarrow[]{\beta^*_1} \gamma^{(Q=3)}|_{\ell=3}   & \Delta<\Delta^*\, .
\end{cases}
\end{align}
As soon as we move away from the repulsive phase into the attractive one, the four partons tend to cluster together  into two separate `blobs' due to the attractive interactions. Since they are charged, if $c_\CJ$ is still small enough, there is some residual repulsion between the partons, so  it is favorable to form two `blobs',  with two partons each, spinning around each other and with internal spin $\ell=2$. This is why at first $a_0$ is given by twice the $Q=2$ spin-2 anomalous dimensions.  Then we pass to the configuration described in Sec.~\ref{Subsec: gravity} with $3$ partons on one side of AdS and one isolated on the opposite side.  Depending on whether $\Delta \lessgtr \Delta^*$, the system can undergo another transition passing from a spin-3 to a scalar `blob' or stop with $a_0= \gamma^{(Q=3)}|_{\ell=3}$, if that is the minimum-twist configuration. 

It is interesting and perhaps surprising that we sometimes see it is favorable at $Q=4$ to form two $Q=2$ blobs rather than a $Q=3$ blob and a $Q=1$ parton. A $Q=3$ blob has 3 different ways to pairwise exchange a graviton, whereas a $Q=2$ blob has only one. So, if we neglect the wavefunction spread of the partons, then the magnitude of the binding energy of a $Q=3$ blob should be at least three times as large as that of a $Q=2$ blob, so that it is never favorable for a $Q=4$ state to form two $Q=2$ blobs.  However, at small values of $\Delta$ the wavefunction spread can be significant; intuitively, there can be Bose enhancement in the $Q=3$ state of regions where the size of the binding energy is diminished. 
\section{Large $Q$ Limit}
\label{sec:LargeQ}
In this section, we consider extending our results to large values of $Q$ in order to make contact with the phase diagram of large $Q$ and large $J$.  We will first consider the `repulsive phase', where the gauge boson exchanges keep the partons far apart at large $J$, so that a perturbative holographic analysis is still reliable.  Then, we will consider the `attractive phase', where a holographic analysis is still useful but must be supplemented with some nonperturbative information.  In the limit of very large $Q$, we describe the qualitative picture that we believe emerges.  At smaller values $Q=3,4$, we believe the same qualitative picture holds but now we can be much more precise by injecting a small amount of nonperturbative data into our bulk computations.  In particular, we include an additional scalar in the bulk corresponding to $\epsilon$, which has $\Delta \approx 1.511$, as well as $\phi^4$ and $\phi^6$ contact terms fixed by matching the $Q=2$ spin-0 operator with dimension $\Delta \approx 1.24$ and the $Q=3$ spin-0 operator with dimension $\Delta \approx 2.1$.  

\subsection{Repulsive Phase}

In the repulsive phase, large $Q$ is still under control as long as $J$ is sufficiently large, because the particles repel and so remain distantly separated from each other. In fact, we can immediately apply our result (\ref{eq:MultQAnomDim}) from our perturbative bulk computation at large $J$ but  $Q$ arbitrary:
\begin{equation}
\gamma^{(d,Q)}(J) \approx \frac{Q}{2} \sum_{n=1}^{Q-1} \gamma^{(d,Q=2)}(2\frac{J}{Q} \left| \sin ( \pi n/Q )\right|).
\end{equation}

At large $J$, we can approximate $\gamma^{(d,Q=2)}(J) \sim \frac{A}{J^{d-2}}$, where $A$ can be read off from (\ref{DoubleTwistLargeJ}). Then, in $d=4$, we obtain
\begin{equation}
\gamma^{(4d,Q)}(J) \approx \frac{A  (Q^2-1)Q^3}{24 J^2}.
\end{equation}
In $d=3$, the sum 
\begin{equation}
\gamma^{(3d,Q)}(Q,J) \approx A \frac{Q^2}{4J} \sum_{n=1}^{Q-1} \frac{1}{  \left| \sin ( \pi n/Q )\right|},
\end{equation}
does not appear to have a simple closed form expression, though it can easily be evaluated numerically, e.g.,
\begin{equation}
\begin{aligned}
\frac{\gamma^{(3d,Q)}(J)}{\gamma^{(3d,Q=2)}(J)} &\approx (1, \frac{4}{\sqrt{3}}, 1+2 \sqrt{2}, 4 \sqrt{1+\frac{2}{\sqrt{5}}}, 5+\frac{4}{\sqrt{3}}) \approx (1,  2.31, 3.83, 5.51, 7.31) \\
& \textrm{ for } Q=(2,3,4,5,6) .
\end{aligned}
\end{equation}
In the limit of large $Q$ we can approximate the sum on $n$ as an integral to obtain
\begin{equation}
\gamma^{(3d,Q)}(J) \stackrel{Q \gg 1}{\approx } A \frac{Q^2}{2J} \int_{\frac{2 \pi}{Q}}^{\pi} \frac{ Q d\theta}{2\pi} \frac{1}{\left|\sin(\theta/2)\right|} \approx \frac{A}{J} \left( \frac{Q^3 \log Q}{2\pi}  +  O(Q^3)\right).
\end{equation}

In summary, we find in $d=3,4$ that
\begin{equation}
\gamma^{(4d,Q)}(J) \approx A \frac{  Q^5}{24 J^2}, \qquad \gamma^{(3d,Q)}(J) \approx A \frac{ Q^3 \log Q}{2\pi J}.
\end{equation}
Because the angular separation between the partons shrinks at large $Q$, there is an enhancement (by $Q$ in $d=4$ and by $\log Q$ in $d=3$) over what one might naively expect by estimating the anomalous dimension to be $\sim Q^2/(J/Q)^{d-2}$. Therefore, this calculation breaks down unless $J\gg Q^3 \log Q$, $J^2 \gg Q^5$ in $d=3,4$, respectively.  In practice, it should be possible to go beyond this limit, and require only $J\gg Q^2$.  The reason is that only the weaker condition is required for the individual photon exchanges to be perturbative. So, the breakdown of the expansion in that regime comes from the sum over a large number of charged particles, each of which individually exerts a negligible force on the others. Therefore, one should be able to resum these interactions by treating the collective gauge field background classically, and solving for the spectrum of the bulk scalar in this background, corresponding to a resummation of multi-stress-tensor operators in an Eikonal limit of the CFT \cite{Cornalba:2007fs,Cornalba:2007zb,Fitzpatrick:2015qma,Fitzpatrick:2019zqz,Fitzpatrick:2019efk,Kulaxizi:2017ixa,Kulaxizi:2018dxo}.  Because of the backreaction of the bulk scalar mode on the gauge field, however, one has to solve a coupled system of equations. We will not attempt such a calculation in this paper.

\subsection{Attractive Phase and O(2) Model}
\subsubsection{Qualitative Picture at Very Large $Q$}
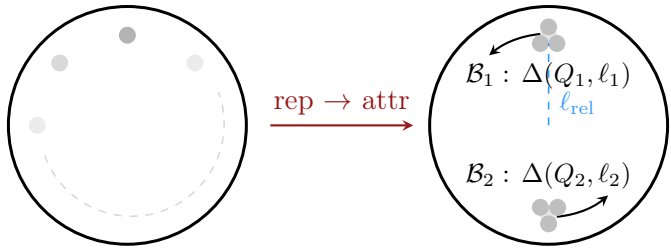
\begin{figure}[t!] \centering
 \begin{tikzpicture}
 \begin{scope}[yshift=1cm]
  \begin{scope}[yshift=1cm]
 \draw[line width=0.4mm] (-3.5,-0.5) circle (45pt);
  \filldraw[gray!60!white] (-3.5,0.7) circle (3pt);
    \filldraw[gray!30!white] (-4.4,0.33) circle (3pt);
  \filldraw[gray!15!white] (-4.7,-0.5) circle (3pt); 
      \filldraw[gray!15!white] (-2.6,0.33) circle (3pt);
           \draw[dashed,semithick,gray!30!white] (-4.6,-0.9) arc[radius=35pt, start angle=200, end angle=380];
           
  \end{scope}

   \begin{scope}[xshift=5.6cm]
     \draw[myRed, thick, ->, -stealth] (-7.2,0.5)--(-5.3, 0.5);
     \node[above=0cm, myRed] at (-6.25, 0.5) {\small rep $\to$ attr};
 \draw[line width=0.4mm] (-3.5,0.5) circle (45pt);
  \draw[->,thick,-stealth] (-3.5,1.72) arc[radius=35pt, start angle=90, end angle=130];
     \draw[->,thick,-stealth] (-3.5,-0.72) arc[radius=35pt, start angle=-90, end angle=-50];
         \filldraw[gray!20!white] (-3.5,-0.7) circle (3pt); 
                  \filldraw[gray!20!white] (-3.5,1.7) circle (3pt); 
  \filldraw[gray!50!white] (-3.5,1.81) circle (3pt);
  \filldraw[gray!50!white] (-3.5,-0.81) circle (3pt);
    \filldraw[gray!50!white] (-3.61,1.59) circle (3pt); 
        \filldraw[gray!50!white] (-3.39,1.59) circle (3pt); 
          \filldraw[gray!50!white] (-3.61,-0.59) circle (3pt);
            \filldraw[gray!50!white] (-3.38,-0.59) circle (3pt);
\node[below=0.1cm] at (-3.5, 1.59) {\footnotesize $\mathcal{B}_1: \, \Delta(Q_1, \ell_1)$};
\node[above=0.1cm] at (-3.5,-0.59) {\footnotesize $\mathcal{B}_2: \, \Delta(Q_2, \ell_2)$};
\draw[blul, semithick, dashed] (-3.5, 0.5) -- (-3.5, 1.59);
\node[right=0cm, blul] at (-3.5, 0.8) {\footnotesize  $\ell_{\mathrm{rel}}$};
  \end{scope}
  \end{scope}
  \end{tikzpicture}
  \caption{Cartoon version of the transition between the repulsive phase to the attractive phase in the O(2) model. In the repulsive phase, the minimal energy configuration is given by evenly distributed partons (\textit{left}). Turning on the attractive interactions, the $Q$ partons form into `$\CB$lobs' $\CB_1$ and $\CB_2$ (\textit{right}). Each of them has some charge $Q_i$ and internal spin $\ell_i$ and they are rotating around the center of AdS with relative spin $\ell_\mathrm{rel}$.  The minimal energy configuration, in this case, is determined by the minimum twist of the individual `$\CB$lobs'.} \label{fig:bloblsCartoon}
\end{figure}
In contrast with the repulsive phase, in the attractive phase the partons at large separation are pulled towards each other, at which point the magnitude of the force between them grows and our perturbative bulk computation is no longer reliable.  Nevertheless, we try to combine the intuition gained from the perturbative analysis of previous sections together with the known phase structure of the O(2) model at large $Q$ and $J$ to obtain a quantitative prediction for the Regge regime at large $Q$ in the attractive phase.

As sketched in Fig.~\ref{fig:bloblsCartoon}, imagine starting with a configuration with the $Q$ partons spread evenly around the equator, which is the minimal energy configuration in the repulsive phase.  In the attractive phase, such a configuration is unstable and the partons will dynamically tend to clump together.  On the other hand, the large value of the conserved spin will prevent the partons from completely clumping together.  Instead, as we saw in the perturbative regime, at large spin they will form `blob's spinning around each other, where the large angular momentum $\ell$ of the state mainly arises from the large radius of rotation between the blobs. 
 More precisely, the charge $Q$ state will be a multi-twist state made from some number of primary states, which are the `blobs', whose combined charge is $Q$.  

Naively, one might expect that the lowest-energy configuration would be to have a multi-twist made from scalar blobs, since for fixed $Q$ the minimum energy increases as a function of spin.  However, total spin $J$ of a multi-twist state gets a contribution $\ell_{\rm rel}$ from the relative motion of the blobs as well as  from the intrinsic spins $\ell_1, \ell_2, \dots, \ell_n$ of the blobs themselves. Therefore, if we increase one of the $\ell_i$s, then we must decrease $\ell_{\rm rel}$, and it is not a priori obvious if the combined effect will be to increase or decrease the total energy of the state.  That is, label the individual blobs by their charges $Q_1,Q_2, \dots, Q_n$ and spins $\ell_1, \ell_2, \dots, \ell_n$, and define their energies to be
\begin{equation}
\Delta(Q_i, \ell_i).
\end{equation}
Then the energy of, for instance, a double-twist state $[\CO_1, \CO_2]_{J}$ will be
\begin{equation}
\Delta_{\rm tot}\!=\!\Delta([\CO_1, \CO_2]_{J})\! \approx \!\Delta(Q_1, \ell_1)+ \Delta(Q_2, \ell_2) + \ell_{\rm rel} .
\end{equation}
We have neglected contributions that vanish at large $\ell_{\rm rel}$. 
The point 
 becomes more transparent if we write everything in terms of the twists $\tau = \Delta-\ell$. Then
\begin{equation}
\tau_{\rm tot} = \Delta_{\rm tot}- J\approx \tau(Q_1, \ell_1) + \tau(Q_2 , \ell_2) ,
\end{equation}
since $\ell_1+\ell_2 + \ell_{\rm rel} = J$.  Consequently, the minimum energy two-blob state will actually be the double-twist state made from the blobs such that each individual blob has  minimum twist, rather than the minimum energy.  More generally, we can consider multi-twist states made from several smaller blobs, with a large relative $\ell_{\rm rel}$ between them, in which case the twists of all the blobs should be individually minimized. In this section   {\it we will assume that the lowest-energy state at large $J \gg Q^2$ can be described as such a multi-twist state made from some number of distantly-separated blobs}.

As we have just argued, the first consideration is to determine, for a blob with fixed $Q_i$, what is the spin $\ell_i$ of the blob that minimizes the twist, since the lowest-energy state in the Regge limit will be constructed from such minimal-twist blobs.  For any fixed value of $Q$, let $\tau_{\rm min}(Q)$ be the twist minimized over all $\ell$:
\begin{equation}
\tau_{\rm min}(Q) \equiv \min_{\ell } \tau(Q,\ell).
\end{equation}
To determine the total number of blobs, we have to ask when it is energetically favorable to have more blobs of less charge each or fewer blobs of more charge each.  Consider for simplicity the case where every blob has charge $q$, and take $Q$ sufficiently large that $Q/q$ can be approximated as a integer.  Then the minimal twist for a  state with $Q/q$ blobs of charge $q$ each, all infinitely separated from each other in AdS, is 
\begin{equation}
\tau = \frac{Q}{q} \tau_{\rm min}(q)  = Q r(q), \qquad r(q) \equiv \tau_{\rm min}(q)/q.
\end{equation}
We have defined the `twist-to-charge' ratio $r(q)$, since by inspection the twist of the full state is minimized by choosing $q$ of the individual blobs to minimize $r(q)$.  

To determine $r(q)$ for various values of $q$, we need some additional nonperturbative information.  By definition, $r(1)=\Delta$.  In the O(2) model, the low-lying $Q=2$ spectrum has been determined accurately and the minimum twist  occurs at $\ell=2$, with the value $\tau_{\rm min}(Q=2) = 1.0154$. Therefore  in this case, $r(2)= 0.5077$, which is slightly smaller than $r(1) = 0.519$.  

Already the value $r(2) = 0.5077$ is quite interesting.  Clearly it puts an upper bound on the minimum value of $r(q)$ over all possible $q$.  What makes this so useful is that the value of the twist-to-charge ratio $r(q)$ plays a similar role to the value of $\Delta$ in the double-twist case for determining whether gravity or gauge interactions are stronger at large distances.  That is, if we consider two charge-$q$ blobs far from each other, whether the attractive force of gravity or the repulsive force of gauge interactions is stronger depends on the relative size of (compare to (\ref{eq:RepulsiveVsAttractive}))
\begin{equation}
\frac{1}{c_\CJ}\gtrless \frac{6  r^2(q)}{ c_\CT} .
\label{eq:RepulsiveVsAttractive2}
\end{equation}
Using $c_\CJ = 0.904c_{J,\rm free}, c_\CT = 0.944c_{T,\rm free}, r(2) = 1.015 \Delta_{\rm free}$, and the fact that the two sides are equal in free theories, $6\Delta_{\rm free}^2 c_{J, \rm free} = c_{T, \rm free}$, one finds that the LHS is slightly larger than the RHS.  This implies that $(q=2)$-blob states repel each other at long distances.  Since $r(2)$ is an upper-bound on the minimum value of $r(q)$, this implies that the optimal-sized blobs (i.e., the ones with $q$ such that they minimize $r(q)$) will also repel each other at long distances.  Thus the results of the previous subsection, on the large $Q$ limit in the repulsive phase, would directly apply to the attractive phase as well, but with the individual partons replaced by blobs of the optimal size $q$!\footnote{An important caveat is that we have been implicitly assuming that $r(q)$ has a minimum, rather than simply decreasing as a function of $q$ to an asymptotic lower limit at infinite $q$.  If this were the case, the optimal value of $q$ for the blobs would simply be the total charge $Q$ of the entire state, and there would only be one such blob, which would contradict our assumption that the state at large $J$ separates into two or more smaller blobs. }

Determining $r(q)$ for $q>2$ requires additional nonperturbative input.  At $q=3$, there is a $\ell=3$ state with $r=0.527$ \cite{Liu:2020tpf}, and at $q=4$ there is a $\ell=4$ state with $r=0.56$. These data points put upper limits on the values of $r(3)$ and $r(4)$, though with less data for fewer values of $\ell$ available it is possible that these are not the best values of $\ell$ and the actual values of $r$ might be lower.  At least, it seems plausible given the available data that $r(q)$ might be minimized at $q=2$. 

\begin{figure}[t!]
\begin{center}
\includegraphics[width=0.6\textwidth]{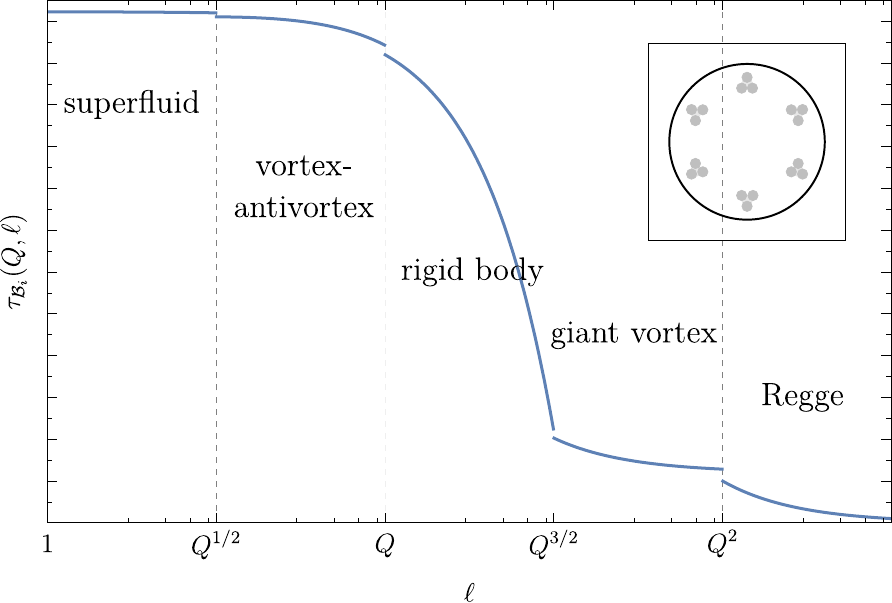}
\caption{Cartoon of the $\ell$-dependence of the twist of a state with charge $Q \gg 1$, determined by considering the corresponding phase that the state will be in. When the state enters  the Regge limit, we expect that it breaks apart into $Q/q$ bound state `blobs' of charge $q$ each for some $q$, which experience a mainly repulsive force between them at large distances.  }
\label{fig:tauBlob}
\end{center}
\end{figure}
Whatever the optimal value of $q$ is, it most likely is an $O(1)$ number, for the following reason.  For large values of $q$, we can turn to  large charge effective theory descriptions.
In particular, $\tau(Q,\ell)$ has already been worked out in the regimes with $\ell \ll Q^2$ --- see also Fig.~\ref{fig:tauBlob}:\\
\begin{itemize}
\item $1 \ll  \ell \ll Q^{1/2} : \frac{\partial}{\partial \ell} \tau(Q,\ell) = \frac{1}{\sqrt{2}} -1  <0 $ , \\

\item $Q^{1/2} \ll \ell \ll Q : \frac{\partial}{\partial \ell} \tau(Q,\ell) = \frac{\sqrt{Q}}{3 c_1 \ell}-1  < 0$ , \\

\item $Q \ll \ell \ll Q^{3/2} : \frac{\partial}{\partial \ell} \tau(Q,\ell) = \frac{\ell}{c_1 Q^{3/2}} -1 <0$ ,  \\

\item $Q^{3/2} \ll \ell \ll Q^2 : \frac{\partial}{\partial \ell} \tau(Q,\ell) = - \frac{9 c_1^2}{4 \pi} \frac{Q^3}{\ell^2} < 0$ . \\

\end{itemize}
In all these regions, the twist is a decreasing function of $\ell$, and so minimizing the twist pushes the spin to larger values.  The last regime is the Regge regime, at which point we assume the state breaks apart into $Q/q$ blobs of charge $q$ each, and the problem is reduced again to the problem of determining the optimal $q$ of the sub-blobs; the value of $r(Q)$ for the full state will then be strictly greater than $r(q)$ for the charge-$q$ sub-blobs, because of the repulsive force between them.  Note that if the minimum twist among the different EFT regimes had turned out to be at spin $\ell$ in one of the other phases, it would instead have implied that the sub-blobs in the Regge limit were described by states in that phase.  Although this analysis strictly applies only in the large $Q$ limit, in the O(2) model the EFT description at spin 0 is accurate already at small $Q$ \cite{Cuomo:2023mxg}; if the same is true of the other regimes, then $r(q)$ will start to grow as a function of $q$ before $q$ becomes large. One cautionary point however is that the minimum twist in the EFT regime occurs at $\ell \gtrsim Q^2$, so to find the minimum twist numerically will require computing the dimensions of states at fairly high spin.

\subsubsection{Quantitative Analysis at $Q=2, 3,4$}
\label{sec:O2}
In the previous section, we have discussed  in detail  our expectations for the dimension of the $Q$-particle state at very large $Q$ and $J$.  Here, we will go back to finite, small $Q$ and we will try to test our results, which we have derived perturbatively in Sec.~\ref{sec:DT} and~\ref{sec:FiniteQJ}, against known data for the spectrum of the  $3d$ O(2) CFT.

In recent years, new information about the spectrum of the O(2) model has become available thanks to the use of confomal bootstrap techniques,  both numerical~\cite{Kos:2013tga,Kos:2015mba,Chester:2019ifh}  as well as analytic~\cite{Alday:2015ewa,Liu:2020tpf}.  The lowest twist scalars at charge $Q=0,1,2,3$ are known to have dimensions, respectively~\cite{Liu:2020tpf}
\eqna{
\Delta_\epsilon=1.51136\, ,  \qquad \Delta_{\Phi}=0.519088\, ,  \qquad \Delta_t=1.23629\, ,  \qquad \Delta_{\chi}=2.1086\, ,
}[deltaO2]
and the conserved current and stress-tensor central charges take  the values
\eqna{
c_{\CJ}=1.809\, , \qquad\qquad c_{\CT}=2.832\, 
\qquad \text{ with }\quad \frac{c_\CJ}{c_\CT}>\frac{1}{6\Delta_{\Phi}^2}\, .
}[centralO2]
Since the ratio of the central charge is greater than the critical value in~\eqref{repToAttr}, the system is in the attractive phase (``gravity is the strongest force''). 

In our setup, we can identify the charge-one scalar with conformal dimension $\Delta_\Phi$ as the CFT dual  of the AdS scalar $\phi$ and use the action  in~\eqref{action}.  Then we can proceed in computing the anomalous dimensions for $Q$-particle states as described in Sec.~\ref{sec:FiniteQJ} with two important modifications.  The first modification is that we have to include a small number of contact terms such as $\phi^4, \phi^6$ to account for the spin-0 multi-twist operators, whose dimensions are not well-approximated by bulk interactions and must be put in by hand to match numeric results.  
The second modification is related to the presence of the neutral scalar $\epsilon(x)$ appearing in the OPE of $\Phi\times \Phi^\dagger$. Although it has a higher twist than the currents, the twist is not that much larger and in practice its contribution is greater than that of the currents until one gets to extremely large $J \gtrsim 10^6$, due to the near cancellation between gauge and gravity with each other.  
Fortunately, taking care of this additional interaction is straightforward.  It is enough to modify the original action in~\eqref{action} as
\eqna{
S_{\mathrm{new}}=S+ \int d^{4} x \sqrt{-g}\left(- \nabla^\mu \epsilon \nabla_\mu \epsilon-m^2_\epsilon \epsilon^2 -\kappa g_\epsilon \lsp \phi^\dagger \phi \lsp \epsilon\right)\, .
}[actionO2]
Similarly to what we have done for the gauge and gravity field,  we can then use the equation of motion for $\epsilon$ 
\eqna{
(\nabla^{\mu}\nabla_\mu -m^2_\epsilon)\epsilon=\frac{\kappa g_\epsilon}{2} \phi^\dagger \phi\,,
}[]
to integrate it out classically. The resulting effective potential gets modified:
\eqna{
V_{\mathrm{eff, new}}[\phi, \phi^\dagger]=V_{\mathrm{eff}}[\phi, \phi^\dagger]+\frac{\kappa g_\epsilon}{2}\epsilon[\phi, \phi^\dagger] \phi^\dagger \phi\,. 
}[]
The OPE coefficient $C_{\phi^\dagger \phi \epsilon}$  is known~\cite{Liu:2020tpf},
\eqna{
C_{\phi^\dagger \phi \epsilon}=0.687126\,  ,
}[]
and it is related to the AdS coupling by 
\eqna{
-\kappa g_\epsilon=C_{\phi^\dagger \phi \epsilon} \frac{\pi ^{3/2} 2^{-\Delta_\epsilon -2 \Delta+\frac{11}{2}} \sqrt{\Gamma (2 \Delta_\epsilon
   -1)} \Gamma (2 \Delta-1)}{\Gamma \left(\frac{\Delta_\epsilon }{2}\right)^2 \Gamma
   \mleft(\frac{\Delta_\epsilon -3}{2}+\Delta\mright) \Gamma \mleft(\Delta-\frac{\Delta_\epsilon }{2}\mright)}\, ,
}[]
where recall $\Delta\equiv\Delta_{\Phi}$.
Using the results is~\cite{DHoker:1999mqo} for a scalar exchange we write
\eqna{
\CE(X, P_1, P_2)&=\langle \Phi(P_1) \epsilon(X) \Phi^*(P_2) \rangle \propto \frac{1}{(-2P_1\cdot P_2)^{\Delta}}f_\epsilon\left( \frac{-P_1\cdot P_2}{2 X \cdot P_1 X \cdot P_2}\right)\,,
}[]
with
\eqna{
f_\epsilon(t)&=\frac{\Gamma\mleft(\frac{\Delta_\epsilon-3}{2}+\Delta\mright)\Gamma\mleft(\Delta-\frac{\Delta_\epsilon}{2}\mright)}{4\Gamma(\Delta)^2}\Big[t^{\frac{\Delta_\epsilon}{2}}\Gamma\mleft(\frac{\Delta_\epsilon}{2}\mright)^2{}_2\tilde{F}_1\left(\frac{\Delta_\epsilon}{2},\frac{\Delta_\epsilon}{2},\Delta_\epsilon-\frac{1}{2},t\right)\\ 
&\quad \, -t^{\Delta}\Gamma(\Delta)^2{}_3\tilde{F}_2 \left( \begin{array}{c}
1, \qquad \Delta, \qquad\Delta \\
 \Delta-\frac{\Delta_\epsilon}{2}+1,\quad \Delta+\frac{\Delta_\epsilon-1}{2}
\end{array}  ; t \right)\Big]\, , 
}[]
where we have defined the regularized Hypergeometric function
\eqna{
{}_p\tilde{F}_q \left( \begin{array}{c}
a_1, \, \cdots, \, a_p\\
b_1, \, \cdots, \, b_q\\
\end{array}  ; t \right) \equiv \frac{1}{\Gamma(b_1)\cdots\Gamma(b_q)}\lsp  {}_p{F}_q \left( \begin{array}{c}
a_1, \, \cdots, \, a_p\\
b_1, \, \cdots, \, b_q\\
\end{array}  ; t \right)\, .
}[]
$f_\epsilon(t)$ admits a  power series expansion 
\eqna{
f_\epsilon(t)=\sum_{k=\frac{\Delta_\epsilon}{2}}^\infty a_k t^k - \sum_{k=\Delta}^\infty a_k t^k \, ,\qquad a_k=
\frac{\Gamma (k)^2 \Gamma \mleft(\Delta -\frac{\Delta_\epsilon }{2}\mright) \Gamma \mleft(\Delta
   +\frac{\Delta_\epsilon }{2}-\frac{3}{2}\mright)}{4 \Gamma (\Delta )^2 \Gamma \mleft(k-\frac{\Delta_\epsilon }{2}+1\mright) \Gamma \mleft(k+\frac{\Delta_\epsilon }{2}-\frac{1}{2}\mright)}.
}[]

As with the gauge and gravity contributions, we have to integrate these exchanges over AdS to calculate their contributions to the anomalous dimensions.  For finite values of $J$, the sum over $k$ converges and the integrals can be done term by term
\eqna{
\CI_\epsilon^{(3d)}&=-\frac{g_\epsilon^2\kappa^2 \pi^{3/2}  \tilde{N}^{(3d)}_{\Delta, \ell_1, \ell_2} \tilde{N}^{(3d)}_{\Delta, \ell_3, \ell_4}\ell_1!\ell_2! \Gamma\mleft( \Delta+k-\frac{3}{2}\mright)\Gamma\mleft(\Delta-\frac{\Delta_\epsilon}{2} \mright)\Gamma\mleft(\Delta+\frac{\Delta_\epsilon-3}{2} \mright)} {16 \Gamma(\Delta-k)\Gamma(\ell_1+\Delta)\Gamma(\ell_2+\Delta)\Gamma\mleft(k-\frac{\Delta_\epsilon}{2}+1 \mright)\Gamma\mleft(k+\frac{\Delta_\epsilon-1}{2} \mright)}\times\\
& \sum_{j=0}^{\ell_1} \frac{\Gamma(j+k) \Gamma(j+k-\ell_1+\ell_2)\Gamma(j-\ell_1+\ell_2+\ell_4+1)\Gamma(\ell_1+\Delta-j-k)}{\Gamma(j+1)\Gamma(\ell_1-j+1)\Gamma(j-\ell_1+\ell_2+1)\Gamma(j+k-\ell_1+\ell_2+\ell_4+\Delta)}\,.
}[matrixEpsilon]
Given the matrix element for the scalar exchange, we can now compute the anomalous dimensions for the $Q$-particle states.
\subsubsection*{Two-particle states}
For $Q=2$, we have already discussed in detail the contributions to the anomalous dimensions from gauge and gravity exchange.  
As for the scalar exchange, the integrals and sums over `monomial' states for each value of $k$ can be done in closed form, with the result 
\begin{align} \label{eq:ScalarExchangeSummand} 
\gamma^{(3d,Q=2)}_J &= \sum_{k=\frac{\Delta_\epsilon}{2}}^\infty \gamma_{J,k} - \sum_{k=\Delta}^\infty \gamma_{J,k}\, , \\  \nonumber
\gamma_{J,k} &= \frac{g_\epsilon^2\kappa^2 (-1)^{J+1} \Gamma (k)^2 \Gamma \left(\Delta -\frac{\Delta _{\epsilon }}{2}\right)
   \Gamma \left(\Delta +\frac{\Delta _{\epsilon }}{2}-\frac{3}{2}\right) \Gamma
   \left(k+\Delta -\frac{3}{2}\right) \Gamma (J-k+\Delta )}{16 \pi ^{3/2} \Gamma
   \left(\Delta -\frac{1}{2}\right)^2 \Gamma (\Delta -k) \Gamma (J+k+\Delta ) \Gamma
   \left(k-\frac{\Delta _{\epsilon }}{2}+1\right) \Gamma \left(k+\frac{\Delta _{\epsilon
   }}{2}-\frac{1}{2}\right)}\, . 
\end{align}
It is important to note that  for generic values of $\Delta$ the large spin $J$ limit does not commute with the sum on $k$. In particular, the second series, which begins with the term $k=\Delta$, in a large $J$ expansion contains powers of the form $J^{-2\Delta-2n}$ with $n=0,1,\dots$, but such terms are not present in a large $J$ expansion of the exact result after first performing the sum on $k$.\footnote{In fact, we expect the double-twist anomalous dimension to decay at large spin as $\sim J^{-\tau_\epsilon}=J^{-\Delta_\epsilon}$.} In Appendix \ref{app:SubtleScalar}, we discuss this issue in more detail. 

\begin{figure}[t!] \centering
\includegraphics[scale=0.6]{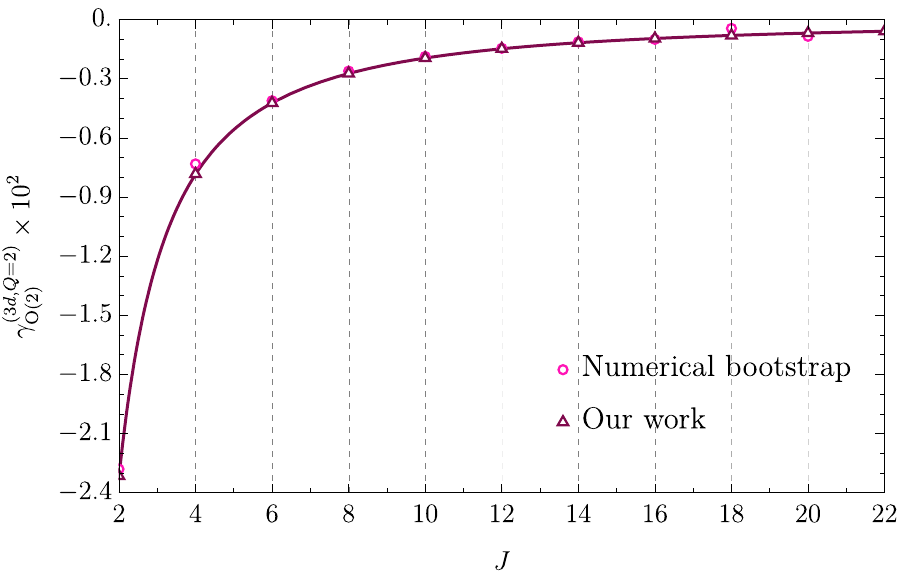} \caption{Anomalous dimension due to graviton, photon and scalar exchange for the two-particle primary in the O(2) model. Notice the accordance between our perturbative computation (violet triangle) and the numerical bootstrap date in~\cite{Liu:2020tpf} (fuchsia circle). The solid line represents the large spin expectation in~\eqref{largeSpinO2Q2}. }\label{fig:O2Q2}
\end{figure}
By carefully taking the large $J$ limit of the scalar exchange and adding the photon and graviton contribution, we can determine the large spin behavior of $[\Phi, \Phi]_J$ in the $3d$ O(2) model\footnote{Here and in the following, for our numerical computations we will set $\Delta_\Phi=0.519\, , \Delta_\epsilon=1.51$.}
\eqna{
\gamma^{(3d, Q=2)}_{\rm{O}(2)} \xrightarrow[]{J\to\infty} -\frac{5\times 10^{-5}}{J}+\frac{9.5 \times 10^{-7}}{J^2}-\frac{0.063}{J^{1.51}}+\frac{0.0018}{J^{2.51}}\, ,
}[largeSpinO2Q2] 
where the first two terms are due to the graviton and photon interactions  and the last two are related to the scalar $\epsilon$-exchange.  This expression is in accordance with previous analytic bootstrap results in~\cite{Alday:2015ewa} and, as they noted, precisely matches  the numerical bootstrap results; for a comparison with recent values extracted in~\cite{Liu:2020tpf} down to spin $J=2$, see Fig.~\ref{fig:O2Q2}.

The correction to the dimension of the spin zero primary deserves a separate discussion.  Based on the discussion in Sec.~\ref{Subsec:contactTerms},  it is not surprising that our perturbative computation fails for $[\Phi, \Phi]_{J=0}$.\footnote{Moreover, recall that  in Sec.~\ref{Sec:smallSpinContact} we have seen that the spin-zero gravity contribution was divergent for $\Delta=\frac{d}{4}$, thus the need to add contact terms for $\Delta\leq \frac{d}{4}$. }  On the other hand, for two-particle states,  adding a $2k$-derivative contact interactions has the effect of changing the anomalous dimensions only for primaries with spin $J\leq k$ \cite{Heemskerk:2009pn}. So at $Q=2$,  to modify only the value of $\gamma^{(3d, Q=2)}_{J=0}$,  it is sufficient  to add to the original action in~\eqref{actionO2} a quartic term\footnote{As discussed in Sec.~\ref{Subsec:contactTerms} for $Q=2$, adding a $\lambda_{1, n}(\partial \phi \partial \phi^\dagger)(\phi \phi^\dagger)^n$ has the same effect on the anomalous dimension of $\lambda_{0, n}(\phi \phi^\dagger)^n$. Hence,  for ease of notation, here and in the following, we will only include $\lambda_n (\phi \phi^\dagger)^n$, bearing in mind that we can always reinterpret the coefficient $\lambda_n$ as a combination of $\lambda_{0,n}$ and $\lambda_{1,n}$.}
\eqna{
S_{\rm new}\to S_{\rm new} -\lambda_4 \kappa^2 (\phi \phi^\dagger)^2\, .
}[]
It is then possible to tune the value of $\lambda_4$ to set $\gamma^{(3d, Q=2)}_{J=0}$ to the desired value.  Considering the lowest-twist charge-2 operator of dimension $\Delta_t$ in~\eqref{deltaO2} as $[\Phi, \Phi]_{J=0}$, we can choose a suitable $\lambda_4$ such that
\eqna{
\gamma^{(3d, Q=2)}_{J=0}(\lambda_4)=\Delta_t-2\Delta_{\Phi}\simeq 0.198 \quad \Rightarrow \quad \lambda_4 \kappa^2\simeq -347.6\, .
}[lamda4]
Adding this contact term, for $Q=2$ and higher,  should be interpreted as an expedient to incorporate all possible non-perturbative effects, which are not captured by our computation. Taken literally, such a large value of $\lambda$ in the bulk Lagrangian would lead to uncontrolled loop corrections. Similarly, the gauge and gravity interactions in the bulk are large at short distances, and would generate large loop corrections.  However, in the regime that we use the AdS action, all tree-level contributions are small and so the large value of $\lambda$ does not pose an obvious problem.  It would be useful if one could formalize a controlled expansion at higher orders, perhaps by defining an effective bulk theory with short distance contributions integrated out, together with a systematic power-counting argument so that interactions like $(\phi \phi^\dagger)^2$ only contribute after being multiplied by inverse powers of a large distance set by the spin $J$. 
\subsubsection*{Three-particle states}
Recall from Sec.~\ref{Sec:smallSpinContact}, and more extensively from Appendix~\ref{App:contact}, that although the $(\phi \phi^\dagger)^2$ contact term affects only the spin $J=0$ at $Q=2$, it nevertheless changes the anomalous dimensions for all the spins  for $Q\geq 3$. Considering this effect together with the contributions from photon, graviton and scalar exchanges, we can determine the corrections to the classical dimensions of three-particle primary states. 
\begin{figure}
\centering
\begin{subfigure}[b]{0.48\textwidth}
    \includegraphics[width=\textwidth]{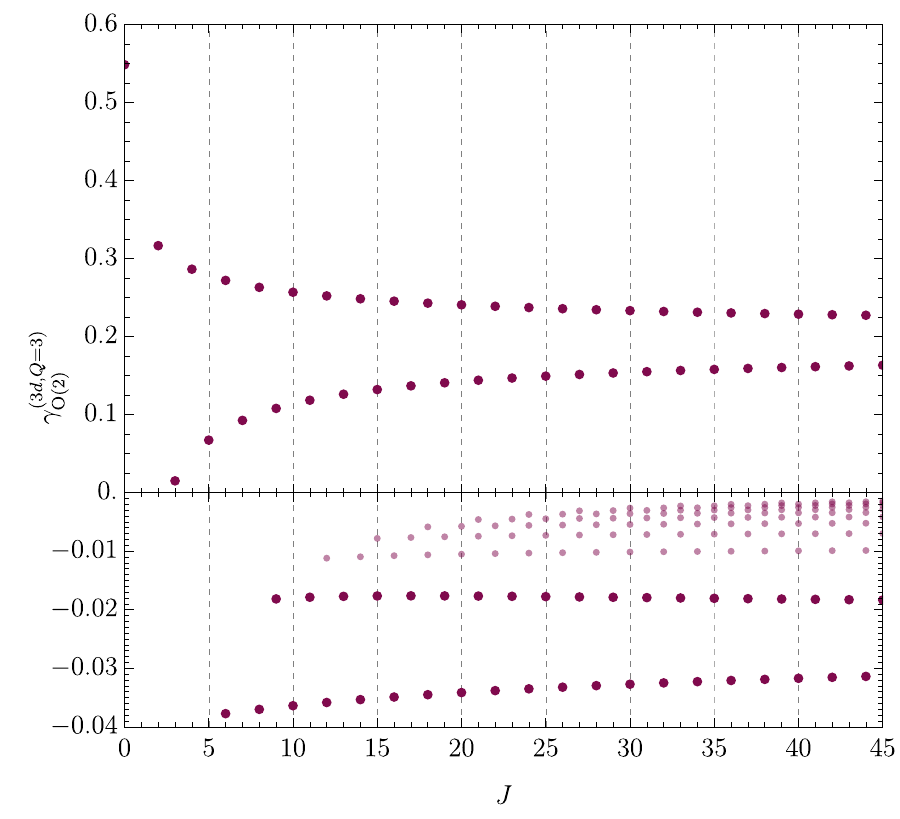}
    \caption{All eigenvalues of $\gamma^{(3d,Q=3)}_{\mathrm{O}(2)}$ matrix}
    \label{fig:Q3O2allEig}
\end{subfigure}
\hfill
\begin{subfigure}[b]{0.48\textwidth} 
    \includegraphics[width=\textwidth]{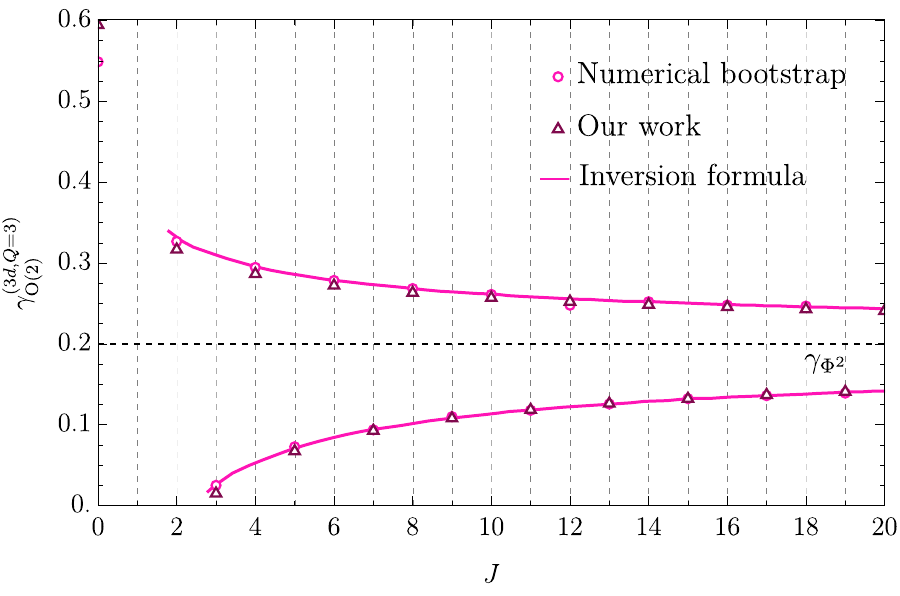}
    \vspace{0.6cm}
    \caption{Highest-energy eigenvalue (HE)}
    \label{fig:Q3O2HE}
\end{subfigure}
\caption{Anomalous dimensions for 3-particle states in the O(2) model as a function of the spin $J$.  \textit{Left}: corrections to all energy eigenvalues. The darker color corresponds to the highest- and lowest-energy states. \textit{Right}: comparison of the highest-energy eigenvalue (violet triangle) with the charge-3 dimensions obtained numerically in~\cite{Liu:2020tpf} (fuchsia circle). The solid fuchsia line corresponds to the inversion formula results in~\cite{Liu:2020tpf}.  The upper trajectory corresponds to even spin, while the lower one to odd $J$.  The dashed line corresponds to the anomalous dimension $\gamma_\phi^2$ of the $[\Phi, \Phi]_J=0$ double-twist primary. }
\label{fig:Q3O2}
\end{figure}
In Fig.~\ref{fig:Q3O2}, we collect all our results at $Q=3$ as a function of the spin $J$.  In Fig.~\ref{fig:Q3O2allEig}, we show all the eigenvalues of
\eqna{
{}^{\qquad r}_{J;Q=3}\!\bra{\Psi} g_{\mathrm{U}(1)}^2\kappa^2\CV_{\CJ}+\kappa^2 \CV_{\CT}+g_\epsilon^2 \kappa^2 \CV_\epsilon+\lambda_4 \kappa^2\CV_{(\phi \phi^\dagger)^2}  \ket{\Psi}_{J;Q=3}^{ s}\, ,
}[Q3O2dims]
where $r,s$ run over the number of degenerate primaries.  The darker points correspond to the lowest- and highest-energy eigenvalues.  By comparing to the results in~\cite{Liu:2020tpf}, we see that these are well approximated by our highest-energy eigenvalues.

In~\cite{Liu:2020tpf}, the authors obtain numerical predictions for the charge-3 OPE data using the extremal functional method, which they then compare with analytic results obtained using the Lorentzian inversion formula (fuchsia circle and solid line respectively in Fig.~\ref{fig:Q3O2HE}).  To obtain this data, they consider the four-point function $\langle \Phi \Phi t t \rangle$ and its crossing symmetric version, so that they are able to extract the twists of $[\Phi, t]_J$ primaries.\footnote{Non-perturbatively,  the anomalous dimension $\gamma$ used here is defined as $\gamma \equiv \tau-Q \Delta$, where $\tau$ is the exact value of the twist.} In Fig. ~\ref{fig:Q3O2HE}, we show these numerical data together with the highest-energy (HE) anomalous dimensions, we compute from~\eqref{Q3O2dims} with $\lambda_4$ fixed as in~\eqref{lamda4}.  The agreement is excellent and improves at larger and larger values of the spin $J$.  As one can infer from the plot, the distribution of the data and the inversion-formula trajectories suggest that the anomalous dimensions of charge-3 particle states approach a constant as $J\to \infty$. We can identify this constant with the value of $\gamma^{(3d, Q=2)}_{\Phi^2}$ in~\eqref{lamda4} ($[\Phi, \Phi]_{J=0}\equiv\Phi^2$).  This  exactly agrees with our  interpretation of the $Q=3$ attractive phase.  In fact we expect that the three-particle state is made by two blobs, one with two partons and one with one, spinning with a very large angular momentum. At large spin, the energy of this system is dominated by the energy of the 2-particle blob.  So for the HE state\footnote{The HE state energy is dominated by the scalar and contact interactions, which are both positive.} we expect the $J\rightarrow \infty$ value of the anomalous dimension at $Q=3$ to be given by the $Q=2$ largest anomalous dimension,  which, from~\eqref{lamda4} and the results in Fig.~\ref{fig:O2Q2}, is indeed $\gamma^{(3d, Q=2)}_{\Phi^2}$
\eqna{
\lim_{J\to \infty }\gamma^{(3d, Q=2)}_{\mathrm{O}(2), \mathrm{max}}&=\max \gamma^{(3d, Q=2)}_{\mathrm{O}(2)}\, , \\
\max \gamma^{(3d, Q=2)}_{\mathrm{O}(2)}&= \gamma^{(3d, Q=2)}_{\mathrm{O}(2)} (\ell=0) = \gamma^{(3d, Q=2)}_{\Phi^2}\,.
}[] 
\begin{figure}[t!]
\begin{subfigure}{0.48\textwidth}
    \includegraphics[scale=0.52]{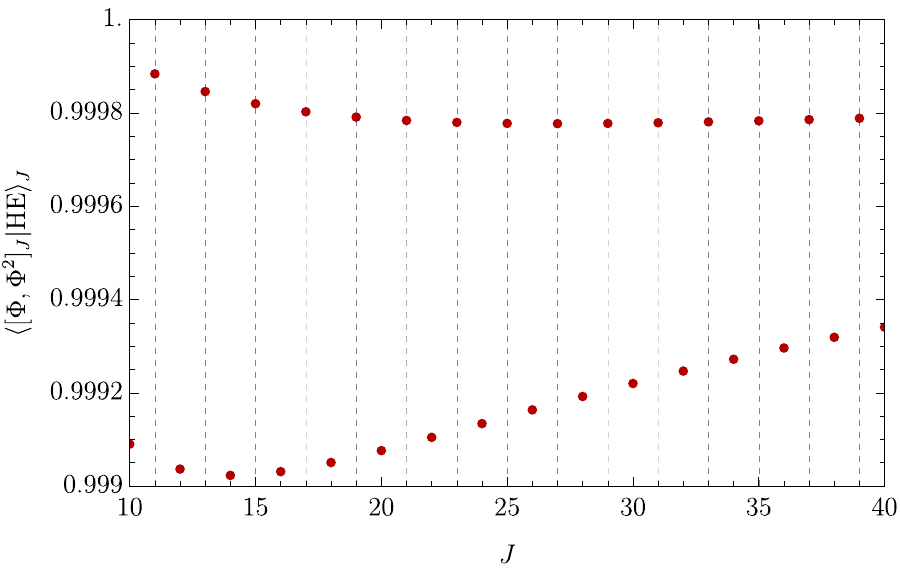}
    \caption{$|\langle[\Phi,\Phi^2]_J|\mathrm{HE}\rangle_J|^2$ }  \label{fig:Q3HEoverlap}
\end{subfigure}
\hfill
\begin{subfigure}{0.48\textwidth}
  \includegraphics[scale=0.5]{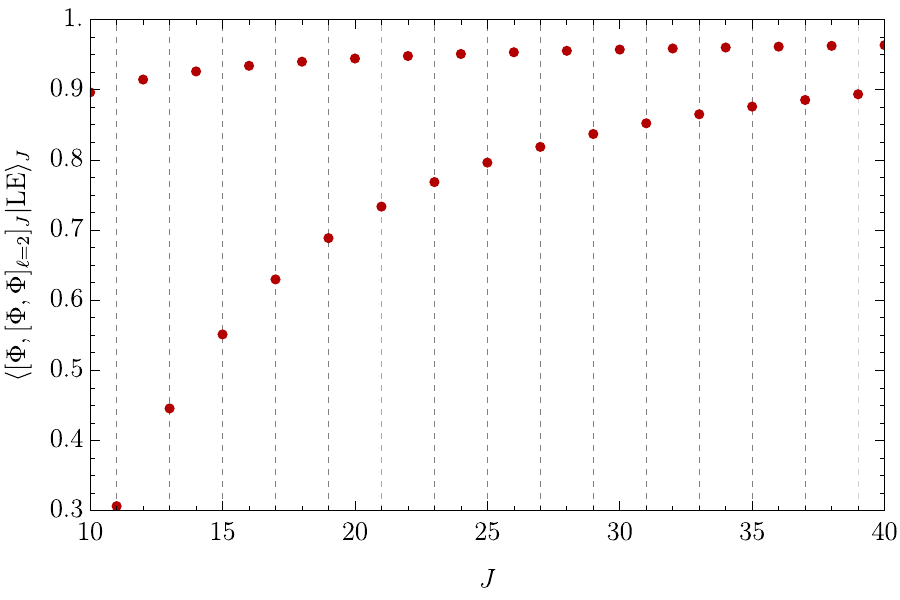}
    \caption{$|\langle[\Phi,[\Phi,\Phi]_{\ell=2}]_J|\mathrm{LE}\rangle_J|^2$ }\label{fig:Q3LEoverlap}
     \end{subfigure}
\caption{Overlap of the Highest-Energy eigenstate (HE) (\textit{left}) and Lowest-Energy eigenstate (LE) (\textit{right}) with the expected triple-twist operators. } \label{fig:Q3overlaps}
\end{figure}

To further validate the identification of the HE state with the charge-3 operators in~\cite{Liu:2020tpf},  we can try to rewrite our largest eigenstate $\ket{\rm HE}_J$ in a basis of states recursively constructed as `double-twists' made from a single $\Phi$ and a double-twist primary  $[\Phi, \Phi]_\ell$, i.e., as $[\Phi, [\Phi, \Phi]_\ell]_J$.  Recall in fact that the authors of~\cite{Liu:2020tpf} are constructing double-twist operators $[\Phi, t]_J=[\Phi, [\Phi, \Phi]_{\ell=0}]_J\equiv [\Phi, \Phi^2]_J$, so we should expect $\ket{\rm HE}_J= [\Phi, \Phi^2]_J$.  In Fig.~\ref{fig:Q3HEoverlap}, we plot the overlap $|\langle [\Phi, \Phi^2]_J |\mathrm{HE}\rangle_J|^2$ for various values of the spin $J$. This function approaches one,  supporting the expected identification.  The deviation from  unity is due to the fact that, at the finite value of $J$ we have access to,  the HE state is actually an admixture of various operators $\ket{\rm HE}_J\sim \beta_0 [\Phi,[\Phi, \Phi]_{\ell=0}]_J+\sum_{\ell \neq 0} \beta_\ell  [\Phi,[\Phi, \Phi]_{\ell}]_J$,  with $\beta_\ell \sim \beta_0/J$, as one can verify by constructing the energy matrix $\bra{\Psi}\CV\ket{\Psi}$ in the triple-twist basis $\ket{\Psi}_J=\ket{[\Phi,[\Phi, \Phi]_{\ell}]_J}$.\footnote{To be precise, the triple-twist basis is redundant, so one should restrict to the linearly independent subset.}

At the opposite side of the spectrum, we expect for the lowest-energy (LE) eigenvalue to have a similar  large spin behavior, except that now what is relevant is the minimum twist for $Q=2$: \eqna{
\lim_{J\to \infty }\gamma^{(3d, Q=2)}_{\mathrm{O}(2), \mathrm{min}}=\min \gamma^{(3d, Q=2)}_{\mathrm{O}(2)}\, .
}[] 
In this case, the large $J$ limit is given by
\eqna{
\min \gamma^{(3d, Q=2)}_{\mathrm{O}(2)}= \gamma^{(3d, Q=2)}_{\mathrm{O}(2)} (\ell=2)\simeq -0.023\,.
}[] 
In Fig.~\ref{fig:Q3O2allEig}, we can see how the odd (even) spins data approach this value from above (below).\footnote{The convergence in $J$ is unfortunately fairly slow, but going to higher $J$ in our numeric calculations is computationally expensive.} Moreover, this suggests that we should identify the corresponding lowest-energy state $\ket{\rm LE}_J$ with the triple-twist  primary $[\Phi, [\Phi, \Phi]_{\ell=2}]_J$, as supported by the computation of the corresponding overlap --- see Fig.~\ref{fig:Q3LEoverlap}.

Let us conclude by discussing 3-particle anomalous dimensions at $J=0$.  Similarly to the $Q=2$ story, our results match quite well to the numerical bootstrap data at small finite spin $J\geq 2$. We can again enforce the match at spin zero by adding a contact interaction.  In analogy with the two-particle case, we would like to add an interaction which modifies the energy of the spin zero three-particle primaries but does not affect $J\geq 2$. This is achieved by further modifying the original action~\eqref{actionO2}
\eqna{
S_{\rm new}\to S_{\rm new}-\lambda_4\kappa^2 (\phi \phi^\dagger)^2-\lambda_6 \kappa^2 (\phi \phi^\dagger)^3\, .
}[actionL6]
Keeping the value of $\lambda_4$ as in~\eqref{lamda4}, we can tune $\lambda_6$ so that
\eqna{
\gamma^{(3d, Q=3)}_{J=0}(\lambda_6)=\Delta_{\chi}-3\Delta_{\Phi}\simeq 0.549 \quad \Rightarrow \quad \lambda_6\kappa^2\simeq -2.8\times 10^3\, ,
}[lamda6]
where we have identified the lowest-twist charge-3 scalar $\chi$ as the triple-twist operator $[\Phi, [\Phi, \Phi]_{\ell=0}]\equiv \Phi^3$.

One could go farther and include even more bulk fields corresponding to known low-dimension CFT operators.  
The next lightest neutral scalar  is $\epsilon^\prime$ ($\Delta_{\epsilon^\prime}\simeq 3.794$ and $\lambda_{\phi\phi^\dagger \epsilon^\prime}\simeq 0.04$~\cite{Liu:2020tpf, private}).  We have estimated the effect of adding such contribution at $Q=2$ and $Q=3$ by retracing the same steps above for the $\epsilon$ case.     The overall result for the anomalous dimensions barely changes, and in particular the difference is not visible by eye in the comparison plots in Fig.~\ref{fig:O2Q2} and~\ref{fig:Q3O2}. The only real difference with  respect to our previous discussion is that  the value of the $\phi^4$ contact term coupling changes, since it is fixed by requiring~\eqref{lamda4} which now also gets contributions from the $\epsilon'$ bulk field, and its new value is significantly decreased to $\lambda_4\kappa^2 \simeq 54$.  This is consistent with our interpretation  of contact terms as a  way of taking into all possible non-perturbative effects, and in EFT language they absorb the contributions of heavy bulk fields that are integrated out, so that `integrating'  back in some of the heavy fields tends to decrease the value of this quartic coupling.
\subsubsection*{Four-particle states}
\begin{figure}[t!]
\centering
    \includegraphics[width=0.5\textwidth]{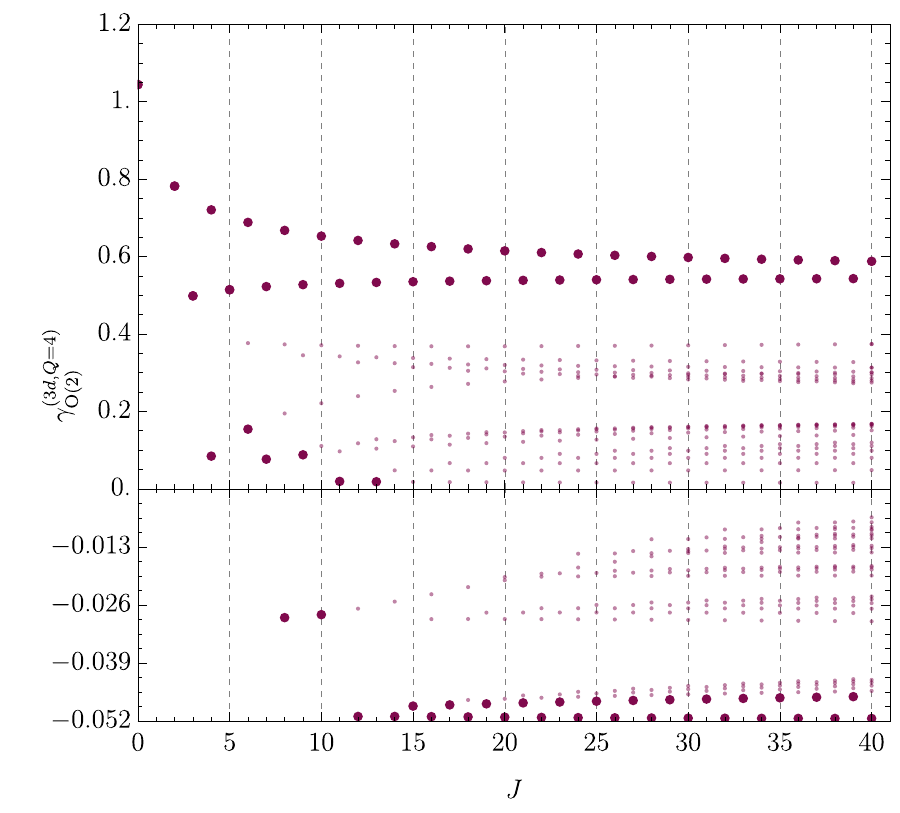}
\caption{Anomalous dimensions for all the energy eigenvalues of 4-particle states in the O(2) model as a function of the spin $J$.  The points with a darker color corresponds to the highest- and lowest-energy states. }
    \label{fig:Q4O2allEig}
\end{figure}
Next we consider $Q=4$.  In Fig.~\ref{fig:Q4O2allEig}, we plot all the eigenvalues of the anomalous dimension matrix
\eqna{
{}^{\qquad r}_{J;Q=4}\!\bra{\Psi} g_{\mathrm{U}(1)}^2\kappa^2\CV_{\CJ}+\kappa^2 \CV_{\CT}+g_\epsilon^2 \kappa^2 \CV_\epsilon+\lambda_4 \kappa^2\CV_{(\phi \phi^\dagger)^2}+\lambda_6 \kappa^2\CV_{(\phi \phi^\dagger)^3}  \ket{\Psi}_{J;Q=4}^{ s}\, ,
}[Q4O2dims]
where $r,s$ run over the number of degenerate primaries.  The darker points correspond to the lowest- and highest-energy eigenvalues.

\begin{figure}
\centering
\begin{subfigure}[b]{0.48\textwidth}
    \includegraphics[width=\textwidth]{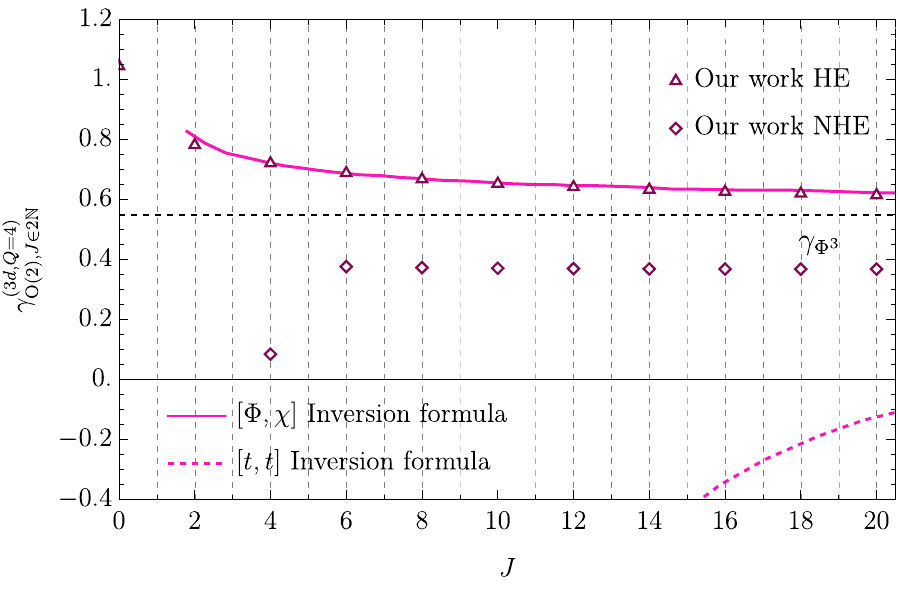}
    \caption{$J \in 2\mathbb{N}$}
    \label{fig:Q4O2HEEven}
\end{subfigure}
\hfill
\begin{subfigure}[b]{0.48\textwidth} 
    \includegraphics[width=\textwidth]{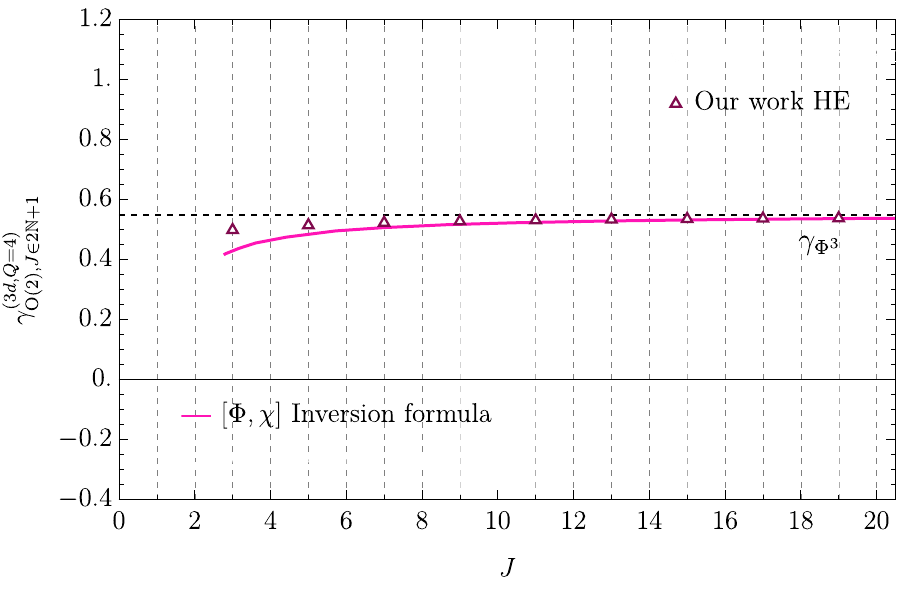}
    \caption{$J \in 2\mathbb{N}+1$}
        \label{fig:Q4O2HEOdd}
\end{subfigure}
\caption{Comparison of the highest-energy eigenvalue (violet triangle)  with $[\Phi,\chi]$-inversion formula results in~\cite{Liu:2020tpf} (solid fuchsia line) for even (\textit{left}) and odd spins (\textit{right}).  The dashed curve shows  their $[t,t]$-inversion formula.  The dashed black line corresponds to the anomalous dimension $\gamma_\Phi^3$ of the $[\Phi, \Phi^2]_J=0\equiv \Phi^3$ triple-twist primary.  We also report numeric results for the Next-to-Highest Energy (NHE) anomalous dimensions for even spins.}
    \label{fig:Q4O2HE}
\end{figure}

Results for primaries at charge four are provided in~\cite{Liu:2020tpf}.\footnote{There the data are obtained by studying a system of mixed correlators: $\langle tttt\rangle,\langle tt\chi\Phi\rangle,\langle \Phi\chi\chi\Phi\rangle$. } In Fig.~\ref{fig:Q4O2HE}, we compare their inversion formula trajectories with our predictions for the highest-energy eigenvalues for even and odd spins separately.  Notice how our data points lie, almost exactly,  on the  $[\Phi, \chi]$-trajectories.

\begin{figure}
\begin{subfigure}{0.48\textwidth}
  \includegraphics[scale=0.5]{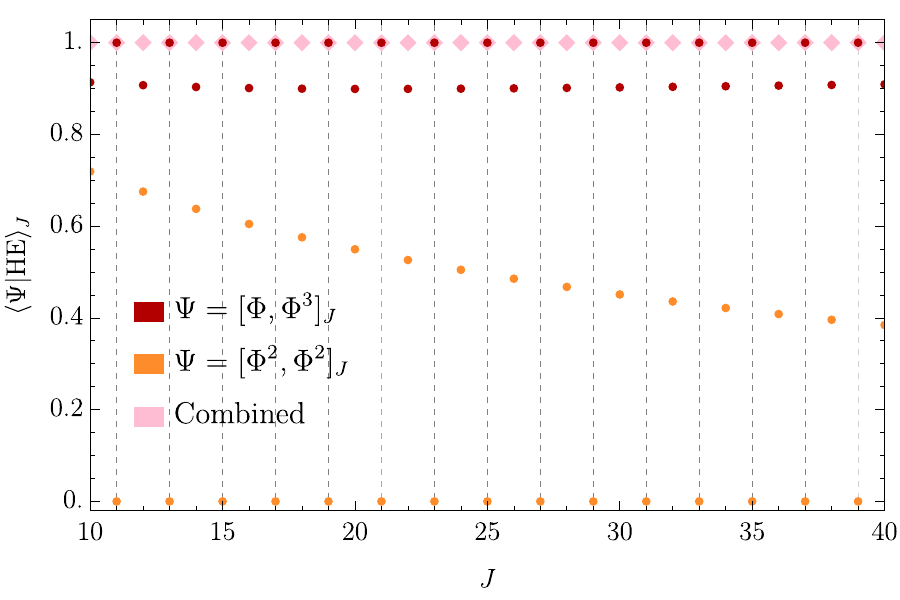}
    \caption{$|\langle \Psi|\mathrm{HE}\rangle_J|^2$}  \label{fig:Q4HEoverlap}
\end{subfigure}
\hfill
\begin{subfigure}{0.48\textwidth}
   \includegraphics[scale=0.5]{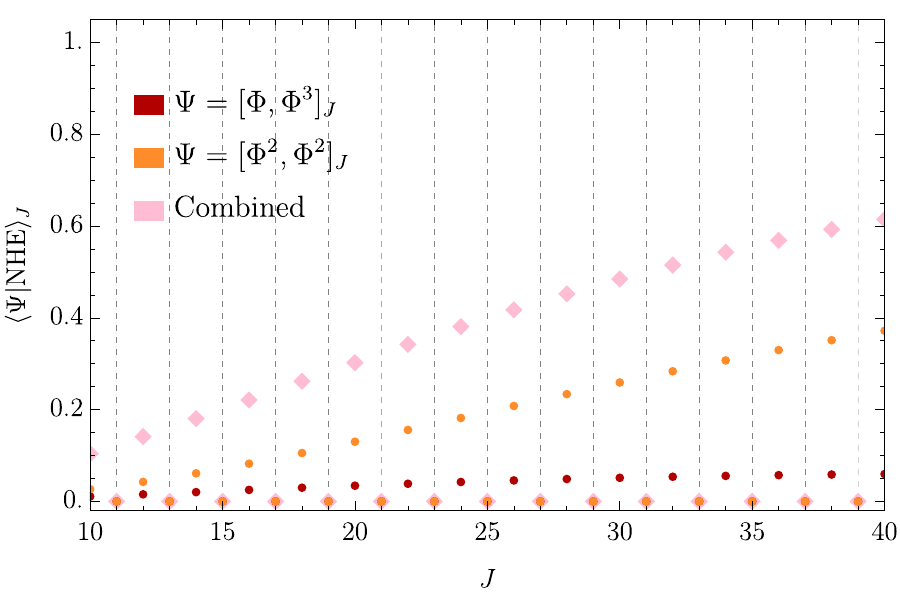}
    \caption{$|\langle \Psi|\mathrm{NHE}\rangle_J|^2$}\label{fig:Q4NHEoverlap}
     \end{subfigure}
\caption{Overlap of the Highest-Energy eigenstate (HE) (\textit{left}) and Next-to-Highest-Energy eigenstate (NHE) (\textit{right}) with two possible quadruple-twist operator: $\ket{\Psi}=[\Phi,\Phi^3]_J$ (red) and $\ket{\Psi}=[\Phi^2,\Phi^2]_J$ (orange) and their sum $|\langle[\Phi,\Phi^3]_J|\mathrm{HE}\rangle_J|^2+|\langle[\Phi^2,\Phi^2]_J|\mathrm{HE}\rangle_J|^2$.} \label{fig:Q4overlapsH}
\end{figure}

As suggested by the dashed black line in the figure,  both our numerical data and the inversion formula curves tend to a constant as the spin increases. The value of the constant is set by 
\eqna{
\lim_{J\to \infty} \gamma^{(3d, Q=4)}_{\mathrm{O}(2), \mathrm{max}}=\max  \gamma^{(3d, Q=3)}_{\mathrm{O}(2)}=\gamma^{(3d, Q=3)}_{\mathrm{O}(2)}(\ell=0)=\gamma_{\Phi_3}^{(3d, Q=3)}\, .
}[]
This result, together with the good agreement with the $[\Phi, \chi]$-trajectories, strongly suggests that we should identify 
 the HE state as the quadruple-twist primary $[\Phi, \Phi^3]_J$. This can be verified by  computing the overlap of this state   with $\ket{\mathrm{HE}}_J$, as in Fig.~\ref{fig:Q4HEoverlap}. As expected, as $J$ increases, $\ket{\mathrm{HE}}$ is better and better approximated by the $[\Phi, \Phi^3]_J$ state.  At lower spins, it receives some minor contributions from other quadruple-twist operators and in particular from the $[\Phi^2, \Phi^2]_J$, but as we can see, this dependence dies off as we increase the spin.\\
In Fig.~\ref{fig:Q4NHEoverlap}, instead, we study the overlap of our Next-to-Highest (NHE) energy eigenstate  with  the operator $[\Phi^2, \Phi^2]_J$. Because of Bose symmetry, there is no overlap for odd spins, while for even spins,  we see how the overlap increases as we go to higher $J$.   Because of the slow convergence seen in the overlap, which is still changing visibly with spin at the highest values we computed,  we suspect that it is necessary to go to very high spins to see the overlap $|\< [ \Phi^2, \Phi^2]_J | {\rm NHE}_J\>|^2$ approach 1. 
Assuming this is in fact what happens, we expect the corresponding NHE eigenvalue to behave at large $J$ as 
\eqna{
\lim_{J\to\infty}\gamma^{(3d, Q=4)}_{\mathrm{O}(2), \mathrm{NHE}}=2 \gamma_{\Phi^2}\simeq 0.396\, .
}[]
As one can see from  Fig.~\ref{fig:Q4O2HEEven}, the NHE data are far from the $[t,t]$ inversion formula expectation of~\cite{Liu:2020tpf}.  As suggested by the authors of~\cite{Liu:2020tpf}, it is likely that the $[t,t]$ curve should not be trusted,  since their analysis did not take into account mixing between $[t,t]$ and all other charge-four operators. 
By contrast, our AdS computations intrinsically take care of all these mixing effects since, at each spin $J$, we are diagonalizing the Hamiltonian in a complete basis of $Q=4$ primaries.

For the lowest-energy (LE) eigenvalue, instead, we expect $\gamma^{(3d, Q=4)}_{\mathrm{O}(2), \mathrm{LE}}$ to approach a constant as $J\to \infty$  set by the energy of the configuration of four partons grouped into two blobs (either two $Q=2$ blobs or one $Q=3$ blob and a $Q=1$ parton), where the nature of the blobs is determined dynamically. By carefully studying the $Q=2$ and $Q=3$ data, we find that the optimal spin for $Q=2$ is $\ell=2$, whereas for $Q=3$ it appears to be $\ell=6$.  Consequently, the  lowest-twist configuration with two $Q=2$ blobs has anomalous dimension
\eqna{
2\gamma^{(3d, Q=2)}_{\mathrm{O}(2), \mathrm{LE}}&=2\gamma_{[\Phi, \Phi]_{\ell=2}} \simeq -0.046\, , \\
}[]
whereas for a  $Q=3$ blob with total spin $\ell=6$ far away from an isolated parton has anomalous dimension
\eqna{
\gamma^{(3d, Q=3)}_{\mathrm{O}(2), \mathrm{LE}}&=\gamma_{\CO_{\ell=6}^{Q=3}} \simeq -0.038\, .
}[]
Thus we would expect the `two $Q=2$ blobs' configuration to set the value of the $J\rightarrow \infty$ anomalous dimension $a_0$.

\begin{figure} \centering
  \includegraphics[scale=0.6]{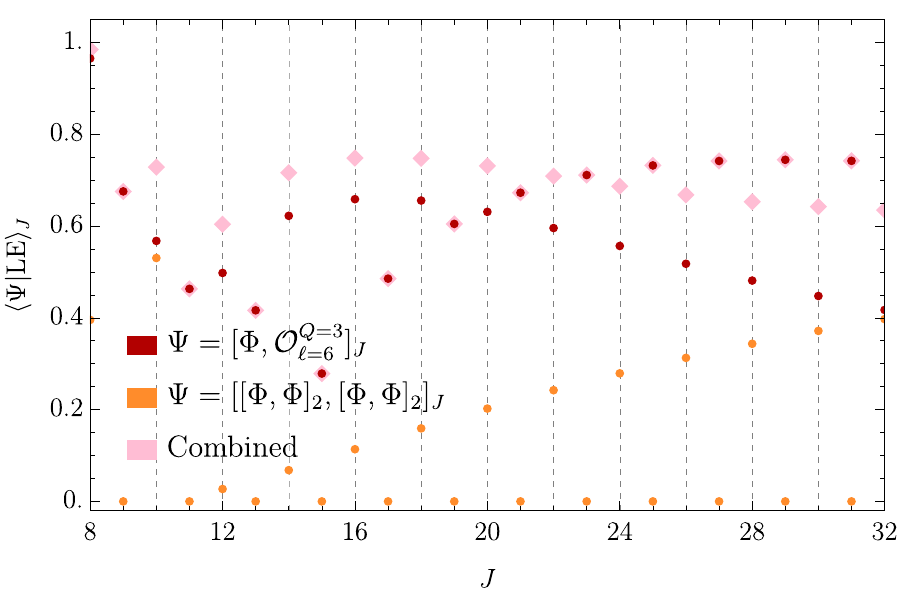}
\caption{Overlap of the Lowest-Energy eigenstate (LE) with two possible quadruple-twist operator: $\ket{\Psi}_1=[\Phi, \CO_{\ell=6}^{Q=3}]_J$ (red) and $\ket{\Psi_2}=[[\Phi, \Phi]_{\ell=2},[\Phi, \Phi]_{\ell=2}]_J$  (orange) and their sum $|\langle \Psi_1|\mathrm{LE}\rangle|^2|+\langle \Psi_2|\mathrm{LE}\rangle|^2$ (pink).} \label{fig:Q4overlapsL}
\end{figure}

Unfortunately, we were not able to verify this prediction with the finite values $J$ that we can access numerically, though we believe this is again due to slow convergence. To better understand the problem, we can evaluate the overlap of the eigenstate corresponding to $\gamma^{(3d, Q=4)}_{\mathrm{O}(2), \mathrm{LE}}$ with the quadruple-twist operators $[[\Phi, \Phi]_{\ell=2},[\Phi, \Phi]_{\ell=2}]_J$ and $[\Phi, \CO_{\ell=3}^{Q=3}]_J$.  As one can see in Fig.~\ref{fig:Q4overlapsL}, for $J$ even the overlap of the LE state with  $[[\Phi, \Phi]_{\ell=2},[\Phi, \Phi]_{\ell=2}]_J$ is slowly increasing and it appears very plausible that it is headed towards the value 1 at large $J$, but it is still only $\sim 0.4$ at $J=32$ (for $J$ odd, the component along this direction is always zero as expected, due to parity and Bose symmetry).  

Finally, as with the three-particle states, we can estimate the effect of adding $\epsilon^\prime$ to the $Q=4$ anomalous dimensions. As before,  the results remain relatively stable after the inclusion of this additional scalar exchange. 

\section{Future Directions}
\label{sec:Future}

The exploration of large charge and large spin phases of CFTs is a rich and fascinating question.  In this paper, we have focused on only a corner of it, in the case of CFTs with a U(1) current in the limit of large spin where states can be described as bulk excitations rotating around the center of AdS. Even in this corner, a complex and interesting picture emerges, with various different possible qualitative behaviors depending on the details of the model.  There is much more that could be done  to explore the physics of this regime.

One obvious extension is to consider more general models.    The simplest extension would be to states in symmetric traceless representations of a global O(N) symmetry, which we expect would be straightforward and could be compared to results for the $3d$ O(N) CFTs, e.g.~\cite{Kos:2015mba,Kos:2013tga,Dey:2016zbg,Henriksson:2022rnm,Bertucci:2022ptt}.  It would also be interesting to include supersymmetry, and in particular one could compare to truly holographic models such as $\CN=4$ SYM or ABJM.  From the bulk side, one could also explore in more detail the effect of adding additional light bulk fields; for instance, a neutral scalar with dimension below $d-2$ would dominate over the gauge and gravity interactions at long distances. More exotic variations include the possibility of multiple light charged species in the bulk, with, say, a charge-2 particle that is lighter than the lightest charge-1 particle. 

One could also try to apply our approach to the $3d$ Ising CFT.  There is no U(1) current in this case, but the $\mathbb{Z}_2$ symmetry separates even-twist and odd-twist states, so that triple-twist states can be distinguished as the lowest-energy multi-particle trajectories with odd twist.  There is already a lot of numeric data available for this model that one could compare to. Moreover,  the lowest-twist double-twist (i.e. $\mathbb{Z}_2$ even) operator is the stress tensor, whose correlators are constrained by Ward identities, which may provide an advantage when using the numeric bootstrap to learn about low spin triple-twist states from mixed correlators.

In terms of the O(2) model, one of our primary motivations for studying the Regge regime was simply the appeal of the beautiful phase diagram at large $Q$ and $J$, and a desire to understand it better.  We have argued that the Regge regime at large $Q$ is likely described by some number of smaller blobs, possibly with charge $q=2$ each, which repel each other at larger distances and therefore can be captured by holography.  In such a state, the calculations done in this paper break down unless $J \gg Q^3 \log Q$, but there is good reason to believe that one could perform a controlled semiclassical calculation down to $J \gtrsim Q^2$. Optimistically, one might even do better and smoothly describe the transition to the giant vortex phase at $J \lesssim Q^2$.  Another interesting direction is to compare to more numeric results at finite values of $Q$ and $J$.  Aside from the bootstrap, it may be possible to get better data from Monte Carlo or numeric diagonalization of microscopic models.  Because of the crucial role played by the conserved spin $J$ quantum number, a promising approach is to use the fuzzy sphere quantization, which exactly preserves rotational invariance.\footnote{See \cite{AndreasTalk} for preliminary results.}  Numerically diagonalizing the dilatation operator as sufficiently large volume should produce the true lightest state in a given sector, which could be compared to holographic results.

Finally, it would be useful to put the approximation used here (tree-level diagrams only) on a more solid ground.  The main reason this approximation should work even for non-holographic CFTs is that interactions universally decay at long distances in AdS when $d\ge 3$.  It would be interesting to systematize this as a long-distance effective theory, and potentially include higher-order corrections. Alternatively, at $Q=2$ one can think of the AdS calculation as a convenient way to implement the inversion formula, supplemented by some data on small values of $J$ where analyticity in $J$ fails.  In light of the new results~\cite{Homrich:2022mmd,Henriksson:2023cnh} of how analyticity in spin should be interpreted in case of multi-twist degeneracy, it may be possible to extend the inversion formula approach to higher-point functions to reproduce our results at $Q\ge 3$.  Especially for the lowest-twist case, where one can consider $(2n)$-point functions with all operators approaching a single null ray, the comb-channel conformal blocks should reduce essentially to a one-dimensional problem, where they have already been worked out \cite{Rosenhaus:2018zqn,Fortin:2020zxw,Alkalaev:2023axo,Fortin:2023xqq}, providing a convenient starting point for deriving a generalized inversion formula.  Along the same lines,  it may also be possible to  directly study the lightcone limit of $(2n)$-point functions  in the comb channel to extract  data about $n$-twist operators of specific form, as suggested in~\cite{Harris:2024nmr}.

\acknowledgments{We are grateful to Gabriel Cuomo, Ami Katz, Lorenzo Quintavalle,  David Simmons-Duffin,  and Ning Su for useful discussions,  and we thank Gabriel Cuomo, Ami Katz, Petr Kravchuk, Jeremy Mann, and  Lorenzo Quintavalle  for comments on a draft.  We especially thank  David Simmons-Duffin for generously sharing some unpublished data for the O(2) model.  GF,  ALF, and WL  are supported by the US Department of Energy Office of Science under Award Number DE-SC0015845, and GL was partially supported by the Simons Collaboration on the Non-perturbative Bootstrap. }

\newpage 
\appendix

\section{$Q=2$ Anomalous Dimensions from 4-point Function}
\label{app:FourPointAnomDim}

In this appendix, as a consistency check on our formulas for  anomalous dimensions of multiparticle states in AdS, we review how to obtain the contribution to the double-twist anomalous dimensions from the Witten diagrams for the gauge and graviton exchange contributions to the four-point functions and compare them to our results in the special case $Q=2$.

The starting point to obtain the anomalous dimensions of double-trace operators $[\Phi, \Phi]_{n,\ell}$ is  the four-point function
\eqna{
\langle \Phi(x_1) \Phi(x_2) \Phi^\dagger(x_3)\Phi^\dagger(x_4) \rangle&=\frac{1}{(x_{12}^ 2 x_{34}^2)^\Delta} \CG(z, \zb) \, ,\\
\CG(z, \zb)&=\sum_{n, \ell} a_{n, \ell}\,  g_{n, \ell}(z, \zb)\, ,
}[blockExp]
where $a_{n, \ell}=c_{\phi \phi [\phi, \phi]_{n, \ell}}c_{\phi^\dagger \phi^\dagger [\phi, \phi]_{n, \ell}}$ is the product of the OPE coefficients involving the exchange of a double-trace operator of spin $\ell$ and twist $2\Delta+2n$,where $\Delta$ is the conformal dimension of $\Phi$.  We have also introduced the cross-rations
\eqna{
u=\frac{x_{12}^2x_{34}^2}{x_{13}^2x_{23}^2}= z \zb\, , \qquad \quad v=\frac{x_{14}^2x_{23}^2}{x_{13}^2x_{23}^2}= (1-z)(1- \zb)\, ,
}[]
with $x_{ij}\equiv (x_i-x_j)^2$. 
Before turning on any interaction, the four-point function is given by the Generalized Free Theory (GFF) result
\eqna{
\CG^{(0)}(z, \zb)=u^\Delta+\left(\frac{u}{v}\right)^\Delta\, .
}[GFF4d] 
The expansion in conformal blocks as in~\eqref{blockExp} of this expression is made easier by the properties of the blocks themselves. For  equal external dimensions, in fact, under the exchange of operators at point $1\leftrightarrow 2$,  they transform as~\cite{Poland:2018epd}
\eqna{
g_{n, \ell}(u,v)=(-1)^\ell g_{n, \ell}\left( \frac{u}{v}, \frac{1}{v}\right)\, .
}[]
Applied to the four-point function in~\eqref{GFF4d}, this implies that we can just expand the first term and we get the second one for free just by multiplying by $(-1)^\ell$. In formula
\eqna{
u^\Delta=\sum_{n, \ell} \tilde{a}_{n, \ell}^{(0)} g_{n, \ell} (z, \zb)\, , \qquad \quad a_{n, \ell}^{(0)}=\left(1+(-1)^\ell\right) \lsp \tilde{a}_{n, \ell}^{(0)}\, ,
}[]
where the superscript $(0)$ stresses to the fact that we are restricting to the GFF result.

As we did in our main discussion, we restrict to the lowest-twist double-trace operators,  \textit{i.e.} $n=0$. The $n=0$ conformal blocks have the same expression in any  $d$ dimension
\eqna{
g^{(\mathrm{coll})}_{\ell}(u,v)=u^{\Delta}\frac{(1-v)^{\ell}}{(-2)^{\ell}}{}_2F_1(\Delta+\ell,\Delta+\ell,2(\Delta+\ell);1-v)\, . 
}[collBlock]
By using these blocks we can determine the GFF OPE coefficients ${a}_{0, \ell}^{(0)}\equiv {a}_{\ell}^{(0)}$
\eqna{
{a}^{(0)}_{\ell}=\left( 1+(-1)^\ell \right)\frac{2^{\ell } \Gamma (\ell +\Delta )^2 \Gamma (\ell +2 \Delta -1)}{\Gamma (\Delta )^2 \Gamma (\ell +1)
   \Gamma (2 \ell +2 \Delta -1)}\, .
}[OPEGFF]

Turning on interactions modify the classical dimensions and the GFF OPE coefficients of the double-trace operators. At tree level
\eqna{
\Delta_{[\Phi, \Phi]_{\ell, n}}&=2\Delta+2n+\ell +\varepsilon \lsp  \gamma^{(1)}_{n,\ell}\, ,\\
a_{n,\ell}&=a_{n,\ell}^{(0)}+\varepsilon \lsp a_{n,\ell}^{(1)}\,,
}[]
where $\varepsilon$ is a generic coupling we will soon identify with either $g_{\mathrm{U}(1)}\kappa^2$ (photon exchange) or $\kappa^2$ (graviton exchange).   When plugging this expansion in the conformal block decomposition~\eqref{blockExp}, the presence of the anomalous dimensions determines the appearance of terms
\eqna{
\frac{1}{2}u^{\Delta+n} \log u \sum_{n, \ell}a_{n, \ell}^{(0)} \gamma_{n, \ell}^{(1)} g_{n, \ell}(z, \zb)\, .
}[]
This gives as a prescription on how to compute the anomalous dimensions starting from the four-point function: single out the $\log u$ coefficient in $\CG(z, \zb)$, expand it in blocks to extract $a^{(0)}_{n, \ell}\gamma^{(1)}_{n, \ell}$, divide by the GFF coefficients in~\eqref{OPEGFF}.

In the following,  we will compute the contributions to $\CG(z, \zb)$ from a tree-level photon and graviton exchange. Then, by following the procedure just described, we will determine the anomalous dimensions for the lowest twist double-trace operators, obtaining a perfect agreement with the AdS results in Sec.~\ref{sec:DT}.
\subsection{Photon exchange}
Starting off with a photon exchange, in terms of Witten diagram we can write the four-point function as 
\begin{center}
\begin{tikzpicture}
\node[left=2cm] at (0,0) {
$\langle \Phi(x_1) \Phi(x_2) \Phi^\dagger (x_3) \Phi^\dagger (x_4) \rangle=$};
\draw (0, 0) circle (1.5cm);
  \draw[snake it] (0,-0.6) -- (0,0.6);
\draw (0, 0.6)--(-0.75, 1.3);
\draw (0, 0.6)--(0.75, 1.3);
\draw (0, -0.6)--(-0.75, -1.3);
\draw (0, -0.6)--(0.75, -1.3);
\node[left=0.1cm, above=0.1cm] at (-0.75, 1.3) {\small $\Phi(x_1)$};
\node[left=0.1cm, below=0.1cm] at (-0.75, -1.3) {\small $\Phi(x_2)$};
\node[right=0.1cm, above=0.1cm] at (0.75, 1.3) {\small $\Phi^\dagger (x_3)$};
\node[right=0.1cm, below=0.1cm] at (0.75, -1.3) {\small $\Phi^\dagger (x_4)$};
\node at (2,0) {+};
\draw (4, 0) circle (1.5cm);
  \draw[snake it] (4,-0.6) -- (4,0.6);
\draw (4, 0.6)--(3.25, 1.3);
\draw (4, -0.6)--(4.96, 1.14);
\draw (4, -0.6)--(3.25, -1.3);
\draw (4, 0.6)--(4.96, -1.14);
\node[left=0.1cm, above=0.1cm] at (3.25, 1.3) {\small $\Phi(x_1)$};
\node[left=0.1cm, below=0.1cm] at (3.25, -1.3) {\small $\Phi(x_2)$};
\node[right=0.2cm, above=0.1cm] at (4.96, 1.3) {\small $\Phi^\dagger (x_3)$};
\node[right=0.2cm, below=0.1cm] at (4.96, -1.3) {\small $\Phi^\dagger (x_4)$};
\end{tikzpicture}
\end{center}
Similarly to the GFF example, because of the $1\leftrightarrow 2$ symmetry, the  results from the second diagram  are simply given by $(-1)^\ell$ the ones coming from the first $t$-channel exchange.  Hence we will analyze in detail only the first contribution, which reads
\eqna{
\frac{\langle \Phi(x_1) \Phi(x_2) \Phi^\dagger (x_3) \Phi^\dagger (x_4) \rangle_t}{(g_{\mathrm{U(1)}}\kappa)^2 (\tilde{N}^{(d)}_{\Delta,0,0})^2}&=\int \frac{d^{d+1} w}{w_0^{d+1}} G_{B\partial}^{(\Delta)}(w,x_2) \lrvec{\nabla}_{\mu} G_{B\partial }^{(\Delta)}(w,x_4) \times \\
&\, \quad  \int  \frac{d^{d+1} z}{z_0^{d+1}}G_{BB}^{\mu \nu} (z, w) G_{B\partial}^{(\Delta)}(z,x_1) \lrvec{\nabla}_{\nu} G_{B\partial }^{(\Delta)}(z,x_3) \, ,
}[tPhoton]
where $G^{(\Delta)}_{B\partial}$ represents the bulk-to-boundary propagator with dimension $\Delta$, while $G^{\mu \nu}_{BB}$ is the photon bulk propagator. The factor of $\tilde{N}^{(d)}_{\Delta, 0,0}$ defined in~\eqref{Ntilde4d} and~\eqref{Ntilde3d} takes into account the normalization of the bulk-to-boundary propagators. 
The integral on the second line can be written as a finite sum\footnote{As in our discussion in Sec.~\ref{sec:DT}, the sum is finite only for specific values of the external dimensions. In our examples we will restrict to these cases,  i.e.  $\Delta \in \mathbb{N}$ in $d=4$ and $\Delta \in \frac{2\mathbb{N}+1}{2}$ for $d=3$.} of contact Witten diagrams~\cite{DHoker:1999mqo,Bissi:2022mrs}
\eqna{
-\sum_{k=\frac{d-2}{2}}^{\Delta-1} \frac{1}{2k}\frac{\Gamma (k) \Gamma \mleft(-\frac{{d}}{2}+\Delta +1\mright)}{2 \Gamma (\Delta ) \Gamma
   \left(k-\frac{\mathit{d}}{2}+2\right)}(x_{13}^2)^{-\Delta+k} g^{\mu\nu} G_{B\partial}^{(k)}(w, x_1)\lrvec{\nabla}_\nu G_{B\partial}^{(k)}(w, x_3)\, .
}[]
If we plug it back to~\eqref{tPhoton} and we use the identity
\eqna{
\nabla^{\mu}G_{B\partial}^{(\Delta_i)}\nabla_{\mu}G_{B\partial}^{(\Delta_j)}=\Delta_i\Delta_j \left(G_{B\partial}^{(\Delta_i)}G_{B\partial}^{(\Delta_j)}-2x_{ij}^2 G_{B\partial}^{(\Delta_i+1)}G_{B\partial}^{(\Delta_j+1)}\right)\, ,
}[]
we get a sum of contact Witten diagrams of the form
\eqna{
D_{\Delta_1\Delta_2\Delta_3\Delta_4}(x_1, \cdots, x_4)=\int \frac{d^{d+1}z}{z_0^{d+1}}\prod_{i=1}^4 G_{B\partial}^{\Delta_i}(z,x_i)\, ,
}[]
which can be written in terms of cross-ratios as
\eqna{
D_{\Delta_1\cdots \Delta_4}=\frac{(x_{14}^2)^{\Sigma-\Delta_1-\Delta_4}(x_{34}^2)^{\Sigma-\Delta_3-\Delta_4}}{(x_{13}^2)^{\Sigma-\Delta_4}(x_{24}^2)^{\Delta_2}}\frac{\pi^{\frac{d}{2}}\Gamma\mleft(\Sigma-\frac{d}{2} \mright)}{2 \prod_{i=1}^4  \Gamma(\Delta_i) } \Db_{\Delta_1\cdots\Delta_4}(u,v)\, ,
}[]
with $\Sigma=\frac{1}{2}\sum \Delta_i$.  All in all~\eqref{tPhoton} becomes
\eqna{
&\frac{(x_{12}^2 x_{34}^2)^\Delta \langle \phi(x_1) \cdots \phi^{\dagger}(x_4)\rangle_t}{{(g_{\mathrm{U(1)}}\kappa)^2 (\tilde{N}^{(d)}_{\Delta,0,0})^2}}=\sum_{k=\frac{d-2}{2}}^{\Delta-1}\frac{\pi ^{\mathit{d}/2} \Gamma \mleft(-\frac{\mathit{d}}{2}+\Delta +1\mright) \Gamma
   \left(k-\frac{\mathit{d}}{2}+\Delta +1\right)}{4 \Gamma (\Delta )^3 \Gamma (k+1) \Gamma
   \left(k-\frac{\mathit{d}}{2}+2\right)}\, \times \\
   &\quad\,  u^\Delta \left(\Db_{k \Delta(k+1) (\Delta+1)}-v \Db_{k (\Delta+1)(k+1)\Delta}-\Db_{(k+1) \Delta k (\Delta+1)}+u \Db_{(k+1) (\Delta+1)k \Delta}\right)\, .
}[Dbexpr]
To extract the $\log u$ from this expression, we can use use the general results about $\Db$ functions in~\cite{Dolan:2004iy}. Let us define
\eqna{
s=\frac{\Delta_1+\Delta_2-\Delta_3-\Delta_4}{2}\, ,
}[]
then for $s=0,1, \cdots$ a generic $\Db$ function can be written as
\eqna{
\Db_{\Delta_1\cdots \Delta_4}=\log u f(u,v)+g(u,v)\, .
}[]
The case for $s$ negative integer can be reduced to the previous one by means of
\eqna{
\Db_{\Delta_1 \Delta_2\Delta_3 \Delta_4}=u^{-s}\Db_{\Delta_4 \Delta_3\Delta_2\Delta_1}\, .
}[]
We can use the fact that $f(u,v)$ admits an expansion in powers of $u$ and~$(1-v)$
\eqna{
f(u,v)&=-\frac{(-1)^s}{s!} \frac{\Gamma(\Delta_1)\Gamma(\Delta_2)\Gamma(\Delta_3+s)\Gamma(\Delta_4+s)}{\Gamma(\Delta_1+\Delta_2)} \times \\
&\quad\, \sum_{m,n=0}^\infty \frac{(\Delta_1)_m (\Delta_1+\Delta_2-\Delta_3-s)_m}{m! (1+s)_m} \frac{(\Delta_2)_{m+n} (\Delta_3+s)_{m+n}}{n! (\Delta_1+\Delta_2)_{2m+n}}u^m (1-v)^n\, .
}[]
As we can see from~\eqref{Dbexpr}, we already have a $u^\Delta$, so to extract the leading twist it is enough to consider the expression above for $m=0$
\eqna{
f(0,v)=\frac{(-1)^{s+1}}{s!} \frac{\Gamma(\Delta_1)\Gamma(\Delta_2)\Gamma(\Delta_3+s)\Gamma(\Delta_4+s)}{\Gamma(\Delta_1+\Delta_2)} {}_2F_1(\Delta_2, \Delta_3+s, \Delta_1+\Delta_2;1-v)\, .
}[DblogLowestTwist]
Comparing with~\eqref{Dbexpr}, it is possible to  extract the anomalous dimensions for $\tau=2\Delta$ in   $d=3$ and $d=4$ 
\twoseqn{
\gamma_{0, \ell}^{(3d)}&={(g_{\mathrm{U(1)}}\kappa)^2 (\tilde{N}^{(3d)}_{\Delta,0,0})^2}\begin{cases}
\frac{2 \pi ^{{3}/2} \Gamma \mleft(-\frac{{3}}{2}+2 \Delta +1\mright)}{ \Gamma
   (2 \Delta )}\, , & \quad \ell=0\, ,\\
   \frac{\pi ^2 \Gamma (\ell +1) \Gamma \mleft(\ell +2 \Delta -\frac{3}{2}\mright)}{\Gamma \mleft(\ell
   +\frac{3}{2}\mright) \Gamma (\ell +2 \Delta -1)}\, , &\quad \ell \geq 2\,,
\end{cases}
}[]
{
\gamma_{0, \ell}^{(4d)}&={(g_{\mathrm{U(1)}}\kappa)^2 (\tilde{N}^{(4d)}_{\Delta,0,0})^2}\begin{cases}
\frac{\pi^2}{2\Delta-1}\, , & \quad \qquad \ell=0\, ,\\
  \frac{ \pi ^2}{(\ell +1) (2 \Delta +\ell -2)}\, , &\quad \qquad \ell \geq 2\,.
\end{cases}
}[]
\subsection{ Graviton exchange}
When a  graviton  is exchanged, the four-point function can be written as
\threeseqn{
&\quad \quad \! \lsp  \frac{\langle \Phi\cdots \Phi^\dagger\rangle_t}{\kappa^2 (\tilde{N}^{(d)}_{\Delta,0,0})^2}=\int \frac{d^{d+1}w}{w_0^{d+1}}A^{\mu\nu}(w, x_1,x_3) T_{\mu \nu}(w, x_2, x_4)\, , 
}[]
{
&A^{\mu \nu}(w, x_1, x_3)=\int \frac{d^{d+1}z}{z_0^{d+1}} G_{BB}^{\mu \nu;\rho \sigma} (z,w) T_{\rho \sigma} (z, x_1, x_3)\, ,
}[]
{
&\begin{aligned}
\, T_{\rho \sigma}(w, x_1, x_3)&=\nabla_\rho G_{B\partial}^{(\Delta)}(z, x_1) \nabla_\sigma G_{B\partial}^{(\Delta)}(z, x_3)-{g_{\rho \sigma}}\Big( m^2   G_{B\partial}^{(\Delta)}(z, x_1) G_{B\partial}^{(\Delta)}(z, x_3) \\
& \quad \, + \nabla^\kappa G_{B\partial}^{(\Delta)}(z, x_1)  \nabla_\kappa G_{B\partial}^{(\Delta)}(z, x_3)\Big)\, , 
\end{aligned}
}[][]
where $m^2=\Delta(\Delta-d)$ and $G_{BB}^{\mu \nu;\rho \sigma}$ is the bulk graviton propagator.
Under the same assumptions, we can rewrite $A^{\mu \nu}$ as a finite sum of bulk-to-boundary propagators\footnote{With the respect to the expression in~\cite{Bissi:2022mrs}, we integrate by part the term $\nabla^\mu \nabla^\nu G_{B\partial}^{(k)}(x_1)$, using the fact we integrate against the conserved tensor and we symmetrize in $x_1 \leftrightarrow x_3$.}
\eqna{
A^{\mu \nu}&=\sum_{k=\frac{d}{2}-1}^{\Delta-1} \frac{\Delta  \Gamma (k) \Gamma \mleft(-\frac{\mathit{d}}{2}+\Delta +1\mright)}{-2 \Gamma (\Delta ) \Gamma
   \mleft(k-\frac{\mathit{d}}{2}+2\mright)} \frac{(x_{13}^2)^{k}}{(x_{13}^2)^{\Delta}} \Bigg( 
   \frac{g^{\mu \nu} \left(d-k-2 \right)}{(k+1)(d-1)} G_{B\partial}^{(k)}(x_1) G_{B\partial}^{(k)}(x_3)\\
   &\quad\, -\frac{1}{2 k(k+1)}\left( \nabla^\mu G_{B\partial}^{(k)}(x_1) \nabla^\nu G_{B\partial}^{(k)}(x_3)+\nabla^\nu G_{B\partial}^{(k)}(x_1) \nabla^\mu G_{B\partial}^{(k)}(x_3)  \right)
   \Bigg)\, .
}[]
All in all
\eqna{
&\frac{(x_{12}^2 x_{34}^2)^\Delta \langle \Phi \cdots \Phi^\dagger \rangle_t}{\kappa^2 (\tilde{N}^{(d)}_{\Delta,0,0})^2} = \frac{\pi ^{\mathit{d}/2} \Delta ^2 \Gamma \mleft(-\frac{\mathit{d}}{2}+\Delta +1\mright)  \Gamma \mleft(k-\frac{\mathit{d}}{2}+\Delta +1\mright)}{2 \Gamma (\Delta )^3 \Gamma
   (k+2) \Gamma \left(k-\frac{\mathit{d}}{2}+2\right)}(z
   \zb)^{\Delta } \times\\
   &\Bigg \lbrace \frac{-\mathit{d}+2 \Delta +2 k+2}{\Delta }(u+v-1) \Db_{k_{+1}\Delta_{+1}k_{+1}\Delta_{+1}}  +\frac{k (-\mathit{d}+2 k+2)}{\Delta }\Db_{k \Delta_{+1}k\Delta_{+1}}\\
   &  -k (\Db_{k \Delta k_{+1}\Delta_{+1}}+v \Db_{k \Delta_{+1}k_{+1}\Delta}+\Db_{k_{+1}\Delta k_{+1}\Delta_{+1}}+u \Db_{k_{+1}\Delta_{+1}k\Delta})\\
   & +(2\Delta-d)\Db_{k_{+1}\Delta k_{+1}\Delta} +\frac{\mathit{d} k \left(\mathit{d}^2-\mathit{d} (2 \Delta +2 k+1)+2 \Delta 
   (k+2)-2\right)}{(\mathit{d}-1) (\mathit{d}-2 \Delta -2 k)}\Db_{k\Delta k\Delta}\Bigg \rbrace\, , 
}[]
where $k_{+1}\equiv k+1$ and analogously for $\Delta$. 

As we have done for the photon exchange, we can isolate the $\log u$ terms using the properties of $\bar{D}$ functions and extract the anomalous dimensions
\twoseqn{
\gamma_{0, \ell}^{(3d)}&=-{\kappa^2 (\tilde{N}^{(3d)}_{\Delta,0,0})^2}\begin{cases}
\frac{2 \pi ^{{3}/2} \Delta ^2 (2 \Delta -3) \Gamma \mleft(2 \Delta
   -\frac{{3}}{2}\mright)}{2 \Gamma (2 \Delta )}\, , & \quad \ell=0\, ,\\
   \frac{\pi ^2 \Delta ^2 \Gamma (\ell +1) \Gamma \mleft(\ell +2 \Delta -\frac{3}{2}\mright)}{2 \Gamma
   \mleft(\ell +\frac{3}{2}\mright) \Gamma (\ell +2 \Delta -1)}\, , &\quad \ell \geq 2\,.
\end{cases}
}[]
{
\gamma_{0, \ell}^{(4d)}&=-{\kappa^2 (\tilde{N}^{(4d)}_{\Delta,0,0})^2}\begin{cases}
\frac{2 \pi ^2 (\Delta -2) \Delta ^2}{3 (\Delta -1) (2 \Delta -1)}\, , & \quad \qquad \quad \ell=0\, ,\\
\frac{2 \pi^2 \Delta^2}{3(\ell+1)(\ell+2\Delta-2)}\,, &\quad \qquad \quad   \ell \geq 2\, .
\end{cases}
}[]
\section{Graviton matrix elements} \label{Appendix:gravitonMatrix}
Similarly to the discussion for $A_\mu$ in Sec.~\ref{subsec:PhotonMatrixEl}, we can extract the spin $\ell_1$ and $\ell_2$ component by taking
\eqna{
(z_1 \cdot \nabla_1)^{\ell_1}(z_2 \cdot \nabla_2)^{\ell_2} h_{\mu\nu}(X; P_1, P_2) 
}[]
and then expand around the points $x_1=0$ and $x_2=\infty$.  To take these derivatives efficiently we can use the parametrization in~\eqref{PPar}, so that 
\eqna{
h_{\mu\nu}\left[ \phi_{0\ell_2\ell_2}^* \phi_{0\ell_1\ell_1}\right] \propto \left[ (\partial_{\zb_1})^{\ell_1}(\partial_{z_2})^{\ell_2} h_{\mu\nu}(X;P_1,P_2)\right]_{\zb_1=0, z_2=0}\, ,
}[]
where the proportionality constant is fixed by requiring to satisfy the equation of motion in~\eqref{EoM}.

For integer (half-integer) values of $\Delta$ in $4d$ ($3d$), the series expansion of $f(t)$ is finite,  so we can evaluate the integrals in~\eqref{IJIT} in closed form for arbitrary $J$.  In $4d$, this takes the form
\eqna{
\CI_{\CT, \Delta \in \mathbb{Z}}^{(4d)}&=\sum_{i=0}^{\min (\ell_1, \ell_2)} \frac{-2\pi^2 \ell_1!\ell_2!(J-i+1)!(\Delta!)^2 \Gamma(\Delta+i-2) \tilde{N}^{(4d)}_{\Delta, \ell_1, \ell_2}\tilde{N}^{(4d)}_{\Delta, \ell_3, \ell_4}}{3 i!(i-J) (1-i+J)  \Gamma (\text{$\ell $1}+\Delta ) \Gamma (\text{$\ell
   $2}+\Delta ) \Gamma (-i+J+\Delta +2)}\\
  &\quad \, \times \Big \lbrace  c_1\,  {}_4F_3 \left( \begin{array}{c}
  1-i+\ell_1, \quad  1-i+\ell_2, \quad 2-\Delta \quad\Delta-1 \\
  2, \quad 2-i-\Delta,\quad 2-i+J+\Delta
\end{array}  ; 1 \right)\\
&\qquad \! +  c_2\,  {}_4F_3 \left( \begin{array}{c}
  2-i+\ell_1, \quad  2-i+\ell_2, \quad 3-\Delta \quad\Delta \\
  3, \quad 3-i-\Delta,\quad 3-i+J+\Delta
\end{array}  ; 1 \right)\\
&\qquad \! +  c_3\,  {}_5F_4 \left( \begin{array}{c}
2, \quad   2-i+\ell_1, \quad  2-i+\ell_2, \quad 3-\Delta \quad\Delta \\
 1, \quad 3, \quad 3-i-\Delta,\quad 3-i+J+\Delta
\end{array}  ; 1 \right)\Big \rbrace\, ,
}[]
where $\ell_1+\ell_3=\ell_2+\ell_4=J$ and we have defined
\eqna{
c_1&=2(\Delta+i-2)(J^2+3\ell_1\ell_2(\Delta+1))-3(\ell_1+\ell_2)(J+i\lsp \Delta)+J(\Delta+4i-2)\\
&\quad\, +i(2-\Delta+i(3\Delta-2))\, ,\\
c_2&=\frac{(i-\ell_1-1)(i-\ell_2-1)(\Delta-2)}{J+\Delta-i+2} \lbrace (J^2+i^2)(\Delta+2)+6\ell_1\ell_2 \Delta\\
&\quad\, -3\Delta(\ell_1+\ell_2)(J+i)+ J (\Delta ^2-2 \Delta +4 \Delta  i-4 i+4)-i((\Delta -2) \Delta +4) \rbrace\, , \\
c_3 &=\frac{3(i-\ell_1-1)(i-\ell_2-1)(\Delta-2)}{J+\Delta-i+2}\lbrace J^2+\ell_1\ell_2 - J (\ell_1+\ell_2)+\frac{\Delta+2}{3}(J-i)\rbrace\, .
}[]
In three dimensions,  $f(t)$ truncates every time $\Delta \in \mathbb{Z}-1/2$ and $\Delta\geq \frac{3}{2}$, so it is convenient to define $\delta=\Delta-\frac{3}{2} \geq 0$. In this case the integral in~\eqref{IJIT} takes the form
\eqna{
&\CI_{\CT, \Delta \in \mathbb{Z}-\frac{1}{2}}^{(3d)}=\sum_{i=0}^{\min(\ell_1, \ell_2)}\sum_{k=0}^{\delta} \frac{\pi^{3/2} (k+1) \ell_1!\ell_2! (i-k+\delta)! (2\delta+1)!\tilde{N}^{(3d)}_{\Delta, \ell_1, \ell_2}\tilde{N}^{(3d)}_{\Delta, \ell_3, \ell_4}}{2^{3-2(k-\delta)} i! (2k+2)!(\ell_1-i)!(\ell_2-i)! (\delta-k)!}\\
&\quad  \, \frac{(3+2\delta)^2\Gamma(J-i)\Gamma\mleft(k+\delta+\frac{1}{2}\mright) \Gamma\mleft(k-i+\ell_1+\frac{1}{2}\mright)\Gamma\mleft(k-i+\ell_2+\frac{1}{2}\mright)}{\Gamma(J+k+\delta-i+3)\Gamma\mleft(\ell_1+\delta+\frac{3}{2}\mright)\Gamma\mleft(\ell_2+\delta+\frac{3}{2}\mright)}\times\\
&\quad  \,\lbrace J^2 (2 k+1) (2 \delta +8 k+1)-4 (\ell_1+\ell_2) (2 \delta +2 k+1) ((2 \delta +3) i+2 J k+J)\\
&\quad\, -(2 k+1) \left(i \left(2 \delta ^2+\delta -6 (2 \delta +1) J+2 \delta  k+9 k\right)-J \left(2 \delta ^2+\delta +2 \delta  k+9 k\right)\right)\\
&\quad  \,+ i^2 ((2 \delta +1) (8 \delta +9)+2 (2 \delta +9) k)+8 {\ell_1} {\ell_2} (\delta +k+2) (2 \delta +2 k+1)\rbrace\,.
}[]

\section{Large Spin Expansion of Exchange Diagram Integrals}
\label{app:SubtleScalar}
When evaluating the exchange diagram contributions to the anomalous dimensions, we have to perform an integration  over AdS. For generic values of the dimensions of operators, this integration cannot be done in closed form and instead we series expand the integrand and perform the integration term-by-term. If one takes the large spin limit of this series, however, one gets spurious contributions to the anomalous dimension because the large spin limit does not commute with the series sum. In this section, we analyze this in detail for a specific choice of the dimensions of operators where we will be able to see this behavior analytically.

We will take the case $d=3, \Delta=1, \Delta_\epsilon=3$, in which case the summand $\gamma_{J,k}$ from (\ref{eq:ScalarExchangeSummand}) simplifies to
\begin{equation}
\gamma_{J,k} = \frac{g_\epsilon^2 \Gamma^2(k)}{(4\pi)^2 k \Gamma(k+J+1)\Gamma(k-J)}, 
\end{equation}
 and we have taken $J$ to be even. We will set $g_\epsilon=4\pi$ to avoid clutter.  The anomalous dimension is 
\begin{equation}
\gamma^{(d,Q)}_J = \sum_{k=\frac{3}{2}}^\infty \gamma_{J,k} - \sum_{k=1}^\infty \gamma_{J,k}.
\end{equation}
The large $J$ limit of each summand is
\begin{equation}
\gamma_{J,k} \sim J^{-2k} \frac{\Gamma^2(k)}{\pi k},
\end{equation}
and therefore naively in the large $J$ limit one can keep only the first few terms in the sum on $k$; in fact, however, doing so would be incorrect. There is already a hint of a problem from the size of the coefficients of $J^{-2k}$, which grow factorially and therefore if we take just the leading large $J$ term at each value of $k$ and sum such terms over $k$, the result has zero radius of convergence.  Another more physical hint of a problem is that the second series, which begins at $k=1$, has a leading contribution of the form $J^{-2}$, whereas the scalar being exchanged in the bulk is dual to $\epsilon$ which has dimension $3$ and so should produce a leading contribution that falls off like $J^{-3}$.\footnote{The bulk Witten diagram for the scalar exchange decomposes into conformal blocks for the scalar operator $\epsilon$ as well as for the double-trace operators $[\Phi \Phi]_{n,J}$, and the latter have dimensions beginning at $2\Delta=2$.  However, these do not produce contributions to the anomalous dimension of $\Phi$, which is perhaps easiest to see by the fact that the Lorentzian inversion formula vanishes when applied to double-trace operators in the exchange channel in the infinite $N$ limit.}

At finite $J$, the summands decay with $k$ like $\sim k^{-2}$ and therefore the series converge.  In this case, the two series each have a closed form expression:
\begin{equation}
\begin{aligned}
\gamma_J^{(1)} &\equiv \sum_{k=\frac{3}{2}}^\infty \gamma_{J,k} = \psi_1\left(\frac{J+1}{2}\right)-\frac{4}{2 J+1}, \\
\gamma_J^{(2)} &\equiv \sum_{k=1}^\infty \gamma_{J,k} = \frac{1}{2} \left(\psi_1\left(\frac{J+1}{2}\right)-\psi_1\left(\frac{J}{2}+1\right)\right),
\end{aligned}
\end{equation}
where $\psi_1(t) \equiv (\frac{d}{dt})^2 \log \Gamma(t)$ is the trigamma function.  The large $J$ series expansion of these is
\begin{equation}
\begin{aligned}
\gamma_J^{(1)} &\sim J^{-2} - \frac{7}{6} J^{-3} + \frac{1}{4} J^{-4} + \frac{97}{120} J^{-5} + \dots, \\
\gamma_J^{(2)}& \sim J^{-2} - J^{-3} +J^{-5} + \dots.
\end{aligned}
\end{equation}
Interestingly, the leading large $J$ behavior of $\gamma_J^{(1)}$ is $J^{-2}$, which is completely invisible in the term-by-term large $J$  expansion of the summands.  Moreover, it is exactly this new $J^{-2}$ term, which is so subtle to see from the original series representation, that cancels the $J^{-2}$ term coming from the second series.  The full resummed answer is
\begin{equation}
\gamma^{(d,Q)}_J  = 2 \psi_1(J+1) - \frac{4}{2J+1} \sim -\frac{1}{6 (J+\frac{1}{2})^3} + \frac{7}{120 (J+\frac{1}{2})^5} - \frac{31}{672 (J+\frac{1}{2})^7} + \dots,
\end{equation}
which indeed begins at $O(J^{-3})$.  Moreover, the series coefficients of $\psi_1(z)$ in powers of $1/z$ at large $z$ are given by Stirling's approximation, which is a canonical example of an asymptotic series.

\section{Fast algorithm for constructing primaries}
\label{app:buildprimary}
In this section we will present the method that recursively builds the \textit{lowest-twist} primary operators made by $\Phi$, which is based on a similar algorithm of \cite[4.1]{Anand:2020gnn}. The only difference is that, in \cite{Anand:2020gnn}, the authors are building primary operators in $d=2$ CFT, which means,  $\partial_-\Phi$ is a primary rather $\Phi$.\footnote{Notice that we can easily generalize the method in~\cite{Anand:2020gnn} because we are just interested in the lowest-twist highest-weight states.}

Let us start by counting the number of independent `monomial' states.  At fixed charge $Q$ and total spin $J$,  define
\begin{equation}
P(Q,J)=\text{ $\#$ of partitions of $J$ objects into exactly $Q$ bins}\, .
\end{equation}
In this counting we are including both primary and descendant states. To single out the first ones, we need to remove from $P(Q,J)$ the number of descendants. But we know that these can be obtained just by acting with $P_\mu$ on the primary states at spin $(J-1)$. So the total number of primaries is given by 
\begin{equation}
\mathcal{N}(Q,J)=P(Q,J)-P(Q,J-1)\, .
\end{equation}
It is straightforward to verify that $\mathcal{N}(Q,J)$ satisfies the following recursion relation,
\begin{equation}\label{Nrec}
\mathcal{N}(Q,J)=\sum_{k=0}^{J/Q}\mathcal{N}(Q-1,J-kQ)\, .
\end{equation}
This relation suggests that we can build a $Q$-primary state  from the $(Q-1)$ one.  In particular, let us call $\mathcal{O}'_{Q-1,\ell}$ the  state at  $(Q-1)$ with total spin $\ell$, then we can construct a $Q$-primary operator  at spin $J$ as
\begin{equation}\label{Ogen}
\mathcal{O}_{Q,J}=\Big[\mathcal{O}'_{Q-1,\ell},\Phi\Big]_{J-\ell}\, , 
\end{equation}
where a double-twist operator made of two generic operators $z \cdot \CO_{\Delta, \ell}$ of dimension $\Delta$ and spin $\ell$ is given by~\cite{Penedones:2010ue},
\eqna{
\Big[z\cdot \CO_{\Delta_A,\ell_A},z\cdot \CO_{\Delta_B,\ell_B}\Big]_\ell&=\sum_{m=0}^\ell c_m^\ell(z\cdot P)^m(z\cdot \CO_{\Delta_A,\ell_A})(z\cdot P)^{\ell-m}(z\cdot \CO_{\Delta_B,\ell_B})\, ,\\
c_m^\ell \equiv c_m^\ell(\Delta_A,\ell_A;\Delta_B,\ell_B)&=\frac{(-1)^m}{m!(\ell-m)!\Gamma(\Delta_A+\ell_A+m)\Gamma(\Delta_B+\ell_B+\ell-m)}
}[]
Moreover, the recursion in~\eqref{Nrec} tells us that to construct the minimal set of independent primaries we do not have to consider any possible $\ell$ in~\eqref{Ogen}, but rather  
\begin{equation}
\Big[\mathcal{O}'_{Q-1,J-kQ},\Phi\Big]_{kQ}\,, \qquad \text{with }\, k=0,1,\cdots,\frac{J}{Q}\, .
\end{equation}

\section{Contact terms contribution at $Q=3$} \label{App:contact}
\begin{figure}[t!] \centering
\includegraphics[scale=0.6]{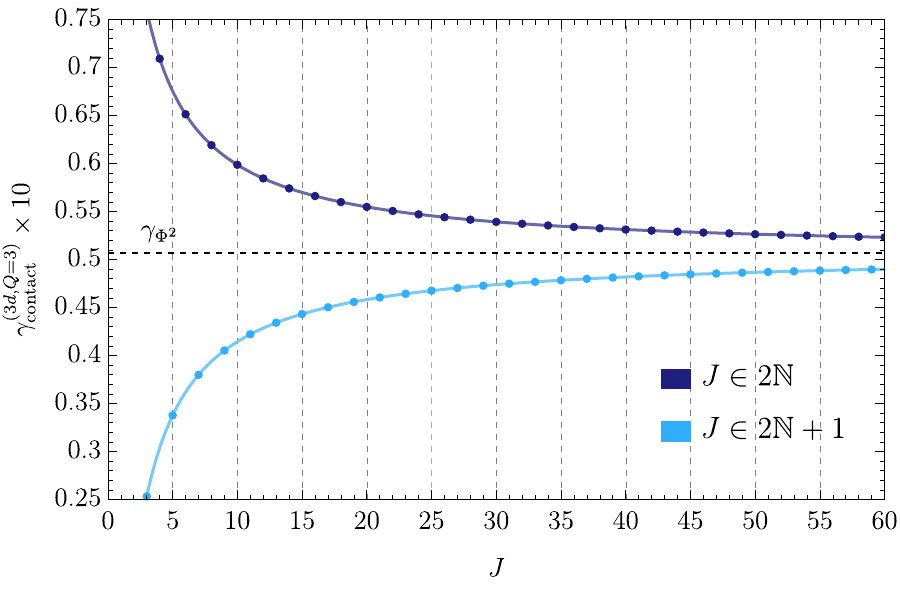}
\caption{Anomalous dimension of the 3-particle state due to the $(\phi \phi^\dagger)^2$ contact interaction in $3d$ for $\Delta=1$.  The two curves correspond to the analytic expression for the even and odd spins and the dots correspond to the values of the anomalous dimensions computed for  integer $J$.  Notice that the curves goes to a constant (dashed line) as $J\to \infty$. The constant value correspond to the anomalous dimension of the $[\Phi, \Phi]_{J=0}\equiv \Phi^2$ primary.}\label{fig:q3contactD13d}
\end{figure}

In this appendix, we explore in detail the effect of a $(\phi \phi^\dagger)^2$ interactions for the anomalous dimensions of the lowest twist three-particle operators. We are interested in studying the matrix 
\eqna{
\int d^d x \sqrt{-\tilde{g}}\lsp {}_{J;Q=3}^{\qquad r}\! \bra{\Psi} (\phi \phi^\dagger)^2 \ket{\Psi}_{J;Q=3}^{s}\, , \qquad r, s=1, \cdots, \CN(Q, J)\, ,
}[]
which at $Q=3$ has only one non-vanishing eigenvalue.  It is easy to verify that the corresponding non trivial eigenvector coincides with the triple-twist operator $[\Phi, [\Phi, \Phi]_{\ell=0}]_J\equiv [\Phi, \Phi^2]_J$. When properly normalized, this primary takes the form
\eqna{
\ket{[\Phi, \Phi^2]_J}&=\frac{1}{\sqrt{N_{[\Phi, \Phi^2]}}}\sum_{\ell=0}^J \sum_{n=0}^\ell  (-1)^\ell C_{n, \ell}\ket{\ell-n, n, J-\ell}\, ,\\
C_{n, \ell}&=\sqrt{\frac{\Gamma(\Delta+n)\Gamma(\Delta+\ell-n)}{n!(\ell-n)!(J-\ell)! \Gamma(\Delta+J-\ell)}} \frac{1}{\Gamma(2\Delta+\ell)}\, ,\\
N_{[\Phi, \Phi^2]}&\!=\frac{2\Gamma(\Delta)\Gamma(3\Delta+2J-1)\!\!\left[2\lsp (-1)^J\Gamma(2\Delta)\Gamma(J+\Delta)+\Gamma(\Delta)\Gamma(2\Delta+J)\right]}{J! \Gamma(2\Delta)\Gamma(\Delta+J) \Gamma(2\Delta+J)^2\Gamma(3\Delta+J-1)}\, .
}[]
Since we have an analytic expression for the primaries and, in~\eqref{lambda0El}, we have computed the two-particle matrix element, we can determine the anomalous dimension of $[\Phi, \Phi^2]$, due to a $(\phi \phi^\dagger)^2$-contact interaction, in closed form
\eqna{
&\gamma^{(d, Q=3)}_{\mathrm{contact}}=\frac{\pi ^{-\mathit{d}/2} \Gamma (\Delta )^2 \Gamma \mleft(2 \Delta -\frac{\mathit{d}}{2}\mright)}{2
   \Gamma (2 \Delta ) \Gamma \mleft(\Delta -\frac{\mathit{d}}{2}+1\mright)^2}+(-1)^J \frac{\pi ^{-\mathit{d}/2} \Gamma (\Delta ) \Gamma \mleft(2 \Delta -\frac{\mathit{d}}{2}\mright) \Gamma
   (J+\Delta )}{\Gamma \mleft(\Delta -\frac{\mathit{d}}{2}+1\mright)^2 \Gamma (J+2 \Delta )}\, , \\
   &\quad\,  \lsp \xrightarrow[\,J\to \infty\,]{}\frac{\pi ^{-\mathit{d}/2} \Gamma (\Delta )^2 \Gamma \mleft(2 \Delta -\frac{\mathit{d}}{2}\mright)}{2
   \Gamma (2 \Delta ) \Gamma \mleft(\Delta -\frac{\mathit{d}}{2}+1\mright)^2}+\frac{(-1)^J}{J^\Delta}\frac{\pi ^{-\mathit{d}/2} \Gamma (\Delta ) \Gamma \mleft(2 \Delta -\frac{\mathit{d}}{2}\mright)}{\Gamma
   \mleft(\Delta-\frac{\mathit{d}}{2} +1\mright)^2}\, .
}[contactAnomDim]
What might seem surprising at first is that the anomalous dimensions decay at large spin as $J^{-\Delta}$. To interpret this result,  it is useful to picture $[\Phi, \Phi^2]$ as three partons in AdS.  The configuration with energy~\eqref{contactAnomDim},  can be imagined as two partons on one side of AdS and an isolated one at the opposite side.  At large spin, the leading correction to the energy is given by the anomalous dimension of the 2-particle blob.  But the only non-vanishing anomalous dimension at $Q=2$, due to the contact interaction, is the spin zero. And indeed the constant term in~\eqref{contactAnomDim} coincides with $\gamma^{(d,Q=2)}_{\Phi^2, \mathrm{contact}}$.  The origin of the  $J^{-\Delta}$ part, instead, should be sought in the non-vanishing overlap of the wavefunctions of the two far-away blobs.  As an example,  in Fig.~\ref{fig:q3contactD13d}, we show the anomalous dimensions of three-particles states due to a $\phi^4$ interaction in $3d$ and for $\Delta=1$.

\section{Background Gauge and Gravity Profiles in Parton Rest Frame}
\label{app:singleparton}
Mapping a parton trajectory to the center of AdS simplifies the analysis of the background gauge and gravity fields that they produce.  In $d=4$, the resulting backgrounds were solved in \cite{Fitzpatrick:2011hh}, which we will briefly review and generalize to arbitrary $d$. This analysis is closely related to the solutions for the background fields given in (\ref{DhokerSolnGauge}) and (\ref{hMNsoln}).

Using the formulas (\ref{JTfromPhi}) and taking $\phi = \phi_0 = \frac{1}{\sqrt{N_{\Delta,0} V_{d-1}}} e^{i \Delta t} \cos^\Delta \rho$,  with $V_{d-1} = \textrm{vol}(S^{d-1}) = \frac{2 \pi^{\frac{d}{2}}}{\Gamma(\frac{d}{2})}$, can easily calculate the resulting current and stress tensor defined in (\ref{JTfromPhi})
\begin{equation}
\begin{aligned}
\CJ_0 &
=  2 g_{\mathrm{U}(1)} \Delta \frac{1}{N_{\Delta,0} V_{d-1}} y^{2\Delta}, \\
 (\CT^0_{\ 0}, \CT^\rho_{ \ \rho},\CT^{\Omega_i}_{ \ \Omega_i}) &= \frac{2\Delta y^{2\Delta} }{N_{\Delta,0} V_{d-1}} (\frac{d}{2}-\Delta, \frac{d}{2}, \frac{d}{2} - \Delta (1-y^2)),
\end{aligned}
\end{equation}
where $y=\cos \rho$, and the remaining components of $\CJ^\mu$ and $\CT^\mu_{\ \nu}$ vanish.  We can take a gauge where the only nonvanishing components of the photon and linearized graviton are $A_0$ and $h_{00}, h_{\rho\rho}$.  The equations of motion for $A$ and $h$ are 
\begin{equation}
\begin{aligned}
& \left( \frac{y^{d+1}}{(1-y^2)^{\frac{d}{2}-1}}\right) \partial_y \left( \frac{(1-y^2)^{\frac{d}{2}}}{y^{d-3}} \partial_y A_0 \right) = 2 g \Delta \frac{1}{N_{\Delta,0} V_{d-1}} y^{2\Delta+2} , \\
& \frac{d-1}{2} \left( \frac{y^{d+1}}{(1-y^2)^{\frac{d}{2}-1}} \right) \partial_y \left( \frac{(1-y^2)^{\frac{d}{2}-1}}{y^{d-2}} h_{\rho \rho}\right) = \frac{2 \Delta y^{2\Delta}}{N_{\Delta,0} V_{d-1}} (\frac{d}{2}-\Delta), \\
& \frac{d-1}{2} y \partial_y (y^2 h_{00} )+ \frac{y^2(d-1) }{1-y^2} (y^2 - \frac{d}{2}) h_{\rho\rho} = \frac{2 \Delta y^{2\Delta}}{N_{\Delta,0} V_{d-1}} \frac{d}{2}.
\end{aligned}
\end{equation}
Fixing the boundary conditions so the solutions are regular at $y=1$ and vanish at $y=0$, the result is 
\begin{equation}
\begin{aligned}
A_0 &=\frac{2 g }{N_{\Delta,0} V_{d-1}}  \frac{1}{4} \left(-\frac{ 
   y^{d-2} \Gamma\left( \frac{d}{2}-1\right) \Gamma \left(\Delta-\frac{d-2}{2} \right)}{\left(1-y^2\right)^{\frac{d}{2}-1} \Gamma (\Delta)}+\frac{2 y^{2 \Delta } \, _2F_1\left(1,\Delta ,\Delta -\frac{d}{2}+2,y^2\right)}{ (2 \Delta -(d-2))}\right) , \\
   h_{00} &= \frac{2\Delta }{N_{\Delta,0} V_{d-1}} \frac{ \left(\frac{  y^{d-2} \Gamma\left( \frac{d}{2} \right)  \Gamma \left(\Delta-\frac{d-2}{2}\right)}{  \left(1-y^2\right)^{\frac{d-2}{2}}\Gamma (\Delta )}-  \frac{(d-2) y^{2 \Delta } \, _2F_1\left(1,\Delta ,\Delta-\frac{d}{2}+2,y^2\right)}{2\Delta -(d-2)}\right)}{(d-1) }, \\
   h_{\rho\rho} & =\frac{2 \Delta}{N_{\Delta,0} V_{d-1}}  \frac{\left(\frac{
   y^{d-2}  \Gamma\left( \frac{d}{2} \right) \Gamma \left(\Delta-\frac{d-2}{2}\right)}{\left(1-y^2\right)^{\frac{d-2}{2}}\Gamma (\Delta )}-  y^{2 \Delta-2 } (1-y^2) \, _2F_1\left(1,\Delta ,\Delta-\frac{d}{2}+1,y^2\right)\right)}{d-1} .
\end{aligned}
\end{equation}
Note that $A_0(y=\sqrt{t})$ is just $f(t)$ from (\ref{DhokerSolnGauge}).

\begin{figure}[t!]
\begin{center}
\includegraphics[width=0.6\textwidth]{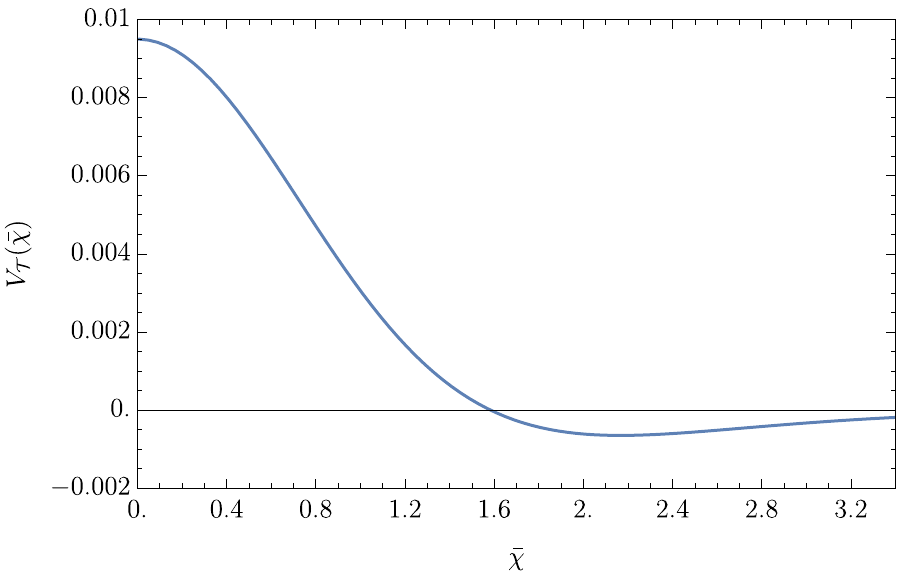}
\caption{Approximate gravitational energy $V_{\CT}(\bar{\chi})$ as a function of the center $\bar{\chi}$ of the rotating parton in the frame where the other parton is at rest in the center of AdS. In this example we restrict to $4d$ and we set the dimension of the external scalars to be $\Delta=\frac{3}{2}$.
}
\label{fig:Vappx}
\end{center}
\end{figure}

An interesting feature of the gravitational background is that, although it is always attractive at large distances, for sufficiently small $\Delta$ it can become repulsive at short distances. 
 To see this,  we can calculate the energy of the two-particle system by evaluating $V(\bar{\chi}) \equiv \< \Phi | \int d^{d} x \sqrt{g} (A_\mu \CJ^\mu + h_{\mu\nu} \CT^{\mu\nu}) | \Phi\>$, where $|\Phi\>$ is the wavefunction for the second parton, and $A_\mu, h_{\mu\nu}$ are the backgrounds created by the first parton. We first normalize the wavefunction $\phi_R$ from (\ref{eq:WvFcnGaussianApprox}) so that its norm $(\Phi, \Phi) = 2i \int d^{d} x \sqrt{g} \Phi^* \partial^0 \Phi$ is one.  Then, the integrals can all be evaluated numerically.   
As an example, in Fig.~\ref{fig:Vappx} we show a numerical result for $\Delta=\frac{3}{2}$ in $d=4$ for  the gravitational part of the potential $V_{\CT}$.

\bibliographystyle{JHEP}
\bibliography{refs}

\providecommand{\href}[2]{#2}\begingroup\raggedright\begin{thebibliography}{10}

\bibitem{Hellerman:2015nra}
S.~Hellerman, D.~Orlando, S.~Reffert and M.~Watanabe, \emph{{On the CFT
  Operator Spectrum at Large Global Charge}},
  \href{http://dx.doi.org/10.1007/JHEP12(2015)071}{\emph{JHEP} {\bf 12} (2015)
  071}, [\href{https://arxiv.org/abs/1505.01537}{{\tt 1505.01537}}].

\bibitem{Cuomo:2017vzg}
G.~Cuomo, A.~de~la Fuente, A.~Monin, D.~Pirtskhalava and R.~Rattazzi,
  \emph{{Rotating superfluids and spinning charged operators in conformal field
  theory}}, \href{http://dx.doi.org/10.1103/PhysRevD.97.045012}{\emph{Phys.
  Rev. D} {\bf 97} (2018) 045012},
  [\href{https://arxiv.org/abs/1711.02108}{{\tt 1711.02108}}].

\bibitem{Cuomo:2022kio}
G.~Cuomo and Z.~Komargodski, \emph{{Giant Vortices and the Regge Limit}},
  \href{http://dx.doi.org/10.1007/JHEP01(2023)006}{\emph{JHEP} {\bf 01} (2023)
  006}, [\href{https://arxiv.org/abs/2210.15694}{{\tt 2210.15694}}].

\bibitem{Su:2022ysc}
J.-H. Su, C.-Y. Xia, W.-C. Yang and H.-B. Zeng, \emph{{Giant vortex in a fast
  rotating holographic superfluid}},
  \href{http://dx.doi.org/10.1103/PhysRevD.107.026006}{\emph{Phys. Rev. D} {\bf
  107} (2023) 026006}, [\href{https://arxiv.org/abs/2208.14172}{{\tt
  2208.14172}}].

\bibitem{Cuomo:2023vvd}
G.~Cuomo, Z.~Komargodski and S.~Zhong, \emph{{Chiral Modes of Giant Superfluid
  Vortices}},  \href{https://arxiv.org/abs/2312.06095}{{\tt 2312.06095}}.

\bibitem{Banerjee:2017fcx}
D.~Banerjee, S.~Chandrasekharan and D.~Orlando, \emph{{Conformal dimensions via
  large charge expansion}},
  \href{http://dx.doi.org/10.1103/PhysRevLett.120.061603}{\emph{Phys. Rev.
  Lett.} {\bf 120} (2018) 061603},
  [\href{https://arxiv.org/abs/1707.00711}{{\tt 1707.00711}}].

\bibitem{Cuomo:2023mxg}
G.~Cuomo, J.~M. V.~P. Lopes, J.~Matos, J.~Oliveira and J.~Penedones,
  \emph{{Numerical tests of the large charge expansion}},
  \href{https://arxiv.org/abs/2305.00499}{{\tt 2305.00499}}.

\bibitem{Komargodski:2012ek}
Z.~Komargodski and A.~Zhiboedov, \emph{{Convexity and Liberation at Large
  Spin}}, \href{http://dx.doi.org/10.1007/JHEP11(2013)140}{\emph{JHEP} {\bf 11}
  (2013) 140}, [\href{https://arxiv.org/abs/1212.4103}{{\tt 1212.4103}}].

\bibitem{Fitzpatrick:2012yx}
A.~L. Fitzpatrick, J.~Kaplan, D.~Poland and D.~Simmons-Duffin, \emph{{The
  Analytic Bootstrap and AdS Superhorizon Locality}},
  \href{http://dx.doi.org/10.1007/JHEP12(2013)004}{\emph{JHEP} {\bf 12} (2013)
  004}, [\href{https://arxiv.org/abs/1212.3616}{{\tt 1212.3616}}].

\bibitem{Fitzpatrick:2014vua}
A.~L. Fitzpatrick, J.~Kaplan and M.~T. Walters, \emph{{Universality of
  Long-Distance AdS Physics from the CFT Bootstrap}},
  \href{http://dx.doi.org/10.1007/JHEP08(2014)145}{\emph{JHEP} {\bf 08} (2014)
  145}, [\href{https://arxiv.org/abs/1403.6829}{{\tt 1403.6829}}].

\bibitem{Alday:2007mf}
L.~F. Alday and J.~M. Maldacena, \emph{{Comments on operators with large
  spin}}, \href{http://dx.doi.org/10.1088/1126-6708/2007/11/019}{\emph{JHEP}
  {\bf 11} (2007) 019}, [\href{https://arxiv.org/abs/0708.0672}{{\tt
  0708.0672}}].

\bibitem{Alday:2015eya}
L.~F. Alday, A.~Bissi and T.~Lukowski, \emph{{Large spin systematics in CFT}},
  \href{http://dx.doi.org/10.1007/JHEP11(2015)101}{\emph{JHEP} {\bf 11} (2015)
  101}, [\href{https://arxiv.org/abs/1502.07707}{{\tt 1502.07707}}].

\bibitem{Alday:2015ewa}
L.~F. Alday and A.~Zhiboedov, \emph{{An Algebraic Approach to the Analytic
  Bootstrap}}, \href{http://dx.doi.org/10.1007/JHEP04(2017)157}{\emph{JHEP}
  {\bf 04} (2017) 157}, [\href{https://arxiv.org/abs/1510.08091}{{\tt
  1510.08091}}].

\bibitem{Alday:2016njk}
L.~F. Alday, \emph{{Large Spin Perturbation Theory for Conformal Field
  Theories}},
  \href{http://dx.doi.org/10.1103/PhysRevLett.119.111601}{\emph{Phys. Rev.
  Lett.} {\bf 119} (2017) 111601},
  [\href{https://arxiv.org/abs/1611.01500}{{\tt 1611.01500}}].

\bibitem{Simmons-Duffin:2016wlq}
D.~Simmons-Duffin, \emph{{The Lightcone Bootstrap and the Spectrum of the 3d
  Ising CFT}}, \href{http://dx.doi.org/10.1007/JHEP03(2017)086}{\emph{JHEP}
  {\bf 03} (2017) 086}, [\href{https://arxiv.org/abs/1612.08471}{{\tt
  1612.08471}}].

\bibitem{Fitzpatrick:2015qma}
A.~L. Fitzpatrick, J.~Kaplan, M.~T. Walters and J.~Wang, \emph{{Eikonalization
  of Conformal Blocks}},
  \href{http://dx.doi.org/10.1007/JHEP09(2015)019}{\emph{JHEP} {\bf 09} (2015)
  019}, [\href{https://arxiv.org/abs/1504.01737}{{\tt 1504.01737}}].

\bibitem{Li:2015rfa}
D.~Li, D.~Meltzer and D.~Poland, \emph{{Non-Abelian Binding Energies from the
  Lightcone Bootstrap}},
  \href{http://dx.doi.org/10.1007/JHEP02(2016)149}{\emph{JHEP} {\bf 02} (2016)
  149}, [\href{https://arxiv.org/abs/1510.07044}{{\tt 1510.07044}}].

\bibitem{Li:2017lmh}
D.~Li, D.~Meltzer and D.~Poland, \emph{{Conformal Bootstrap in the Regge
  Limit}}, \href{http://dx.doi.org/10.1007/JHEP12(2017)013}{\emph{JHEP} {\bf
  12} (2017) 013}, [\href{https://arxiv.org/abs/1705.03453}{{\tt 1705.03453}}].

\bibitem{Caron-Huot:2017vep}
S.~Caron-Huot, \emph{{Analyticity in Spin in Conformal Theories}},
  \href{http://dx.doi.org/10.1007/JHEP09(2017)078}{\emph{JHEP} {\bf 09} (2017)
  078}, [\href{https://arxiv.org/abs/1703.00278}{{\tt 1703.00278}}].

\bibitem{Kaviraj:2015cxa}
A.~Kaviraj, K.~Sen and A.~Sinha, \emph{{Analytic bootstrap at large spin}},
  \href{http://dx.doi.org/10.1007/JHEP11(2015)083}{\emph{JHEP} {\bf 11} (2015)
  083}, [\href{https://arxiv.org/abs/1502.01437}{{\tt 1502.01437}}].

\bibitem{Dey:2016zbg}
P.~Dey, A.~Kaviraj and K.~Sen, \emph{{More on analytic bootstrap for O(N)
  models}}, \href{http://dx.doi.org/10.1007/JHEP06(2016)136}{\emph{JHEP} {\bf
  06} (2016) 136}, [\href{https://arxiv.org/abs/1602.04928}{{\tt 1602.04928}}].

\bibitem{Pal:2022vqc}
S.~Pal, J.~Qiao and S.~Rychkov, \emph{{Twist Accumulation in Conformal Field
  Theory: A Rigorous Approach to the Lightcone Bootstrap}},
  \href{http://dx.doi.org/10.1007/s00220-023-04767-w}{\emph{Commun. Math.
  Phys.} {\bf 402} (2023) 2169--2214},
  [\href{https://arxiv.org/abs/2212.04893}{{\tt 2212.04893}}].

\bibitem{Bertucci:2022ptt}
F.~Bertucci, J.~Henriksson and B.~McPeak, \emph{{Analytic bootstrap of mixed
  correlators in the O(n) CFT}},
  \href{http://dx.doi.org/10.1007/JHEP10(2022)104}{\emph{JHEP} {\bf 10} (2022)
  104}, [\href{https://arxiv.org/abs/2205.09132}{{\tt 2205.09132}}].

\bibitem{Henriksson:2022rnm}
J.~Henriksson, \emph{{The critical O(N) CFT: Methods and conformal data}},
  \href{http://dx.doi.org/10.1016/j.physrep.2022.12.002}{\emph{Phys. Rept.}
  {\bf 1002} (2023) 1--72}, [\href{https://arxiv.org/abs/2201.09520}{{\tt
  2201.09520}}].

\bibitem{Albayrak:2019gnz}
S.~Albayrak, D.~Meltzer and D.~Poland, \emph{{More Analytic Bootstrap:
  Nonperturbative Effects and Fermions}},
  \href{http://dx.doi.org/10.1007/JHEP08(2019)040}{\emph{JHEP} {\bf 08} (2019)
  040}, [\href{https://arxiv.org/abs/1904.00032}{{\tt 1904.00032}}].

\bibitem{Fitzpatrick:2010zm}
A.~L. Fitzpatrick, E.~Katz, D.~Poland and D.~Simmons-Duffin, \emph{{Effective
  Conformal Theory and the Flat-Space Limit of AdS}},
  \href{http://dx.doi.org/10.1007/JHEP07(2011)023}{\emph{JHEP} {\bf 07} (2011)
  023}, [\href{https://arxiv.org/abs/1007.2412}{{\tt 1007.2412}}].

\bibitem{Fitzpatrick:2011hh}
A.~L. Fitzpatrick and D.~Shih, \emph{{Anomalous Dimensions of Non-Chiral
  Operators from AdS/CFT}},
  \href{http://dx.doi.org/10.1007/JHEP10(2011)113}{\emph{JHEP} {\bf 10} (2011)
  113}, [\href{https://arxiv.org/abs/1104.5013}{{\tt 1104.5013}}].

\bibitem{Costa:2014kfa}
M.~S. Costa, V.~Gon\c{c}alves and J.~a. Penedones, \emph{{Spinning AdS
  Propagators}}, \href{http://dx.doi.org/10.1007/JHEP09(2014)064}{\emph{JHEP}
  {\bf 09} (2014) 064}, [\href{https://arxiv.org/abs/1404.5625}{{\tt
  1404.5625}}].

\bibitem{Buric:2021kgy}
I.~Buric, S.~Lacroix, J.~A. Mann, L.~Quintavalle and V.~Schomerus,
  \emph{{Gaudin models and multipoint conformal blocks III: comb channel
  coordinates and OPE factorisation}},
  \href{http://dx.doi.org/10.1007/JHEP06(2022)144}{\emph{JHEP} {\bf 06} (2022)
  144}, [\href{https://arxiv.org/abs/2112.10827}{{\tt 2112.10827}}].

\bibitem{Harris:2024nmr}
S.~Harris, A.~Kaviraj, J.~A. Mann, L.~Quintavalle and V.~Schomerus, \emph{{Comb
  Channel Lightcone Bootstrap II: Triple-Twist Anomalous Dimensions}},
  \href{https://arxiv.org/abs/2401.10986}{{\tt 2401.10986}}.

\bibitem{Costa:2011mg}
M.~S. Costa, J.~Penedones, D.~Poland and S.~Rychkov, \emph{{Spinning Conformal
  Correlators}}, \href{http://dx.doi.org/10.1007/JHEP11(2011)071}{\emph{JHEP}
  {\bf 11} (2011) 071}, [\href{https://arxiv.org/abs/1107.3554}{{\tt
  1107.3554}}].

\bibitem{Penedones:2010ue}
J.~Penedones, \emph{{Writing CFT correlation functions as AdS scattering
  amplitudes}}, \href{http://dx.doi.org/10.1007/JHEP03(2011)025}{\emph{JHEP}
  {\bf 03} (2011) 025}, [\href{https://arxiv.org/abs/1011.1485}{{\tt
  1011.1485}}].

\bibitem{DHoker:1999mqo}
E.~D'Hoker, D.~Z. Freedman and L.~Rastelli, \emph{{AdS / CFT four point
  functions: How to succeed at z integrals without really trying}},
  \href{http://dx.doi.org/10.1016/S0550-3213(99)00526-X}{\emph{Nucl. Phys. B}
  {\bf 562} (1999) 395--411}, [\href{https://arxiv.org/abs/hep-th/9905049}{{\tt
  hep-th/9905049}}].

\bibitem{Andriolo:2022hax}
S.~Andriolo, M.~Michel and E.~Palti, \emph{{Self-binding energies in AdS}},
  \href{http://dx.doi.org/10.1007/JHEP02(2023)078}{\emph{JHEP} {\bf 02} (2023)
  078}, [\href{https://arxiv.org/abs/2211.04477}{{\tt 2211.04477}}].

\bibitem{Alday:2017gde}
L.~F. Alday, A.~Bissi and E.~Perlmutter, \emph{{Holographic Reconstruction of
  AdS Exchanges from Crossing Symmetry}},
  \href{http://dx.doi.org/10.1007/JHEP08(2017)147}{\emph{JHEP} {\bf 08} (2017)
  147}, [\href{https://arxiv.org/abs/1705.02318}{{\tt 1705.02318}}].

\bibitem{Maldacena:2015waa}
J.~Maldacena, S.~H. Shenker and D.~Stanford, \emph{{A bound on chaos}},
  \href{http://dx.doi.org/10.1007/JHEP08(2016)106}{\emph{JHEP} {\bf 08} (2016)
  106}, [\href{https://arxiv.org/abs/1503.01409}{{\tt 1503.01409}}].

\bibitem{Antunes:2021kmm}
A.~Antunes, M.~S. Costa, V.~Goncalves and J.~V. Boas, \emph{{Lightcone
  bootstrap at higher points}},
  \href{http://dx.doi.org/10.1007/JHEP03(2022)139}{\emph{JHEP} {\bf 03} (2022)
  139}, [\href{https://arxiv.org/abs/2111.05453}{{\tt 2111.05453}}].

\bibitem{Kaviraj:2022wbw}
A.~Kaviraj, J.~A. Mann, L.~Quintavalle and V.~Schomerus, \emph{{Multipoint
  lightcone bootstrap from differential equations}},
  \href{http://dx.doi.org/10.1007/JHEP08(2023)011}{\emph{JHEP} {\bf 08} (2023)
  011}, [\href{https://arxiv.org/abs/2212.10578}{{\tt 2212.10578}}].

\bibitem{Anand:2020gnn}
N.~Anand, A.~L. Fitzpatrick, E.~Katz, Z.~U. Khandker, M.~T. Walters and Y.~Xin,
  \emph{{Introduction to Lightcone Conformal Truncation: QFT Dynamics from CFT
  Data}},  \href{https://arxiv.org/abs/2005.13544}{{\tt 2005.13544}}.

\bibitem{Mikhailov:2002bp}
A.~Mikhailov, \emph{{Notes on higher spin symmetries}},
  \href{https://arxiv.org/abs/hep-th/0201019}{{\tt hep-th/0201019}}.

\bibitem{Fitzpatrick:2011dm}
A.~L. Fitzpatrick and J.~Kaplan, \emph{{Unitarity and the Holographic
  S-Matrix}}, \href{http://dx.doi.org/10.1007/JHEP10(2012)032}{\emph{JHEP} {\bf
  10} (2012) 032}, [\href{https://arxiv.org/abs/1112.4845}{{\tt 1112.4845}}].

\bibitem{Derkachov:2010zza}
S.~E. Derkachov and A.~N. Manashov, \emph{{Anomalous dimensions of composite
  operators in scalar field theories}},
  \href{http://dx.doi.org/10.1007/s10958-010-0032-9}{\emph{J. Math. Sci.} {\bf
  168} (2010) 837--855}.

\bibitem{PetrInProgress}
P.~Kravchuk and J.~A. Mann, \emph{{\textit{in progress}}}, .

\bibitem{Breitenlohner:1982bm}
P.~Breitenlohner and D.~Z. Freedman, \emph{{Positive Energy in anti-De Sitter
  Backgrounds and Gauged Extended Supergravity}},
  \href{http://dx.doi.org/10.1016/0370-2693(82)90643-8}{\emph{Phys. Lett. B}
  {\bf 115} (1982) 197--201}.

\bibitem{Cornalba:2007fs}
L.~Cornalba, \emph{{Eikonal methods in AdS/CFT: Regge theory and multi-reggeon
  exchange}},  \href{https://arxiv.org/abs/0710.5480}{{\tt 0710.5480}}.

\bibitem{Cornalba:2007zb}
L.~Cornalba, M.~S. Costa and J.~Penedones, \emph{{Eikonal approximation in
  AdS/CFT: Resumming the gravitational loop expansion}},
  \href{http://dx.doi.org/10.1088/1126-6708/2007/09/037}{\emph{JHEP} {\bf 09}
  (2007) 037}, [\href{https://arxiv.org/abs/0707.0120}{{\tt 0707.0120}}].

\bibitem{Fitzpatrick:2019zqz}
A.~L. Fitzpatrick and K.-W. Huang, \emph{{Universal Lowest-Twist in CFTs from
  Holography}}, \href{http://dx.doi.org/10.1007/JHEP08(2019)138}{\emph{JHEP}
  {\bf 08} (2019) 138}, [\href{https://arxiv.org/abs/1903.05306}{{\tt
  1903.05306}}].

\bibitem{Fitzpatrick:2019efk}
A.~L. Fitzpatrick, K.-W. Huang and D.~Li, \emph{{Probing universalities in d
  \ensuremath{>} 2 CFTs: from black holes to shockwaves}},
  \href{http://dx.doi.org/10.1007/JHEP11(2019)139}{\emph{JHEP} {\bf 11} (2019)
  139}, [\href{https://arxiv.org/abs/1907.10810}{{\tt 1907.10810}}].

\bibitem{Kulaxizi:2017ixa}
M.~Kulaxizi, A.~Parnachev and A.~Zhiboedov, \emph{{Bulk Phase Shift, CFT Regge
  Limit and Einstein Gravity}},
  \href{http://dx.doi.org/10.1007/JHEP06(2018)121}{\emph{JHEP} {\bf 06} (2018)
  121}, [\href{https://arxiv.org/abs/1705.02934}{{\tt 1705.02934}}].

\bibitem{Kulaxizi:2018dxo}
M.~Kulaxizi, G.~S. Ng and A.~Parnachev, \emph{{Black Holes, Heavy States, Phase
  Shift and Anomalous Dimensions}},
  \href{http://dx.doi.org/10.21468/SciPostPhys.6.6.065}{\emph{SciPost Phys.}
  {\bf 6} (2019) 065}, [\href{https://arxiv.org/abs/1812.03120}{{\tt
  1812.03120}}].

\bibitem{Liu:2020tpf}
J.~Liu, D.~Meltzer, D.~Poland and D.~Simmons-Duffin, \emph{{The Lorentzian
  inversion formula and the spectrum of the 3d O(2) CFT}},
  \href{http://dx.doi.org/10.1007/JHEP09(2020)115}{\emph{JHEP} {\bf 09} (2020)
  115}, [\href{https://arxiv.org/abs/2007.07914}{{\tt 2007.07914}}].

\bibitem{Kos:2013tga}
F.~Kos, D.~Poland and D.~Simmons-Duffin, \emph{{Bootstrapping the $O(N)$ vector
  models}}, \href{http://dx.doi.org/10.1007/JHEP06(2014)091}{\emph{JHEP} {\bf
  06} (2014) 091}, [\href{https://arxiv.org/abs/1307.6856}{{\tt 1307.6856}}].

\bibitem{Kos:2015mba}
F.~Kos, D.~Poland, D.~Simmons-Duffin and A.~Vichi, \emph{{Bootstrapping the
  O(N) Archipelago}},
  \href{http://dx.doi.org/10.1007/JHEP11(2015)106}{\emph{JHEP} {\bf 11} (2015)
  106}, [\href{https://arxiv.org/abs/1504.07997}{{\tt 1504.07997}}].

\bibitem{Chester:2019ifh}
S.~M. Chester, W.~Landry, J.~Liu, D.~Poland, D.~Simmons-Duffin, N.~Su et~al.,
  \emph{{Carving out OPE space and precise $O(2)$ model critical exponents}},
  \href{http://dx.doi.org/10.1007/JHEP06(2020)142}{\emph{JHEP} {\bf 06} (2020)
  142}, [\href{https://arxiv.org/abs/1912.03324}{{\tt 1912.03324}}].

\bibitem{Heemskerk:2009pn}
I.~Heemskerk, J.~Penedones, J.~Polchinski and J.~Sully, \emph{{Holography from
  Conformal Field Theory}},
  \href{http://dx.doi.org/10.1088/1126-6708/2009/10/079}{\emph{JHEP} {\bf 10}
  (2009) 079}, [\href{https://arxiv.org/abs/0907.0151}{{\tt 0907.0151}}].

\bibitem{private}
{Private Communication with David Simmons-Duffin.}

\bibitem{AndreasTalk}
A.~Lauchli, \emph{{\textbf{Fuzzy Sphere Study of the 3D O(2) CFT: Spectrum,
  Finite Size Corrections and some OPE coefficients}, talk given at Fuzzy
  Sphere Meets Bootstrap}}, .

\bibitem{Homrich:2022mmd}
A.~Homrich, D.~Simmons-Duffin and P.~Vieira, \emph{{Complex Spin: The Missing
  Zeroes and Newton's Dark Magic}},
  \href{https://arxiv.org/abs/2211.13754}{{\tt 2211.13754}}.

\bibitem{Henriksson:2023cnh}
J.~Henriksson, P.~Kravchuk and B.~Oertel, \emph{{Missing local operators,
  zeros, and twist-4 trajectories}},
  \href{https://arxiv.org/abs/2312.09283}{{\tt 2312.09283}}.

\bibitem{Rosenhaus:2018zqn}
V.~Rosenhaus, \emph{{Multipoint Conformal Blocks in the Comb Channel}},
  \href{http://dx.doi.org/10.1007/JHEP02(2019)142}{\emph{JHEP} {\bf 02} (2019)
  142}, [\href{https://arxiv.org/abs/1810.03244}{{\tt 1810.03244}}].

\bibitem{Fortin:2020zxw}
J.-F. Fortin, W.-J. Ma and W.~Skiba, \emph{{All Global One- and Two-Dimensional
  Higher-Point Conformal Blocks}},
  \href{https://arxiv.org/abs/2009.07674}{{\tt 2009.07674}}.

\bibitem{Alkalaev:2023axo}
K.~Alkalaev, A.~Kanoda and V.~Khiteev, \emph{{Wilson networks in AdS and global
  conformal blocks}},
  \href{http://dx.doi.org/10.1016/j.nuclphysb.2023.116413}{\emph{Nucl. Phys. B}
  {\bf 998} (2024) 116413}, [\href{https://arxiv.org/abs/2307.08395}{{\tt
  2307.08395}}].

\bibitem{Fortin:2023xqq}
J.-F. Fortin, W.-J. Ma, S.~Parikh, L.~Quintavalle and W.~Skiba, \emph{{One- and
  two-dimensional higher-point conformal blocks as free-particle wavefunctions
  in $ {\textrm{AdS}}_3^{\otimes m} $}},
  \href{http://dx.doi.org/10.1007/JHEP01(2024)031}{\emph{JHEP} {\bf 01} (2024)
  031}, [\href{https://arxiv.org/abs/2310.08632}{{\tt 2310.08632}}].

\bibitem{Poland:2018epd}
D.~Poland, S.~Rychkov and A.~Vichi, \emph{{The Conformal Bootstrap: Theory,
  Numerical Techniques, and Applications}},
  \href{http://dx.doi.org/10.1103/RevModPhys.91.015002}{\emph{Rev. Mod. Phys.}
  {\bf 91} (2019) 015002}, [\href{https://arxiv.org/abs/1805.04405}{{\tt
  1805.04405}}].

\bibitem{Bissi:2022mrs}
A.~Bissi, A.~Sinha and X.~Zhou, \emph{{Selected topics in analytic conformal
  bootstrap: A guided journey}},
  \href{http://dx.doi.org/10.1016/j.physrep.2022.09.004}{\emph{Phys. Rept.}
  {\bf 991} (2022) 1--89}, [\href{https://arxiv.org/abs/2202.08475}{{\tt
  2202.08475}}].

\bibitem{Dolan:2004iy}
F.~A. Dolan and H.~Osborn, \emph{{Conformal partial wave expansions for N=4
  chiral four point functions}},
  \href{http://dx.doi.org/10.1016/j.aop.2005.07.005}{\emph{Annals Phys.} {\bf
  321} (2006) 581--626}, [\href{https://arxiv.org/abs/hep-th/0412335}{{\tt
  hep-th/0412335}}].

\end{thebibliography}\endgroup

\end{document}